\documentclass[10pt]{article}
\usepackage{epsfig}
\usepackage{amssymb}
\usepackage{amsmath}
\usepackage{amsfonts}
\usepackage{mathrsfs}
\usepackage[dvips]{color}

\newcommand{\R}{\mathbb{R}}
\newcommand{\C}{\mathbb{C}}
\newcommand{\Z}{\mathbb{Z}}
\newcommand{\N}{\mathbb{N}}

\newcommand{\fa}{\mathfrak{a}}
\newcommand{\fb}{\mathfrak{b}}
\newcommand{\fc}{\mathfrak{c}}
\newcommand{\fg}{\mathfrak{g}}
\newcommand{\fh}{\mathfrak{h}}
\newcommand{\bfh}{\mbox{\large$\fh$}}

\newcommand{\fp}{\mathfrak{p}}

\newcommand{\ft}{\mathfrak{t}}

\newcommand{\fz}{\mathfrak{z}}
\newcommand{\fD}{\mathfrak{D}}

\newcommand{\fB}{\mathfrak{B}}

\newcommand{\fM}{\mathfrak{M}}
\newcommand{\fP}{\mathfrak{P}}

\newcommand{\fS}{\mathfrak{S}}
\newcommand{\fT}{\mathfrak{T}}

\newcommand{\fU}{\mathfrak{U}}

\newcommand{\fK}{\mathfrak{K}}

\newcommand{\fH}{\mathfrak{H}}
\newcommand{\fN}{\mathfrak{N}}

\newcommand{\rx}{{\mbox{\large x}}}
\newcommand{\ry}{{\mbox{\large y}}}
\newcommand{\rz}{{\mbox{\large z}}}
\newcommand{\rp}{{\mbox{\large p}}}

\newcommand{\be}{\begin{equation}}
\newcommand{\ee}{\end{equation}}
\newcommand{\bea}{\begin{eqnarray}}
\newcommand{\eea}{\end{eqnarray}}

\newcommand{\nn}{\nonumber}
\newcommand{\kt}{\rangle}
\newcommand{\br}{\langle}
\newcommand{\cun}{\mbox{\footnotesize${\cal N}$}}

\newcommand{\ed}{\end{document}}

\newcommand{\pbr}{\prec\!}
\newcommand{\pkt}{\!\succ}

\newcommand{\cH}{\mathcal{H}}
\newcommand{\cJ}{\mathcal{J}}
\newcommand{\cK}{\mathcal{K}}
\newcommand{\cL}{\mathcal{L}}
\newcommand{\cF}{\mathcal{F}}
\newcommand{\cG}{\mathcal{G}}

\newcommand{\pH}{{\cal H}_{\rm phys}}

\newcommand{\cP}{{\cal P}}

\newcommand{\cT}{\mathcal{T}}
\newcommand{\cC}{\mathcal{C}}
\newcommand{\cD}{\mathcal{D}}

\newcommand{\cV}{\mathcal{V}}
\newcommand{\cU}{\mathcal{U}}
\newcommand{\cS}{\mathcal{S}}

\newcommand{\psigma}{\mbox{\bf{\large{$\sigma$}}}}
\newcommand{\sg}{{\rm sgn}}
\newcommand{\lpb}{\{\!\!\{}
\newcommand{\rpb}{\}\!\!\}}
\newcommand{\etap}{{\eta_{_+}}}
\newcommand{\rd}{{\mbox{\large{\rm d}}}}
\newcommand{\tr}{{\rm tr}}
\newcommand{\Ep}{{\stackrel{\leftrightarrow}{\mbox{\large{$\varepsilon$}}}}}
\newcommand{\MU}{{\stackrel{\leftrightarrow}{\mbox{\large{$\mu$}}}}}

\newcommand{\ru}{{\mbox{\Large u}}}
\newcommand{\sF}{\mathscr{F}}

\begin{document}

\title{Pseudo-Hermitian Representation of Quantum Mechanics}

\author{\\
Ali Mostafazadeh
\\
\\
Department of Mathematics, Ko\c{c} University,\\
34450 Sariyer, Istanbul, Turkey\\ amostafazadeh@ku.edu.tr}
\date{ }
\maketitle

\begin{abstract}
A diagonalizable non-Hermitian Hamiltonian having a real spectrum
may be used to define a unitary quantum system, if one modifies the
inner product of the Hilbert space properly. We give a comprehensive
and essentially self-contained review of the basic ideas and
techniques responsible for the recent developments in this subject.
We provide a critical assessment of the role of the geometry of the
Hilbert space in conventional quantum mechanics to reveal the basic
physical principle motivating our study. We then offer a survey of
the necessary mathematical tools, present their utility in
establishing a lucid and precise formulation of a unitary quantum
theory based on a non-Hermitian Hamiltonian, and elaborate on a
number of relevant issues of fundamental importance. In particular,
we discuss the role of the antilinear symmetries such as ${\cal
PT}$, the true meaning and significance of the so-called charge
operators $\cC$ and the ${\cal CPT}$-inner products, the nature of
the physical observables, the equivalent description of such models
using ordinary Hermitian quantum mechanics, the pertaining duality
between local-non-Hermitian versus nonlocal-Hermitian descriptions
of their dynamics, the corresponding classical systems, the
pseudo-Hermitian canonical quantization scheme, various methods of
calculating the (pseudo-) metric operators, subtleties of dealing
with time-dependent quasi-Hermitian Hamiltonians and the
path-integral formulation of the theory, and the structure of the
state space and its ramifications for the quantum Brachistochrone
problem. We also explore some concrete physical applications
and manifestations of the abstract concepts and tools that have been
developed in the course of this investigation. These include
applications in nuclear physics, condensed matter physics,
relativistic quantum mechanics and quantum field theory, quantum
cosmology, electromagnetic wave propagation, open quantum systems,
magnetohydrodynamics, quantum chaos, and biophysics.

\vspace{5mm}

\vspace{2mm}


\end{abstract}

\tableofcontents
\contentsline{section}{Appendix}{\pageref{appendix-1}}



\section{Introduction and Overview}
\label{sec1}

General Relativity (GR) and Quantum Mechanics (QM) are the most
important achievements of the twentieth century theoretical physics.
Their discovery has had an enormous impact on our understanding of
Nature. Ironically, these two pillars of modern physics are
incompatible both conceptually and practically. This has made their
unification into a more general physical theory the most fundamental
problem of modern theoretical physics. The unification of Special
Relativity and QM, which is by far an easier task, has been the
subject of intensive research since late 1920's. It has led to the
formulation of various quantum field theories. A most successful
example is the Standard Model which provides a satisfactory
description of all available observational data in high energy
particle physics. In spite of the immensity of the amount of
research activity on the subject and the fact that this is conducted
by the most capable theoretical physicists of our times, the
attempts at quantizing gravity have not been as successful. In fact,
one can claim with confidence that these attempts have so far failed
to produce a physical theory offering concrete experimentally
verifiable predictions. This state of affairs has, over the years,
motivated various generalizations of GR and QM. Although none of
these generalizations could be developed into a consistent physical
theory capable of replacing GR or QM, the hope that they might
facilitate the discovery of a unified theory of quantum gravity
still motivates research in this direction.

The development of the special relativistic quantum theories has
also involved attempts at generalizing QM. Among these is an idea
initially put forward by Dirac in 1942 \cite{Dirac-1942} and
developed by Pauli \cite{pauli-1943} into what came to be known as
the \emph{indefinite-metric} quantum theories
\cite{sudarshan,nagy,nakanishi}. This is a rather conservative
generalization of QM in which one considers in addition to the
physical states of the system a set of hypothetical states, called
\emph{ghosts}, whose function is mainly to improve the regularity
properties of the mathematical description of the physical model.
The indefinite-metric quantum theories lost their interest by mid
1970's and perhaps unfortunately never found a detailed coverage in
standard textbooks on relativistic quantum mechanics and quantum
field theory.\footnote{In fact, Pauli who had made fundamental
contributions to the very foundations of the subject had developed
strong critical views against it. For example in his Nobel Lecture
of 1946, he identifies a ``correct theory'' with one that does not
involve ``a hypothetical world'', \cite{pauli-nobel}. For a
comprehensive critical assessment of indefinite-metric quantum field
theories, see \cite{nakanishi}.}

A more recent attempt at generalizing QM is due to Bender and his
collaborators \cite{bbj,bbj-ajp} who adopted all its axioms except
the one that restricted the Hamiltonian to be Hermitian. They
replaced the latter condition with the requirement that the
Hamiltonian must have an exact ${\cal PT}$-symmetry. Here ${\cal P}$
and ${\cal T}$ are the parity and time-reversal operators whose
action on position wave functions $\psi(x)$ is given by $({\cal
P}\psi)(x):=\psi(-x)$ and $({\cal
T}\psi)(x):=\psi(x)^*$.\footnote{We use asterisk to denote
complex-conjugation.} The exact ${\cal PT}$-symmetry of a
Hamiltonian operator $H$ means that it has a complete set of ${\cal
PT}$-invariant eigenvectors $\psi_n$, i.e., ${\cal
PT}\psi_n=a_n\psi_n$ for some complex numbers $a_n$. This condition
assures the reality of the spectrum of $H$. A class of thoroughly
studied examples is provided by the ${\cal PT}$-symmetric
Hamiltonians of the form
    \be
    H_\nu=\frac{1}{2}\,p^2-(ix)^\nu,
    \label{pt-sym-nu}
    \ee
where $\nu$ is a real number not less than 2, and the eigenvalue
problem for $H_\nu$ is defined using an appropriate contour $\Gamma$
in the complex plane $\C$ which for $\nu<4$ may be taken as the real
line $\R$. A correct choice for $\Gamma$ assures that the spectrum
of $H$ is discrete, real, and positive,
\cite{bender-prl,bender-jmp,dorey,shin,jpa-2005a}. Another example
with identical spectral properties is the ${\cal PT}$-symmetric
cubic anharmonic oscillator,
    \be
    H=\frac{1}{2}\,p^2+\frac{1}{2}\,\mu^2 x^2+i\epsilon x^3,
    \label{pt-sym-3}
    \ee
whose coupling constants $\mu$ and $\epsilon$ are real and its
eigenvalue problem is defined along the real axis ($\Gamma=\R$),
\cite{bender-prd-2004,jpa-2005b}.

This ${\cal PT}$-symmetric modification of QM leads to an
indefinite-metric quantum theory
\cite{japaridze,tanaka-2006,cjp-2006}, if one endows the Hilbert
space with the indefinite inner product,
    \be
    \br\cdot|\cdot\kt_{_{\cal P}}:=\br\cdot|{\cal P}\cdot\kt,
    \label{inn-PT}
    \ee
known as the ${\cal PT}$-inner product \cite{bbj}. The symbol
$\br\cdot|\cdot\kt$ that appears in (\ref{inn-PT}) stands for the
$L^2$-inner product:
$\br\phi|\psi\kt:=\int_\Gamma\phi(z)^*\psi(z)dz$ that defines the
Hilbert space $L^2(\Gamma)$ of square-integrable functions
$\psi:\Gamma\to\C$, where $\Gamma$ is the contour in complex plane
that specifies the ${\cal PT}$-symmetric model \cite{jpa-2005a}.

The main advantage of the indefinite inner product (\ref{inn-PT})
over the positive-definite inner product $\br\cdot|\cdot\kt$ is that
the former is invariant under the time-evolution generated by the
Schr\"odinger equation \cite{znojil-pr-2001,p1,jmp-2004}, i.e., if
$\phi(t)$ and $\psi(t)$ are solutions of the Schr\"odinger equation
for the ${\cal PT}$-symmetric Hamiltonian $H$,
$\br\phi(t)|\psi(t)\kt_{_{\cal P}}$ does not depend on time.

In order to employ the standard formulation of indefinite metric
quantum theories \cite{nagy} for a ${\cal PT}$-symmetric model we
proceed as follows \cite{japaridze}.
    \begin{enumerate}
    \item
    We split the space ${\cal H}$ of state-vectors into the
    subspaces ${\cal H}_\pm:=\{\psi\in{\cal H}\,|\,{\rm sgn}(
    \br\psi|\psi\kt_{_{\cal P}})=\pm\}$,
    where sgn$(a):=a/|a|$ if $a$ is a nonzero real
    number and sgn$(0):=0$. ${\cal H}_\pm$ are orthogonal subspaces
    in the sense that for all $\psi_\pm\in{\cal H}_\pm$,
    $\br\psi_-|\psi_+\kt_{_{\cal P}}=0$.\footnote{The assumption of the
    existence of such an orthogonal decomposition is referred to as
    ``decomposability'' of the model in indefinite-metric theories,
    \cite{nagy}. For ${\cal PT}$-symmetric systems considered in the
    literature, this assumption is valid if there is a complete basis of
    common eigenvectors of the Hamiltonian and ${\cal PT}$.}

    \item We impose a superselection rule that
    forbids interactions mixing the elements of $\cH_-$ and $\cH_+$ and
    try to devise a solution for the difficult problem of providing a
    physical interpretation of the theory \cite{nagy,nakanishi}. The
    simplest way of dealing with this problem is to
    identify the elements of $\cH_+$ with physical state-vectors
    \cite{wald,itzikson-zuber} and effectively discard the rest of
    state-vectors as representing unphysical or ghost states.

    \end{enumerate}

An alternative formulation of the theory that avoids the
interpretational difficulties of indefinite-metric theories is the
one based on the construction of a genuine positive-definite inner
product on $\cH$. This construction was initially obtained in
\cite{p2} as a by-product of an attempt to derive a necessary and
sufficient condition for the reality of the spectrum of a general
Hamiltonian operator $H$ that possesses a complete set of
eigenvectors \cite{p1,p2,p3}. Under the assumption that the spectrum
of $H$ is discrete, one can show that it is real if and only if
there is a positive-definite inner product $\br\cdot|\cdot\kt_+$
that makes it Hermitian, i.e., $\br\cdot|H\cdot\kt_+=\br
H\cdot|\cdot\kt_+$, \cite{p2,p3}. The proof of this statement
involves an explicit construction of $\br\cdot|\cdot\kt_+$. This
inner product depends on the choice of the Hamiltonian through its
eigenvectors. Hence it is not unique
\cite{p4,jmp-2003,geyer-cjp,jpa-2006a}.

In \cite{bbj} the authors propose a different approach to the
problem of identifying an appropriate inner product for the
$\cP\cT$-symmetric Hamiltonians such as (\ref{pt-sym-nu}). They
introduce a generic symmetry of these Hamiltonians which they term
as $\cC$-symmetry and construct a class of positive-definite inner
products, called the \emph{${\cal CPT}$-inner products}, that, as we
show in Section~3.4, turn out to coincide with the inner products
$\br\cdot|\cdot\kt_+$, \cite{jmp-2003,jpa-2005a}. The approach of
\cite{bbj} may be related to a much older construction originally
proposed in the context of the indefinite-metric quantum theories
\cite{nagy,nakanishi}. It involves the following two steps.
    \begin{enumerate}
    \item Suppose that $\cH=\cH_+\oplus\cH_-$ where $\cH_\pm$ are the
    orthogonal subspaces we defined above, and that both $\cH_\pm$
    admit a basis consisting of the eigenvectors of $H$.

    \item Let $\Pi:{\cal H}\to{\cal H}$ be the projection operator
    onto $\cH_+$, so that for all $\psi\in{\cal H}$,
    $\psi_+:=\Pi\psi\in{\cal H}_+$
    and $\psi_-:=\psi-\psi_+\in{\cal H}_-$. Clearly $\Pi^2=\Pi$ and
    $\Pi\psi_-=0$.

    \item Endow ${\cal H}$ with the positive-definite inner product:
    $(\phi,\psi):=\br\phi_+|\psi_+\kt_{_{\cal P}}-
    \br\phi_-|\psi_-\kt_{_{\cal P}}.$

    \item Let $\cC:\cH\to\cH$ be defined by ${\cal
    C}\psi:=\psi_+-\psi_-$. Then, in view of the orthogonality of ${\cal
    H}_\pm$,
    \be
    (\phi,\psi)=\br\phi|{\cal C}\psi\kt_{_{\cal P}}=
    \br{\cal C}\phi|\psi\kt_{_{\cal P}},
    \label{inn-cpt-1}
    \ee
\end{enumerate}
It is not difficult to see that ${\cal C}=2\Pi-I$, where $I$ stands
for the identity operator acting in ${\cal H}$, i.e., the operator
that leaves all the elements of ${\cal H}$ unchanged. Obviously for
all $\psi\in{\cal H}_\pm$, ${\cal C}\psi=\pm\psi$. Hence, ${\cal C}$
is a grading operator associated with the direct sum decomposition
${\cal H}={\cal H}_+\oplus{\cal H}_-$ of ${\cal H}$. Furthermore, in
view of the assumption~1 above, the eigenvectors of $H$ have a
definite grading. This is equivalent to the condition that $\cC$
commutes with the Hamiltonian operator, $[{\cal C},H]=0$,
\cite{AGK}. It turns out that the ${\cal CPT}$-inner product
introduced in \cite{bbj} coincides with the inner product
(\ref{inn-cpt-1}).

In \cite{bbj,bbj-ajp}, the authors use the ${\cal CPT}$-inner
product to formulate a unitary quantum theory based on the ${\cal
PT}$-symmetric Hamiltonians (\ref{pt-sym-nu}). They identify the
observables $O$ of the theory with the ${\cal CPT}$-symmetric
operators\footnote{To ensure that the spectrum of such a ${\cal
CPT}$-symmetric operator $O$ is real, one demands that $O$ has an
exact ${\cal CPT}$-symmetry, i.e., its eigenstates are left
invariant under the action of ${\cal CPT}$. This does not however
ensure the completeness of the eigenvectors of $O$.}, in particular
    \be
    {\cal CPT}\: O\: {\cal CPT}=O.
    \label{spt-sym-add}
    \ee
This definition is motivated by the demand that the structure of the
theory must not involve mathematical operations such as Hermitian
conjugation and be determined only using physical conditions. The
definition (\ref{spt-sym-add}) does fulfil this demand\footnote{The
authors of \cite{bbj} emphasize this point by stating that: ``In
effect, we replace the mathematical condition of Hermiticity, whose
physical content is somewhat remote and obscure, by the physical
condition of space-time and charge-conjugation symmetry.'' One must
note however that in this theory the usual coordinate operator $x$
no longer represents a physical observable. As a result ${\cal P}$
does not affect a reflection in the physical space, and there is no
reason why one should still refer to ${\cal PT}$-symmetry as the
physical ``space-time reflection symmetry'' as done in
\cite{bbj,bbj-ajp}.}, but it suffers from a serious dynamical
inconsistency in the sense that in the Heisenberg picture an
operator $O(t):=e^{itH}O(0)e^{-itH}$ that commutes with $\cC\cP\cT$
at $t=0$ may not commutes with this operator at $t>0$. Therefore, in
general, under time-evolution an observable can become unobservable
\cite{critique,comment}! This inconsistency rules out
(\ref{spt-sym-add}) as an acceptable definition of a physical
observable. As noticed in \cite{bbj-erratum,bbrr}, it can be
avoided, if one replaces the symmetry condition (\ref{spt-sym-add})
with:
    \be
    {\cal CPT}\: O\:{\cal CPT}=O^T,
    \label{revised}
    \ee
where all operators are identified with their matrix representation
in the coordinate-basis and $O^T$ stands for the transpose of $O$.
In particular, $\br x|O^T|x'\kt:=\br x'|O|x\kt$.

The presence of the mathematical operation of transposition in
(\ref{revised}) shows that apparently the theory could not be
defined just using ``the physical condition of space-time and
charge-conjugation symmetry'' as was initially envisaged,
\cite{bbj,bbj-ajp}. Note also that (\ref{revised}) puts an implicit
and difficult-to-justify restriction on the Hamiltonian $H$. Being
an observable commuting with ${\cal CPT}$, $H$ must satisfy $H^T=H$,
i.e., it is necessarily symmetric!\footnote{In mathematical
literature the term ``symmetric operator'' is usually used for a
different purpose, as discussed in footnote~\ref{symmetric-op}
below. To avoid possible confusion we will not adopt this
terminology.} Therefore (\ref{revised}) cannot be used to determine
the observables of a theory that has a nonsymmetric Hamiltonian. The
restriction to symmetric Hamiltonians may be easily lifted, if one
is willing to adopt the conventional definition of the observables,
namely identifying them with the operators that are Hermitian with
respect to the ${\cal CPT}$-inner product \cite{critique,cjp-2004b}.
    \be
    (\cdot,O\cdot)=(O\cdot,\cdot),
    \label{Hermitian-cpt}
    \ee
Indeed this definition is forced upon us by a well-known
mathematical theorem that we present a detailed proof of in the
appendix. It states that \emph{if a linear operator $O$ has real
expectation values computed using a given inner product, then $O$ is
necessarily Hermitian with respect to this inner product}. This
shows that the requirement of the Hermiticity of observables and in
particular the Hamiltonian has a simple ``physical''
justification.\footnote{What is however not dictated by this
requirement is the choice of the inner product.}

An important motivation for considering this so-called \emph{${\cal
PT}$-symmetric Quantum Mechanics} is provided by an interesting idea
that is rooted in special relativistic local quantum field theories
(QFT). Among the most celebrated results of QFT is the ${\cal {\rm
C}PT}$-theorem. It states that every field theory satisfying the
axioms of QFT is ${\cal {\rm C}PT}$-invariant, \cite{haag}, where
${\rm C}$ is the charge-conjugation operator. Clearly replacing the
axiom that the Hamiltonian is Hermitian with the statement of the
${\cal {\rm C}PT}$-theorem might lead to a generalization of QFT.
The implementation of this idea in a nonrelativistic setting
corresponds to the requirement that the Hamiltonian possesses an
exact ${\cal PT}$-symmetry. This in turn allows for the construction
of a ${\cal C}$-operator that similarly to the charge-conjugation
operator ${\rm C}$ of QFT squares to identity and generates a
symmetry of the system. The idea that this establishes a
nonrelativistic analog of the ${\cal {\rm C}PT}$-theorem is quite
tempting. Yet there are clear indications that this is not really
the case. For example, unlike the charge-conjugation operator of
QFT, ${\cal C}$ depends on the choice of the Hamiltonian. It turns
out that in fact this operator does not play the role of the
relativistic charge-conjugation operator, it is merely a useful
grading operator for the Hilbert space.\footnote{One can more
appropriately compare ${\cal C}$ with the chirality operator
($\gamma^5$) for the Dirac spinors.} In this sense the adopted
terminology is rather unfortunate.

One of the aims of the present article is to show that ${\cal
PT}$-symmetric QM is an example of a more general class of theories,
called \emph{Pseudo-Hermitian Quantum Mechanics}, in which ${\cal
PT}$-symmetry does not play a basic role and one does not need to
introduce a ${\cal C}$-operator to make the theory well-defined. The
analogs of ${\cal PT}$ and ${\cal C}$ operators can nevertheless be
defined in general \cite{jmp-2003}, but they do not play a
fundamental role. All that is needed is to determine the class of
non-Hermitian Hamiltonians that are capable of generating a unitary
evolution and a procedure that associates to each member of this
class a positive-definite inner product that renders it Hermitian.
It turns out that there are always an infinite class of
positive-definite inner products satisfying this condition. Each of
these defines a separate physical Hilbert space with a complete set
of observables. In this way one obtains a set of quantum systems
that have the same Hamiltonian but different Hilbert spaces.
Therefore, they are dynamically equivalent but kinematically
distinct.

In order to elucidate the conceptual foundations of Pseudo-Hermitian
QM we will next examine some of the basic properties of the
mathematical notions of the ``transpose'' and
``Hermitian-conjugate'' of a linear operator. For clarity of the
presentation we will only consider the operators that act in the
space of square-integrable functions $L^2(\R)$. The discussion may
be generalized to square-integrable functions defined on a complex
contour \cite{jpa-2005a}.

In the literature on ${\cal PT}$-symmetric QM, notably
\cite{bbj,bbj-ajp,bbrr}, the transpose $O^T$ and Hermitian-conjugate
$O^\dagger$ of a linear operator $O$ are respectively defined with
respect to the coordinate-basis, $\{|x\kt\}$, according to
    \bea
    \br x_1|O^T|x_2\kt&:=&\br x_2|O|x_1\kt,
    \label{transpose}\\
    \br x_1|O^\dagger|x_2\kt&:=&\br x_2|O|x_1\kt^*,
    \label{Hermitian}
    \eea
where $x_1,x_2$ are arbitrary real numbers. Therefore the terms
``symmetric'' and ``Hermitian'' respectively refer to the conditions
$\br x_1|O|x_2\kt=\br x_2|O|x_1\kt$ and $\br x_1|O|x_2\kt^*=\br
x_2|O|x_1\kt$. These definitions reflect the inclination to treat
operators as matrices. This is certainly permissible provided that
one specifies the particular basis one uses for this purpose. In
this sense the following equivalent definitions are more preferable.
    \be
    O^T:=\int dx_1\int dx_2\: \br x_2|O|x_1\kt~ |x_1\kt\br x_2|,~~~~~
    O^\dagger:=\int dx_1\int dx_2\: \br x_2|O|x_1\kt^*~
    |x_1\kt\br x_2|.
    \label{spec}
    \ee

A nice feature of (\ref{Hermitian}) that is not shared with
(\ref{transpose}) is that it is invariant under the basis
transformations that map $\{|x\kt\}$ onto any \emph{orthonormal
basis}. For example one can easily show that if (\ref{Hermitian})
holds, so do
    \be
    \br p_1|O^\dagger|p_2\kt=\br p_2|O|p_1\kt^*~~~~~{\rm and}~~~~~
    O^\dagger=\int dp_1\int dp_2 \br p_2|O|p_1\kt^*
    |p_1\kt\br p_2|.
    \label{Hermitian-p}
    \ee
This invariance under orthonormal basis transformations stems from
the fact that $O^\dagger$ admits a basis-independent definition: It
is the linear operator satisfying
    \be
    \br\phi|O^\dagger\psi\kt=\br O\phi|\psi\kt,
    \label{adj}
    \ee
i.e., the \emph{adjoint operator} for $O$.\footnote{A rigorous
definition of the adjoint operator is given in
Subsection~\ref{sec-Hermitian}.}

The notions of ``transpose'' and ``symmetric operator'' introduced
above and employed in \cite{bbj,bbj-ajp,bbj-erratum,bbrr} do not
share this invariance property of ``Hermitian-conjugate'' and
``Hermitian operator''. For example, it is easy to see that $\br
x_1|(ip)|x_2\kt=-\br x_2|(ip)|x_1\kt$ while $\br p_1|(ip)|p_2\kt=\br
p_2|(ip)|p_1\kt$. Therefore, $ip$ is represented by a symmetric
matrix in the $p$-basis while it is represented by an antisymmetric
matrix in the $x$-basis. This shows that there is no
basis-independent notion of the transpose of an operator or a
symmetric operator.\footnote{One can define a notion of the
transpose of a linear operator $O$ acting in a complex inner product
space ${\cal V}$ in terms of an arbitrary antilinear involution
$\tau:{\cal V}\to{\cal V}$ according to $O^T=\tau O^\dagger \tau$,
\cite{akhiezer-glazman}. A function $\tau:{\cal V}\to{\cal V}$ is
called an \emph{antilinear operator} if
$\tau(a\psi+b\phi)=a^*\tau\psi+b^*\tau\phi$ for all complex numbers
$a,b$ and all elements $\psi,\phi$ of ${\cal V}$. It is called an
\emph{involution} if $\tau^2=I$, where $I:{\cal V}\to{\cal V}$ is
the identity operator. The choice of a basis to define $O^T$ is
equivalent to the choice of an antilinear involution $\tau$. The
notion of the transpose used in \cite{bbj,bbj-ajp,bbrr} corresponds
to choosing the time-reversal operator ${\cal T}$ as
$\tau$.\label{antilinear}}

Obviously once we specify a basis, there is no danger of using
definition (\ref{transpose}). But we must keep in mind that any
theory in which one uses the notion of transposition in the sense of
(\ref{transpose}) involves the implicit assumption that the
coordinate-basis is a preferred basis. The use of the notion of
Hermitian-conjugation as defined by (\ref{Hermitian}) does not rely
on such an assumption. As we will explain in the following section
the choice of an orthonormal basis is equivalent to the choice of an
inner product. This is why one can define $O^\dagger$ using its
basis-independent defining relation (\ref{adj}) which only involves
the inner product $\br\cdot|\cdot\kt$. In summary, while the use of
the terms ``transpose'' and ``symmetric operator'' involves making a
particular choice for a preferred basis, the use of the term
``Hermitian-conjugate'' and ``Hermitian operator'' involves making a
particular choice for an inner product.

In conventional QM the inner product is fixed from the outset. Hence
the notions of Hermitian-conjugation and Hermitian operator are
well-defined. The opposite is true about the notions of
transposition and symmetric operator. This does not cause any
difficulty, because they never enter into quantum mechanical
calculations, and in principle one does not need to introduce them
at all. We will see that the same is the case in Pseudo-Hermitian
QM. In particular, in the discussion of ${\cal PT}$-symmetric
systems, there is no need to identify physical observables using
(\ref{revised}).

The main reason for making a universal and preassigned choice for
the inner product in QM is the curious fact that up to
unitary-equivalence there is a unique inner product.\footnote{This
will be explained in detail in Subsection~\ref{sec-unitary}.} This
means that using different inner products leads to physically
identical theories, or more correctly to different representations
of a single physical theory. In conventional QM, one eliminates the
chance of employing these alternative representations by adopting
the usual ($L^2$-) inner product as the only viable choice. The
situation resembles a gauge theory in which one fixes a gauge from
the outset and then forgets about the gauge symmetry. This will have
no effect on the physical quantities computed using such a theory,
but it is clearly not recommended. It is quite possible that an
alternative choice of gauge would facilitate a particular
calculation.

We wish to argue that because \emph{no one has ever made an
independent measurement of the inner product of the Hilbert space},
it must be kept as a degree of freedom of the formulation of the
theory. This is the basic principle underlying Pseudo-Hermitian QM.

We will see that any inner product may be defined in terms of a
certain linear operator $\eta_+$. It is this so-called \emph{metric
operator} that determines the kinematics of pseudo-Hermitian quantum
systems. The Hamiltonian operator $H$ that defines the dynamics is
linked to the metric operator via the pseudo-Hermiticity relation,
    \be
    H^\dagger=\eta_+ H\eta_+^{-1}.
    \label{ph}
    \ee
We will explore some of the consequences of this equation whose
significance has not been fully noticed or appreciated in its
earlier investigations, notably in the context of the
indefinite-metric quantum theories \cite{pauli-1943}.

We wish to point out that there is a very large number of
publications on the subject of this article. Many of these focus on
the mathematical issues related to the investigation of the spectrum
of various non-Hermitian operators or formalisms developed to study
such problems. Here we will not deal with these issues. The
interested reader may consult the review article
\cite{dorey-review}. Another related line of research is the
mathematical theory of linear spaces with an indefinite metric and
its applications \cite{bognar,azizov}. This is also beyond the aim
and the scope of the present article. There are a series of review
articles by Bender and his collaborators
\cite{bbj-ajp,bbrr,b-cp-2005,b-ijmpa-2005,b-rpp-2007} that also
address the physical aspects of the subject. The approach pursued in
these articles is based on the use of the $\cC\cP\cT$-inner products
and the definition of observables given by (\ref{revised}). This
restricts the domain of validity of their results to symmetric
Hamiltonians. The discussion of the classical limit of
$\cP\cT$-symmetric systems offered in these articles is also not
satisfactory, because it does not involve a quantization scheme that
would relate classical and quantum observables.

It is our opinion that to gain a basic understanding of the subject
demands a careful study of the underlying mathematical structures
without getting trapped in the physically irrelevant mathematical
details and technicalities. The need for a comprehensive and
readable treatment of basic mathematical notions and their physical
consequences has not been met by any of the previously published
reviews. In the first part of the present article (Sections~2 and
3), we intend to address this need. Here we only discuss the
mathematical tools and results that are necessary for addressing the
conceptual issues of direct relevance to the physical aspects of the
subject. In section~3, we use the mathematical machinery developed
in Section~2 to present a complete formulation of pseudo-Hermitian
QM and its connection with $\cP\cT$- and $\cC$-symmetries. The
second part of the article (Sections 4-9) aims to survey various
recent developments. In Section~4 we survey different methods of
computing metric operators. In Sections 5-8 we explore systems
defined on complex contours, the classical limit of pseudo-Hermitian
quantum systems, the subtleties involving time-dependent
Hamiltonians and the path-integral formulation of the theory, and
the quantum Brachistochrone problem, respectively. In Section~9 we
discuss some of the physical applications of pseudo-Hermitian
Hamiltonians, and in Section~10 we present our concluding remarks.

\section{Mathematical Tools and Conceptual Foundations} \label{sec2}

In this section we survey the necessary mathematical tools and
elaborate on a number of conceptual issues that are helpful in
clarifying various existing misconceptions on the subject. We also
offer a through discussion of the motivation for considering a more
general formulation of QM.

One of the axioms of QM is that pure physical states of a quantum
system are rays in a Hilbert space ${\cal H}$. Each ray may be
determined in a unique manner by a nonzero element $\psi$ of ${\cal
H}$ which we call a \emph{state-vector}. The physical quantities
associated with a pure state is computed using a corresponding
state-vector and the inner product of the Hilbert space. We begin
our discussion by a precise description of Hilbert spaces, inner
products, bases, Hermitian and unitary operators, biorthonormal
systems, and their relevance to our investigation.

We will use the following notations and conventions: $\R$, $\R^+$,
$\C$, $\Z$, $\Z^+$, $\N$ denote the sets of real numbers, positive
real numbers, complex numbers, integers, positive integers, and
nonnegative integers (natural numbers), respectively. The symbol
``$~:=~$'' means that the left-hand side is defined to be the
right-hand side, ``$~=:~$'' means that the converse is true, and
``$~\in~$'' and ``$~\subseteq~$'' respectively stand for ``is an
element of'' and ``is a subset of''. Throughout this paper we will
only consider time-independent Hamiltonian operators unless
otherwise is explicitly stated.

\subsection{Hilbert Spaces and Riesz Bases}
\label{sec-inn}

Consider a complex vector space ${\cal V}$ and a function
$\br\cdot|\cdot\kt:{\cal V}\times{\cal V}\to\C$ that assigns to any
pair $\psi,\phi$ of elements of ${\cal V}$ a complex number
$\br\psi|\phi\kt$. Suppose that  $\br\cdot|\cdot\kt$ has the
following properties.
    \begin{itemize}
    \item[]({\em i}) It is \emph{positive-definite}, i.e., for all
nonzero elements $\psi$ of ${\cal V}$, $\br\psi|\psi\kt$ is a
positive real number, and it vanishes if and only if $\psi=0$, where
we use $0$ also for the zero vector.
    \item[]({\em ii}) It is Hermitian, i.e., for any pair
$\psi,\phi$ of elements of ${\cal V}$,
$\br\psi|\phi\kt^*=\br\phi|\psi\kt$.
    \item[]({\em iii}) It is linear in its second slot, i.e., for all
$\psi,\phi,\chi\in{\cal V}$ and all $a,b\in\C$,
$\br\psi|a\phi+b\chi\kt=a\br\psi|\phi\kt+b\br\psi|\chi\kt$.
    \end{itemize}
Then $\br\cdot|\cdot\kt$ is called an \emph{inner
product}\footnote{We use the terms ``inner product'' and
``positive-definite inner product'' synonymously.} on ${\cal V}$,
and the pair $({\cal V},\br\cdot|\cdot\kt)$ is called an \emph{inner
product space.}

An inner product $\br\cdot|\cdot\kt$ on ${\cal V}$ assigns to each
element $\psi$ of ${\cal V}$ a nonnegative real number,
$\parallel\psi\parallel:=\sqrt{\br\psi|\psi\kt}$, that is called the
\emph{norm} of $\psi$. We can use the norm to define a notion of
distance between elements of ${\cal V}$, according to $\parallel
\psi-\phi\parallel$, and develop analysis and geometry on the inner
product space $({\cal V},\br\cdot|\cdot\kt)$.

A \emph{Hilbert space} ${\cal H}$ is an inner product space which
fulfills an additional technical condition, namely that its norm
defines a \emph{complete metric space}, i.e., for any infinite
sequence $\{\psi_k\}$ of elements $\psi_k$ of ${\cal H}$, the
condition that $\lim_{j,k\to\infty}\parallel
\psi_j-\psi_k\parallel=0$ implies that $\{\psi_k\}$ converges to an
element $\psi$ of ${\cal H}$;
$\lim_{k\to\infty}\parallel\psi-\psi_k\parallel=0$. In other words,
a Hilbert space is a complete inner product space.

A subset ${\cal S}$ of a Hilbert space ${\cal H}$ is said to be
\emph{dense}, if every element of ${\cal H}$ may be obtained as the
limit of a sequence of elements of ${\cal S}$. A Hilbert space is
said to be \emph{separable}, if it has a countable dense subset. It
turns out that ${\cal H}$ is separable if and only if it has a
countable \emph{basis}. The latter is a sequence $\{\chi_n\}$ of
linearly independent elements of ${\cal H}$ such that the set of its
finite linear combinations,
    \be
    {\cal L}(\{\chi_n\})=\left\{\sum_{n=1}^K c_n\chi_n~|~
    K\in\Z^+,~c_n\in\C\right\},
    \label{finite-span}
    \ee
is a dense subset of ${\cal H}$. For an infinite-dimensional Hilbert
space ${\cal H}$, $\{\chi_n\}$ is an infinite sequence and the
assertion that ${\cal L}(\{\chi_n\})$ is a dense subset means that
every element $\psi$ of ${\cal H}$ is the limit of a convergent
series of the form $\sum_{n=1}^\infty c_n\chi_n$ whose coefficient
$c_n$ are assumed to be uniquely determined by $\psi$.\footnote{This
notion of basis is sometimes called a \emph{Schauder basis}
\cite{young}. It is not to be confused with the algebraic or Hamel
basis which for infinite-dimensional Hilbert spaces is always
uncountable \cite{halmos}.}

It is not difficult to show that any finite-dimensional inner
product space is both complete and separable. In this sense
infinite-dimensional separable Hilbert spaces are natural
generalizations of the finite-dimensional inner product spaces. In
the following we will use the label $N$ to denote the dimension of
the Hilbert space in question. $N=\infty$ will refer to an
infinite-dimensional separable Hilbert space.

An important difference between finite and infinite-dimensional
Hilbert spaces is that the definition of a basis in a
finite-dimensional Hilbert space does not involve the inner product
while the opposite is true about the infinite-dimension Hilbert
spaces. The requirement that (\ref{finite-span}) be a dense subset
makes explicit use of the norm. Therefore whether a given sequence
of linearly independent vectors forms a basis of an
infinite-dimensional Hilbert space ${\cal H}$ depends in a crucial
manner on the inner product of ${\cal H}$.

Given a basis $\{\chi_n\}$ of a separable Hilbert space ${\cal H}$,
one can apply the Gram-Schmidt process \cite{roman,cheney} to
construct an \emph{orthonormal basis}, i.e., a basis $\{\xi_n\}$
satisfying
    \be
    \br\xi_m|\xi_n\kt=\delta_{mn}~~~~\mbox{for all
    $m,n\in\{1,2,3,\cdots,N\}$,}
    \label{ortho}
    \ee
where $\delta_{mn}$ denotes the Kronecker delta symbol:
$\delta_{mn}:=0$ if $m\neq n$ and $\delta_{nn}:=1$ for all $n$. For
an orthonormal basis $\{\xi_n\}$, the coefficients $c_n$ of the
basis expansion,
    \be
    \psi=\sum_{n=1}^N c_n\xi_n,
    \label{expand-1}
    \ee
of the elements $\psi$ of ${\cal H}$ are given by
    \be
    c_n=\br\xi_n|\psi\kt.
    \label{expand-2}
    \ee
Furthermore, in view of (\ref{ortho}) -- (\ref{expand-2}),
    \be
    \br\phi|\psi\kt=\sum_{n=1}^N
    \br\phi|\xi_n\kt\br\xi_n|\psi\kt~~~~
    \mbox{for all $\phi,\psi\in{\cal H}$}.
    \label{norm}
    \ee
In particular,
    \be
    \parallel\psi\parallel^2=\sum_{n=1}^N |\br\xi_n|\psi\kt|^2
    ~~~~\mbox{for all $\psi\in{\cal H}$}.
    \label{norm2}
    \ee
Eq.~(\ref{norm}) implies that the operator ${\cal I}$ defined by
${\cal I}\,\psi:=\sum_{n=1}^N \br\xi_n|\psi\kt\,\xi_n$ equals the
identity operator $I$ acting in ${\cal H}$; ${\cal
I}=I$.\footnote{Let $\psi_1,\psi_2\in{\cal H}$ be such that for all
$\phi\in{\cal H}$, $\br\phi|\psi_1\kt=\br\phi|\psi_2\kt$. Then
$\br\phi|\psi_1-\psi_2\kt=0$ for all $\phi\in{\cal H}$. Setting
$\phi=\psi_1-\psi_2$, one then finds
$\parallel\psi_1-\psi_2\parallel^2=0$ which implies $\psi_1=\psi_2$.
In view of this argument, $\br\phi|{\cal
I}\,\psi\kt=\br\phi|\psi\kt$ for all $\psi,\phi\in{\cal H}$ implies
${\cal I}\,\psi=\psi$ for all $\psi\in{\cal H}$. Hence ${\cal I}=I$.
\label{trick}} This is called the \emph{completeness relation} which
in Dirac's bra-ket notation takes the familiar form: $\sum_{n=1}^N
|\xi_n\kt\br\xi_n|=I$.

Next, we wish to examine if \emph{given a basis $\{\zeta_n\}$ of a
separable Hilbert space ${\cal H}$, there is an inner product
$(\cdot|\cdot)$ on ${\cal H}$ with respect to which $\{\zeta_n\}$ is
orthonormal}. Because $\{\zeta_n\}$ is a basis, for all
$\psi,\phi\in{\cal H}$ there are unique complex numbers $c_n,d_n$
such that  $\psi=\sum_{n=1}^Nc_n\zeta_n$ and
$\phi=\sum_{n=1}^Nd_n\zeta_n$. We will attempt to determine
$(\psi|\phi)$ in terms of $c_n$ and $d_n$.

First, consider the finite-dimensional case, $N<\infty$. Then, in
view of Eq.~(\ref{norm}), the condition that $\{\zeta_n\}$ is
orthonormal with respect to $(\cdot|\cdot)$ defines the latter
according to
    \be
    (\psi|\phi):=\sum_{n=1}^N c_n^*d_n.
    \label{inn-basis}
    \ee
We can easily check that $(\cdot|\cdot)$ possesses the defining
properties (\emph{i}) -- (\emph{iii}) of an inner product and
satisfies $(\zeta_m|\zeta_n)=\delta_{mn}$. It is also clear from
Eq.~(\ref{norm}) that any other inner product with this property
must satisfy (\ref{inn-basis}). This shows that $(\cdot|\cdot)$ is
the unique inner product that renders $\{\zeta_n\}$ orthonormal.

The case $N=\infty$ may be similarly treated, but in general there
is no guarantee that the right-hand side of (\ref{inn-basis}) is a
convergent series. In fact, it is not difficult to construct
examples for which it is divergent. Therefore, an inner product that
makes an arbitrary basis $\{\zeta_n\}$ orthonormal may not exist.
The necessary and sufficient condition for the existence of such an
inner product $(\cdot|\cdot)$ is that $\psi=\sum_{n=1}^\infty
c_n\zeta_n$ implies $\sum_{n=1}^\infty |c_n|^2<\infty$ for all
$\psi\in{\cal H}$.\footnote{If $(\cdot|\cdot)$ exists,
(\ref{inn-basis}) must hold because $\{\zeta_n\}$ is orthonormal
with respect to $(\cdot|\cdot)$. Hence the left-hand side of
(\ref{inn-basis}) is well-defined and its right-hand must be
convergent. In particular, for $\phi=\psi$ we find that
$(\psi|\psi)=\sum_{n=1}^\infty |c_n|^2$ must be finite. This
establishes the necessity of the above condition. Its sufficiency
follows from the inequality: For all $K\in\Z^+$, $|\sum_{n=1}^K
c_n^*d_n|^2\leq  \sum_{n=1}^\infty |c_n|^2 \sum_{m=1}^\infty
|d_m|^2<\infty$.} Furthermore, we shall demand that the inner
product space ${\cal H}'$ obtained by endowing the underlying vector
space of ${\cal H}$ with the inner product $(\cdot|\cdot)$ is a
Hilbert space. As we will discuss in Subsection~\ref{sec-unitary},
any two infinite-dimensional separable Hilbert spaces, in particular
${\cal H}$ and ${\cal H}'$, have the same topological properties,
i.e., the set of open subsets of ${\cal H}$ coincide with those of
${\cal H}'$. This restricts $(\cdot|\cdot)$ to be
\emph{topologically equivalent} to $\br\cdot|\cdot\kt$, i.e., there
are positive real numbers $c_1$ and $c_2$ satisfying
$c_1\br\psi|\psi\kt\leq (\psi|\psi)\leq c_2 \br\psi|\psi\kt$ for all
$\psi\in{\cal H}$. It turns out that the inner product
(\ref{inn-basis}) that renders the basis $\{\zeta_n\}$ orthonormal
and is topologically equivalent to $\br\cdot|\cdot\kt$ exists and is
unique if and only if it is obtained from an orthonormal basis
$\{\xi_n\}$ through the action of an everywhere-defined bounded
invertible linear operator $A:{\cal H}\to{\cal H}$, i.e.,
$\zeta_n=A\xi_n$. A basis $\{\zeta_n\}$ having this property is
called a \emph{Riesz basis}, \cite{gohberg-krein,young}. In summary,
we can construct a new separable Hilbert space ${\cal H}'$ in which
$\{\zeta_n\}$ is orthonormal if and only if it is a Riesz basis. We
will give a derivation of a more explicit form of the inner product
of ${\cal H}'$ in Subsection~\ref{sec-biortho}.

\subsection{Bounded, Invertible, and Hermitian Operators}
\label{sec-Hermitian}

Consider two Hilbert spaces ${\cal H}_1$ and ${\cal H}_2$ with inner
products $\br\cdot|\cdot\kt_{_1}$ and$\br\cdot|\cdot\kt_{_2}$,
respectively, and a linear operator $A$ that maps ${\cal H}_1$ to
${\cal H}_2$. The \emph{domain} ${\cal D}(A)$ of $A$ is the subset
of ${\cal H}_1$ such that the action of $A$ on any element of ${\cal
D}(A)$ yields a unique element of ${\cal H}_2$. The range ${\cal
R}(A)$ of $A$ is the subset of ${\cal H}_2$ consisting of elements
of the form $A\psi_1$ where $\psi_1$ belongs to ${\cal D}(A)$. If
${\cal D}(A)={\cal H}_1$, one says that $A$ has \emph{full domain},
or that it is \emph{everywhere-defined}. If ${\cal R}(A)={\cal
H}_2$, one says that $A$ has \emph{full range}, i.e., it is an
\emph{onto} function. As an example consider the momentum operator
$p$ acting in ${\cal H}_1={\cal H}_2=L^2(\R)$,
$(p\psi)(x):=-i\hbar\frac{d}{dx}\,\psi(x)$. Then ${\cal D}(p)$
consists of the square-integrable functions that have a
square-integrable derivative, and ${\cal R}(p)$ is the set of
square-integrable functions $\psi_2$ such that
$\psi_1(x)=\int_{-\infty}^x \psi_2(u)du$ is also square-integrable.
In particular $p$ is not everywhere-defined, but its domain is a
dense subset of $L^2(\R)$. Such an operator is said to be
\emph{densely-defined}. All the operators we encounter in this
article and more generally in QM are densely-defined.

A linear operator $A:{\cal H}_1\to{\cal H}_2$ is said to be
\emph{bounded} if there is a positive real number $M$ such that for
all $\psi\in{\cal D}(A)$, $\parallel A\psi\parallel_2\:\leq M
\parallel\psi\parallel_1$, where $\parallel\cdot\parallel_1$ and
$\parallel\cdot\parallel_2$ are respectively the norms defined by
the inner products $\br\cdot|\cdot\kt_{_1}$ and
$\br\cdot|\cdot\kt_{_2}$. The smallest $M$ satisfying this
inequality is called the norm of $A$ and denoted by $\parallel
A\parallel$. A characteristic feature of a bounded operator is that
all its eigenvalues $a$ are bounded by its norm, $|a|\leq\,\parallel
A\parallel$. Furthermore, a linear operator is bounded if and only
if it is continuous \cite{reed-simon}.\footnote{A function
$f:\cH_1\to\cH_2$ is said to be continuous if for all
$\psi\in\cD(f)$ and every sequence $\{\psi_n\}$ in $\cD(f)$ that
converges to $\psi$, the sequence $\{f(\psi_n)\}$ converges to
$f(\psi)$.} Linear operators relating finite-dimensional Hilbert
spaces are necessarily bounded. Therefore the concept of boundedness
is only important for infinite-dimensional Hilbert spaces.

$A:{\cal H}_1\to{\cal H}_2$ is called an \emph{invertible operator}
if it satisfies both of the following conditions
\cite{hislop-sigal}.\footnote{Some authors do not require the second
condition. The above definition is the most convenient for our
purposes.}
    \begin{enumerate}
    \item $A$ is one-to-one and onto, so $A^{-1}:{\cal H}_2\to{\cal H}_1$
exists and has a full domain;
    \item $A^{-1}$ is a bounded operator.
    \end{enumerate}
If $A$ is bounded, one-to-one and onto, then according to a theorem
due to Banach its inverse is also bounded; it is invertible,
\cite{kolmogorov-fomin}.\footnote{A continuous one-to-one onto
function with a continuous inverse is called a \emph{homeomorphism}.
The domain and range of a homeomorphism have the same topological
properties. If $f:{\cal H}_1\to{\cal H}_2$ is a homeomorphism
relating two Hilbert spaces ${\cal H}_1$ and ${\cal H}_1$, a
sequence $\{x_n\}$ of elements $x_n$ of ${\cal H}$ converges to an
element $x\in{\cal H}$ if and only if the $\{f(x_n)\}$ converges to
$f(x)$ in ${\cal H}_2$.\label{homeo1}} An important class of bounded
invertible operators that play a fundamental role in QM is the
unitary operators. We will examine them in
Subsection~\ref{sec-unitary}.

Next, consider a linear operator $A:{\cal H}_1\to{\cal H}_2$ that
has a dense domain ${\cal D}(A)$. Let ${\cal D}'$ be the subset of
${\cal H}_2$ whose elements $\psi_2$ satisfy the following
condition: For all $\psi_1\in{\cal D}(A)$, there is an element
$\phi_1$ of ${\cal H}_1$ such that
$\br\psi_2|A\psi_1\kt_{_2}=\br\phi_1|\psi_1\kt_{_1}$. Then there is
a unique linear operator $A^\dagger:{\cal H}_2\to{\cal H}_1$
fulfilling \cite{reed-simon}
    \be
    \br\psi_1|A^\dagger\psi_2\kt_{_1}=\br A\psi_1|\psi_2\kt_{_2}~~~~
    \mbox{for all}~\psi_1\in{\cal D}(A)~{\rm and}~
    \psi_2\in{\cal D'}.
    \label{dagger}
    \ee
This operator is called the \emph{adjoint} or
\emph{Hermitian-conjugate} of $A$. By construction, ${\cal D}'$ is
the domain of $A^\dagger$; ${\cal D}(A^\dagger)={\cal D}'$.

For the case ${\cal H}_2={\cal H}_1=:{\cal H}$, where ${\cal H}$ is
a Hilbert space with inner product $\br\cdot|\cdot\kt$, a linear
operator $A:{\cal H}\to{\cal H}$ having a dense domain ${\cal D}(A)$
is called a \emph{self-adjoint} or \emph{Hermitian}
operator\footnote{In some mathematics texts, e.g.,
\cite{reed-simon}, the term ``Hermitian'' is used for a more general
class of operators which satisfy (\ref{self-adjoint}) but have
${\cal D}(A)\subseteq{\cal D}(A^\dagger)$. A more commonly used term
for such an operator is ``symmetric operator''. We will avoid using
this terminology which conflicts with the terminology used in the
literature on ${\cal PT}$-symmetric QM, \cite{bbj,bbj-ajp,bbrr}.
\label{symmetric-op}} if $A^\dagger=A$. In particular, ${\cal
D}(A^\dagger)={\cal D}(A)$ and
    \be
    \br\psi_1|A\psi_2\kt=\br A\psi_1|\psi_2\kt~~~~
    \mbox{for all}~\psi_1,\psi_2\in{\cal D}(A).
    \label{self-adjoint}
    \ee
An occasionally useful property of Hermitian operators is that
\emph{every Hermitian operator having a full domain is necessarily
bounded}. This is known as the \emph{Hellinger-Toeplitz theorem}
\cite{reed-simon}.

Hermitian operators play an essential role in QM mainly because of
their spectral properties \cite{von-neumann}. In particular, their
spectrum\footnote{For a precise definition of the spectrum of a
linear operator see Subsection~\ref{sec-spec-prop}.} is real, their
eigenvectors with distinct eigenvalues are orthogonal, and they
yield a \emph{spectral resolution of the identity operator} $I$. For
a Hermitian operator $A$ with a discrete spectrum the latter takes
the following familiar form, if we use the Dirac bra-ket notation.
    \be
    I=\sum_{n=1}^N |\alpha_n\kt\br\alpha_n|,
    \label{resolution-I}
    \ee
where $\{\alpha_n\}$ is an orthonormal basis consisting of the
eigenvectors $\alpha_n$ of $A$ whose eigenvalues $a_n$ are not
necessarily distinct,
    \be
    A\alpha_n=a_n\alpha_n,~~~~~~~\mbox{for all $n\in\{1,2,3,\cdots,N\}$}.
    \label{eg-va}
    \ee

Eqs.~(\ref{resolution-I}) and (\ref{eg-va}) imply that $A$ is
\emph{diagonalizable} and admits the following \emph{spectral
representation}
    \be
    A=\sum_{n=1}^N a_n |\alpha_n\kt\br\alpha_n|.
    \label{resolution-A}
    \ee
Well-known analogs of (\ref{resolution-I}) and (\ref{resolution-A})
exist for the cases that the spectrum is not discrete, \cite{kato}.

An important property that makes Hermitian operators indispensable
in QM is the fact that for a given
densely-defined\footnote{Observables must have dense domains, for
otherwise one can construct a state in which an observable cannot be
measured!} linear operator $A$ with $\cD(A)=\cD(A^\dagger)$, the
expectation value $\br\psi|A\psi\kt$ is real-valued for all unit
state-vectors $\psi\in{\cal D}(A)$ if and only if the Hermiticity
condition (\ref{self-adjoint}) holds, \cite{kato}.\footnote{A simple
proof of this statement is given in the appendix.} This shows that
in a quantum theory that respects von Neumann's measurement
(projection) axiom, the observables cannot be chosen from among
non-Hermitian operators even if they have a real spectrum. The same
conclusion may be reached by realizing that the measurement axiom
also requires the eigenvectors of an observable with distinct
eigenvalues to be orthogonal, for otherwise the reading of a
measuring device that is to be identified with an eigenvalue of the
observable will not be sufficient to determine the state of the
system immediately after the measurement \cite{jpa-2004b}. This is
because an eigenvector that is the output of the measurement may
have a nonzero component along an eigenvector with a different
eigenvalue. This yields nonzero probabilities for the system to be
in two different physical states, though one measures the eigenvalue
of one of them only!

Let $\{\xi_n\}$ be an orthonormal basis of ${\cal H}$ and $A$ be a
Hermitian operator acting in ${\cal H}$, then according to property
(\emph{ii}) of the inner product and the Hermiticity condition
(\ref{self-adjoint}) we have
    \be
    A_{mn}:=\br\xi_m|A\xi_n\kt=\br\xi_n|A\xi_m\kt^*=A_{nm}^*.
    \label{matrix-hermit}
    \ee
This shows that $A$ is represented in the basis $\{\xi_n\}$,
according to
    \be
    A=\sum_{n=1}^N A_{mn}|\xi_m\kt\br\xi_n|,
    \label{matrix-rep}
    \ee
using the $N\times N$ \emph{Hermitian matrix}\footnote{A square
matrix $M$ is called Hermitian if its entries $M_{mn}$ satisfy
$M_{mn}=M_{nm}^*$ for all $m$ and $n$.} $\underline{A}:=(A_{mn})$.

It is essential to realize that a Hermitian operator can be
represented by a non-Hermitian matrix in a non-orthonormal basis.
This implies that having the expression for the matrix
representation of an operator and knowing the basis used for this
representation are not sufficient to decide if the operator is
Hermitian. One must in addition know the inner product and be able
to determine if the basis is orthonormal. \emph{Referring to an
operator as being Hermitian or non-Hermitian (using its matrix
representation) without paying attention to the inner product of the
space it acts in is a dangerous practice.}

For example, it is not difficult to check that the Hermitian matrix
$\psigma_1=\mbox{\scriptsize
$\left(\begin{array}{cc}0&1\\1&0\end{array}\right)$}$ represents the
operator $L:{\C}^2\to{\C}^2$ defined by $L${\scriptsize$
\left(\begin{array}{c} z_1\\
z_2\end{array}\right):=\left(\begin{array}{c} z_1\\
z_1-z_2\end{array}\right)$} in the basis ${\cal
B}:=${\scriptsize$\left\{ \left(\begin{array}{c}
1\\0\end{array}\right), \left(\begin{array}{c}
1\\1\end{array}\right)\right\}$}. The same operator is represented
in the standard basis ${\cal B}_0:=${\scriptsize$\left\{
\left(\begin{array}{c} 1\\0\end{array}\right),
\left(\begin{array}{c} 0\\1\end{array}\right)\right\}$} using the
non-Hermitian matrix {\scriptsize
$\left(\begin{array}{cc}1&0\\1&-1\end{array}\right)$}. We wish to
stress that this information is, in fact, not sufficient to
ascertain if $L$ is a Hermitian operator, unless we fix the inner
product on the Hilbert space $\C^2$. For instance, if we choose the
standard Euclidean inner product which is equivalent to requiring
${\cal B}_0$ to be orthonormal, then $L$ is a non-Hermitian
operator. If we choose the inner product that makes the basis ${\cal
B}$ orthonormal, then $L$ is a Hermitian operator. We can use
(\ref{inn-basis}) to construct the latter inner product. It has the
form $(\vec z|\vec w ):=z_1^*(w_1-w_2)+z_2^*(-w_1+2w_2)$, where $
\vec z=\mbox{\scriptsize$\left(\begin{array}{c}
    z_1\\z_2\end{array}\right)$}$, and $
    \vec w=\mbox{\scriptsize$
    \left(\begin{array}{c}
    w_1\\w_2\end{array}\right)$}$.

The above example raises the following natural question. Given a
linear operator $H$ that is not represented by a Hermitian matrix in
an orthonormal basis, is there another (non-orthonormal) basis in
which it is represented by a Hermitian matrix? This is equivalent to
asking if one can modify the inner product so that $H$ becomes
Hermitian. The answer to this question is: ``No, in general.'' As we
will see in the sequel, there is a simple necessary and sufficient
condition on $H$ that ensures the existence of such an inner
product.

\subsection{Unitary Operators and Unitary-Equivalence}
\label{sec-unitary}

Equations (\ref{expand-2}) and (\ref{norm}) may be employed to
derive one of the essential structural properties of separable
Hilbert spaces, namely that up to unitary-equivalence they are
uniquely determined by their dimension. To achieve this we first
explain how one compares inner product spaces. Two inner product
spaces ${\cal H}_1$ and ${\cal H}_2$ with inner products
$\br\cdot|\cdot\kt_{_1}$ and $\br\cdot|\cdot\kt_{_2}$ are said to be
\emph{unitary-equivalent}, if there is an everywhere-defined onto
linear operator $U:{\cal H}_1\to{\cal H}_2$ such that for every
$\phi_1,\psi_1$ in ${\cal H}_1$ we have
    \be
    \br U\phi_1|U\psi_1\kt_{_2}=\br\phi_1|\psi_1\kt_{_1}.
    \label{isometry}
    \ee
Such an operator is called a {\em unitary operator}.  In view of
(\ref{adj}) and (\ref{isometry}), we have
    \be
    U^\dagger U=I_1,
    \label{UU=I}
    \ee
where $I_1$ denotes the identity operator acting on ${\cal H}_1$.
One can use (\ref{isometry}) and the ontoness property of $U$ to
show that $U$ is an invertible operator and its inverse $U^{-1}$,
that equals $U^\dagger$, is also unitary.

Unitary-equivalence is an equivalence relation.\footnote{This means
that every inner product space is unitary-equivalent to itself; if
${\cal H}_1$ is unitary-equivalent to ${\cal H}_2$, so is ${\cal
H}_2$ to ${\cal H}_1$; if ${\cal H}_1$ is unitary-equivalent to
${\cal H}_2$ and ${\cal H}_2$ is unitary-equivalent to ${\cal H}_3$,
then ${\cal H}_1$ is unitary-equivalent to ${\cal H}_3$.} Therefore
to establish the unitary-equivalence of all $N$-dimensional
separable Hilbert spaces it suffices to show that all of them are
unitary-equivalent to a chosen one. The most convenient choice for
the latter is the Hilbert space
    \[{\cal H}_0^N=\left\{\begin{array}{ccc}
    \C^N &{\rm for}& N\neq\infty\\
    \ell^2 &{\rm for}& N=\infty,\end{array}\right.\]
where $\C^N$ is the set of $N$-dimensional complex column vectors
endowed with the standard Euclidean inner product $\br\vec w|\vec
z\kt:=\vec w^*\cdot\vec z$, a dot denotes the usual dot product,
$\ell^2$ is the set of square-summable sequences,
$\ell^2:=\left\{~\{c_n\}\left|~ c_n\in\C~,~
    \sum_{n=1}^\infty|c_n|^2 <\infty\right.\right\}$,
equipped with the inner product:
    $\br\,\{\tilde c_n\}\,|\,\{c_n\}\,\kt:=\sum_{n=1}^\infty
    \tilde c_n^*c_n$, and $\{\tilde c_n\},\{c_n\}\in\ell^2$.
Now, let ${\cal H}$ be any $N$-dimensional separable Hilbert space
with inner product $(\cdot|\cdot)$, $\{\xi_n\}$ be an orthonormal
basis of ${\cal H}$, and $U:{\cal H}\to{\cal H}_0^N $ be defined by
$U(\psi):=\{\,(\xi_n|\psi)\,\}$ for all $\psi$ in ${\cal H}$. It is
not difficult to see that $U$ is an everywhere-defined and onto
linear map. Furthermore in view of (\ref{norm}) it satisfies, $\br
U\phi|U\psi\kt=\sum_{n=1}^N(\xi_n|\phi)^*(\xi_n|\psi)= (\phi|\psi)$,
for all $\phi,\psi\in{\cal H}$. Hence, (\ref{isometry}) holds, $U$
is a unitary operator, and ${\cal H}$ is unitary-equivalence to
${\cal H}_0^N$.

For a quantum system having $\R$ as its configuration space, one
usually uses the coordinate wave functions $\psi(x)$ to represent
the state-vectors. The latter are elements of the Hilbert space
$L^2(\R)$ of square-integrable complex-valued functions
$\psi:\R\to\C$. The inner product of $L^2(\R)$ has the form
    \be
    \br\phi|\psi\kt=\int_{-\infty}^\infty dx~\phi(x)^*\psi(x).
    \label{L2-inn}
    \ee
A concrete example for an orthonormal basis $\{\xi_n\}$ for
$L^2(\R)$ is the basis consisting of the standard normalized
eigenfunctions $\xi_n=\psi_{n-1}$ of the unit simple harmonic
oscillator Hamiltonian \cite{messiah}, or equivalently that of the
operator $-\frac{d^2}{dx^2}+x^2$, i.e., $\psi_m(x)=\pi^{-1/4}(m!
2^m)^{-1/2} e^{-x^2/2}H_m(x),$ where
$H_m(x)=e^{x^2/2}(x-\frac{d}{dx})^me^{-x^2/2}$ are Hermite
polynomials and $m\in\N$. For this example, switching from the
coordinate-representation of the state-vectors to their
representation in terms of the energy eigenbasis of the above
harmonic oscillator corresponds to affecting the unitary operator
$U$.

Unitary operators have a number of important properties that follow
from (\ref{isometry}). For example if $U:{\cal H}_1\to{\cal H}_2$ is
a unitary operator relating two separable Hilbert spaces ${\cal
H}_1$ and ${\cal H}_2$, then
    \begin{enumerate}
    \item $U$ is an everywhere-defined, bounded, and invertible
    operator.
    \item If $\{\xi_n\}$ is an orthonormal basis of
    ${\cal H}_1$, then $\{U\xi_n\}$ is an orthonormal basis of
    ${\cal H}_2$.\footnote{The converse is also true in the
sense that given an orthonormal basis $\{\xi'_n\}$ of ${\cal H}_2$
there is a unique unitary operator $U:{\cal H}_1\to{\cal H}_2$ such
that $\xi'_n=U\xi_n$ for all $n\in\{1,2,\cdots,N\}$.}
    \item Let $A_1:{\cal H}_1\to{\cal H}_1$ be a Hermitian
    operator with domain ${\cal D}(A_1)$. Then $U A_1 U^\dagger:
    {\cal H}_2\to{\cal H}_2$ is a Hermitian operator with domain
    $U(\cD(A_1))=
    \{U\psi_1\in{\cal H}_2|\psi_1\in{\cal D}(A_1)\}$.
    \end{enumerate}
A direct consequence of statement~1 above and the fact that the
inverse of a unitary operator is unitary is that unitary operators
are homeomorphisms\footnote{See footnote~\ref{homeo1} for the
definition.}. As a result, all $N$-dimensional separable Hilbert
spaces have identical topological properties. In particular, if two
separable Hilbert spaces ${\cal H}_1$ and ${\cal H}_2$ share an
underlying vector space, a sequence $\{x_n\}$ converges to $x$ in
${\cal H}_1$ if and only if it converges to $x$ in ${\cal H}_2$.

Next, we recall that every quantum system $s$ is uniquely determined
by a separable Hilbert space $\cH$ that determines the kinematic
structure of $s$ and a Hamiltonian operator $H:\cH\to\cH$ that gives
its dynamical structure via the Schr\"odinger equation. Let $s_1$
and $s_2$ be quantum systems corresponding to the Hilbert spaces
${\cal H}_1$, ${\cal H}_2$ and the Hamiltonians $H_1:{\cal
H}_1\to{\cal H}_1$, $H_2:{\cal H}_2\to{\cal H}_2$. By definition the
observables of the system $s_i$, with $i\in\{1,2\}$, are Hermitian
operators $O_i:{\cal H}_i\to{\cal H}_i$. $s_1$ and $s_2$ are
physically equivalent, if there is a one-to-one correspondence
between their states and observables in such a way that the physical
quantities associated with the corresponding states and observables
are identical. Such a one-to-one correspondence is mediated by a
unitary operator $U:{\cal H}_1\to{\cal H}_2$ according to
    \be
    \psi_1\to \psi_2:=U\psi_1~~~~~~O_1\to O_2:=U\:O_1U^\dagger.
    \label{unitary-eq2}
    \ee
In particular, if $\psi_i(t)$ is an evolving state-vector of the
system $s_i$, i.e., it is a solution of the Schr\"odinger equation
$i\hbar\frac{d}{dt}\psi_i(t)=H_i\psi_i(t)$, we have
$\psi_2(t)=U\psi_1(t)$. The necessary and sufficient condition for
the latter is $H_2=U\:H_1U^\dagger$. This observation motivates the
following theorem.
    \begin{center}
    \parbox{15.5cm}{\textbf{Theorem~1:} \emph{As physical systems
    $s_1=s_2$, if there is a unitary operator $U:\cH_1\to\cH_2$
    satisfying $H_2=U\:H_1U^\dagger$.}}
    \end{center}
To prove this assertion we recall that all physical quantities in QM
may be expressed as expectation values of observables. Suppose that
such a unitary operator exists, and let us prepare a state of $s_1$
that is represented by $\psi_1\in{\cal H}_1$ and measure the
observable $O_1$. The expectation value of this measurement is given
by
    \be
    \frac{\br \psi_1|O_1\psi_1\kt_1}{\br \psi_1|\psi_1\kt_1}=
    \frac{\br U^\dagger\psi_2|U^\dagger
    O_2U\psi_1\kt_1}{\br U^\dagger\psi_2|U^\dagger\psi_2\kt_1}=
    \frac{\br \psi_2|
    O_2\psi_2\kt_2}{\br \psi_2|\psi_2\kt_2},
    \label{ex-va=}
    \ee
where $\br\cdot|\cdot\kt_i$ is the inner product of ${\cal H}_i$,
and we have used (\ref{unitary-eq2}) and the fact that $U^\dagger$
is also a unitary operator. Eqs.~(\ref{ex-va=}) show that the above
measurement is identical with measuring $O_2$ in a state of $s_2$
represented by the state-vector $\psi_2$. This argument is also
valid for the case that $\psi_1$ is an evolving state-vector. It
shows that the existence of $U$ implies the physical equivalence of
$s_1$ and $s_2$.\footnote{The converse of Theorem~1 can be
formulated similarly to the Wigner's symmetry theorem
\cite[p91]{weinberg}.}

If the hypothesis of Theorem~1 holds, i.e., there is a unitary
operator $U:\cH_1\to\cH_2$ satisfying $H_2=U\:H_1U^\dagger$, we say
that the pairs $(\cH_1,H_1)$ and $(\cH_2,H_2)$ and the quantum
systems they define are \emph{unitary-equivalent}.

In conventional QM one mostly considers unitary operators that act
in a single Hilbert space ${\cal H}$. These generate linear
transformations that leave the inner product of the state-vectors
invariant. They form the unitary group ${\cal U}({\cal H})$ of the
Hilbert space which includes the time-evolution operator
$e^{-itH/\hbar}$ as a one-parameter subgroup.\footnote{All possible
symmetry, dynamical, and kinematic groups are also subgroups of
${\cal U}({\cal H})$, \cite{nova}.} Given a quantum system $s$ with
Hamiltonian $H$, one may use the unitary operators $U\in{\cal
U}({\cal H})$ to generate unconventional unitary-equivalent systems
$s_{_U}$ having the same Hilbert space. If $x$ and $p$ are the
standard position and momentum operators of $s$, the position,
momentum, and Hamiltonian operators for $s_{_U}$ are respectively
given by $x_{_U}:=U\:x\:U^\dagger$, $p_{_U}:=U\:p\:U^\dagger$, and
$H_{_U}:=U\:H\:U^\dagger$. The transformation $x\to x_{_U}$, $p\to
p_{_U}$, and $H\to H_{_U}$ is the quantum analog of a classical
time-independent canonical transformation. Therefore, unitary
transformations generated by the elements of $\cU(\cH)$ play the
role of canonical transformations.

\subsection{Biorthonormal Systems}
\label{sec-biortho}

Let $\{\psi_n\}$ be a basis of an $N$-dimensional separable Hilbert
space ${\cal H}$, with $N\leq\infty$, and $\{\xi_n\}$ be the
orthonormal basis obtained by performing the Gram-Schmidt process on
$\{\psi_n\}$. Because $\{\psi_n\}$ is a basis, there are unique
complex numbers $B_{mn}\in\C$ such that for all
$m\in\{1,2,3,\cdots,N\}$
    \be
    \xi_m=\sum_{n=1}^N B_{nm}\psi_n,
    \label{xi-zeta}
    \ee
and because $\{\xi_n\}$ is an orthonormal basis,
    \bea
    \psi_n&=&\sum_{k=1}^N \br\xi_k|\psi_n\kt\:\xi_k,
    \label{zeta-xi}
    \eea
for all $n\in\{1,2,3,\cdots,N\}$. If we substitute (\ref{zeta-xi})
in (\ref{xi-zeta}) and use the orthonormality of $\xi_n$, we find
$\sum_{n=1}^N B_{nm}\br\xi_j|\psi_n\kt=\delta_{mj}$ for all
$m,j\in\{1,2,3,\cdots,N\}$. This shows that the $N\times N$ matrix
$\underbar{B}=(B_{mn})$ is invertible and the entries of $\underbar
B^{-1}$ are given by $B^{-1}_{mn}=\br\xi_m|\psi_n\kt$. We can use
this relation to express (\ref{zeta-xi}) in the form
    \be
    \psi_n=\sum_{k=1}^N B^{-1}_{kn}\:\xi_k.
    \label{zeta-xi-2}
    \ee
This equation suggests that the vectors $\phi_m$ defined by
    \be
    \phi_m:=\sum_{j=1}^N B_{mj}^*\:\xi_j
    ~~~~\mbox{for all $m\in\{1,2,3,\cdots,N\}$},
    \label{phi-m=}
    \ee
fulfil
    \be
    \br\phi_m|\psi_n\kt=\delta_{mn}~~~~
    \mbox{for all $m,n\in\{1,2,3,\cdots,N\}$}.
    \label{biortho}
    \ee
Furthermore, employing (\ref{expand-1}), (\ref{zeta-xi-2}), and
(\ref{phi-m=}), we can show that for all $\psi\in{\cal H}$,
$\sum_{n=1}^N\br\phi|\psi\kt\psi_n=\psi$. We can use Dirac's bra-ket
notation to express this identity as
    \be
    \sum_{n=1}^N |\psi_n\kt\br\phi_n|=I.
    \label{bi-complete}
    \ee
This is a generalization of the more familiar completeness relation
(\ref{resolution-I}).

A sequence $\{(\psi_n,\phi_n)\}$ of ordered pairs of elements of
${\cal H}$ that satisfy (\ref{biortho}) is called a {\em
biorthonormal system} \cite{sadovnichii,schmeidler,young}. A
biorthonormal system satisfying (\ref{bi-complete}) is said to be
\emph{complete}.

Let $\{\psi_n\}$ be a basis of a separable Hilbert space ${\cal H}$,
and $\{\phi_n\}$  be a sequence in $\cH$ such that
$\{(\psi_n,\phi_n)\}$ is a complete biorthonormal system. Then, one
can show that $\{\phi_n\}$ is the unique sequence with this property
and that it is necessarily a basis of ${\cal H}$, \cite{young}.
$\{\phi_n\}$ is called the \emph{biorthonormal basis} associated
with $\{\psi_n\}$, and the biorthonormal system
$\{(\psi_n,\phi_n)\}$ is called a \emph{biorthonormal extension} of
$\{\psi_n\}$.

If $N<\infty$, the right-hand side of (\ref{phi-m=}) is well-defined
and the above construction yields the biorthonormal basis
$\{\phi_n\}$ associated with every basis $\{\psi_n\}$ of ${\cal H}$.
If $N=\infty$, $\{\phi_m\}$ can be constructed provided that the
right-hand side of (\ref{phi-m=}) converges. This is the case if and
only if
    \be
    \sum_{j=1}^\infty |B_{mj}|^2<\infty~~~~
    \mbox{for all $m\in\Z^+$}.
    \label{bounded-B}
    \ee
A theorem due to Bari \cite{gohberg-krein} states that given a basis
$\{\psi_n\}$, the biorthonormal system $\{(\psi_n,\phi_n)\}$ exists
and $\sum_{n=1}^\infty |\br\psi_n|\psi\kt|^2$ and $\sum_{n=1}^\infty
|\br\phi_n|\psi\kt|^2$ both converge for all $\psi\in{\cal H}$, if
and only if $\{\psi_n\}$ is a Riesz basis, i.e., there are an
orthonormal basis $\{\chi_n\}$ and an everywhere-defined bounded
invertible operator $A:{\cal H}\to{\cal H}$ such that
$\psi_n=A\chi_n$.\footnote{As any pair of orthonormal bases are
related by a unitary operator $U:{\cal H}\to{\cal H}$ which is an
everywhere-defined bounded invertible operator, one can take
$\chi_n=\xi_n$ without loss of generality. This allows for the
identification of the infinite matrix $\underbar B$ with the matrix
representation of $A^{-1}$ in the basis $\{\xi_n\}$, for we have
$B_{mn}=\br\xi_m|A^{-1}\xi_n\kt$.} In this case $\{\phi_n\}$ is also
a Riesz basis, and $\{(\psi_n,\phi_n)\}$ is the unique biorthonormal
extension of $\{\psi_n\}$, \cite{gohberg-krein,sadovnichii,young}.

The coefficients of the expansion of a vector $\psi\in{\cal H}$ in a
Riesz basis $\{\psi_n\}$ can be expressed in terms of its
biorthonormal basis $\{\phi_n\}$. Given $\psi=\sum_{n=1}^Nc_n\psi_n$
we have, in light of (\ref{biortho}), $c_n=\br\phi_n|\psi\kt$.
Hence, for all $\psi\in{\cal H}$,
    \be
    \psi=\sum_{n=1}^N\br\phi_n|\psi\kt\:\psi_n.
    \label{biortho-exp}
    \ee
Clearly the roles of $\{\phi_n\}$ and $\{\psi_n\}$ are
interchangeable. In particular, $\sum_{n=1}^N
|\phi_n\kt\br\psi_n|=I$ and
$\psi=\sum_{n=1}^N\br\psi_n|\psi\kt\:\phi_n$ for all $\psi\in{\cal
H}$.

Let $\{\psi_n\}$ be a Riesz basis and $\{(\psi_n,\phi_n)\}$ be its
biorthonormal extension. As we explained in
Subsection~\ref{sec-inn}, we can construct a unique inner product
$(\cdot|\cdot)$ on ${\cal H}$ that makes a Riesz basis orthonormal.
We can use (\ref{inn-basis}) and (\ref{biortho-exp}) to obtain the
following simplified expression for this inner product.
    \be
    (\psi|\phi):=\sum_{n=1}^N \br\psi|\phi_n\kt\br\phi_n|\phi\kt=
    \br\psi|\eta_+\phi\kt,~~~~\mbox{for all $\psi,\phi\in
    {\cal H}$},
    \label{inn-basis-1}
    \ee
where we have introduced the operator $\eta_+:{\cal H}\to{\cal H}$
according to
    \be
    \eta_+\psi:=\sum_{n=1}^N \br\phi_n|\psi\kt\:\phi_n.
    \label{eta-plus=}
    \ee
Using Dirac's bra-ket notation we can write it in the form
    \be
    \eta_+=\sum_{n=1}^N |\phi_n\kt\br\phi_n|.
    \label{eta-plus=2}
    \ee
Because $\{\psi_n\}$ is a Riesz basis, the inner
product~(\ref{inn-basis-1}) is defined for all $\psi,\phi\in{\cal
H}$. This shows that $\eta_+$ is everywhere-defined. Furthermore, it
is not difficult to see, by virtue of (\ref{biortho}), that it has
an everywhere-defined inverse given by
    \be
    \eta_+^{-1}:=\sum_{n=1}^N |\psi_n\kt\br\psi_n|.
    \label{eta-plus-inverse=}
    \ee
This shows that $\eta_+$ is a one-to-one onto linear operator. As
suggested by (\ref{eta-plus=2}) it is also Hermitian, which in
particular implies that both $\eta_+$ and $\eta_+^{-1}$ are bounded.
Finally, in view of (\ref{eta-plus=2}),
$\br\psi|\eta_+\psi\kt=\sum_{n=1}^N|\br\phi_n|\psi\kt|^2$, for all
$\psi\in\cH$. Therefore $\eta_+$ is a positive operator. Moreover
because it is an invertible operator, its spectrum is strictly
positive. A Hermitian operator with this property is called a
\emph{positive-definite operator}. The operator $\eta_+$ constructed
above is an everywhere-defined, bounded, positive-definite,
invertible operator. A linear operator with these properties is
called a \emph{metric operator}.

\subsection{Metric Operators and Conventional QM}
\label{sec-metric}

Consider a pair of separable Hilbert spaces ${\cal H}_1$ and ${\cal
H}_2$ that are identical as vector spaces but have different inner
products. We will denote the inner products of ${\cal H}_1$ and
${\cal H}_2$ by $\br\cdot|\cdot\kt_{_1}$ and
$\br\cdot|\cdot\kt_{_2}$, respectively, and view
$\br\cdot|\cdot\kt_{_2}$ as an alternative inner product on ${\cal
H}_1$. Our aim is to find a way to express $\br\cdot|\cdot\kt_{_2}$
in terms of $\br\cdot|\cdot\kt_{_1}$. We will first consider the
case that the underlying vector space ${\cal V}$ of both ${\cal
H}_1$ and ${\cal H}_2$ is finite-dimensional, i.e., $N<\infty$.

Let $\{\xi_n\}$ be an orthonormal basis of ${\cal H}_1$. As we
argued above, it satisfies the completeness relation:
$\sum_{n=1}^N|\xi_n\kt_{_1}\, _{_1}\!\br\xi_n|=I$, where $I$ is the
identity operator of ${\cal V}$ . In general, $\{\xi_n\}$ is not an
orthonormal basis of ${\cal H}_2$ and the operator
    \be
    \eta_+:=\sum_{n=1}^N|\xi_n\kt_{_2}\, _{_2}\!\br\xi_n|,
    \label{eta-def-1}
    \ee
does not coincide with $I$. Eq.~(\ref{eta-def-1}), which we can
express in the more conventional form:
    \be
    \eta_+\,\psi:=
    \sum_{n=1}^N\br\xi_n|\psi\kt_{_2}~\xi_n,~~~~
    \mbox{for all $\psi\in{\cal V}$},
    \label{eta-def}
    \ee
defines $\eta_+$ as a linear operator $\eta_+:{\cal V}\to{\cal V}$
having a full domain.

Now, let $\phi$ be an arbitrary element of ${\cal V}$. In view of
(\ref{expand-1}), we can express it as
    \be
    \phi=\sum_{m=1}^N\br\xi_m|\phi\kt_{_1}\,\xi_m.
    \label{expand-3}
    \ee
Using Eqs.~(\ref{eta-def}), (\ref{expand-3}) and the properties
(\emph{ii}) and (\emph{iii}) shared by both the inner products, we
can easily show that $\eta_+$ fulfils
    \be
    \br\phi|\psi\kt_{_2}=\br\phi|\eta_+\,\psi\kt_{_1}~~~~~~~
    \mbox{for all $\psi,\phi\in{\cal V}$.}
    \label{eta-prop1}
    \ee
Employing properties (\emph{i}) and (\emph{ii}) of
$\br\cdot|\cdot\kt_2$, we can also verify that
    \bea
    &&\br\phi|\eta_+\,\psi\kt_{_1}^*=\br\psi|\eta_+\,\phi\kt_{_1}
    ~~~~\mbox{for all $\psi,\phi\in{\cal V}$,}
    \label{eta-prop3}\\
    &&\br\psi|\eta_+\,\psi\kt_{_1}>0~~~~~~~~~~~~\mbox{for all
    nonzero $\psi\in{\cal V}$}.
    \label{eta-prop2}
    \eea
Eq.~(\ref{eta-prop3}) shows that $\eta_+$ is a Hermitian operator
acting in ${\cal H}_1$, in particular it has a real spectrum.
Eq.~(\ref{eta-prop2}) implies that indeed the spectrum of $\eta_+$
is strictly positive; $\eta_+$ is a positive-definite operator.

If ${\cal H}_1$ and ${\cal H}_2$ are infinite-dimensional, the sum
appearing in (\ref{eta-def}) stands for an infinite series and we
must address the issue of its convergence. The convergence of this
series is equivalent to the requirement that $\sum_{n=}^\infty
|\br\xi_n|\psi\kt_{_2}|^2<\infty$ for all $\psi\in{\cal V}$.
According to the above-mentioned theorem of Bari this requirement is
fulfilled provided that $\{\xi_n\}$ is a Riesz basis of ${\cal
H}_2$. Under this assumption (\ref{eta-def}) defines a linear
operator $\eta_+$ acting in ${\cal H}_1$ and satisfying
(\ref{eta-prop1}). It is a positive-definite (in particular
Hermitian) operator with a full domain. Being Hermitian and
everywhere-defined it is also necessarily bounded.

Next, we exchange the roles of ${\cal H}_1$ and ${\cal H}_2$. This
gives rise to an everywhere-defined bounded positive-definite
operator $\eta_+'$ acting in ${\cal H}_2$ such that
$\br\phi|\psi\kt_{_1}=\br\phi|\eta_+'\,\psi\kt_{_2}$ for all
$\psi,\phi$ in ${\cal V}$. Combining this equation and
(\ref{eta-prop1}) yields
$\br\phi|\psi\kt_{_1}=\br\phi|\eta_+\eta_+'\,\psi\kt_{_1}$ for all
$\psi,\phi$ in ${\cal V}$. This in turn implies that
$\eta_+'\eta_+=I$.\footnote{The proof uses the argument given in
footnote~\ref{trick}.} Because $\eta_+'$ is a bounded operator with
a full domain, $\eta_+$ is an invertible operator with inverse
$\eta_+^{-1}=\eta_+'$.\footnote{If we use the prescription of
Subsection~\ref{sec-inn} to obtain the inner product $(\cdot|\cdot)$
on ${\cal H}_2$ that renders the Riesz basis $\{\xi_1\}$
orthonormal, we find by its uniqueness that
$(\cdot|\cdot)=\br\cdot|\cdot\kt_{_1}$. According to
(\ref{inn-basis-1}), there is an everywhere-defined bounded
invertible operator $\tilde\eta_+$ such that
$\br\cdot|\cdot\kt_{_1}=\br\cdot|\tilde\eta_+\cdot\kt_{_2}$. It is
given by $\tilde\eta_+=\eta_+^{-1}$.} Therefore $\eta_+$ is a metric
operator.

The above construction of the metric operator $\eta_+$ is based on
the choice of an orthonormal basis $\{\xi_n\}$ of ${\cal H}_1$. We
will next show that indeed $\eta_+$ is independent of this choice.
Let $\eta'_+:{\cal V}\to{\cal V}$ be an everywhere-defined linear
operator satisfying
    \be
    \br\phi|\psi\kt_{_2}=\br\phi|\eta_+'\,\psi\kt_{_1}~~~~~~~
    \mbox{for all $\psi,\phi\in{\cal V}$.}
    \label{eta-prop1-prime}
    \ee
Eqs.~(\ref{eta-prop1}) and (\ref{eta-prop1-prime}) show that
$\br\phi|(\eta'_+-\eta_+)\psi\kt_{_1}=0$ for all $\psi,\phi\in{\cal
V}$. In view of the argument presented in footnote~\ref{trick}, this
implies $\eta'_+\psi=\eta_+\psi$ for all $\psi\in{\cal V}$, i.e.,
$\eta'_+=\eta_+$. This establishes the uniqueness of the metric
operator $\eta_+$ which in turn means that \emph{the inner products
that make a complex vector space ${\cal V}$ into a separable Hilbert
space are in one-to-one correspondence with the metric operators
$\eta_+$ acting in one of these Hilbert spaces}.

To employ the characterization of the inner products in terms of
metric operators we need to select a Hilbert space ${\cal H}$. We
will call this Hilbert space a \emph{reference Hilbert space}. We
will always fix a reference Hilbert space and use its inner product
$\br\cdot|\cdot\kt$ to determine if a given linear operator acting
in ${\cal V}$ is Hermitian or not. The following are some typical
examples.
    \begin{itemize}
    \item
    For systems having a finite number ($N$) of linearly
    independent state-vectors, i.e., when ${\cal V}=\C^N$,
    ${\cal H}$ is defined by the Euclidean inner product,
    $\br\vec\phi|\vec\psi\kt:=\vec\phi^*\cdot\vec\psi$, for all
    $\vec\phi,\vec\psi\in\C^N$.

    \item For systems whose configuration space is a differentiable
    manifold $M$ with an integral measure $d\mu(x)$, ${\cal V}$ is the
    space $L^2(M)$ of all square-integrable functions $\psi:M\to\C$
    and ${\cal H}$ is defined by the $L^2$-inner product:
    $\br\phi|\psi\kt:=\int_{M}\phi(x)^*\psi(x)\,d\mu(x)$,
    for all $\phi,\psi\in L^2(M)$. The systems whose configuration
    space is a Euclidean space $(M=\R^d)$ or a complex contour
    $(M=\Gamma)$ belong to this class. We will discuss the latter
    systems in Section~\ref{sec-contour}.
    \end{itemize}

In summary, given a separable Hilbert space ${\cal H}$ with inner
product $\br\cdot|\cdot\kt$, we can characterize every other inner
product on ${\cal H}$ by a metric operator $\eta_+:{\cal H}\to{\cal
H}$ according to
    \be
    \br\cdot|\cdot\kt_{_{\eta_+}}:=\br\cdot|\eta_+\,\cdot\kt.
    \label{inn}
    \ee
Each choice of $\eta_+$ defines a unique separable Hilbert space
${\cal H}_{_{\eta_+}}$. Because $\eta_+$ is a positive-definite
operator, we can use its spectral representation to construct its
positive square root $\rho:=\sqrt\eta_+$. As a linear operator
acting in ${\cal H}$, $\rho$ is a Hermitian operator satisfying
$\rho^2=\eta_+$. We can use this observation to establish
    \be
    \br\phi|\psi\kt_{\eta_+}=\br\phi|\eta_+\psi\kt=
    \br\rho^\dagger\phi|\rho\,\psi\kt=
    \br\rho\,\phi|\rho\,\psi\kt,~~~~~~\mbox{for all $\phi,\psi\in\cH$}.
    \label{rho}
    \ee
This relation shows that as a linear operator mapping ${\cal
H}_{_{\eta_+}}$ onto ${\cal H}$, $\rho$ is a unitary
operator.\footnote{Strictly speaking (\ref{rho}) shows that $\rho$
is an isometry. However, in view of the fact that $\eta_+$ is
invertible, so does $\rho$. This implies that $\rho:{\cal
H}_{_{\eta_+}}\to{\cal H}$ is a genuine unitary operator.} It
provides a realization of the unitary-equivalence of the Hilbert
spaces ${\cal H}_{_{\eta_+}}$ and ${\cal H}$.

In ordinary QM one fixes the physical Hilbert space of the system to
be one of the reference Hilbert spaces listed above and develops a
theory based on this preassigned Hilbert space. The argument that
the unitary-equivalence of all separable Hilbert spaces justifies
this convention is not quite acceptable. For example, although for
all $d\in\Z^+$, $L^2(\R^d)$ is unitary-equivalent to $L^2(\R)$, we
never use $L^2(\R)$ to describe a system having more than one real
degree of freedom. \emph{The choice of the particular Hilbert space
should in principle be determined by physical considerations or left
as a freedom of the formulation of the theory.} In view of lack of a
direct measurement of the inner product or the associated metric
operator, we propose to adopt the second option. We will see that
this does not lead to a genuine generalization of QM, but it reveals
a set of alternative and equally useful representations of QM which
could not be utilized within its conventional formulation.
Furthermore, the introduction of the freedom in choosing the metric
operator may be used as an interesting method of extending QM by
relaxing some of the restrictions put on the metric operator. Indeed
the indefinite-metric quantum theories are examples of such a
generalization.

\section{Pseudo-Hermitian QM: Ingredients and Formalism}
\label{sec-phqm}
\subsection{Quasi-Hermitian versus Pseudo-Hermitian QM}
\label{sec-quasi}

To the best of our knowledge, the first publication investigating
the consequences of the freedom in the choice of the metric operator
is the article by Scholtz, Geyer, and Hahne \cite{geyer} in which
the choice of the metric operator is linked to that of an
\emph{irreducible} set of linear operators. The latter is any
(minimal) set ${\cal S}$ of operators $O_\alpha$ acting in a vector
space ${\cal V}$ that do not leave any proper subspace of ${\cal V}$
invariant, i.e., the only subspace ${\cal V}'$ of ${\cal V}$ that
satisfies: ``~$\mbox{$O_\alpha\in{\cal S}$ and $\psi\in{\cal V}'$
imply $O_\alpha\psi\in{\cal V}'$,}$~'' is ${\cal V}$.

The approach pursued in \cite{geyer} involves using the physical
characteristics of a given system to determine an irreducible set of
operators (that are to be identified with the observables of the
theory) and employing the latter to fix a metric operator and the
associated inner product of the Hilbert space. We will call this
formalism \emph{Quasi-Hermitian Quantum Mechanics}. The main problem
with this formalism is that it is generally very difficult to
implement. This stems from the fact that the operators belonging to
an irreducible set must in addition be compatible, i.e., there must
exist an inner product with respect to which all the members of the
set are Hermitian. In order to use this formalism to determine the
inner product, one must in general employ a complicated iterative
scheme.
    \begin{itemize}
    \item Select a linear operator $O_1$ with a complete set of eigenvectors
    and a real spectrum;
    \item Find the set $\fU_1$ of all possible metric operators
that make $O_1$ Hermitian, and select a linear operator $O_2$,
linearly independent of $O_1$, from among the linear operators that
are Hermitian with respect to the inner product defined by some
$\eta_+\in\fU_1$;
    \item Find the set $\fU_2$ of all possible metric
operators that make $O_2$ Hermitian, and select a linear operator
$O_3$, linearly independent of $O_1$ and $O_2$, from among the
linear operators that are Hermitian with respect to the inner
product defined by some $\eta_+\in\fU_1\cap \fU_2$, where $\cap$
stands for the intersection of sets;
    \item Repeat this procedure until the inner product
(respectively metric operator $\eta_+$) is fixed up to a constant
coefficient.
    \end{itemize}
As we see, in trying to employ Quasi-Hermitian QM, one needs a
procedure to compute the most general metric operator associated
with a given diagonalizable linear operator with a real spectrum.
This is the main technical tool developed within the framework of
Pseudo-Hermitian QM.

Pseudo-Hermitian QM differs from Quasi-Hermitian QM in that in the
former one chooses $O_1$ to be the Hamiltonian, finds $\fU_1$ and
leaves the choice of $O_2$ arbitrary, i.e., identifies all the
operators $O$ that are Hermitian with respect the inner product
associated with some unspecified metric operator $\eta_+$ belonging
to $\fU_1$. The metric operator $\eta_+$ fixes a particular inner
product and defines the ``\emph{physical Hilbert space}'' ${\cal
H}_{\rm phys}$ of the system. The ``\emph{physical observables}''
are the Hermitian operators $O$ acting in ${\cal H}_{\rm phys}$. We
can use the unitary-equivalence of ${\cal H}_{\rm phys}$ and ${\cal
H}$ realized by $\rho:=\sqrt\eta_+$ to construct the physical
observables $O$ using those of the conventional QM, i.e., Hermitian
operators $o$ acting in the reference Hilbert space ${\cal H}$. This
is done according to \cite{critique,jpa-2004b}
    \be
    O=\rho^{-1}o\rho.
    \label{O=ror}
    \ee
Note that as an operators mapping ${\cal H}_{\rm phys}$ onto ${\cal
H}$, $\rho$ is a unitary operator. Hence $O$ is a Hermitian operator
acting in ${\cal H}_{\rm phys}$ if and only if $o$ is a Hermitian
operator acting in ${\cal H}$.

For instance, let ${\cal H}=L^2(\R^d)$ for some $d\in\Z^+$. Then we
can select the usual position $x_i$ and momentum $p_i$ operators to
substitute for $o$ in (\ref{O=ror}). This defines a set of basic
physical observables,
    \be
    X_i:=\rho^{-1}x_i\rho,~~~~~~~~P_i:=\rho^{-1}p_i\rho,
    \label{X-P}
    \ee
which we respectively call the \emph{$\eta_+$-pseudo-Hermitian
position and momentum operators}. They furnish an irreducible
unitary representation of the Weyl-Heisenberg algebra,
    \be
    [X_i,X_j]=[P_i,P_j]=0,~~~~~~[X_i,P_j]=\hbar\delta_{ij}I,~~~~
    \mbox{for all $i,j\in\{1,2,\cdots,d$\}}.
    \label{W-H}
    \ee

In principle, we can express the Hamiltonian $H$ as a function of
$X_i$ and $P_i$ and attempt to associate a physical meaning to it by
devising a quantum-to-classical \emph{correspondence principle}. One
way of doing this is to define the underlying classical Hamiltonian
$H_c$ for the system as
    \be
    H_c(\vec x_c,\vec p_c):=\left.\lim_{\hbar\to 0}
    H(\vec X,\vec P)\right|_{\mbox{\tiny$\begin{array}{c}
    \vec X\to\vec x_c\\
    \vec P\to\vec p_c\end{array}$}},
    \label{class-H}
    \ee
where $\vec w:=(w_1,w_2,\cdots,w_d)^T$ for $\vec w=\vec X,\vec
P,\vec x_c,\vec p_c$; and $\vec x_c$ and $\vec p_c$ stand for
classical position and momentum variables. Supposing that the
right-hand side of (\ref{class-H}) exists, one may reproduce the
quantum system described by the Hilbert space ${\cal H}_{\rm phys}$
and Hamiltonian $H$ by quantizing the classical system corresponding
to $H_c$ according to
    \be
    \vec x_c\to \vec X,~~~~~~~
    \vec p_c\to \vec P,~~~~~~~
    \{\cdot,\cdot\}_c \to -i\hbar^{-1}[\cdot,\cdot],
    \label{ph-quantize}
    \ee
where $\{\cdot,\cdot\}_c$ and $[\cdot,\cdot]$ stand for the Poisson
bracket and the commutator, respectively. This is called the
\emph{$\eta_+$-pseudo-Hermitian canonical quantization} scheme
\cite{jpa-2004b,jpa-2005b}.

The quantum system described by ${\cal H}_{\rm phys}$ and $H$ admits
a representation in conventional QM in which the Hilbert space is
${\cal H}=L^2(\R^d)$, the observables are Hermitian operators acting
in ${\cal H}$, and the Hamiltonian is given by
    \be
    h:=\rho H\rho^{-1}.
    \label{h-Hermitian}
    \ee
Because $\rho:{\cal H}_{\rm phys}\to{\cal H}$ is unitary, so is its
inverse $\rho^{-1}:{\cal H}\to{\cal H}_{\rm phys}$. This in turn
implies that $h$ is a Hermitian operator acting in ${\cal H}$. We
will call the representation of the quantum system that is based on
the Hermitian Hamiltonian $h$ the \emph{Hermitian representation}.
In this representation, we can proceed employing the usual
prescription for identifying the underlying classical Hamiltonian,
namely as
    \be
    H_c(\vec x_c,\vec p_c):=\left.\lim_{\hbar\to 0}
    h(\vec x,\vec p)\right|_{\mbox{\tiny$\begin{array}{c}
    \vec x\to\vec x_c\\
    \vec p\to\vec p_c\end{array}$}}.
    \label{class-H-2}
    \ee
Note that this relation is consistent with (\ref{class-H}), because
in view of (\ref{X-P}) and (\ref{h-Hermitian}), $h=f(\vec x,\vec p)$
if and only if $H=f(\vec X,\vec P)$, where we suppose that
$f:\R^{2d}\to\C$ is a piecewise real-analytic function.

Each choice of a metric operator $\eta_+\in\fU_1$ corresponds to a
particular quantum system with an associated Hermitian
representation. One can in principle confine his (her) attention to
this representation which can be fully understood using the
conventional QM. The main disadvantage of employing the Hermitian
representation is that in general the Hamiltonian $h$ is a terribly
complicated nonlocal operator. Therefore, the computation of the
energy levels and the description of the dynamics are more
conveniently carried out in the pseudo-Hermitian representation. In
contrast, it is the Hermitian representation that facilitates the
computation of the expectation values of the physical position and
momentum operators as well as that of the localized states in
physical position or momentum spaces. See
\cite{jpa-2004b,jpa-2005b,jmp-2005,jpa-2006a} for explicit examples.

\subsection{Pseudo-Hermitian and Pseudo-Metric Operators}
\label{sec-ph-metric}

\begin{center}
\parbox{15.5cm}{\textbf{Definition~1:} \emph{A linear operator $A:{\cal
H}\to{\cal H}$ acting in a separable Hilbert space ${\cal H}$ is
said to be {pseudo-Hermitian} if ${\cal D}(A)$ is a dense subset of
${\cal H}$, and there is an everywhere-defined invertible Hermitian
linear operator $\eta:{\cal H}\to{\cal H}$ such that
    \be
    A^\dagger=\eta A \eta^{-1}.
    \label{ph-2}
    \ee}}
    \end{center}
We will refer to an operator $\eta$ satisfying (\ref{ph-2}) as a
\emph{pseudo-metric operator associated with the operator $A$} and
denote the set of all such operators by $\fM_A$. Clearly, $A$ is
pseudo-Hermitian if and only if $\fM_A$ is nonempty. Furthermore,
for every linear operator $A$, $\fM_A\subseteq \fM_I$ where $I$ is
the identity operator. We will call elements of $\fM_I$
\emph{pseudo-metric operators}. Because they are Hermitian and have
full domain, they are necessarily bounded.\footnote{The definition
of a pseudo-Hermitian operator given in \cite{p1} requires $\eta$ to
be a Hermitian automorphism. This is equivalent to the definition
given above because of the following. Firstly, an automorphism is,
by definition, everywhere-defined. Hence, if it is Hermitian, it
must be bounded. Secondly because the inverse of every Hermitian
automorphism is a Hermitian automorphism, $\eta^{-1}$ is
everywhere-defined and bounded, i.e., $\eta$ is invertible. The fact
that every everywhere-defined invertible operator is an automorphism
is obvious.}

Clearly if $\eta_1$ belongs to $\fM_A$, then so does
$\eta_r:=r\eta_1$, for every nonzero real number $r$. The scaling
$\eta_1\to\eta_r$ of the pseudo-metric operators has no physical
significance. It signifies a spurious symmetry that we will
eliminate by restricting our attention to pseudo-metric operators
$\eta$ whose spectrum $\sigma(A)$ is bounded above by $1$, i.e.,
max$[\sigma(A)]=1$.\footnote{For a given $\eta_1\in\fM_A$ there
always exists $r_\star\in\R$ such that the spectrum of
$\eta_{r_\star}$ is bounded above by $1$ This follows from the fact
that because $\eta_1$ is an invertible bounded self-adjoint
operator, its spectrum $\sigma(\eta_1)$ is a compact subset of $\R$
excluding zero, \cite{kolmogorov-fomin}. Let $\alpha_1$ and
$\alpha_2$ be respectively the minimum and maximum values of
$\sigma(\eta_1)$. If $\alpha_2>0$, we take $r_\star:=\alpha_2^{-1}$;
if $\alpha_2<0$, we take $r_\star:=-\alpha_1^{-1}$.\label{scale}}
The latter form a subset of $\fM_A$ which we will denote by $\fU_A$.

In general, either $\fU_A$ is the empty set and $A$ is not
pseudo-Hermitian or $\fU_A$ is an infinite set\footnote{This is true
unless the Hilbert space is one-dimensional.}; \emph{the
pseudo-metric operator associated with a pseudo-Hermitian operator
is not unique.} To see this, let $\eta\in\fU_A$, $B:{\cal H}\to{\cal
H}$ be an everywhere-defined invertible bounded operator commuting
with $A$, and $\tilde\eta:=B^\dagger\eta B$. Then $B^\dagger$ is an
everywhere-defined bounded operator \cite{yosida} commuting with
$A^\dagger$. These in turn imply that $\tilde\eta$ is an
everywhere-defined invertible Hermitian operator which in view of
(\ref{ph-2}) satisfies $\tilde\eta A\tilde\eta^{-1}=B^\dagger\eta B
A B^{-1}\eta^{-1} B^{-1\dagger}=B^\dagger A^\dagger
B^{-1\dagger}=A^\dagger$, i.e., $\tilde\eta\in\fM_A$. Clearly there
is an infinity of choices for $B$ defining an infinite set of
pseudo-metric operators of the form $B^\dagger\eta B$. It is not
difficult to observe that upon making appropriate scaling of these
pseudo-metric operators one can construct an infinite set of
elements of $\fU_A$, i.e., $\fU_A$ is an infinite set.

The non-uniqueness of pseudo-metric operators associated with a
pseudo-Hermitian operator motivates the following definition.
\begin{center}
\parbox{15.5cm}{
\textbf{Definition~2:} \emph{Let $A$ be a pseudo-Hermitian operator
acting in ${\cal H}$ and $\eta\in \fM_I$ be a pseudo-metric
operator. Then $A$ is said to be {$\eta$-pseudo-Hermitian} if
$\eta\in\fM_A$.}} \end{center} Clearly, in order to determine
whether a pseudo-Hermitian operator $A$ is $\eta$-pseudo-Hermitian
one must know both $A$ and $\eta$. It is quite possible that a
pseudo-Hermitian operator fails to be $\eta$-pseudo-Hermitian for a
given $\eta\in \fM_I$.\footnote{The term ``$\eta$-pseudo-Hermitian
operator'' is used to emphasize that one works with a particular
pseudo-metric operator. It coincides with the notion of a
``$J$-Hermitian operator'' used by mathematicians
\cite{pease,azizov} and the old notion of a ``pseudo-Hermitian
operator'' used occasionally in the context of indefinite-metric
quantum theories \cite{case-1954,sudarshan}.}

Given a pseudo-Hermitian operator $A$, the set $\fM_A$ may or may
not include a positive-definite element $\eta_+$. If such a
positive-definite element exists, we can use it to construct the
inner product
    \be
    \br\cdot|\cdot\kt_{\eta_+}:=\br\cdot|\eta_+\cdot\kt
    \label{inn-eta2}
    \ee
that renders $A$ Hermitian,
    \be
    \br\cdot|A\cdot\kt_{\eta_+}=\br A \cdot|\cdot\kt_{\eta_+}.
    \label{eta-self-adjoint-1}
    \ee
This means that if we endow the underlying vector space ${\cal V}$
of ${\cal H}$ with the inner product (\ref{inn-eta2}), we find a
separable Hilbert space ${\cal H}_{_{\eta_+}}$ such that $A:{\cal
H}_{_{\eta_+}}\to {\cal H}_{_{\eta_+}}$ is Hermitian. In particular
the spectrum $\sigma(A)$ of $A$ is real. If $\sigma(A)$ happens to
be discrete, we can construct an orthonormal basis $\{\psi_n\}$ of
${\cal H}_{_{\eta_+}}$ consisting of the eigenvectors of $A$. As a
sequence of elements of ${\cal H}$, $\{\psi_n\}$ is a Riesz basis.
Hence $A:{\cal H}\to{\cal H}$ is diagonalizable. This shows that for
a densely-defined operator having a discrete spectrum, the condition
that it is a diagonalizable operator having a real spectrum is
necessary for the existence of a metric operator $\eta_+$ among the
elements of $\fM_A$. The existence of $\eta_+$, in particular,
implies that $\fM_A$ is non-empty. Hence, $A$ is necessarily
pseudo-Hermitian.

It is not difficult to show that the same conditions are also
sufficient for the inclusion of a metric operator in $\fM_A$,
\cite{p2,p3}. Suppose that $A:{\cal H}\to{\cal H}$ is a
densely-defined diagonalizable operator having a real spectrum. Let
$\{\psi_n\}$ be a Riesz basis consisting of the eigenvectors of $A$
and $\{(\psi_n,\phi_n)\}$ be its biorthonormal extension. Then, in
view of the spectral representation of $A$, i.e., $A=\sum_{n=1}^N
a_n|\psi_n\kt\br\phi_n|$ where $a_n$ are eigenvalues of $A$, and the
basic properties of the biorthonormal system $\{(\psi_n,\phi_n)\}$,
we can easily show that $\eta_+$, as defined by
    \be
    \eta_+:=\sum_{n=1}^N|\phi_n\kt\br\phi_n|,
    \label{eta=quasi}
    \ee
is a positive-definite operator belonging to $\fM_A$. As we
explained in Subsection~\ref{sec-biortho}, this is the unique metric
operator whose inner product makes $\{\psi_n\}$ orthonormal.

Again if $\fM_A$ includes a metric operator $\eta_+$, then
$\tilde\eta_+=B^\dagger\eta_+ B$ for any everywhere-defined,
bounded, invertible operator $B$ commuting with $A$ is also a metric
operator belonging to $\fM_A$. This shows that the subset $\fM^+_A$
of $\fM_A$ that consists of metric operators is either empty or has
an infinity of elements, \cite{p4,jmp-2003}. The same holds for
$\fU^+_A:=\fM^+_A\cap \fU_A$. In summary, \emph{for a Hilbert space
with dimension $N>1$, either there is no metric operator $\eta_+$
satisfying (\ref{eta-self-adjoint-1}) or there is an infinite set of
such metric operators that in addition fulfil ${\rm
max}[\sigma(\eta_+)]=1$.}

One may generalize the notion of the inner product by replacing
Condition~(\emph{i}) of Subsection~\ref{sec-inn} by the following
weaker condition.
    \begin{itemize}
    \item[](\'{\em i}\,) $\br\cdot|\cdot\kt$ is \emph{nondegenerate},
i.e., given $\psi\in{\cal H}$ the condition ``$\br\phi|\psi\kt=0$
for all $\phi\in{\cal H}$'' implies ``$\psi=0$''.
    \end{itemize}
A function $\pbr\cdot|\cdot\pkt:{\cal H}\times{\cal H}\to\C$ which
satisfies conditions (\'{\em i}\,), (\emph{ii}) and (\emph{iii}) is
called a \emph{pseudo-inner product}. Clearly every inner product on
${\cal H}$ is a pseudo-inner product. The converse is not true,
because in general there are pseudo-inner products
$\pbr\cdot|\cdot\pkt$ that fail to satisfy (\emph{i}). This means
that there may exist nonzero $\psi\in{\cal H}$ such that
$\pbr\psi|\psi\pkt\leq 0$. Such a pseudo-inner product is called an
\emph{indefinite inner product}. It is not difficult to see that
given a pseudo-metric operator $\eta\in\fM_I$, the following
relation defines a pseudo-inner product on ${\cal H}$.
    \be
    \pbr\cdot|\cdot\pkt=\br\cdot|\eta\cdot\kt=:
    \br\cdot|\cdot\kt_{_\eta}.
    \label{inn-3}
    \ee

Let $A:\cH\to\cH$ be a densely-defined operator, $\eta:\cH\to\cH$ be
a pseudo-metric operator, and
$A_\eta^\dagger:=\eta^{-1}A^\dagger\eta$. Then $A_\eta^\dagger$ for
all $\psi_1\in{\cal D}(A)$ and $\psi_2\in{\cal D}(A_\eta^\dagger)$,
we have $\br\psi_1|A_\eta^\dagger\psi_2\kt_{_\eta}=\br
A\psi_1|\psi_2\kt_{_\eta}$. In particular if $A$ is
$\eta$-pseudo-Hermitian, $A_\eta^\dagger=A$ and
    \be
    \br\psi_1|A\psi_2\kt_{_\eta}=\br A\psi_1|\psi_2\kt_{_\eta}.
    \label{ph-3}
    \ee
This means that every pseudo-Hermitian operator $A$ is Hermitian
with respect to the pseudo-inner product $\br\cdot|\cdot\kt_\eta$
defined by an arbitrary element $\eta$ of $\fM_A$. It is not
difficult to see that the converse is also true: If $\eta A$ and
$A^\dagger\eta$ have the same domains and $A$ satisfies (\ref{ph-3})
for some $\eta\in\fM_I$, then $A$ is pseudo-Hermitian and
$\eta\in\fM_A$.

An \emph{indefinite-metric space} is a complex vector space ${\cal
V}$ endowed with a function $\pbr\cdot|\cdot\pkt:{\cal H}\times{\cal
H}\to\C$ satisfying (\'{\em i}\,), (\emph{ii}) and (\emph{iii}),
\cite{bognar,azizov}. One can turn a Hilbert space into an
indefinite-metric space by endowing the underlying vector space with
an indefinite inner product $\br\cdot|\cdot\kt_\eta$ whose
pseudo-metric operator $\eta$ is not positive-definite. The latter
is called an \emph{indefinite metric operator}. It is important to
realize that the study of general indefinite-metric spaces is not
the same as the study of consequences of endowing a given Hilbert
space with an indefinite inner product. The latter, which is known
as the $\eta$-formalism, avoids a host of subtle questions such as
the nature of the topology of indefinite-metric spaces
\cite{nagy,nakanishi}.

The indefinite-metric quantum theories
\cite{pauli-1943,sudarshan,nagy,nakanishi} involve the study of
particular indefinite-metric spaces having a fixed indefinite inner
product. In this sense they share the philosophy adopted in
conventional QM; \emph{the indefinite-inner product is fixed from
the outset and the theory is built upon this choice.} The situation
is just the opposite in pseudo-Hermitian QM where the space of
state-vectors is supposed to have the structure of a (separable)
Hilbert space with an inner product which is neither indefinite nor
fixed \emph{a priori}.\footnote{Failure to pay attention to this
point is responsible for confusing pseudo-Hermitian QM with
indefinite-metric quantum theories. See for example
\cite{kleefeld}.}

In pseudo-Hermitian QM, the physical Hilbert space is constructed
using the following prescription. First one endows the vector space
of state-vectors with a fixed auxiliary (positive-definite) inner
product. This defines the reference Hilbert space ${\cal H}$ in
which all the relevant operators act. Next, one chooses a
Hamiltonian operator that acts in ${\cal H}$, is diagonalizable, has
a real spectrum, but needs not be Hermitian. Finally, one determines
the (positive-definite) inner products on ${\cal H}$ that render the
Hamiltonian Hermitian. Because there is an infinity of such inner
products, one obtains an infinite class of kinematically different
but dynamically equivalent quantum systems. The connection to
indefinite-metric theories is that for the specific ${\cal
PT}$-symmetric models whose study motivated the formulation of
pseudo-Hermitian QM, there is a simple and universal choice for an
indefinite inner product, namely the $\cP\cT$-inner product
(\ref{inn-PT}), which makes the Hamiltonian Hermitian. But this
indefinite inner product does not define the physical Hilbert space
of the theory.

Clearly the basic ingredient in both the indefinite-metric and
pseudo-Hermitian QM is the pseudo-metric operator. In general the
spectrum of a pseudo-metric operator need not be discrete. However,
for simplicity, we first consider a pseudo-metric operator
$\eta\in\fM_{I}$ that has a discrete spectrum. We can express it
using its spectral representation as
    \be
    \eta=\sum_{n=1}^N e_n\:
    |\varepsilon_n\kt\br\varepsilon_n|,
    \label{spec-rep-eta}
    \ee
Because $\eta$ is a bounded invertible Hermitian operator, its
eigenvalues $e_n$ are nonzero and its eigenvectors
$|\varepsilon_n\kt$ form a complete orthonormal basis of the Hilbert
space ${\cal H}$. Therefore, we can define
    \be
    B:=\sum_{n=1}^N |e_n|^{-1/2}\:
    |\varepsilon_n\kt\br\varepsilon_n|=|\eta|^{-1/2},
    \label{B=}
    \ee
and use it to obtain a new pseudo-metric operator, namely
    \be
    \tilde\eta:=B^\dagger\eta B=
    \sum_{n=1}^N {\rm sgn}(e_n)\:
    |\varepsilon_n\kt\br\varepsilon_n|.
    \label{tilde-eta=}
    \ee
The presence of a continuous part of the spectrum of $\eta$ does not
lead to any difficulty as far as the above construction is
concerned. Because $\eta$ is Hermitian, one can always define
$|\eta|:=\sqrt{\eta^2}$ and set $B:=|\eta|^{-1/2}$. Both of these
operators are bounded, positive-definite, and invertible. Hence
$\tilde\eta:=B^\dagger\eta B$ is an element of $\fM_A$ whose
spectrum is a subset of $\{-1,1\}$.

If we perform the transformation $\eta\to\tilde\eta$ on a
(positive-definite) metric operator $\eta_+$, we find $\tilde\eta=I$
and $\br\cdot|\cdot\kt_{_{\tilde\eta}}=\br\cdot|\cdot\kt$. This
observation is used by Pauli to argue that we would only gain
``something essentially new if we take into consideration indefinite
bilinear forms $\cdots$,'' \cite{pauli-1943}.\footnote{Pauli uses
the ``term bilinear form'' for what we call a ``pseudo-inner
product''.} To provide a precise justification for this assertion,
let $\eta_+$ be a (positive-definite) metric operator and ${\cal
H}_{\eta_+}$ be the Hilbert space having the inner product
$\br\cdot|\cdot\kt_{_{\eta_+}}$. Then, for all
$\psi_1,\psi_2\in{\cal H}$,
    \be
    \br B\psi_1|B\psi_2\kt_{_{\eta_+}}=
    \br B\psi_1|\eta_+ B\psi_2\kt=\br \psi_1|B^\dagger
    \eta B\psi_2\kt=\br \psi_1|\tilde\eta_+\psi_2\kt=
    \br \psi_1|\psi_2\kt,
    \label{unitary-B}
    \ee
where we have used the identities $B:=|\etap|^{-1/2}=\etap^{-1/2}$
and $\tilde\eta_+:=B^\dagger\eta_+ B=I$. Eq.~(\ref{unitary-B}) shows
that $B$ is a unitary operator mapping ${\cal H}$ onto ${\cal
H}_{\eta_+}$. As a result, the quantum system $s_{_{\eta_+}}$ whose
state-vectors belong to ${\cal H}_{\eta_+}$ is unitary-equivalent to
the quantum system $s_{_I}$ whose Hilbert space is the reference
Hilbert space ${\cal H}$. They describe the same physical system.
This is the conclusion reached by Pauli in 1943, \cite{pauli-1943}.
There is a simple objection to this argument. It ignores the
dynamical aspects of the theory. As we will see below, the
description of the Hamiltonian and the time-evolution operator can
be very complicated in the ``\emph{Hermitian representation}'' of
the physical system. Therefore, although considering ${\cal
H}_{\eta_+}$ with a (positive-definite) metric operator ${\eta_+}$
generally yields an equivalent ``\emph{pseudo-Hermitian
representation}'' of the conventional QM, a clever choice of
$\eta_+$ may be of practical significance in deriving the physical
properties of the system under investigation. As we discuss in
Subsection~\ref{rqm-cq-qft}, it turns out that indeed these new
representations play a key role in the resolution of one the oldest
problems of modern physics, namely the problem of negative
probabilities in relativistic QM of Klein-Gordon fields
\cite{cqg,ap,ijmpa-2006,ap-2006a} and certain quantum field theories
\cite{bender-lee-model}.

We end this subsection with the following remarks.
\begin{itemize}
\item Strictly speaking Pauli's above-mentioned argument does not
hold, if one keeps $\etap$ to be positive-definite but does not
require it to be invertible or bounded \cite{kato,shubin}. For
example one might consider the case that $\etap^{-1}$ exists but is
unbounded. In this case, $\etap$ is not onto and $B$ fails to be a
unitary operator. This type of generalized metric operators and the
corresponding non-unitary transformations $B$ have found
applications in the description of resonances
\cite{lowdin,antoniou,reed-simon-4}. They also appear in the
application of pseudo-Hermitian quantum mechanics for typical
$\cP\cT$-symmetric and non-$\cP\cT$-symmetric models. For these
models the Hamiltonian operator is a second order differential
operator $H$ acting in an appropriate function space $\cF$ that
renders the eigenvalue problem for $H$ well-posed. As discussed in
great detail in \cite{cjp-2006}, if $H$ is to serve as the
Hamiltonian operator for a unitary quantum system, one must
construct an appropriate reference Hilbert space $\cH$ in which $H$
acts as a quasi-Hermitian operator. This in particular implies the
existence of an associated metric operator that satisfies the
boundedness and other defining conditions of the metric operators.

Suppose $H'$ is a differential operator acting in a function space
$\cF$ and having a discrete spectrum, i.e., there is a countable set
of linearly-independent eigenfunctions of $H'$ with isolated
non-degenerate or finitely degenerate eigenvalues. We can use $H'$
and $\cF$ to define a unitary quantum system as follows
\cite{jpa-2003,jpa-2004b}.

First, we introduce $\sF$ to be the subset of $\cF$ that contains
the eigenfunctions of $H'$ with real eigenvalues, and let $\cL$ be
the span of $\sF$, i.e., $\cL:=\left\{\:\sum_{m=1}^M
c_m\psi_m\:\big|\: M\in\Z^+, c_m\in\C,\psi_m\in\sF\:\right\}$. Next,
we endow $\cL$ with the inner product \cite{kretschmer-szymanowski}
    \be
    \br\: \sum_{j=1}^J c_j\psi_j\,\big|\,
    \sum_{k=1}^K d_k\psi_k\:\kt:=\sum_{m=1}^{{\rm min}(J,K)}
    c_m^*d_m,
    \label{KS-inn}
    \ee
and Cauchy-complete\footnote{Every separable inner product space $N$
can be extended to a separable Hilbert space $\cK$, called its
Cauchy completion, in such a way that $N$ is dense in $\cK$ and
there is no proper Hilbert subspace of $\cK$ with the same
properties, \cite{reed-simon}.} the resulting inner product space
into a Hilbert space $\cK$. We can then identify the restriction of
$H'$ onto $\cL$, that we denote by $H$, with the Hamiltonian
operator of a quantum system. It is a densely-defined operator
acting in $\cK$, because its domain has a subset $\cL$ that is dense
in $\cK$, \cite{reed-simon}. In fact, in view of (\ref{KS-inn}) and
the fact that $\cL$ is dense in $\cK$, the eigenvectors of $H$ form
an orthonormal basis of $\cK$. Moreover, $H:\cK\to\cK$ has, by
construction, a real spectrum. Therefore, it is a Hermitian
operator.

This construction is quite difficult to implement in practice.
Instead, one takes the reference Hilbert space $\cH$ to be an
$L^2$-space, ensures that the given differential operator that is
now denoted by $H$ has a real spectrum, and that the set of its
eigenfunctions $\sF$ is dense in $\cH$. Then, one constructs an
invertible positive operator $\eta_+$ satisfying the
pseudo-Hermiticity condition,
    \be
    H^\dagger=\eta_+H\,\eta_+^{-1},
    \label{ph-new5}
    \ee
and uses it to construct the physical Hilbert space and the
Hermitian representation of the system.

For most of the concrete models that have so far been studied the
obtained $\eta_+$ turns out not to be everywhere-defined or bounded.
But, these qualities are highly sensitive to the choice of the
reference Hilbert space that may also be considered as a degree of
freedom of the formalism. The mathematical data that have physical
content are the eigenfunctions of $H$ and their linear combinations,
i.e., elements of $\cL$. Therefore, the only physical restriction on
the reference Hilbert spaces ${\cal H}$ is that $\cL$ be a dense
subset of $\cH$. This means that the question of the existence of a
genuine metric operator associated with $H$ requires addressing the
problem of the existence of a (reference) Hilbert space $\cH$ such
that
    \begin{enumerate}
    \item $\cL$ is a dense subset of $\cH$, and
    \item there is metric operator
    $\eta_+:\cH\to\cH$ satisfying (\ref{ph-new5}).
    \end{enumerate}
It is also possible that given a Hilbert space $\cH$ fulfilling the
first of these conditions and an unbounded invertible positive
operator $\eta_+:\cH\to\cH$ satisfying (\ref{ph-new5}), one can
construct a genuine bounded metric operator fulfilling the latter
condition. These mathematical problems require a systematic study of
their own. Following physicists' tradition, we shall ignore
mathematical subtleties related to these problems when we deal with
specific models that allow for an explicit investigation.

\item Let $A$ be a densely-defined linear operator with a
non-empty $\fU^+_A$ and $\eta_+\in \fU^+_A$. Because
$B^{-1}:=\eta_+^{1/2}:{\cal H}_{_{\eta_+}}\to{\cal H}$ is a unitary
operator and $A:{\cal H}_{_{\eta_+}}\to{\cal H}_{_{\eta_+}}$ is
Hermitian, the operator {\large$a$}$:=B^{-1} A B$ is a Hermitian
operator acting in ${\cal H}$. This shows that $A:{\cal H}\to{\cal
H}$ is related to a Hermitian operator {\large$a$}$:{\cal H}\to{\cal
H}$ via a similarity transformation,
    \be
    A=B\,\mbox{\large$a$}\,B^{-1}.
    \label{A=rar}
    \ee
Such an operator is called {\em quasi-Hermitian},
\cite{geyer}.\footnote{A linear densely-defined operator $A:{\cal
H}\to{\cal H}$ acting in a Hilbert space ${\cal H}$ is said to be
\emph{quasi-Hermitian} if there exists an everywhere-defined,
bounded, invertible linear operator $B:{\cal H}\to{\cal H}$ and a
Hermitian operator {\large$a$}$:{\cal H}\to{\cal H}$ such that
$A=B\mbox{\large$a$}B^{-1}$. The above analysis shows that $A$ is
quasi-Hermitian if and only if it is pseudo-Hermitian and $\fU^+_A$
is nonempty. In mathematical literature, the term quasi-Hermitian is
used for bounded operators $A$ satisfying $A^\dagger T=TA$ for a
positive but possibly non-invertible linear operator $T$,
\cite{dieudonne}. These and their various generalizations and
special cases have been studied in the context of symmetrizable
operators \cite{reid-1951,lax-1954,silberstein,zannen,kharazov}. For
a more recent review see \cite{istratescu}.}

\item Let $B:=\eta_+^{-1/2}$, $B':{\cal H}\to{\cal H}_{_{\eta_+}}$
be an arbitrary unitary operator, and
$\mbox{\large$a$}':={B'}^{-1}AB'$. Then in view of (\ref{A=rar}),
$\mbox{\large$a$}'=U^{-1}\mbox{\large$a$}\,U$, where $U:{\cal
H}\to{\cal H}$ is defined by $U:=B^{-1}B'$. Because both $B$ and
$B'$ are unitary operators mapping ${\cal H}$ onto ${\cal
H}_{_{\eta_+}}$, $\mbox{\large$a$}'$ and $U$ are respectively
Hermitian and unitary operators acting in ${\cal H}$. Conversely,
for every unitary operator $U:{\cal H}\to{\cal H}$ the operator
$B':=BU$ is a unitary operator mapping ${\cal H}$ onto ${\cal
H}_{_{\eta_+}}$ and $\mbox{\large$a$}':={B'}^{-1}AB'$ is a Hermitian
operator acting in ${\cal H}$. These observations show that the most
general Hermitian operator $\mbox{\large$a$}':{\cal H}\to{\cal H}$
that is related to $A$ via a similarity transformation,
    \be
    A=B'\,\mbox{\large$a$}'\,{B'}^{-1},
    \label{A=rar-gen}
    \ee
has the form
    \be
    B'=BU=\eta_+^{-1/2}U,
    \label{B-prime=}
    \ee
where $U$ is an arbitrary unitary transformation acting in ${\cal
H}$, i.e., $U\in{\cal U}({\cal H})$. If we identify $A$ with the
Hamiltonian operator for a quantum system and employ the formalism
of pseudo-Hermitian QM, the metric operator $\eta_+$ defines the
physical Hilbert space as ${\cal H}_{\rm phys}:={\cal
H}_{_{\eta_+}}$. Being Hermitian operators acting in ${\cal H}_{\rm
phys}$, the observables $O$ can be constructed using the unitary
operator $B:{\cal H}\to{\cal H}_{_{\eta_+}}$ and Hermitian operators
$o:{\cal H}\to{\cal H}$ according to
    \be
    O=B\,o{B}^{-1}=\eta_+^{-1/2}o\,\eta_+^{1/2}.
    \label{observables-construct}
    \ee
One can use any other unitary operator $B':{\cal H}\to{\cal
H}_{_{\eta_+}}$ for this purpose. Different choices for $B'$
correspond to different one-to-one mappings of the observables $O$
to Hermitian operators $o$. According to (\ref{B-prime=}) if
$o=B^{-1}O\,B$, then $o':={B'}^{-1}O\,B'=U^{-1}o\,U$. Therefore
making different choices for $B'$ corresponds to performing quantum
canonical transformations in ${\cal H}$. This in turn means that,
without loss of generality, we can identify the physical observables
of the system in its pseudo-Hermitian representation using
(\ref{observables-construct}).

\end{itemize}

\subsection{Spectral Properties of Pseudo-Hermitian Operators}
\label{sec-spec-prop}

Consider a pseudo-Hermitian operator $A$ acting in an
$N$-dimensional separable Hilbert space ${\cal H}$, with
$N\leq\infty$, and let $\eta\in\fM_A$. The spectrum of $A$ is the
set $\sigma(A)$ of complex numbers $\lambda$ such that the operator
$A-\lambda I$ is not invertible. Let $\lambda\in \sigma(A)$, then
$A-\lambda I$ is not invertible and because $\eta$ is invertible,
$\eta(A-\lambda I)\eta^{-1}=A^\dagger-\lambda I$ must not be
invertible. This shows that $\lambda\in\sigma(A^\dagger)$. But the
spectrum of $A^\dagger$ is the complex-conjugate of the spectrum of
$A$, i.e., $\lambda\in\sigma(A^\dagger)$ if and only if
$\lambda^*\in\sigma(A)$, \cite{kato}. This argument shows that as a
subset of the complex plane $\C$, the spectrum of a pseudo-Hermitian
operator is symmetric under the reflection about the real axis,
\cite{azizov,bognar}. In particular, the eigenvalues $a_n$ of $A$
(for which $A-a_nI$ is not one-to-one) are either real or come in
complex-conjugate pairs, \cite{pauli-1943,p1}.

Let $A$ be a diagonalizable pseudo-Hermitian operator with a
discrete spectrum \cite{reed-simon}.\footnote{This in particular
implies that the eigenvalues of $A$ have finite multiplicities.}
Then, one can use a Riesz basis $\{\psi_n\}$ consisting of a set of
eigenvectors of $A$ and the associated biorthonomal basis
$\{\phi_n\}$ to yield the following spectral representation of $A$
and a pseudo-metric operator $\eta\in\fM_A$, \cite{p1,p4}.
    \bea
    A&=&\sum_{n_0=1}^{N_0} a_{n_0} |\psi_{n_0}\kt\br\phi_{n_0}|+
    \sum_{\nu=1}^{\cun} \left(\alpha_{\nu} |\psi_{\nu}\kt\br\phi_{\nu}|+
    \alpha_{\nu}^* |\psi_{-\nu}\kt\br\phi_{-\nu}|\right),
    \label{ph-sp-decom}\\
    \eta&:=&\sum_{n_0=1}^{N_0} \sigma_{n_0}\,|\phi_{n_0}\kt\br\phi_{n_0}|+
    \sum_{\nu=1}^{\cun} \left(  |\phi_{\nu}\kt\br\phi_{-\nu}|+
    |\phi_{-\nu}\kt\br\phi_{\nu}|\right),
    \label{eta-sp-decom}
    \eea
where $n_0$ labels the real eigenvalues $a_{n_0}$ (if any), $\nu$
labels the complex eigenvalues $\alpha_{\nu}$ with positive
imaginary part (if any), $-\nu$ labels the complex eigenvalues
$\alpha_{-\nu}=\alpha_{\nu}^*$ with negative imaginary part, the
eigenvalues with different spectral labels $n\in\{n_0,\nu,-\nu\}$
need not be distinct, $0\leq N_0,\cun\leq\infty$,
$\sigma_{n_0}\in\{-1,1\}$ are arbitrary, and we have
    \bea
    A|\psi_{n_0}\kt=a_{n_0}|\psi_{n_0}\kt,&&
    A|\psi_{\pm\nu}\kt=\alpha_{\pm\nu}|\psi_{\pm\nu}\kt,
    \label{eg-va-A}\\
    \br\phi_{m_0}|\psi_{n_0}\kt=\delta_{m_0,n_0}&&
    \mbox{for all $m_0,n_0\in\{1,2,3,\cdots, N_0\}$},
    \label{biortho-A-1}\\
    \br\phi_{\fg\mu}|\psi_{\fh\,\nu}\kt=
    \delta_{\fg,\fh}\, \delta_{\mu,\nu}&&
    \mbox{for all $\fg,\fh\in\{-,+\}$ and
    $\mu,\nu\in\{1,2,3,\cdots,\cun\}$}.
    \label{biortho-A-2}
    \eea

Consider the set ${\cal L}(\{\psi_n\})$ of finite linear
combinations of $\psi_n$'s, as defined by (\ref{finite-span}).
According to (\ref{ph-sp-decom}), elements of ${\cal L}(\{\psi_n\})$
belong to the domain of $A$, i.e., ${\cal L}(\{\psi_n\})\subseteq
{\cal D}(A)$. But because $\{\psi_n\}$ is a basis, ${\cal
L}(\{\psi_n\})$ is a dense subset of ${\cal H}$. This implies that
${\cal D}(A)$ is a dense subset of ${\cal H}$.

Next, we show that the operator $\eta$ defined by
(\ref{eta-sp-decom}) does actually define a pseudo-metric operator,
i.e., it is an everywhere-defined, bounded, invertible, Hermitian
operator.

Let $\{\chi_n\}$ be an orthonormal basis of ${\cal H}$ and $B:{\cal
H}\to{\cal H}$ be the everywhere-defined, bounded, invertible
operator that maps $\{\chi_n\}$ onto the Riesz basis $\{\psi_n\}$,
i.e., $\psi_n=B\chi_n$ for all $n$. It is not difficult to see that
the biorthonormal basis $\{\phi_n\}$ may be mapped onto $\{\chi_n\}$
by $B^\dagger$, $\chi_n=B^\dagger\phi_n$ for all $n$. We can use
this relation and (\ref{eta-sp-decom}) to compute
    \be
    \tilde\eta:=B^\dagger\eta B=
    \sum_{n_0=1}^{N_0} \sigma_{n_0}\,|\chi_{n_0}\kt\br\chi_{n_0}|+
    \sum_{\nu=1}^{\cun} \left(  |\chi_{\nu}\kt\br\chi_{-\nu}|+
    |\chi_{-\nu}\kt\br\chi_{\nu}|\right).
    \label{tilde-eta-sp-decom}
    \ee

It is not difficult to show that $\tilde\eta$ is a pseudo-metric
operator, $\tilde\eta\in\fM_I$. To see this, let $\psi\in{\cal H}$
be arbitrary. Then because $\{\chi_n\}$ is orthonormal,
    \be
    \psi=\sum_{n=1}^N\br\chi_n|\psi\kt\:\chi_n=
    \sum_{n_0=1}^{N_0}\br\chi_{n_0}|\psi\kt\:\chi_{n_0}+
    \sum_{\nu=1}^{\cun}\left(\br\chi_{\nu}|\psi\kt\:\chi_{\nu}+
    \br\chi_{-\nu}|\psi\kt\:\chi_{-\nu}\right).
    \label{psi-expand-1}
    \ee
In view of (\ref{tilde-eta-sp-decom}) and (\ref{psi-expand-1}), we
have
    \be
    \tilde\eta\,\psi=\sum_{n_0=1}^{N_0}
    \sigma_{n_0}\br\chi_{n_0}|\psi\kt\:\chi_{n_0}+
    \sum_{\nu=1}^{\cun}\left(\br\chi_{-\nu}|\psi\kt\:\chi_{\nu}+
    \br\chi_{\nu}|\psi\kt\:\chi_{-\nu}\right).
    \label{tilde-eta-psi}
    \ee
In particular,
    \be
    \parallel\tilde\eta\,\psi\parallel^2=
    \sum_{n_0=1}^{N_0} |\br\chi_{n_0}|\psi\kt|^2+
    \sum_{\nu=1}^{\cun}\left(|\br\chi_{-\nu}|\psi\kt|^2+
    |\br\chi_{\nu}|\psi\kt|^2\right) = \:\parallel\psi\parallel^2.
    \label{norm=norm}
    \ee
This shows that not only $\tilde\eta$ is everywhere-defined but it
is bounded. Indeed, we have $\parallel\tilde\eta\parallel=1$.
Because $\tilde\eta$ is a bounded everywhere-defined operator,
$\tilde\eta^\dagger$ is also everywhere-defined and as is obvious
from (\ref{tilde-eta-sp-decom}), it coincides with $\tilde\eta$,
i.e., $\tilde\eta$ is Hermitian. Finally, in view of
(\ref{tilde-eta-sp-decom}), we can easily show that
$\tilde\eta^2=I$. In particular, $\tilde\eta^{-1}=\tilde\eta$ is
bounded and $\tilde\eta$ is invertible. This completes the proof of
$\tilde\eta\in\fM_I$.

Next, we observe that $\tilde\eta\in\fM_I$ implies $\eta\in\fM_I$.
This is because according to (\ref{tilde-eta-sp-decom}),
$\eta={B^{\dagger}}^{-1}\tilde\eta B^{-1}$ and $B^{-1}$ and
${B^{\dagger}}^{-1}$ are bounded everywhere-defined invertible
operators. Therefore, $\eta$ as defined by (\ref{eta-sp-decom}) is a
pseudo-metric operator. Let us also note that the inverse of $\eta$
is given by
    \be
    \eta^{-1}=B\tilde\eta B^\dagger=\sum_{n_0=1}^{N_0} \sigma_{n_0}\,|\psi_{n_0}\kt\br\psi_{n_0}|+
    \sum_{\nu=1}^{\cun} \left(  |\psi_{\nu}\kt\br\psi_{-\nu}|+
    |\psi_{-\nu}\kt\br\psi_{\nu}|\right).
    \label{eta-inverse-sp-decom}
    \ee
We can easily show that $\eta$ belongs to $\fM_A$ by substituting
(\ref{ph-sp-decom}), (\ref{eta-sp-decom}), and
(\ref{eta-inverse-sp-decom}) in $\eta A\eta^{-1}$ and checking that
the result coincides with $A^\dagger$.

In Ref.~\cite{p4}, it is shown that any element $\eta\in\fM_A$ can
be expressed in the form (\ref{eta-sp-decom}) where $\{\psi_n\}$ is
the biorthonormal basis associated with some (Riesz) basis
$\{\psi_n\}$ consisting of the eigenvectors of $A$. As any two Riesz
bases are related by a bounded everywhere-defined invertible
operator $L:{\cal H}\to{\cal H}$ one may conclude that the elements
of $\fM_A$ have the following general form.
    \be
    \eta'=L^\dagger \eta\, L,
    \label{general-eta}
    \ee
where $\eta$ is the pseudo-metric operator (\ref{eta-sp-decom}) that
is defined in terms of a fixed (but arbitrary) (Riesz) basis
$\{\psi_n\}$ consisting of the eigenvectors of $A$ and
$\sigma_{n_0}$ are a set of arbitrary signs. The operator $L$
appearing in (\ref{general-eta}) maps eigenvectors $\psi_n$ of $A$
to eigenvectors $L\psi_n$ of $A$ in such a way that $\psi_n$ and
$L\psi_n$ have the same eigenvalue, \cite{p4}. This in particular
implies that $L$ commutes with $A$.

Suppose $A$ has a complex-conjugate pair of nonreal eigenvalues
$\alpha_{\pm\nu}$, let $\psi_{\pm\nu}$ be a corresponding pair of
eigenvectors, $\eta'\in\fM_A$ be an arbitrary pseudo-metric operator
associated with $A$, $L$ be an everywhere-defined, bounded,
invertible operator commuting with $A$ and satisfying
(\ref{general-eta}), and $\xi:=L^{-1}\psi_\nu$. Then $\br
\xi|\eta'\xi\kt=\br\psi_\nu|\eta\,\psi_\nu\kt=0$. Because $L$ is
invertible $\xi\neq 0$, this is an indication that $\eta'$ is not a
positive-definite operator. Similarly suppose that one of the signs
$\sigma_{n_0}$ appearing in (\ref{eta-sp-decom}) is negative and let
$\zeta:=L^{-1}\psi_{n_0}$. Then
$\br\zeta|\eta'\zeta\kt=\br\psi_{n_0}|\eta\,\psi_{n_0}\kt=
\sigma_{n_0}=-1$, and again $\eta'$ fails to be positive-definite.
These observations show that in order for $\eta'$ to be a
positive-definite operator the spectrum of $A$ must be real and all
the signs $\sigma_{n_0}$ appearing in (\ref{eta-sp-decom}) must be
positive. In this case, we have $\eta'=\eta'_+$, where
    \be
    \eta'_+:=L^\dagger\eta_+\,L,~~~~~~~~~~
    \eta_+=\sum_{n=1}^N |\phi_n\kt\br\phi_n|.
    \label{eta-old-positive}
    \ee

The choice of the signs $\sigma_{n_0}$ is not dictated by the
operator $A$ itself. Therefore it is the reality of the spectrum of
$A$ that ensures the existence of positive-definite elements of
$\fM_A$. By definition, such elements are metric operators belonging
to $\fM^+_A$. Their general form is given by
(\ref{eta-old-positive}).

\subsection{Symmetries of Pseudo-Hermitian Hamiltonians}
\label{sec-sym}

Consider a pseudo-Hermitian operator $A:\cH\to\cH$ and let $\eta_1$
and $\eta_2$ be a pair of associated pseudo-metric operators;
$\eta_1 A \eta_1^{-1}=A^\dagger=\eta_2 A \eta_2^{-1}$. Then it is a
trivial exercise to show that the invertible linear operator
$S:=\eta_2^{-1}\eta_1$ commutes with $A$, \cite{p1}. If we identify
$A$ with the Hamiltonian of a quantum system, which we shall do in
what follows, $S$ represents a linear symmetry of $A$.

Next, consider a diagonalizable pseudo-Hermitian operator $A$ with a
discrete spectrum, $\psi_n$ be eigenvectors of $A$, and
$\{(\psi_n,\phi_n)\}$ be the complete biorthonormal extension of
$\{\psi_n\}$ so that $A$ admits a spectral representation of the
form (\ref{ph-sp-decom}):
    \be
    A=\sum_{n_0=1}^{N_0} a_{n_0} |\psi_{n_0}\kt\br\phi_{n_0}|+
    \sum_{\nu=1}^{\cun} \left(\alpha_{\nu} |\psi_{\nu}\kt\br\phi_{\nu}|+
    \alpha_{\nu}^* |\psi_{-\nu}\kt\br\phi_{-\nu}|\right).
    \label{ph-sp-decom-2}
    \ee
Moreover, for every sequence $\sigma=(\sigma_{n_0})$ of signs
($\sigma_{n_0}\in\{-1,+1\}$), let
    \bea
    \eta_{ \sigma }&:=&\sum_{n_0=1}^{N_0} \sigma_{n_0}\,|\phi_{n_0}\kt\br\phi_{n_0}|+
    \sum_{\nu=1}^{\cun} \left(  |\phi_{\nu}\kt\br\phi_{-\nu}|+
    |\phi_{-\nu}\kt\br\phi_{\nu}|\right),
    \label{eta-sp-decom-2}\\
    \cC_{ \sigma }&:=&\sum_{n_0=1}^{N_0} \sigma_{n_0} |\psi_{n_0}\kt\br\phi_{n_0}|+
    \sum_{\nu=1}^{\cun} \left( |\psi_{\nu}\kt\br\phi_{\nu}|+
    |\psi_{-\nu}\kt\br\phi_{-\nu}|\right),
    \label{gen-C=}
    \eea
and $\eta_1,\fS:\cH\to\cH$ be defined by
    \bea
    \eta_1&:=&\sum_{n_0=1}^{N_0} |\phi_{n_0}\kt\br\phi_{n_0}|+
    \sum_{\nu=1}^{\cun} \left(  |\phi_{\nu}\kt\br\phi_{-\nu}|+
    |\phi_{-\nu}\kt\br\phi_{\nu}|\right),
    \label{eta-1=}\\
    \fS&:=&\sum_{n_0=1}^{N_0} |\psi_{n_0}\kt\star\br\phi_{n_0}|+
    \sum_{\nu=1}^{\cun} \left( |\psi_{\nu}\kt\star\br\phi_{\nu}|+
    |\psi_{-\nu}\kt\star\br\phi_{-\nu}|\right),
    \label{gen-S=}
    \eea
where for every $\psi,\phi\in\cH$ the symbol $|\psi\kt\star\br\phi|$
denotes the following antilinear operator acting in $\cH$.
    \be
    |\psi\kt\star\br\phi|\:\zeta:=\br\zeta|\phi\kt\:\psi=
    \br\phi|\zeta\kt^*\,\psi,~~~~~~~~
    \mbox{for all $\zeta\in\cH$}.
    \label{star=}
    \ee
As we discussed in Section~\ref{sec-spec-prop}, $\eta_\sigma$ and
$\eta_1$ are pseudo-metric operators associated with $A$. The
operators $\cC_\sigma$ and $\fS$ have the following remarkable
properties \cite{jmp-2003}.
    \begin{itemize}

\item In view of the fact that
$\eta_1^{-1}=\sum_{n_0=1}^{N_0} |\psi_{n_0}\kt\br\psi_{n_0}|+
\sum_{\nu=1}^{\cun} \left(  |\psi_{\nu}\kt\br\psi_{-\nu}|+
|\psi_{-\nu}\kt\br\psi_{\nu}|\right)$, we have
    \be
    \cC_\sigma=\eta_1^{-1}\eta_\sigma.
    \label{C-sigma}
    \ee
Therefore, $\cC_\sigma$ is a linear invertible operator that
generates a symmetry of $A$:
    \be
    [\cC_{\sigma},A]=0.
    \label{C-sigma-sym}
    \ee

\item Using (\ref{star=}) and the biorthonormality and
completeness properties of $\{(\psi_n,\phi_n)\}$, we can check that
$\fS$ is an invertible antilinear operator that also commutes with
$A$,
    \be
    [\fS,A]=0.
    \label{gen-pt-sym}
    \ee

\item $\cC_\sigma$ and $\fS$ are commuting involutions, i.e.,
    \be
    [\cC_\sigma,\fS]=0,~~~~~~~~~~\cC_\sigma^2=\fS^2=I.
    \label{involutions}
    \ee

\end{itemize}
In summary, we have constructed an involutive antilinear symmetry
generator $\fS$ and a class of involutive linear symmetry generators
$\cC_\sigma$ that commute with $\fS$.

It turns out that if a given diagonalizable operator $A$ with a
discrete spectrum commutes with an invertible antilinear operator,
then $A$ is necessarily pseudo-Hermitian. Therefore, for such
operators pseudo-Hermiticity and the presence of (involutive)
antilinear symmetries are equivalent conditions
\cite{p3,solombrino-2002}. Furthermore, each of these conditions is
also equivalent to the \emph{pseudo-reality} of the spectrum of $A$.
The latter means that the complex-conjugate of every eigenvalue of
$A$ is an eigenvalue with the same multiplicity
\cite{p3,solombrino-2002}. These observations are the key for
understanding the role of ${\cal PT}$ symmetry in the context of our
study. They admit extensions for a certain class of
non-diagonalizable operators with discrete spectrum
\cite{jmp-2002d,scolarici-solombrino-2003,sac-jpa-2006,cgs-jpa-2007}
and some operators with continuous spectrum
\cite{cqg,jmp-2005,jpa-2006b}.

The reality of the spectrum of $A$ is the necessary and sufficient
condition for the existence of an associated metric operator and the
corresponding positive-definite inner product that renders $A$
Hermitian \cite{p2}. For the case that the spectrum of $A$ is real,
the expressions for $\eta_{\sigma},\eta_1,\cC_\sigma,$ and $\fS$
simplify:
    \bea
    \eta_{ \sigma }&:=&\sum_{n=1}^{N} \sigma_{n }\,
    |\phi_{n }\kt\br\phi_{n}|,~~~~~~~~~
    \eta_1=\sum_{n=1 }^{N } |\phi_{n }\kt\br\phi_{n }|=\eta_+,
    \label{etas-real}\\
    \cC_{ \sigma }&=&\sum_{n=1 }^{N } \sigma_{n }
    |\psi_{n }\kt\br\phi_{n }|,~~~~~~~~~
    \fS=\sum_{n=1 }^{N } |\psi_{n }\kt\star\br\phi_{n }|,
    \label{gen-C-S-real}
    \eea
and we find
    \be
    \cC_\sigma\psi_n=\sigma_n\psi_n,~~~~~~~~\fS\psi_n=\psi_n.
    \label{sec-6-exact-sym}
    \ee
Hence, $\cC_\sigma$ and $\fS$ generate exact symmetries of $A$.

In order to make the meaning of $\cC_\sigma$ more transparent, we
use the basis expansion of an arbitrary $\psi\in\cH$, namely
$\psi=\sum_{n=1}^N c_n\psi_n$, to compute
    \be
    \cC_\sigma\psi=\sum_{n=1}^Nc_n~\cC\psi_n=
    \sum_{n=1}^N \sigma_n c_n\psi_n=\psi_+-\psi_-,
    \label{c-sigma-action}
    \ee
where $\psi_\pm:=\sum_{n\in\fN_\pm}c_n\psi_n$ and $
\fN_\pm:=\Big\{n\in\{1,2,3,\cdots,N\}~\big|~\sigma_n=\pm 1\Big\}$.
According to (\ref{c-sigma-action}), for every $\psi\in{\cal H}$
there are unique state-vectors $\psi_\pm$ belonging to the
eigenspaces ${\cal H}_\pm:=\{\psi\in{\cal
H}~|~\cC_\sigma\psi=\pm\psi~\}$ of $\cC_\sigma$ such that
$\psi=\psi_+-\psi_-$. This identifies $\cC_\sigma$ with a ($\Z_2$-)
grading operator for the Hilbert space.

If $\sigma_n=\pm 1$ for all $n$, we find $\cC_\sigma=\pm I$. In the
following we consider the nontrivial cases: $\cC\neq\pm I$. Then
$\cH_\pm$ are proper subspaces of $\cH$ satisfying
    \be
    \cH=\cH_+\oplus\cH_-,
    \label{oplus}
    \ee
$\cC_\sigma$ is a genuine grading operator, both $\eta_\sigma$ and
$-\eta_\sigma$ fail to be positive-definite, and
$\br\cdot|\cdot\kt_{\eta_\sigma}$ is indefinite. Furthermore, in
view of (\ref{c-sigma-action}), the operators
    $
    \Pi_\pm:=\frac{1}{2}~(I\pm\cC_\sigma),
    $
satisfy $\Pi_\pm\psi=\psi_\pm$, i.e., they are projection operators
onto $\cH_\pm$.

Next, consider computing $\br\psi_\fg|\phi_\fh\kt_{\eta_\sigma}$ for
arbitrary $\fg,\fh\in\{-,+\}$ and $\psi_\pm,\phi_\pm\in\cH_\pm$.
Using the basis expansion of $\psi_\pm$ and $\phi_\pm$, i.e.,
$\phi_\pm=\sum_{n\in\fN_\pm}d_n\psi_n$ and
$\psi_\pm=\sum_{n\in\fN_\pm}c_n\psi_n$, and (\ref{etas-real}), we
have $\eta_\sigma\phi_\pm=\pm\sum_{n\in\fN_\pm}d_n\phi_n$ and
    \be
    \br\psi_\fg|\phi_\fh\kt_{\eta_\sigma}=
    \br\psi_\fg|\eta_\sigma\phi_\fh\kt=
    \fg\:\delta_{\fg,\fh}\sum_{n\in\fN_\fg}c_n^{*}d_n.
    \label{eq-123z}
    \ee
Therefore, with respect to the indefinite inner product
$\br\cdot|\cdot\kt_{\eta_\sigma}$, the subspaces $\cH_+$ and $\cH_-$
are orthogonal, and (\ref{oplus}) is an orthogonal direct sum
decomposition.

Another straightforward implication of (\ref{eq-123z}) is that, for
all $\psi,\phi\in{\cal H}$,
    \be
    \br\psi|\phi\kt_{\eta_\sigma}=\br\psi_+|\phi_+\kt_{\eta_\sigma}+
    \br\psi_-|\phi_-\kt_{\eta_\sigma}=
    \sum_{n\in\fN_+}c_n^{*}d_n
    -\sum_{n\in\fN_-}c_n^{*}d_n=
    \br\psi_+|\phi_+\kt_{\eta_+}-\br\psi_-|\phi_-\kt_{\eta_+},
    \label{sec6-eq1}
    \ee
where $\psi_\pm:=\Pi_\pm\psi$, $\phi_\pm:=\Pi_\pm\phi$, $c_n:=
\br\phi_n|\psi\kt$ and $d_n:= \br\phi_n|\phi\kt$. Similarly, we have
    \be
    \br\psi|\cC_\sigma\phi\kt_{\eta_\sigma}=
    \br\psi_+|\phi_+\kt_{\eta_\sigma}-
    \br\psi_-|\phi_-\kt_{\eta_\sigma}=
    \sum_{n\in\fN_+}c_n^{*}d_n
    +\sum_{n\in\fN_-}c_n^{*}d_n=
    \sum_{n=1}^N c_n^{*}d_n=\br\psi|\phi\kt_{\eta_+}.
    \label{sec6-eq2}
    \ee
This calculation shows that the positive-definite inner product
$\br\cdot|\cdot\kt_{\eta_+}$ can be expressed in terms of the
indefinite inner product $\br\cdot|\cdot\kt_{\eta_\sigma}$ and the
grading operator $\cC_\sigma$ according to
    \be
    \br\cdot|\cdot\kt_{\eta_+}=\br\cdot|\cC_\sigma
    \cdot\kt_{\eta_\sigma}.
    \label{sec6-inn=inn}
    \ee
Conversely, one can use (\ref{sec6-inn=inn}) to define a
positive-definite inner product that makes $H$ Hermitian. The latter
scheme may be traced back to a similar construction developed in the
1950's in the context of indefinite-metric quantum theories
\cite{nevanlinna}. See \cite{nagy,nakanishi} for reviews. In the
context of ${\cal PT}$-symmetric quantum mechanics, it was proposed
(with a specific choice for the sequence $\sigma$) in \cite{bbj} and
coined the name ${\cal CPT}$-inner product.

To see the connection with the treatment of \cite{bbj}, consider the
case that $A$, which is now viewed as the Hamiltonian operator for a
quantum system, is symmetric, i.e.,
    $
    A^T:={\cal T}A^\dagger{\cal T}=A,
    $
where $\cT$ is the time-reversal operator.\footnote{Recall that
$\cT$ is an antilinear Hermitian (and unitary) involution
\cite{wigner-1960,messiah}.} Then given the spectral representation
of $A$ and $A^\dagger$, we can easily choose a biorthonormal system
$\{(\psi_n,\phi_n)\}$ such that
    \be
    \phi_n=\sigma_n\cT\psi_n.
    \label{sec6-e11}
    \ee
Using this relation together with the biorthonormality and
completeness properties of $\{(\psi_n,\phi_n)\}$, we can obtain the
following spectral representation of $\cT$.
    \be
    \cT=\sum_{n=1}^N\sigma_n|\phi_n\kt\star\br\phi_n|.
    \label{sec6-e12}
    \ee
In view Eqs.~(\ref{etas-real}), (\ref{gen-C-S-real}),
(\ref{sec6-e12}), and $\cT^2=I$, we have
    \bea
    &&\br\psi_m|\psi_n\kt=\sigma_m\sigma_n\br\phi_n|\phi_m\kt,
    \label{sec6-e13}\\
    &&\fS=\cT\eta_\sigma.
    \label{sec6-e14}
    \eea
Clearly, we could use (\ref{sec6-e12}) to define an invertible
antilinear operator satisfying (\ref{sec6-e11}) for an arbitrary
possibly non-symmetric $A$, namely
    \be
    \cT_\sigma:=\sum_{n=1}^N\sigma_n|\phi_n\kt\star\br\phi_n|.
    \label{sec6-e12n}
    \ee
But in this more general case, (\ref{sec6-e13}) may not hold and
$\cT_\sigma$ may not be an involution.

In fact, condition (\ref{sec6-e13}) is not only a necessary
condition for $\cT_\sigma^2=I$ but it is also sufficient
\cite{jmp-2003}. An analogous necessary and sufficient condition for
$\eta_\sigma^2=I$ is \cite{jmp-2003}
    \be
    \br\psi_m|\psi_n\kt=\sigma_m\sigma_n\br\phi_m|\phi_n\kt,
    \label{sec6-e15}
    \ee
If both (\ref{sec6-e13}) and (\ref{sec6-e15}) hold,
    \be
    \cT_\sigma^2=\eta_\sigma^2=I,
    \label{sec6-e16}
    \ee
and as a result
    \be
    \fS=\cT_\sigma\eta_\sigma.
    \label{S=TP}
    \ee
By virtue of this relation and $\fS^2=I$,
    \be
    [\eta_\sigma,\cT_\sigma]=0.
    \label{sec6-e17}
    \ee
Therefore,
    \be
    \fS=\eta_\sigma \cT_\sigma.
    \label{S=PT}
    \ee

For the symmetric and $\cP\cT$-symmetric Hamiltonians
(\ref{pt-sym-nu}) that are considered in \cite{bbj}, we can find a
biorthonormal system $\{(\psi_n,\phi_n)\}$ satisfying
(\ref{sec6-e13}) -- (\ref{sec6-e15}), \cite{jmp-2003}. Moreover,
setting $\sigma_n:=(-1)^{n+1}$ for all $n\in\Z^+$, we have
    \be
    \eta_\sigma=\cP.
    \label{sec6-e18}
    \ee
Therefore, in light of (\ref{sec6-e14}) and (\ref{sec6-e17}), the
antilinear symmetry generator $\fS$ coincides with $\cP\cT$,
    \be
    \fS=\cP\cT,
    \label{S=PTn}
    \ee
and the linear symmetry generator $\cC_\sigma$ is the
``charge-conjugation'' operator $\cC$ of \cite{bbj}. In view of
(\ref{C-sigma}), (\ref{gen-pt-sym}), (\ref{etas-real}),
(\ref{involutions}), and (\ref{sec6-e18}), it satisfies
    \be
    \cC^2=I,~~~~~~[\cC,A]=[\cC,{\cal PT}]=0,
    ~~~~~~\cC=\eta_+^{-1}\cP.
    \label{sec6-C-op}
    \ee
Furthermore, because in the position representation of the state
vectors $(\cT\psi)(x)=\psi(x)^*$, the positive-definite inner
product (\ref{sec6-inn=inn}) coincides with the $\cC\cP\cT$-inner
product.

This completes the demonstration that the $\cC\cP\cT$-inner product
is an example of the positive-definite inner products
$\br\cdot|\cdot\kt_\etap$ that we explored earlier.

We conclude this subsection by noting that although in general we
can introduce a pair \emph{generalized time-reversal and parity
operators}, $\eta_\sigma$ and $\cT_\sigma$, they may fail to be
involutions.

\subsection{A Two-Level Toy Model} \label{sec-two-level}

In this subsection we demonstrate the application of our general
results in the study of a simple two-level model which, as we will
see in Subsection~\ref{rqm-cq-qft}, admits physically important
infinite-dimensional generalizations~\cite{cqg,ijmpa-2006,ap}.

Let ${\cal H}$ be the (reference) Hilbert space obtained by endowing
$\C^2$ with the Euclidean inner product and $\{e_1,e_2\}$ be the
standard basis of $\C^2$, i.e., $e_1:=${\scriptsize$
\left(\begin{array}{c}1\\0\end{array}\right)$}, $e_2:=${\scriptsize$
\left(\begin{array}{c}0\\1\end{array}\right)$}. Then we can
represent every linear operator $K$ acting in ${\cal H}$ in the
basis $\{e_1,e_2\}$ by a $2\times 2$ matrix which we denote by
$\underline{K}$; the entries of $\underline{K}$ have the form
$\underline{K}_{ij}:=\br e_i|Ke_j\kt$ where $i,j\in\{1,2\}$.

Now, consider a linear operator $A:\C^2\to\C^2$ represented by
    \be
    \underbar A:=\frac{1}{2}\,\left(\begin{array}{cc}
    D+1 & D-1\\
    -D+1 & -D-1\end{array}\right),
    \label{matrix-A=}
    \ee
where $D$ is a real constant. $A:{\cal H}\to{\cal H}$ is a Hermitian
operator if and only if $D=1$. We can easily solve the eigenvalue
problem for $A$. Its eigenvalues $a_n$ and eigenvectors $\psi_n$
have the form
    \be
    a_1=-a_2= D^{1/2},~~~~~~~
    \psi_1=c_1\left(\begin{array}{c}
    1+D^{1/2}\\
    1-D^{1/2}\end{array}\right),~~~~~
    \psi_2=c_2\left(\begin{array}{c}
    1-D^{1/2}\\
    1+D^{1/2}\end{array}\right),
    \label{eg-va-A=}
    \ee
where $c_1,c_2$ are arbitrary nonzero complex numbers. Clearly, for
$D=0$, $a_1=a_2=0$, the eigenvectors become proportional, and $A$ is
not diagonalizable. For $D\neq 0$, $A$ has two distinct eigenvalues
and $\{\psi_1,\psi_2\}$ forms a basis of $\C^2$. This shows that
$D=0$ marks an exceptional spectral point \cite{kato,heiss}, for
$D>0$ the eigenvalues are real and for $D<0$ they are imaginary.

It is an easy exercise to show that $A$ is
$\Sigma_3$-pseudo-Hermitian where $\Sigma_3:\C^2\to\C^2$ is the
linear operator represented in the standard basis by the Pauli
matrix $\psigma_3:=${\scriptsize$
\left(\begin{array}{cc}1&0\\0&-1\end{array}\right)$}. Hence $\fM_A$
includes $\Sigma_3$, and $A$ is $\Sigma_3$-pseudo-Hermitian for all
$D\in\R$. This is consistent with the fact that $\Sigma_3$ is not a
positive-definite operator, because otherwise $A$ could not have
imaginary eigenvalues. According to our general results, for $D>0$,
$\fM_A$ must include positive-definite operators. To construct these
we first construct the biorthonormal basis $\{\phi_1,\phi_2\}$
associated with $\{\psi_1,\psi_2\}$. For $D>0$, the basis vectors
$\phi_n$ are given by
    \be
    \phi_1=(4c_1^*)^{-1}\left(\begin{array}{c}
    1+D^{-1/2}\\
    1-D^{-1/2}\end{array}\right),~~~~~
    \phi_2=(4c_2^*)^{-1}\left(\begin{array}{c}
    1-D^{-1/2}\\
    1+D^{-1/2}\end{array}\right).
    \label{phi-eg-va-A=}
    \ee
Inserting these relations in (\ref{eta-sp-decom}) we find the
following expression for the matrix representation of the most
general pseudo-metric operator $\eta\in\fM_A$.
    \be
    \mbox{\underbar{$\eta$}}=r_1\sigma_1
    \left(\begin{array}{cc}
    (1+D^{-1/2})^2 & 1-D^{-1}\\
    1-D^{-1} & (1-D^{-1/2})^2\end{array}\right)+
    r_2 \sigma_2
    \left(\begin{array}{cc}
    (1-D^{-1/2})^2 & 1-D^{-1}\\
    1-D^{-1} & (1+D^{-1/2})^2\end{array}\right),
    \label{gen-eta-example}
    \ee
where $r_1:=|4c_1|^{-2}$ and $r_2:=|4c_2|^{-2}$ are arbitrary
positive real numbers and $\sigma_1,\sigma_2$ are arbitrary signs.

The choice $\sigma_1=-\sigma_2=1$ and $r_1=r_2=D^{1/2}/4$ yields
$\eta=\Sigma_3$. The choice $\sigma_1=\sigma_2=1$ yields the form of
the most general positive-definite element of $\fM_A$. A
particularly simple example of the latter is obtained by taking
$c_1=c_2=D^{-1/4}/2$ which implies $r_1=r_2=D^{1/2}/4$. It has the
form
    \be
    \mbox{\underbar{$\eta$}}_+=\frac{1}{2}\,
    \left(\begin{array}{cc}
    D^{1/2}+D^{-1/2} & D^{1/2}-D^{-1/2}\\
    D^{1/2}-D^{-1/2} & D^{1/2}+D^{-1/2}\end{array}\right).
    \label{eta-plus-example}
    \ee
We can simplify this expression by introducing \footnote{This was
pointed out to me by Professor Haluk Beker.} $\theta:=\frac{1}{2}\ln
D$, and using the fact that the Pauli matrix
$\psigma_1:=${\scriptsize$
\left(\begin{array}{cc}0&1\\1&0\end{array}\right)$} squares to the
identity matrix $\underbar{\mbox{$I$}}$. This yields
    \be
    \mbox{\underbar{$\eta$}}_+=
    \left(\begin{array}{cc}
    \cosh\theta & \sinh\theta\\
    \sinh\theta &
    \cosh\theta\end{array}\right)=
    \cosh\theta\, \underbar{\mbox{$I$}}+\sinh\theta\,\psigma_1=
    e^{\theta\,\psigma_1}.
    \label{eta-plus-example-2}
    \ee

The metric operator represented by (\ref{eta-plus-example}) and
(\ref{eta-plus-example-2}) defines the following (positive-definite)
inner product on $\C^2$ with respect to which $A$ is a Hermitian
operator.
    \be
    \br\vec z|\vec w\kt_{_{\eta_+}}:=
    \br\vec z|\eta_+\vec w\kt=
    (z_1^*w_1+z_2^*w_2)\cosh\theta+(z_1^*w_2+z_2^*w_1)\sinh\theta,
    \label{inn-prod-example}
    \ee
where $\vec z=(z_1,z_2)^T,\vec w=(w_1,w_2)^T\in\C^2$ are arbitrary.
The inner product (\ref{inn-prod-example}) has a more complicated
form than both the reference (Euclidean) inner product, $\br\vec
z|\vec w\kt=z_1^*w_1+z_2^*w_2$, and the indefinite inner product
defined by $\Sigma_3$,
    \be
    \br\vec z|\vec w\kt_{_{\Sigma_3}}:=
    \br\vec z|\Sigma_3\vec w\kt=z_1^*w_1-z_2^*w_2.
    \label{sigam3-inn-prod}
    \ee
Furthermore, unlike $\br\cdot|\cdot\kt$ and
$\br\cdot|\cdot\kt_{_{\Sigma_3}}$, the inner product $\br\cdot
|\cdot\kt_{_{\eta_+}}$ depends on $\theta$ and consequently $D$. In
particular, as $D\to 0$ it degenerates. In fact, a quick inspection
of Eq.~(\ref{gen-eta-example}) shows that every $D$-independent
pseudo-metric operator is proportional to $\Sigma_3$ and hence
necessarily indefinite; $\pm\Sigma_3$ are the only $D$-independent
elements of $\fU_A$.

Now, suppose that $A$ is the Hamiltonian operator of a two-level
quantum system. If we employ the prescription provided by the
indefinite-metric quantum theories, we should endow $\C^2$ with an
indefinite inner-product from the outset. The simplest choice that
is historically adopted and viewed, according to the above-mentioned
argument due to Pauli, as being the unique choice is
$\eta=\Sigma_3$. In this case the system has a physical state
corresponding to the state-vector $e_1$ and a hypothetical state or
ghost corresponding to $e_2$. Unfortunately, the subspace of
physical state-vector, i.e., the span of $\{e_1\}$, is not invariant
under the action of $A$ unless $D=1$. Hence, for $D\neq 1$, such an
indefinite-metric quantum theory suffers from interpretational
problems and is inconsistent. In contrast, pseudo-Hermitian quantum
mechanics provides a consistent description of a unitary quantum
theory based on the Hamiltonian $A$. This is done by endowing $\C^2$
with the (positive-definite) inner product
$\br\cdot|\cdot\kt_{\eta_+'}$, where $\eta_+'$ is given by the
right-hand side of (\ref{gen-eta-example}) with
$\sigma_1=\sigma_2=1$. It involves the free parameters $r_1$ and
$r_2$ that can be fixed from the outset or left as degrees of
freedom of the formulation of the theory. We can represent this most
general metric operator by
    \be
    \mbox{\underbar{$\eta'$}}_+= r
    \left(\begin{array}{cc}
    \cosh\theta+s & \sinh\theta\\
    \sinh\theta & \cosh\theta-s
    \end{array}\right),
    \label{gen-eta-plus-example}
    \ee
where $r:=2(r_1+r_2)D^{-1/2}\in\R^+$ and
$s:=\frac{r_1-r_2}{r_1+r_2}\in (-1,1)$ are arbitrary. We can use
(\ref{gen-eta-plus-example}) to determine the most general inner
product on $\C^2$ that makes $A$ Hermitian. This is given by
    \be
    \br\vec z|\vec w\kt_{_{\eta'_+}}:=
    \br\vec z|\eta'_+\vec w\kt=r\left[\br\vec z|\vec w\kt_{_{\eta_+}}
    +s\, \br\vec z|\vec w\kt_{_{\Sigma_3}}\right],
    \label{inn-prod-example-general}
    \ee
where $\br\cdot|\cdot\kt_{_{\eta_+}}$ and
$\br\cdot|\cdot\kt_{_{\Sigma_3}}$ are respectively defined by
(\ref{inn-prod-example}) and (\ref{sigam3-inn-prod}).

We can relate $\eta_+'$ to $\eta_+$ using a linear operator
$L:\C^2\to\C^2$ commuting with $A$ via $\eta_+'=L^\dagger\eta_+L$.
This operator has the following general form $L=2
D^{-1/4}\left[\sqrt{r_1}e^{i\varphi_1}|\psi_1\kt\br\phi_1|+
    \sqrt{r_2}e^{i\varphi_2}|\psi_1\kt\br\phi_1|\right]$,
where $\varphi_1,\varphi_2\in[0,2\pi)$ are arbitrary. We can
represent it by
    \be
    \mbox{\underbar{$L$}}=
    \left(\begin{array}{cc}
    \lambda_-\cosh\theta+\lambda_+ & \lambda_-\sinh\theta\\
    -\lambda_-\sinh\theta & -\lambda_-\cosh\theta+\lambda_+
    \end{array}\right),
    \label{B-matrix-rep=}
    \ee
where $\lambda_\pm:=D^{-1/4}(\sqrt r_1 e^{i\varphi_1}\pm \sqrt r_2
e^{i\varphi_2})$. With the help of (\ref{eta-plus-example-2}),
(\ref{gen-eta-plus-example}), and (\ref{B-matrix-rep=}), we have
checked that indeed
$\mbox{\underbar{$\eta'$}$_+=$\underbar{$L$}$^\dagger$
\underbar{$\eta$}$_+$\underbar{$L$}}$.

Next, we wish to establish the quasi-Hermiticity of $A$ for $D>0$,
i.e., show that it can be expressed as $A=B\mbox{\large$a$}B^{-1}$
for an invertible operator $B:\C^2\to\C^2$ and a Hermitian operator
$\mbox{\large$a$}:{\cal H}\to{\cal H}$. As we explained in
Section~\ref{sec-ph-metric}, we can identify $B$ with the inverse of
the positive square root of a metric operator belonging to
$\fM_A^+$. A convenient choice is $B=\eta_+^{-1/2}$, for in light of
(\ref{eta-plus-example-2}) we have
    \[\mbox{\underbar{$B$}}^{\pm 1}=e^{\mp\frac{\theta}{2}\,
    \psigma_1}=
    \left(\begin{array}{cc}
    \cosh\frac{\theta}{2} & \mp\sinh\frac{\theta}{2}\\
    \mp\sinh\frac{\theta}{2} &
    \cosh\frac{\theta}{2}\end{array}\right).\]
This leads to the following remarkably simple expression for the
matrix representation of {\large$a$}.
    \be
    \mbox{\large\underbar{$a$}}=
    \mbox{\underbar{$B$}}^{-1}\,
    \mbox{\underbar{$A$}}\,
    \mbox{\underbar{$B$}}=
    \left(\begin{array}{cc}
    D^{1/2} & 0\\
    0 & -D^{1/2}\end{array}\right)= D^{1/2}\psigma_3.
    \label{Hermitian-A=}
    \ee
Hence, $\mbox{\large$a$}=D^{1/2}\Sigma_3$.

Because every Hermitian operator $o:{\cal H}\to{\cal H}$ is a linear
combination, with real coefficients, of the identity operator $I$
and the operators $\Sigma_1$, $\Sigma_2$, and $\Sigma_3$ that are
respectively represented by Pauli matrices $\psigma_1$, $\psigma_2$,
and $\psigma_3$, we can express every physical observable $O:{\cal
H}_{_{\eta_+}}\to{\cal H}_{_{\eta_+}}$ as
    $O=\fa_0 I+\sum_{j=1}^3 \fa_j\,S_j$,
where $\fa_0,\fa_1,\fa_2,\fa_3$ are some real numbers and
$S_j:=B\Sigma_jB^{-1}=\eta_+^{-1/2}\Sigma_j\,\eta_+^{1/2}$ for all
$j\in\{1,2,3\}$. The observables $S_j$ are represented by
    $\mbox{\underbar{$S_1$}}=\psigma_1$, $
    \mbox{\underbar{$S_2$}}=
    \cosh\theta\,\psigma_1-i\sinh\theta\,\psigma_3$, and
    $\mbox{\underbar{$S_3$}}=i\sinh\theta\,\psigma_2+
    \cosh\theta\,\psigma_3$.

Next, we repeat the calculation of the Hermitian Hamiltonian and the
physical observables for the case that we choose the general metric
operator $\eta_+'$ to construct the physical Hilbert space, i.e.,
set ${\cal H}_{\rm phys}={\cal H}_{_{\eta'_+}}$. The matrix
representation of the Hermitian Hamiltonian {\large$a'$}:${\cal
H}\to{\cal H}$ is then given by
    \be
    \mbox{\large\underbar{$a'$}}=
    \mbox{\underbar{$\eta_+'$}}^{1/2}\,
    \mbox{\underbar{$A$}}\,
    \mbox{\underbar{$\eta_+'$}}^{-1/2}=D^{1/2}
    \left(\begin{array}{cc}
    u(\theta,s) & v(\theta,s)\\
    v(\theta,s) & -u(\theta,s)\end{array}\right)=
    D^{1/2}[v(\theta,s)\,\psigma_1+u(\theta,s)\,\psigma_3],
    \label{Hermitian-A-prime=}
    \ee
where
    $u(\theta,s):=\frac{\sqrt{1-s^2}\,\sinh^2\theta+s^2\cosh\theta}{
    \sinh^2\theta+s^2}$ and $
    v(\theta,s):=\frac{s(\cosh\theta-\sqrt{1-s^2})\sinh\theta}{
    \sinh^2\theta+s^2}$.
We can similarly express the physical observables in the form
    \be
    O'=\fa_0 I+\sum_{j=1}^3 \fa_j\,S'_j,
    \label{gen-obs-prime=}
    \ee
where $S'_j:={\eta'_+}^{-1/2}\Sigma_j\,{\eta'_+}^{1/2}$.

As seen from (\ref{Hermitian-A-prime=}) the Hermitian Hamiltonian
{\large$a'$} describes the interaction of a spin $\frac{1}{2}$
particle with a magnetic field that is aligned along the unit vector
$(u,0,v)^T$ in $\R^3$. As one varies $s$ the magnetic field rotates
in the $x$-$z$ plane. It lies on the $z$-axis for $s=0$ which
corresponds to using $\eta_+$ to define the physical Hilbert space.
Clearly, there is no practical advantage of choosing $s\neq 0$.
Furthermore, for all $s\in(-1,1)$ and in particular for $s=0$, the
Hermitian representation of the physical system is actually less
complicated than its pseudo-Hermitian representations. This seems to
be a common feature of a large class of two-level systems
\cite{jpa-2003}.

Next, we compute the symmetry generator $\cC_\sigma$ for
$\sigma_1=-\sigma_2=1$. Denoting this operator by $\cC$ for
simplicity, realizaing that $
\cC=|\psi_1\kt\br\phi_1|-|\psi_2\kt\br\phi_2|$, and using
(\ref{eg-va-A=}), (\ref{phi-eg-va-A=}), and $D=e^{2\theta}$, we find
    \be
    \underline{\cC}=\left(\begin{array}{cc}
    \cosh\theta & \sinh\theta\\
    -\sinh\theta & -\cosh\theta\end{array}\right).
    \label{sec7-C=}
    \ee
Observing that in view of (\ref{matrix-A=}), $A^2=D I$, and making
use of this relation and (\ref{sec7-C=}) we are led to the curious
relation \cite{ijmpa-2006}
    \be
    \cC=\frac{A}{\sqrt{A^2}}.
    \label{sec7-C=AA}
    \ee

It is important to note that in performing the above calculation we
have not fixed the normalization constants $c_1$ and $c_2$ appearing
in (\ref{eg-va-A=}) and (\ref{phi-eg-va-A=}). Therefore, up to an
unimportant sign, $\cC$ is unique.

As we mentioned above setting $c_1=c_2=D^{-1/4}/2$ and
$\sigma_1=-\sigma_2=1$, we find the pseudo-metric operator
$\eta_\sigma=|\phi_1\kt\br\phi_1|-|\phi_2\kt\br\phi_2|=\Sigma_3$.
Because $\Sigma_3$ is a linear involution we can identify it with
$\cP$. It is a simple exercise to show that $\cP,\cC$ and $\eta_+$
actually satisfy
    \be
    \cC=\eta_+^{-1}\cP.
    \label{sec7-e31}
    \ee

We can similarly construct the antilinear symmetry generator $\fS$.
It turns out that unlike $\cC$, $\fS$ depends on (the phase of)
normalization constants $c_1$ and $c_2$ that appear in
(\ref{eg-va-A=}) and (\ref{phi-eg-va-A=}). Setting
$c_1=c_2=D^{-1/4}/2$, we find
    \be
    \fS=\cT,
    \label{sec7-e32}
    \ee
where $\cT$ denotes complex conjugation, $\cT\vec z=\vec z^*$. In
view of (\ref{sec7-e32}), the symmetry condition $[\fS,A]=0$
corresponds to the statement that $A$ is a real operator, i.e.,
$\underline{A}$ is a real matrix, which is a trivial observation.

Similarly, we can introduce an antilinear operator $\cT_\sigma$
according to (\ref{sec6-e12n}). This yields
$\cT_\sigma:=|\phi_1\kt\star\br\phi_1|-|\phi_2\kt\star\br\phi_2|$,
which in general depends on the choice of $c_1$ and $c_2$. For
$c_1=c_2=D^{-1/4}/2$, we have $\cT_\sigma=\cP\cT$. Combining this
relation with (\ref{sec7-e32}) yields $\fS=\cP\cT_\sigma$. It is not
difficult to see that indeed $\cT_\sigma$ is an involution. But it
differs from the usual time-reversal operator $\cT$. Let us also
point out that we could construct the pseudo-Hermitian quantum
system defined by $A$ without going through the computation of
$\cC$, $\cP$ and $\cT_\sigma$ operators. What is needed is a metric
operator that defines the inner product of the physical Hilbert
space.

The construction of the $\cC$ operator for two-level systems with a
symmetric Hamiltonian has been initially undertaken in
\cite{bender-jpa-2003}. The $\cC$ operator for general two-level
systems and its relation to the metric operators of pseudo-Hermitian
quantum mechanics are examined in \cite{jpa-2003}. See also
\cite{znojil-geyer-2006}. A comprehensive treatment of the most
general pseudo-Hermitian two-level system that avoids the use of the
$\cC$ operator is offered in \cite{tjp}.

\section{Calculation of Metric Operator}

A pseudo-Hermitian quantum system is defined by a (quasi-Hermitian)
Hamiltonian operator and an associated metric operator $\eta_+$.
This makes the construction of $\eta_+$ the central problem in
pseudo-Hermitian quantum mechanics. There are various methods of
calculating a metric operator. In this section, we examine some of
the more general and useful of these methods.

\subsection{Spectral Method}\label{sec-spectral}
The spectral method, which we employed in
Section~\ref{sec-two-level}, is based on the spectral representation
of the metric operator:
    \be
    \eta_+=\sum_1^N|\phi_n\kt\br\phi_n|.
    \label{sec8-q1}
    \ee
It involves the construction of a complete set of eigenvectors
$\phi_n$ of $A^\dagger$ and summing the series appearing in
(\ref{sec8-q1}) (or performing the integrals in case that the
spectrum is continuous).

\subsubsection{$\cP\cT$-symmetric infinite square well}
\label{sec-pt-well}

The first pseudo-Hermitian and $\cP\cT$-symmetric model with an
infinite-dimensional Hilbert space that has been treated within the
framework of pseudo-Hermitian quantum mechanics is the one
corresponding to the $\cP\cT$-symmetric square well potential
\cite{znoji-well,bmq}:
    \be
    v(x)=\left\{\begin{array}{ccc}
    -i\zeta ~{\rm sgn}(x)&{\rm for}&|x|<L/2,\\
    \infty&{\rm for}&|x|\geq L/2,\end{array}\right.
    \label{sec8-p2}
    \ee
where $\zeta$ and $L$ are real parameters, $L$ is positive, and $x$
takes real values. This was achieved in \cite{jpa-2004b} using the
spectral method combined with a certain approximation scheme that
allowed for a reliable approximate evaluation of a metric operator
as well as the corresponding equivalent Hermitian Hamiltonian and
pseudo-Hermitian position and momentum operators. A more recent
treatment of this model that makes use of the spectral method and
obtains a perturbative expansion for a $\cC$ operator and the
corresponding metric operator $\eta_+$ in powers of $\zeta$ is given
in \cite{bender-well}.

\subsubsection{$\cP\cT$-symmetric barrier}
\label{sec-pt-barrier}

In \cite{jmp-2005}, the spectral method has been used for treating a
pseudo-Hermitian quantum system defined by the scattering potential:
    \be
    v(x)=\left\{\begin{array}{ccc}
    -i\zeta ~{\rm sgn}(x)&{\rm for}&|x|<L/2,\\
    0&{\rm for}&|x|\geq L/2,\end{array}\right.
    \label{sec8-p3}
    \ee
where again $\zeta,L,x\in\R$ and $L$ is positive. This potential was
originally used in \cite{ruschhaupt} as a phenomenological tool for
describing the propagation of electromagnetic waves in certain
dielectric wave guides.\footnote{The use of complex potential in
constructing various phenomenological models and effective theories
has a long history. For a discussion that is relevant to complex
scattering potentials see the review article \cite{muga-2004} and
\cite{ahmed-pra-2001,dkr-1,cannata}.} It is the first example of a
$\cP\cT$-symmetric potential with a continuous spectrum that could
be studied thoroughly within the context of pseudo-Hermitian quantum
mechanics.

Application of the spectral method for this potential involves
replacing the sum in (\ref{sec8-q1}) with an integral over the
spectral parameter and taking into account the double degeneracy of
the energy levels. The extremely lengthy calculation of a metric
operator for this potential yields the following remarkably simple
expression \cite{jmp-2005}.
    \be
    \br x|\eta_{+}|y\kt=\delta(x-y)+
    {\mbox{$\frac{imL^2\zeta}{16\hbar^2}$}}\,
    (2L+2|x+y|-|x+y+L|-|x+y-L|)\,
    {\rm sgn}(x-y)+{\cal O}(\zeta^2),
    \label{sec8-eta=}
    \ee
where $x,y\in\R$, and ${\cal O}(\zeta^2)$ stands for terms of order
$\zeta^2$ and higher.

An unexpected feature of the scattering potential (\ref{sec8-p3}) is
that the corresponding equivalent Hermitian Hamiltonian has an
effective interaction region that is three times larger than that of
the potential (\ref{sec8-p3}). In other words, in the physical
space, which is represented by the spectrum of the pseudo-Hermitian
position operator, the interaction takes place in the interval
$[-\frac{3L}{2},\frac{3L}{2}]$ rather than
$[-\frac{L}{2},\frac{L}{2}]$.

\subsubsection{Delta-function potential with a complex coupling}

Another complex scattering potential for which the spectral method
could be successfully applied is the delta-function potential
\cite{jpa-2006b}:
    \be
    v(x)=\fz\:\delta(x),
    \label{delta-po}
    \ee
where $\fz$ is a complex coupling constant with a non-vanishing real
part. For this system it has been possible to compute a metric
operator and show that it is actually a bounded operator up to and
including third order terms in the imaginary part of $\fz$. It is
given by
    \be
    \br x|\eta_{+}|y\kt=\delta(x-y)+
    \frac{im\zeta}{2\hbar^2}\,\left[
    \theta( x y)\;e^{-\kappa| x- y|}+
    \theta(- x y)\;e^{-\kappa| x+ y|}\right]\,
    {\rm sgn}(y^2-x^2)+{\cal O}(\zeta^2),
    \label{sec8-delta-metric}
    \ee
where $\zeta:=\Im(\fz)$, $\kappa:=m\hbar^{-2}\Re(\fz)$, $\Re$ and
$\Im$ denote the real and imaginary parts of their arguments,
$\theta$ is the step function defined by $\theta(x):=[1+{\rm
sgn}(x)]/2$ for all $x\in\R$, and we have omitted the quadratic and
cubic terms for brevity.

In order to determine the physical meaning of the quantum system
defined by the potential (\ref{delta-po}) and the metric operator
(\ref{sec8-delta-metric}), we should examine the Hermitian
representation of the system. The equivalent Hermitian Hamiltonian
is given by \cite{jpa-2006b}
    \be
    h=\frac{p^2}{2m}+\Re(\fz)\,\delta(x)+\frac{m\zeta^2}{8\hbar^2}
    \;h_2+
    {\cal O}(\zeta^3),
    \label{sec8-h=phys}
    \ee
where
    \be
    (h_2\psi)(x):=\fa_\psi
    \,e^{-\kappa|x|}+ \fb_\psi\delta(x),
    \label{sec8-h2=phys}
    \ee
$\psi\in L^2(\R)$ and $x\in\R$ are arbitrary, $\fa_\psi:=\psi(0)$,
and $\fb_\psi:=\int_{-\infty}^\infty e^{-\kappa|y|}\psi(y)\,dy$. As
seen from (\ref{sec8-h=phys}) and (\ref{sec8-h2=phys}) the nonlocal
character of the Hermitian Hamiltonian $h$ is manifested in the
$\psi$-dependence of the coefficients $\fa_\psi$ and $\fb_\psi$.

A generalization of the delta-function potential (\ref{delta-po})
that allows for a similar analysis is the double-delta function
potential: $v(x)=\fz_-\delta(x+a)+\fz_+\delta(x-a)$, where $\fz_\pm$
and $a$ are complex and real parameters, respectively
\cite{jpa-2009,jpa-2010}. Depending on the values of the coupling
constants $\fz_\pm$, this potential may develop spectral
singularities \cite{naimark,ljance}. These are the points where the
eigenfunction expansion for the corresponding Hamiltonian breaks
down \cite{jpa-2009}. This is a well-known mathematical phenomenon
\cite{naimark,ljance} with an interesting and potentially useful
physical interpretation: A spectral singularity is a real energy
where both the reflection and transmission coefficients diverge.
Therefore it corresponds to a peculiar type of scattering states
that behave exactly like resonances: They are resonances with
zero-width \cite{prl-2009,pra2-2009}.\footnote{The single
delta-function potential (\ref{delta-po}) develops a spectral
singularity for imaginary values of $\fz$. For other examples of
complex potentials with a spectral singularity, see
\cite{samsonov,prl-2009,pra2-2009}.}

\subsubsection{Other Models}

The application of the spectral method for systems with an
infinite-dimensional Hilbert space is quite involved. If the system
has a discrete spectrum it requires summing complicated series, and
if the spectrum is continuous it involves evaluating difficult
integrals. This often makes the use of certain approximation scheme
necessary and leads to approximate expressions for the metric
operator. A counterexample to this general situation is the quantum
system describing a free particle confined within a closed interval
on the real line and subject to a set of $\cP\cT$-symmetric Robin
boundary conditions \cite{KBZ}. For this system the spectral method
may be employed to yield a closed formula for a metric operator.
Other systems for which the spectral method could be employed to
give an explicit and exact expression for the metric operator are
the infinite-dimensional extensions of the two-level system
considered in Section~\ref{sec-two-level} where $D$ is identified
with a positive-definite operator acting in an infinite-dimensional
Hilbert space \cite{cqg,ap}. These quantum systems appear in a
certain two-component representation of the Klein-Gordon
\cite{jpa-1998} and (minisuperspace) Wheeler-DeWitt fields
\cite{jmp-1998}.

\subsection{Perturbation Theory}
\label{sec-pert}

The standard perturbation theory has been employed in the
determination of the spectrum of various complex potentials since
long ago \cite{caliceti-1980}.\footnote{For more recent
developments, see
\cite{bender-dunne-1999,caliceti-2005,caliceti-2006} and references
therein.} In the present discussion we use the term ``perturbation
theory'' to mean a particular perturbative method of constructing a
metric operator for a given quasi-Hermitian Hamiltonian operator.
This method involves the following steps.
    \begin{enumerate}
    \item Decompose the Hamiltonian $H$ in the form
        \be
        H=H_0+\epsilon H_1,
        \label{sec8-q2}
        \ee
where $\epsilon$ is a real (perturbation) parameter, and $H_0$ and
$H_1$ are respectively Hermitian and anti-Hermitian
$\epsilon$-independent operators.
    \item Use the fact that $\eta_+$ (being a positive-definite
operator) has a unique Hermitian logarithm to introduce the
Hermitian operator $Q:=-\ln\eta_+$, so that
        \be
        \eta_+=e^{-Q},
        \label{sec8-q3}
        \ee
and express the pseudo-Hermiticity relation
$H^\dagger=\eta_+H\eta_+^{-1}$ in the form
        \be
        H^\dagger=e^{-Q}H\,e^{Q}.
        \label{sec8-q3-ph}
        \ee
In view of the Backer-Campbell-Hausdorff identity \cite{ryder},
        \be
        e^{-Q}H\,e^{Q}=H+\sum_{\ell=1}^\infty
        \frac{1}{\ell!}[H,Q]_\ell=
        H+[H,Q]+\frac{1}{2!}[[H,Q],Q]+\frac{1}{3!}[[[H,Q],Q],Q]+
        \cdots,
        \label{bch-identity}
        \ee
where $[H,Q]_\ell:=[[\cdots[[H,Q],Q],\cdots],Q]$ and $\ell$ is the
number of copies of $Q$ appearing on the right-hand side of this
relation, (\ref{sec8-q3-ph}) yields
        \be
        H^\dagger=H+\sum_{\ell=1}^\infty
        \frac{1}{\ell!}[H,Q]_\ell.
        \label{ph-bch-identity}
        \ee
    \item Expand $Q$ in a power series in $\epsilon$ of the form
        \be
        Q=\sum_{j=1}^\infty Q_j\:\epsilon^j,
        \label{sec8-q4}
        \ee
    where $Q_j$ are $\epsilon$-independent Hermitian operators.
    \item Insert (\ref{sec8-q2}) and (\ref{sec8-q4}) in (\ref{ph-bch-identity})
    and equate terms of the same order in powers
    of $\epsilon$ that appear on both sides of this equation. This
    leads to a set of operator equations for $Q_j$ which have the form
    \cite{jpa-2006a}
        \be
        [H_0,Q_j]=R_j.
        \label{sec8-q5}
        \ee
    Here $j\in\Z^+$ and $R_j$ is determined in terms of $H_1$ and
    $Q_k$ with $k<j$ according to
        \bea
        &&R_j:=\left\{\begin{array}{ccc}
        -2H_1&{\rm for}& j=1,\\
        \sum_{k=2}^j q_k Z_{kj}&{\rm for}& j\geq 2,\end{array}
        \right.~~~~~
        q_k:=\sum_{m=1}^k\sum_{n=1}^m
        \frac{(-1)^nn^k m!}{k!2^{m-1}n!(m-n)!}\,,~~
        \label{sec8-q6}\\
        &&Z_{kj}:=
        \sum_{\stackrel{s_1,\cdots,s_k\in\Z^+}{s_1+\cdots+s_k=j}}
        [[[\cdots[H_0,Q_{s1}],Q_{s_2}],\cdots,],Q_{s_k}].
        \label{sec8-q7}
        \eea
    More explicitly we have
    {\small
    \bea
    \left[H_0,Q_1\right]&=&-2H_1,
    \label{e1}\\
    \left[H_0,Q_2\right]&=& 0,
    \label{e2}\\
    \left[H_0,Q_3\right]&=&-\frac{1}{6}[H_1,Q_1]_{_2},
    \label{e3}\\
    \left[H_0,Q_4\right]&=&
    -\frac{1}{6}\left([[H_1,Q_1],Q_2]+[[H_1,Q_2],Q_1]\right),
    \label{e4}\\
    \left[H_0,Q_5\right]&=&
    \frac{1}{360}\,[H_1,Q_1]_{_4}-\frac{1}{6}\,
    \left([H_1,Q_2]_{_2}+[[H_1,Q_1],Q_3]+[[H_1,Q_3],Q_1]\right).
    \label{e5}
    \eea}%
    \item Solve the above equations for $Q_j$ iteratively by
    making an appropriate ansatz for their general form.

    \end{enumerate}

A variation of this method was originally developed in
\cite{bender-prd-2004} to compute the $\cC$ operator for the
following $\cP\cT$-symmetric Hamiltonians and some of their
multidimensional and field-theoretical generalizations.
    \bea
    H&=&\frac{1}{2m}\,p^2+\frac{1}{2}\,\mu^2 x^2+i\epsilon\, x^3,
    \label{sec8-pt-sym-3}\\
    H&=&\frac{1}{2m}\,p^2+i\epsilon\, x^3,
    \label{sec8-pt-cubic}
    \eea
where $\mu$ and $\epsilon$ are nonzero real coupling constants.

\subsubsection{$\cP\cT$-symmetric cubic anharmonic oscillator}
\label{sec-cubic-osc}

A perturbative calculation of a metric operator and the
corresponding equivalent Hermitian Hamiltonian and pseudo-Hermitian
position and momentum operators for the Hamiltonian
(\ref{sec8-pt-sym-3}) has been carried out in
\cite{jpa-2005b,jones-2005}.

Following \cite{bender-prd-2004} one can satisfy the operator
equations for $Q_j$ by taking $Q_{2i}=0$ for all $i\in\Z^+$ and
adopting the ansatz
    \be
    Q_{2i+1}=\sum_{j,k=0}^{i+1} c_{ijk}~\{x^{2j},p^{2k+1}\},
    \label{ansatz}
    \ee
where $\{\cdot,\cdot\}$ stands for the anticommutator and $c_{ijk}$
are real constants. Inserting (\ref{ansatz}) in (\ref{sec8-q5}), one
can determine $c_{ijk}$ for small values of $i$ \cite{jpa-2005b}.
See also \cite{bender-prd-2004,jones-2005}.

Again, to determine the physical content of the system defined by
the Hamiltonian (\ref{sec8-pt-sym-3}) and the  metric operator
$\eta_+=e^{-Q}$, we need to inspect the associated Hermitian
Hamiltonian operator \cite{jpa-2005b}:
    \bea
    h&=&\frac{p^2}{2m}+\frac{1}{2}\mu^2x^2+
    \frac{1}{m\mu^4}\left(\{x^2,p^2\}+p\,x^2p+
    \frac{3m\mu^2}{2}\,x^4\right)
    \epsilon^2+\frac{2}{\mu^{12}}
    \left(\frac{p^6}{m^3}-\frac{63\mu^2}{16m^2}\{x^2,p^4\}\right.
    \nn\\
    &&\left.
    -\frac{81\mu^2}{8m^2}\,p^2x^2p^2-\frac{33\mu^4}{16m}
    \{x^4,p^2\}-\frac{69\mu^4}{8m}\,x^2p^2x^2
    -\frac{7\mu^6}{4}\,x^6
    \right)\epsilon^4+
    {\cal O}(\epsilon^6),
    \label{h-pert=4b}
    \eea
and the underlying classical Hamiltonian (\ref{class-H-2}):
    \bea
    H_c&=&\frac{p_c^2}{2m}+\frac{1}{2}\mu^2x_c^2+
    \frac{3}{2\mu^4}\left(\frac{2}{m}\,x_c^2p_c^2+\mu^2x_c^4\right)
    \epsilon^2+\nn\\
    &&\frac{2}{\mu^{12}}
    \left(\frac{p_c^6}{m^3}-\frac{18\mu^2}{m^2}\,x_c^2p_c^4-
    \frac{51\mu^4}{4m}
    \,x_c^4p_c^2-\frac{7\mu^6}{4}\,x_c^6\right)\epsilon^4+
    {\cal O}(\epsilon^6).
    \label{H-class=}
    \eea
If we only consider the terms of order $\epsilon^2$ and lower, we
can express (\ref{H-class=}) in the form
    \be
    H_c=\frac{p_c^2}{2M(x_c)}+\frac{\mu^2}{2}\,x_c^2+
    \frac{3\epsilon^2}{2\mu^2}\,x_c^4+{\cal O}(\epsilon^4),
    \label{H-class=2}
    \ee
where $ M(x_c):=m(1+3\mu^{-4}\epsilon^2\,x_c^2)^{-1}=
    m(1-3\mu^{-4}\epsilon^2\,x_c^2)+{\cal O}(\epsilon^4)$.
This shows that for small values of $\epsilon$, the
$\cP\cT$-symmetric Hamiltonian (\ref{sec8-pt-sym-3}) describes a
position-dependent-mass quartic anharmonic oscillator
\cite{jpa-2005b}. This observation has motivated the use of
non-Hermitian constant-mass standard Hamiltonians,
$H=p^2/(2m)+v(x)$, in the perturbative description of a class of
position-dependent-mass standard Hamiltonians \cite{BQR}.

As seen from (\ref{h-pert=4b}), the 4-th order (in $\epsilon$)
contribution to the equivalent Hermitian Hamiltonian $h$ involves
$p^6$. It is not difficult to show that $h=\sum_{\ell=0}^\infty
h_\ell \:\epsilon^{2\ell}$, where $h_\ell$ is a polynomial in $p$
whose degree is an increasing function of $\ell$. Therefore, the
perturbative expansion of $h$ includes arbitrarily large powers of
$p$. This confirms the expectation that $h$ is a nonlocal operator.
The same holds for the pseudo-Hermitian position and momentum
operators \cite{jpa-2005b,jones-2005}.

\subsubsection{Imaginary cubic potential}
\label{sec-cubic-imag}

Ref.~\cite{jpa-2006a} gives a perturbative treatment of the
Hamiltonian
    \be
    H=\frac{p^2}{2m}+i\,\epsilon\, x^3,
    \label{sec8-cubic-pot}
    \ee
in which the operator equations (\ref{sec8-q5}) are turned into
certain differential equations and solved iteratively. This method
relies on the observation that for this Hamiltonian, $H_0=p^2/2m$.
Therefore,
    $\br x|[H_0,Q_j]|y\kt=$\linebreak$
    -\frac{\hbar^2}{2m}\left(\partial_x^2-\partial_y^2\right)\br
    x|Q_j|y\kt$.
In view of this identity and (\ref{sec8-q5}), we find
    \be
    (-\partial_x^2+\partial_y^2)\,\br x|Q_j|y\kt=\frac{2m}{\hbar^2}
    \:\br x|R_j|y\kt.
    \label{sec8-wave}
    \ee
Because $R_j$ is given in terms of $H_1$ and $Q_i$ with $i<j$, one
can solve (\ref{sec8-wave}) iteratively for $\br x|Q_j|y\kt$. Note
also that this equation is a non-homogeneous $(1+1)$-dimensional
wave equation which is exactly solvable.

This approach has two important advantages over the earlier
perturbative calculation of the metric operator for the imaginary
cubic potential \cite{bender-prd-2004}. Firstly, it involves solving
a well-known differential equation rather than dealing with
difficult operator equations. Secondly, it is not restricted by the
choice of an ansatz, i.e., it yields the most general expression for
the metric operator. In particular, it reveals large classes of
$\cC\cP\cT$ and non-$\cC\cP\cT$-inner products that were missed in
an earlier calculation given in \cite{bender-prd-2004}. Here we give
the form of the equivalent Hermitian Hamiltonian associated with the
most general admissible metric operator:
    \bea
    h&=&\frac{p^2}{2m}+\frac{3m}{16}
    \left(\{x^6,\frac{1}{p^2}\}
    +22\hbar^2\{x^4,\frac{1}{p^4}\}+\alpha_2\,\hbar^4
    \{x^2,\frac{1}{p^6}\}+\right.\nn\\
    &&\left.
    \hspace{.7cm}
    \frac{(14\alpha_2+1680)\hbar^6}{p^8}
    +\beta_2\,\hbar^3\{x^3,\frac{1}{p^5}\}\:{\cal P}\right)
    \epsilon^2+\nn\\
    &&\hspace{1.4cm}\hbar^6\left(
    \alpha_3 (\hbar\{x^2,\frac{1}{p^{11}}\}
    +\frac{44\hbar^3}{p^{13}} )+i\beta_3\,\{x^3,\frac{1}{p^{10}}\}\:
    {\cal P}\right)\epsilon^3+
    {\cal O}(\epsilon^4),
    \label{h=man}
    \eea
where $\alpha_2,\alpha_3,\beta_2,\beta_3$ are free real parameters
characterizing the nonuniqueness of the metric operator, and $\cP$
is the parity operator \cite{jpa-2006a}.

A remarkable feature of the Hamiltonian~(\ref{sec8-cubic-pot}) is
that the underlying classical Hamiltonian is independent of the
choice of the metric operator (to all orders of perturbation). Up to
terms of order $\epsilon^3$ it is given by the following simple
expression.
    \be
    H_c=\frac{p^2_c}{2m}+\frac{3}{8}\,m\epsilon^2\,
    \frac{x_c^6}{p_c^2}
    +{\cal O}(\epsilon^4).
    \label{cubic-H-classical}
    \ee
The presence of $p_c^2$ in the denominator of the second term is a
clear indication that the equivalent Hermitian Hamiltonian is a
nonlocal operator (and that this is the case regardless of the
choice of the metric operator.) Again the classical Hamiltonian
(\ref{cubic-H-classical}) clarifies the meaning of the imaginary
cubic potential $i\,\epsilon\,x^3$.

\subsubsection{Other Models}

The perturbation theory usually leads to an infinite series
expansion for the metric operator whose convergence behavior is
difficult to examine. There are however very special models for
which this method gives exact expressions for $Q$ and consequently
the metric operator $\eta_+=e^{-Q}$. Examples of such models are
given in \cite{bender-lee-model,BJR,jm,bender-mannheim}. The
simplest example is the free particle Hamiltonian studied in
\cite{jpa-2006a}.

For other examples of the perturbative calculation of a metric
operator and the corresponding equivalent Hermitian Hamiltonian, see
\cite{banerjee} and particularly \cite{FF}.

\subsection{Differential Representations of Pseudo-Hermiticity}

In the preceding subsection we show how one can turn the operator
equations appearing in the perturbative calculation of the metric
operator into certain differential equations. In this subsection we
outline a direct application of differential equations in the
computation of pseudo-metric operators for a large class of
pseudo-Hermitian Hamiltonian operators $H$ acting in the reference
Hilbert space $L^2(\R)$.

In the following we outline two different methods of identifying a
differential representation of the pseudo-Hermiticity condition,
    \be
    H^\dagger=\eta H\eta^{-1}.
    \label{sec8-ph-pre}
    \ee

\subsubsection{Field equation for the metric from
Moyal product}

Consider expressing (\ref{sec8-ph-pre}) in the form
    \be
    \eta H=H^\dagger\eta,
    \label{sec8-ph}
    \ee
and viewing $\eta$ and $H$ as complex-valued functions of $x$ and
$p$ that are composed by the Moyal $\mbox{\large$*$}$-product:
    \be
    f(x,p)\:\mbox{\large$*$}\:g(x,p):=f(x,p)\:
    e^{i\hbar\stackrel{\leftarrow}{\partial_x}
    \stackrel{\rightarrow}{\partial_p}}\:g(x,p)=
    \sum_{k=0}^\infty \frac{(i\hbar)^k}{k!}\:[\partial^k_x f(x,p)]
    \,\partial^k_p g(x,p).
    \label{moyal}
    \ee
This yields: $\eta(x,p)\:\mbox{\large$*$}\:H(x,p)=
H(x,p)^*\:\mbox{\large$*$}\:\eta(x,p)$, \cite{scholtz-geyer-2006}.
With the help of (\ref{moyal}) we can express this equation more
explicitly as
    \be
    \sum_{k=0}^\infty \frac{(i\hbar)^k}{k!}\:\Big\{
    [\partial_p^k H(x,p)]\,\partial_x^k-
    [\partial_x^k H(x,p)^*]\,\partial_p^k\Big\}\,\eta(x,p)=0.
    \label{ph-moyal-eq}
    \ee
This is a linear homogeneous partial differential equation of finite
order only if $H(x,p)$ is a polynomial in $x$ and $p$. For example,
for the imaginary cubic potential, i.e., the Hamiltonian
$H=\frac{p^2}{2m}+i\,\epsilon\, x^3$, it reads
    \be
    \left[\epsilon\,\hbar^3\partial_p^3
    -3\,i\,\epsilon\,\hbar^2 x\,\partial_p^2-(2m)^{-1}\hbar^2\partial_x^2
    -3\,\epsilon\,\hbar x^2\partial_p+im^{-1}\hbar\, p\,\partial_x+
    2\,i\,\epsilon\, x^3\right]\eta(x,p)=0.
    \label{sec8-moy-e1}
    \ee
The presence of variable coefficients in this equation is an
indication that it is not exactly solvable. Particular perturbative
solutions can however be constructed. This applies more generally
for other polynomial Hamiltonians. Explicit examples are given in
\cite{scholtz-geyer-2006,FF-moyal,assis-fring}.

We should like to note however that not every solution of
(\ref{ph-moyal-eq}) defines a pseudo-metric (respectively metric)
operator. We need to find solutions that correspond to Hermitian
(respectively positive-definite) and invertible operators $\eta$.
Ref.~\cite{scholtz-geyer-2006} suggests ways to address this
problem.

The above-described method that is based on the Moyal product has
two important shortcomings.
    \begin{enumerate}
    \item If the Hamiltonian is not a polynomial of $x$ and $p$, then the
resulting equation (\ref{ph-moyal-eq}) is not a differential
equation with a finite order. This makes its solution extremely
difficult. This is true unless $H(x,p)$ has a particularly simple
form. A typical example is the exponential potential $e^{ix}$
treated in Ref.~\cite{curtright-jmp-2007}. This potential is
actually one of the oldest $\cP\cT$-symmetric potentials whose
spectral problem has been examined thoroughly \cite{gasymov}. For
$x\in\R$, its spectrum includes an infinity of spectral
singularities that prevent this potential from defining a genuine
unitary evolution.\footnote{For a discussion of biorthonormal
systems for this potential with $x$ taking values on a circle (a
closed interval with periodic boundary condition on the
eigenfunctions), see \cite{CM-jmp-2007}.}

    \item For the polynomial Hamiltonians, for which (\ref{ph-moyal-eq})
is a differential equation, the general form and even the order of
this equation depends on the Hamiltonian. In particular, for the
standard Hamiltonians of the form
    \be
    H=\frac{p^2}{2m}+v(x),
    \label{sec8-st-H}
    \ee
they depend on the choice of the potential $v(x)$.

\end{enumerate}

We shall next discuss a differential representation of the
pseudo-Hermiticity that does not suffer from any of these
shortcomings.

\subsubsection{Universal field equation for the metric}

Consider a pseudo-Hermitian Hamiltonian of standard form
(\ref{sec8-st-H}). Substituting (\ref{sec8-st-H}) in the
pseudo-Hermiticity relation $\eta H=H^\dagger\eta$ and evaluating
the matrix elements of both sides of the resulting equation in the
coordinate basis $\{|x\kt\}$, we find \cite{jmp-2006a}
    \be
    \left[-\partial_x^2+\partial_y^2+\mu^2(x,y)\right]
    \eta(x,y)=0,
    \label{sec-8-diff-1}
    \ee
where $\mu^2(x,y):=\frac{2m}{h^2}\,[v(x)^*-v(y)]$,
and $\eta(x,y):=\br x|\eta|y\kt$.
Eq.~(\ref{sec-8-diff-1}) is actually a Klein-Gordon equation for
$\br x|\eta|y\kt$ (with a variable mass term.) As such, it is much
easier to handle than the equation obtained for the pseudo-metric in
the preceding subsection, i.e., (\ref{ph-moyal-eq}). Moreover, it
applies to arbitrary polynomial and non-polynomial potentials.

It turns out that one can actually obtain a formal series expansion
for the most general solution of (\ref{sec-8-diff-1}) that satisfies
the Hermiticity condition: $\eta(x,y)=\eta(y,x)^*$.
This solution has the form \cite{jmp-2006a}
    \be
    \eta(x,y)=\sum_{\ell=0}^\infty {\cal K}^\ell u(x,y),
    \label{sec-8-per}
    \ee
where ${\cal K}$ is the integral operator defined by
    \be
    {\cal K}\,f(x,y):=\frac{m}{\hbar^2}\,\left[
    \int^ydr \int_{x-y+r}^{x+y-r}
    ds~v(r)\,f(s,r)+
    \int^xds \int_{-x+y+s}^{x+y-s}
    dr~v(s)^*\,f(s,r)\right],
    \label{sec8-K=}
    \ee
$f:\R^2\to\C$ is an arbitrary test function, $u:\R^2\to\C$ is
defined by $u(x,y):=u_+(x-y)+u_-(x+y)$, and $u_\pm$ are arbitrary
complex-valued (piecewise) smooth (generalized) functions satisfying
$u_\pm(x)^*=u_\pm(\mp x)$.

For imaginary potentials, the series solution (\ref{sec-8-per})
provides an extremely effective perturbative method for the
construction of the most general metric operator. For example, the
application of this method for the $\cP\cT$-symmetric square well
potential that we discussed in Subsection~\ref{sec-pt-well} yields,
after a page-long straightforward calculation, \cite{jmp-2006a}
    \be
    \eta(x,y)=\delta(x-y)+\zeta\left[w_+(x-y)+w_-(x+y)+
    \frac{im}{2\hbar^2}\,|x+y|\,{\rm sgn}(x-y)\right]+{\cal
    O}(\zeta^2),
    \label{gen-eta-sw}
    \ee
where $w_\pm:[-\frac{L}{2},\frac{L}{2}]\to\C$ are arbitrary
functions satisfying $w_\pm(x)^*=w_\pm(\mp x)$ and $w_\pm(\pm L)=0$.
The metric operator associated with the $\cC\cP\cT$-inner product
that is computed using the spectral method in \cite{bender-well}
turns out to correspond to a particular choice for $w_\pm$ in
(\ref{gen-eta-sw}).

A probably better evidence of the effectiveness of this method is
its application in the construction of a metric operator for the
$\cP\cT$-symmetric barrier potential that we examined in
Subsection~\ref{sec-pt-barrier}. Again, a two-pages-long calculation
yields \cite{jmp-2006a}
    \bea
    \eta(x,y)&=&\delta(x-y)+\zeta\big[~w_+(x-y)+w_-(x+y)+\nn\\
    &&\hspace{.2cm}
    \frac{im}{4\hbar^2}
    \left(2|x+y|-|x+y+L|-|x+y-L|\right){\rm sgn}(x-y)\big]+
    {\cal O}(\zeta^2).
    \label{eta-pde-gen}
    \eea
This expression reproduces the result obtained in \cite{jmp-2005}
using the spectral method (after over a hundred pages of
calculations), namely (\ref{sec8-eta=}), as a special case.

In Ref.~\cite{jmp-2006a} this differential representation of
pseudo-Hermiticity has been used to obtain a perturbative expression
for the metric operator associated with the imaginary delta-function
potentials of the form\footnote{The spectral properties of
$\cP\cT$-symmetric potentials of this form and their consequences
have been studied in
\cite{jones-1,ahmed-1,albaverio,dkr-1,demiralp2}. In particular, see
\cite{jpa-2009}.}
    \be
    v(x)=i\sum_{n=1}^N \zeta_n\,\delta(x-a_n),
    \label{sec8-delta}
    \ee
where $\zeta_n,a_n\in\R$. The result is
    \be
    \eta(x,y)=\delta(x-y)+\sum_{n=1}^N \frac{2m\zeta_n}{\hbar^2}\!\!
    \left[w_{n+}(x-y)+w_{n-}(x+y)+\frac{i}{2}\theta(x+y-2a_n)\,
    {\rm sgn}(y-x)\right]
    +{\cal O}(\zeta_n^2).
    \label{sec8-eta-delta-gen}
    \ee
For the special case: $N=2$, $a_1=-a_2>0$ and $\zeta_1=-\zeta_2>0$,
where (\ref{sec8-delta}) is a $\cP\cT$-symmetric potential, a
careful application of the spectral method yields a
positive-definite perturbatively bounded metric operator
\cite{batal} that turns out to be a special case of
(\ref{sec8-eta-delta-gen}). The general $N=2$ case, that depending
on the choice of $\zeta_k$ may or may not posses $\cP\cT$-symmetry,
has been examined in \cite{jpa-2009,jpa-2010}.

The main difficulty with the approaches presented in this section
(and its subsections) is that they may lead to a ``metric'' operator
that is unbounded or non-invertible.\footnote{One must also restrict
the free functions appearing in the formula for $\eta(x,y)$ so that
the operator $\eta$ they define is at least densely-defined.} For
example setting $N=1$ in (\ref{sec8-delta}), one finds a delta
function potential with an imaginary coupling that gives rise to a
spectral singularity \cite{jpa-2006b}. Therefore, the corresponding
Hamiltonian is not quasi-Hermitian, and there is actually no genuine
(bounded, invertible, positive-definite) metric operator for this
potential. Yet, one can use (\ref{sec8-eta-delta-gen}) to obtain a
formula for a ``metric operator''! This observation suggests that
one must employ this method with extra care.

\subsection{Lie Algebraic Method}

In Subsection~\ref{sec-pert}, we described a perturbative scheme for
solving the pseudo-Hermiticity relation,
    \be
    H^\dagger=e^{-Q}H\, e^{Q},
    \label{ph-Lie}
    \ee
for the operator $Q$ that yields a metric operator upon
exponentiation, $\etap=e^{-Q}$. In this section we explore a class
of quasi-Hermitian Hamiltonians and corresponding metric operators
for which (\ref{ph-Lie}) reduces to a finite system of numerical
equations, although the Hilbert space is infinite-dimensional. The
key idea is the use of an underlying Lie algebra. In order to
describe this method we first recall some basic facts about Lie
algebras and their representations.

\subsubsection{Lie algebras and their representations}

Consider a matrix Lie group $G$, i.e., a subgroup of the general
linear group $GL(N,\C)$ for some $N\in\Z^+$, and let $\cG$ denote
its Lie algebra, \cite{elliott-dawber,isham}. A unitary
representation of $G$ is a mapping $\cU$ of $G$ into the group of
all unitary operators acting in a separable Hilbert space $\cH$ such
that the identity element of $G$ is mapped to the identity operator
acting in $\cH$ and for all $g_1,g_2\in G$,
$\cU(g_1g_2)=\cU(g_1)\cU(g_2)$. Such a unitary representation
induces a unitary representation for $\cG$, i.e., a linear mapping
$\fU$ of $\cG$ into the set of anti-Hermitian linear operators
acting in $\cH$ such that for all $X_1,X_2\in\cG$, $\fU([X_1,X_2])=
[\fU(X_1),\fU(X_2)]$, \cite{elliott-dawber,fell-doran}. The mappings
$\cU$ and $\fU$ are related according to: $\cU(e^X)=e^{\fU(X)}$, for
all $X\in\cG$. Furthermore, because for all $X\in\cG$, $\fU(X)$ is
an anti-Hermitian operator acting in $\cH$, there is a Hermitian
operator $K:\cH\to\cH$ such that $\fU(X)=iK$.

Let $\{K_1,K_2,\cdots, K_d\}$ be a set of Hermitian operators acting
in $\cH$ such that $\{iK_1,iK_2,\cdots,iK_d\}$ is a basis of
$\fU(\cG)$. Then $K_a$ with $a\in\{1,2,\cdots,d\}$ are called
\emph{generators} of $G$ in the representation $\cU$. If $\cU$ is a
faithful representation, i.e., it is a one-to-one mapping, the same
holds for $\fU$ and $d$ coincides with the dimension of $G$. In this
case, we refer to the matrices
    \be
    \underline{K_a}:=\fU^{-1}(K_a)
    \label{sec4-generators=}
    \ee
as generators of $G$ in its standard representation.

Next, consider the set of complex linear combinations of
$\underline{K_a}$, i.e., the \emph{complexification}  of $\cG$:
$\cG_{_\C}:=\left\{\sum_{a=1}^d
\fc_a\,\underline{K_a}~|~\fc_a\in\C~\right\}$. We can extend the
domain of definition of $\fU$ to $\cG_{_\C}$ by linearity: For all
$\fc_a\in\C$, $\fU\left(\sum_{a=1}^d \fc_a\,\underline{K_a}\right):=
    \sum_{a=1}^d \fc_a\:\fU(\underline{K_a})=
    \sum_{a=1}^d \fc_a\, K_a$.
Similarly, we extend the definition of $\cU$ to the set of elements
of $GL(N,\C)$ that are obtained by exponentiation of those of
$\cG_{_\C}$. This is done according to
    \be
    \cU(e^X)=e^{\fU(X)},~~~~~~~\mbox{for all $X\in\cG_{_\C}$}.
    \label{sec4-pre-cbh-id}
    \ee
Next, we recall that according to Backer-Campbell-Hausdorff identity
(\ref{bch-identity}), for all $X,Y\in\cG_\C$, $e^{-X}Y
e^{X}\in\cG_{_\C}$. Furthermore, (\ref{bch-identity}) and
(\ref{sec4-pre-cbh-id}) imply
    \be
    \cU(e^{-X}Y\,e^{X})=e^{-\fU(X)}\fU(Y)\,e^{\fU(X)}=\cU(e^{-X})\,
    \fU(Y)\,\cU(e^{X})~~~~~\mbox{for all
    $X,Y\in\cG_\C$.}
    \label{sec4-main-id}
    \ee
This completes our mathematical digression.

\subsubsection{General outline of the method}

Suppose that $H:\cH\to\cH$ can be expressed as a polynomial in the
Hermitian generates $K_a$ of $G$ in a faithful unitary
representation $\cU$,\footnote{In mathematical terms, one says that
$H$ is an element of the enveloping algebra of $\cG$ in the
representation $\cU$.} i.e.,
    \be
    H=\sum_{k=1}^n\sum_{a_1,a_2,\cdots,a_k=1}^d
    \lambda_{a_1,a_2,\cdots a_k}~K_{a_1}\, K_{a_2}\cdots K_{a_k},
    \label{sec4-Lie1}
    \ee
where $n\in\Z^+$, $d$ is the dimension of $G$, and
$\lambda_{a_1,a_2,\cdots a_k}\in\C$. Demand that $H$ admits a metric
operator of the form $\etap=e^{-Q}$ with $Q$ given by
    \be
    Q=\sum_{a=1}^d r_a~ K_a,~~~~~\mbox{for some $r_a\in\R$}.
    \label{sec4-Lie4-Q}
    \ee
Then as we will show below the right-hand side of (\ref{ph-Lie}) can
be evaluated using the standard representation of $\cG$ and readily
expressed as a polynomial in $K_a$ with the same order as $H$. Upon
imposing (\ref{ph-Lie}), we therefore obtain a (finite) set of
numerical equations involving the coupling constants
$\lambda_{a_1,a_2,\cdots a_k}$ and the parameters $r_a$ that
determine the metric operator via
    \be
    \etap:=\exp\left(-\sum_{a=1}^d
    r_a~ K_a\right).
    \label{sec4-Lie4}
    \ee

In order to demonstrate how this method works, we introduce the
matrix $\underline{\etap}:=\exp\left(-\sum_{a=1}^d
    r_a~\underline{K_a}\right)$,
that belongs to $\exp(\cG_{_\C})$ and satisfies $
    \etap=\cU(\underline{\etap})$.
This together with (\ref{sec4-generators=}) and (\ref{sec4-main-id})
imply
    \bea
    \etap H\,\etap^{-1}&=&\sum_{k=1}^n\sum_{a_1,a_2,\cdots,a_k=1}^d
    \lambda_{a_1,a_2,\cdots a_k}~
    (\etap K_{a_1}\etap^{-1}) (\etap K_{a_2}\etap^{-1})\cdots (\etap
    K_{a_k}\etap^{-1})\nn\\
    &=&\sum_{k=1}^n\sum_{a_1,a_2,\cdots,a_k=1}^d\lambda_{a_1,a_2,\cdots a_k}~
    \fU(\underline{\etap}\:\underline{K_{a_1}}\:\underline{\etap}^{-1})\cdots
    \fU(\underline{\etap}\:\underline{K_{a_k}}\:\underline{\etap}^{-1})
    \label{sec4-ph-Lie-exp1}
    \eea
Because for all $a\in\{1,2,\cdots,d\}$,
$\underline{\etap}\:\underline{K_{a}}\:\underline{\etap}^{-1}$
belongs to $\cG_{_\C}$, there are complex coefficients $\kappa_{ab}$
depending on the structure constants $C_{abc}$ of the Lie algebra
$\cG$ and the coefficients $r_a$ such that
    \be
    \underline{\etap}\:\underline{K_{a}}\:\underline{\etap}^{-1}=
    \sum_{b=1}^d\kappa_{ab}\underline{K_b}.
    \label{sec4-kappa-exp1}
    \ee
As a result, $\fU(\underline{\etap}\:\underline{K_{a}}\:
\underline{\etap}^{-1})=\sum_{b=1}^d\kappa_{ab}\:\fU(\underline{K_b})=
\sum_{b=1}^d\kappa_{ab}\,K_b$.
Inserting this relation in (\ref{sec4-ph-Lie-exp1}), we find
    \be
    \etap H\,\etap^{-1}=
    \sum_{k=1}^n\sum_{b_1,b_2,\cdots,b_k=1}^d
    \tilde\lambda_{b_1b_2\cdots b_k}K_{b_1}K_{b_2}\cdots K_{b_k},
    \label{sec4-ph-Lie-exp2}
    \ee
where $\tilde\lambda_{b_1b_2\cdots
b_k}:=\sum_{a_1,a_2,\cdots,a_k=1}^d
    \lambda_{a_1,a_2,\cdots a_k}\kappa_{a_1b_1}\kappa_{a_2b_2}
    \cdots\kappa_{a_k b_k}$.
Note that the coefficients $\tilde\lambda_{b_1b_2\cdots b_k}$ depend
on the parameters $r_a$ of the metric operator (\ref{sec4-Lie4}).

In view of (\ref{sec4-Lie1}) and (\ref{sec4-ph-Lie-exp2}), the
pseudo-Hermiticity relation $H^\dagger=\etap\,H\etap^{-1}$ takes the
form
    \be
    \sum_{k=1}^n\sum_{c_1,c_2,\cdots,c_{k-1},c_k=1}^d
    (\lambda^*_{c_k c_{k-1} \cdots c_2c_1}-\tilde\lambda_{c_1c_2\cdots
    c_{k-1}c_k})~K_{c_1}K_{c_2}\cdots K_{c_k}=0.
    \label{sec4-main-eq}
    \ee
We can use the commutation relations for the generators $K_a$,
namely $[K_a,K_b]=i\sum_{c=1}^d C_{abc}K_c$,
to reorder the factors $K_{c_1}K_{c_2}\cdots K_{c_k}$ and express
the left-hand side of (\ref{sec4-main-eq}) as a sum of linearly
independent operators. Consequently, the coefficients of this sum
must identically vanish. This yields a system of equations for
$r_a$. In general this system is over-determined and a solution
might not exist. However, there is a class of Hamiltonians of the
form (\ref{sec4-Lie1}) for which this system has solutions. In this
case, each solution determines a metric operator.

For the particular case that $n=1$, so that
    \be
    H=\sum_{a=1}^d\lambda_aK_b,
    \label{sec4-H-linear}
    \ee
we may employ a more direct method of deriving the system of
equations for $r_a$. This is based on the observation that in this
case we can obtain a representation of the pseudo-Hermiticity
relation $H^\dagger=\etap H\etap^{-1}$ in $\cG_\C$, namely
    \be
    \underline{H^\maltese}=\underline{\etap}\;\underline{H}\;
    \underline{\etap}^{-1},
    \label{sec4-ph-matrix-form}
    \ee
where\footnote{Unless $G$ is a unitary group, $\fU$ is not a
$*$-representation \cite{fell-doran}, and
$\underline{H^\maltese}\neq \underline{H}^\dagger$.}
    \be
    \underline{H}:=\sum_{a=1}^d\lambda_a\underline{K_a},~~~~~~~~~~~
    \underline{H^\maltese}:=\sum_{a=1}^d\lambda^*_a\underline{K_a}.
    \label{sec4-H-linear-under}
    \ee
The matrix equation~(\ref{sec4-ph-matrix-form}) is equivalent to a
system of $d$ complex equations for $d$ real variables $r_a$.
Therefore, it is generally over-determined.

We can use the above Lie algebraic method to compute the equivalent
Hermitian Hamiltonian $h$ for the quasi-Hermitian Hamiltonians of
the form (\ref{sec4-Lie1}). In view of the definitions: $h:=\rho
H\rho^{-1}$ and $\rho:=\sqrt{\etap}=\exp\left(\sum_{a=1}^d
\frac{r_a}{2}\,K_a\right),$ $h$ is given by the right-hand side of
(\ref{sec4-kappa-exp1}) provided that we use $\frac{r_a}{2}$ in
place of $r_a$.

An alternative Lie algebraic approach of determining metric operator
and the equivalent Hermitian Hamiltonian is the following. First, we
use the argument leading to (\ref{sec4-ph-Lie-exp1}) to obtain
    \be
    h=\rho H\rho^{-1}=\sum_{k=1}^n
    \sum_{a_1,a_2,\cdots,a_k=1}^d\lambda_{a_1,a_2,\cdots a_k}~
    \fU(\underline{\rho}\:\underline{K_{a_1}}\:\underline{\rho}^{-1})\:
    \fU(\underline{\rho}\:\underline{K_{a_2}}\:\underline{\rho}^{-1})\cdots
    \fU(\underline{\rho}\:\underline{K_{a_k}}\:\underline{\rho}^{-1}),
    \label{sec4-Hermitian-h}
    \ee
where $\underline{\rho}:=\exp\left(\sum_{a=1}^d
\frac{r_a}{2}\,\underline{K_a}\right)$. Then, we evaluate
$\underline{\rho}\:\underline{K_{a}}\:\underline{\rho}^{-1}$ and
express it as a linear combination of $\underline{K_{a}}$ with
$r_a$-dependent coefficients.\footnote{This is possible, because
$\underline{\rho}\in\exp(\cG_{_\C})$.} Substituting the result in
(\ref{sec4-Hermitian-h}) and using the linearity of $\fU$ and
$K_a=\fU(\underline{K_a})$ give
$h=\sum_{k=1}^n\sum_{a_1,a_2,\cdots,a_k=1}^d
    \varepsilon_{a_1,a_2,\cdots a_k}~K_{a_1}\, K_{a_2}\cdots K_{a_k}$,
where $\varepsilon_{a_1,a_2,\cdots a_k}$ are $r_a$-dependent complex
coefficients. In this approach, we obtain the desired system of
equations for $r_a$ by demanding that $h=h^\dagger$. This is the
root taken in \cite{quesne-jpa-2007} where the Lie algebraic method
was originally used for the construction of the metric operators and
equivalent Hermitian Hamiltonians for a class of quasi-Hermitian
Hamiltonians of the linear form~(\ref{sec4-H-linear}) with
underlying $su(1,1)$ algebra. For an application of this approach to
Hamiltonians that are quadratic polynomials in generators of
$SU(1,1)$, see \cite{assis-fring-2008}.

\subsubsection{Swanson Model: $\cG=su(1,1)$}

In this section we explore the application of the Lie algebraic
method to construct metric operators for Swanson's Hamiltonian
\cite{swanson}:
    \be
    H=\hbar\omega(a^\dagger
    a+\frac{1}{2})+\alpha\:a^2+\beta\:{a^\dagger}^2,
    \label{sec4-swanson}
    \ee
where $a:=\frac{\rx+i\rp}{\sqrt 2}$,
$\rx:=\sqrt{\frac{m\omega}{\hbar}}~x$,
$\rp:=\frac{p}{\sqrt{m\hbar\omega}}$, $\alpha,\beta,\omega,m$ are
real parameters, $m>0$, $\omega>0$, and $
    \hbar^2\omega^2>4\alpha\beta$.
The latter condition ensures the reality and discreteness of the
spectrum of (\ref{sec4-swanson}).

The problem of finding metric operators for the
Hamiltonian~(\ref{sec4-swanson}) is addressed in
\cite{jones-2005,scholtz-geyer-2006,mgh-jpa-2007}. The use of the
properties of Lie algebras for solving this problem was originally
proposed in \cite{quesne-jpa-2007}.

Swanson's Hamiltonian (\ref{sec4-swanson}) is an example of a rather
trivial class of quasi-Hermitian Hamiltonians of the standard form
    \be
    H=\frac{[p+A(x)]^2}{2M}+v(x),
    \label{sec9-HN-3}
    \ee
where $A$ and $v$ are respectively a complex-valued vector potential
and a real-valued scalar potential, and $M\in\R^+$ is the mass. It
is easy to see that these Hamiltonians admit the $x$-dependent
metric operator: $\etap=\exp\left(-\frac{2}{\hbar}\int
dx~\Im[A(x)]\right)$.
This in turn yields the equivalent Hermitian Hamiltonian:
$h=\frac{1}{2m}\left(p+\Re[A(x)]\right)^2+v(x)$.
The subclass of the Hamiltonians (\ref{sec9-HN-3}) corresponding to
imaginary vector potentials has been considered in
\cite{ahmed-pla-2002}. The Swanson Hamiltonian~(\ref{sec4-swanson})
is a special case of the latter. It corresponds to the choice:
$M=\frac{m}{1-\tilde\alpha-\tilde\beta}$,
$A(x)=i\left(\frac{m\omega(\tilde\alpha-
\tilde\beta)}{1-\tilde\alpha-\tilde\beta}\right)x$, and
$v(x)=\frac{1}{2}\left(\frac{1-4\tilde\alpha
\tilde\beta}{1-\tilde\alpha-\tilde\beta}\right)m\omega^2x^2$, where
    \be
    \tilde\alpha:=\frac{\alpha}{\hbar\omega},~~~~~~
    \tilde\beta:=\frac{\beta}{\hbar\omega}.
    \label{sec4-dimensionless-swanson}
    \ee

As shown in \cite{jones-2005,mgh-jpa-2007}, the
Hamiltonian~(\ref{sec4-swanson}) admits other exactly constructible
metric operators. The Lie algebraic method considered in this
section offers a systematic approach for constructing metric
operators for this Hamiltonian. In order to describe the details of
this construction we begin by recalling that the operators $a$ and
$a^\dagger$ are the usual harmonic oscillator annihilation and
creation operators that satisfy
    \be
    [a,a^\dagger]=1.
    \label{boson-algebra}
    \ee
A well-known consequence of this relation is the possibility of
constructing a unitary representation of the Lie algebra $su(1,1)$
using quadratic polynomials in $a$ and $a^\dagger$. To see this,
consider the Hermitian operators:
$K_1:=\frac{1}{4}(a^2+{a^\dagger}^2)$,
$K_2:=\frac{i}{4}(a^2-{a^\dagger}^2)$, and
$K_3:=\frac{1}{4}(aa^\dagger+a^\dagger a)=\frac{1}{2}(a^\dagger
a+\frac{1}{2})$ that act in $\cH:=L^2(\R)$, \cite{nova}. In view of
(\ref{boson-algebra}), they satisfy the $su(1,1)$ algebra:
$[K_1,K_2]=-iK_3$, $[K_2,K_3]=iK_1$, $[K_3,K_1]=iK_2$.
Clearly the Hamiltonian~(\ref{sec4-swanson}) can be expressed as a
linear combination of $K_1,K_2$ and $K_3$:
    \be
    H=2\left[(\alpha+\beta)K_1+i(\alpha-\beta)K_2+\hbar\omega\,K_3\right].
    \label{sec4-swanson-2}
    \ee
This relation identifies the Hamiltonian (\ref{sec4-swanson-2}) as a
special case of the Hamiltonians of the form (\ref{sec4-H-linear})
with $G=SU(1,1)$, $d=3$, $\lambda_1=2(\alpha+\beta)$,
$\lambda_2=2i(\alpha-\beta)$, and $\lambda_3=2\hbar\omega$. We can
further simplify (\ref{sec4-swanson-2}) by introducing non-Hermitian
generators $K_\pm:=K_1\pm iK_2$. In terms of these, we have
$H=2\left(\alpha K_+ +\beta K_- +\hbar\omega\,K_3\right)$.

In order to apply the above method of constructing a metric operator
for (\ref{sec4-swanson-2}), we need to find a set of generators
$\underline{K_a}$ of $SU(1,1)$ in its standard representation and a
faithful unitary representation $\fU$ of the Lie algebra $su(1,1)$
such that $K_a=\fU(\underline{K_a})$ for all $a\in\{1,2,3\}$. A
simple choice is
    \be
    \underline{K_1}:=\frac{i}{2}\,\mbox{\large$\sigma_1$}=
        \frac{1}{2}\left(\begin{array}{cc}
        0&i\\
        i& 0\end{array}\right),~~~
    \underline{K_2}:=\frac{i}{2}\,\mbox{\large$\sigma_2$}=
        \frac{1}{2}\left(\begin{array}{cc}
        0&1\\
        -1& 0\end{array}\right),~~~
    \underline{K_3}:=\frac{1}{2}\,\mbox{\large$\sigma_3$}=
        \frac{1}{2}\left(\begin{array}{cc}
        1&0\\
        0&-1\end{array}\right),
    \label{su1-1-st}
    \ee
where $\mbox{\large$\sigma_1$}$, $\mbox{\large$\sigma_2$}$, and
$\mbox{\large$\sigma_3$}$ are the Pauli matrices. We also have
    \be
    \underline{K_+}:=\underline{K_1}+i\underline{K_2}=
        \left(\begin{array}{cc}
        0&i\\
        0&0\end{array}\right),~~~~~~
    \underline{K_-}:=\underline{K_1}-i\underline{K_2}=
        \left(\begin{array}{cc}
        0&0\\
        i&0\end{array}\right),
        \label{sec4-Kpm=}
    \ee
that fulfill $\fU(\underline{K_\pm})=K_\pm$.

Comparing~(\ref{su1-1-st}) and (\ref{sec4-Kpm=}), we see that it is
more convenient to work with the generators $K_\pm$ and $K_3$ rather
than $K_a$ with $a\in\{1,2,3\}$. This in particular suggests the
following alternative parametrization of the metric operator
(\ref{sec4-Lie4}).
    \be
    \etap=\exp(z\,K_+)\:\exp(2rK_3)\:\exp(z^*\,K_-),~~~~~~\mbox{with
    $z\in\C, r\in\R$}.
    \label{sec4-rep-etap}
    \ee
Clearly,
    \bea
    \underline{\etap}&=&\exp(z\,\underline{K_+})\;
    \exp(2r\underline{K_3})\;\exp(z^*\,\underline{K_-})=
    \left(\begin{array}{cc}
    e^r-e^{-r}|z|^2 & i e^{-r}z\\
    ie^{-r}z^* & e^{-r}\end{array}\right),
    \label{sec4-rep-etap-under}\\
    \underline{\etap}^{-1}&=&\exp(-z^*\,\underline{K_-})\;
    \exp(-2r\underline{K_3})\;\exp(-z\,\underline{K_+})=
    \left(\begin{array}{cc}
     e^{-r} & -i e^{-r}z\\
    -ie^{-r}z^* &  e^r-e^{-r}|z|^2\end{array}\right),
    \label{sec4-rep-etap-under-inv}
    \eea
where we have made use of (\ref{su1-1-st}) and (\ref{sec4-Kpm=}).
Also in view of (\ref{sec4-H-linear-under}),
(\ref{sec4-dimensionless-swanson}), (\ref{sec4-swanson-2}), and
$K_\pm:=K_1\pm iK_2=K_\mp^\dagger$, we have
    \bea
    \underline{H}&=&2\hbar\omega[\tilde\alpha\underline{K_+}+\tilde\beta
    \underline{K_-}+\underline{K_3}]=\hbar\omega\left(\begin{array}{cc}
    1 & 2i\tilde\alpha\\
    2i\tilde\beta & -1\end{array}\right),
    \label{sec4-H-under-swanson}\\
    \underline{H^\maltese}&=&2\hbar\omega[\tilde\alpha\underline{K_-}+\tilde\beta
    \underline{K_+}+\underline{K_3}]=\hbar\omega\left(\begin{array}{cc}
    1 & 2i\tilde\beta\\
    2i\tilde\alpha & -1\end{array}\right).
    \label{sec4-Hp-under-swanson}
    \eea

Next, we insert (\ref{sec4-rep-etap-under}) --
(\ref{sec4-Hp-under-swanson}) in (\ref{sec4-ph-matrix-form}). This
yields the following three independent complex equations that are
more conveniently expressed in terms of $s:=e^r$ and $w:=e^{-r}z$.
    \bea
    &&s(\tilde\alpha s-w)+\tilde\beta(w^2-1)+
    s|w|^2[w+\tilde\alpha s(|w|^2-2)]=0,
    \label{sec4-e1=z}\\
    &&\tilde\beta-\tilde\alpha s^2+sw^*(1+\tilde\alpha s w^*)=0,
    \label{sec4-e2=z}\\
    &&\tilde\beta w+s w^*[w+\tilde\alpha s(|w|^2-1)]=0.
    \label{sec4-e3=z}
    \eea
To solve these equations, we rewrite (\ref{sec4-e3=z}) as:
$w+\tilde\alpha s(|w|^2-2)=-\tilde\alpha s-\frac{\tilde\beta
w}{sw^*}$, and use this relation in (\ref{sec4-e1=z}) to obtain
    \be
    \tilde\alpha s(1-|w|^2)=\frac{\tilde\beta}{s}+w.
    \label{sec4-4=z}
    \ee
Substituting this equation back into (\ref{sec4-e3=z}), we find
$\tilde\beta(w-w^*)=0$. Therefore, either $\tilde\beta=0$ or
$w\in\R$. It is easy to show using (\ref{sec4-e2=z}) that the
condition $\tilde\beta=0$ implies $w\in\R$ as well. Hence, $w$ is
real and (\ref{sec4-4=z}) reduces to a quadratic equation
whose solution is
    \be
    w=\frac{-1\pm\sqrt{4\tilde\alpha^2s^2+1-4\tilde\alpha\tilde\beta}}{2\tilde\alpha
    s}.
    \label{sec4-w=sol}
    \ee
It turns out that (\ref{sec4-e1=z}) -- (\ref{sec4-e3=z}) do not
impose any further restriction on $s$. Therefore, (\ref{sec4-w=sol})
is the solution of the system (\ref{sec4-e1=z}) --
(\ref{sec4-e3=z}). In terms of the original parameters $r$ and $z$,
it reads
$z=(-1\pm\sqrt{4\tilde\alpha^2e^{2r}+1-4\tilde\alpha\tilde\beta})/
    (2\tilde\alpha)
    =(-\hbar\omega\pm\sqrt{4\alpha^2e^{2r}+\hbar\omega-4\alpha\beta})/(
    2\alpha)$.
Substituting this formula in (\ref{sec4-rep-etap}), we find two
one-parameter families of metric operators for Swanson's
Hamiltonian.

\section{Systems Defined on a Complex Contour}
\label{sec-contour}

\subsection{Spectral problems defined on a contour}
\label{sec9-spec}

Consider the Schr\"odinger operator $-\frac{d^2}{dx^2}+V(x)$,
where $V:\R\to\C$ is a complex-valued piecewise real-analytic
potential. The study of the spectral problem for this operator and
its complex generalization, $-\frac{d^2}{dz^2}+V(z)$,
with $z$ taking values along a contour\footnote{Here by the term
`contour' we mean (the graph of) a piecewise smooth simple curve
that needs not be closed.} $\Gamma$ in $\C$, predates the discovery
of quantum mechanics.\footnote{Hermann Weyl's dissertation of 1909
provides a systematic approach to this problem. For a detailed
discussion of Weyl's results, see \cite[\S 10]{hille}.} The case of
polynomial potentials have been studied thoroughly in \cite{sibuya}.
For a more recent discussion, see \cite{shin-jpa-2005}.

The spectrum of $-\frac{d^2}{dz^2}+V(z)$ depends on the choice of
the contour $\Gamma$. In the case that $\Gamma$ visits the point at
infinity, the spectrum is essentially determined by the boundary
condition imposed on the solutions $\Psi:\Gamma\to\C$ of the
following eigenvalue equation at infinity.
    \be
    \left[-\frac{d^2}{dz^2}+V(z)\right]\Psi(z)=E\,\Psi(z).
    \label{sec9-eg-va}
    \ee

For an extended contour $\Gamma$, that is obtained by a continuous
invertible deformation of the real axis in $\C$, the eigenvalue
problem (\ref{sec9-eg-va}) is well-posed provided that we demand
$\Psi(z)$ to decay exponentially as $|z|\to\infty$ along $\Gamma$.
To make this condition more explicit, we identify $\Gamma$ with the
graph of a parameterized curve $\zeta:\R\to\C$ in $\C$, i.e.,
    \be
    \Gamma=\{\zeta(s)~|~s\in\R~\}.
    \label{sec9-param}
    \ee
The assumption that $\Gamma$ is simple implies that $\zeta$ is a
one-to-one function, and we can express the above-mentioned boundary
condition as
    \be
    |\Psi(\zeta(s))|\to 0~~~{\rm exponentially~as}~~~
    s\to\pm\infty.
    \label{sec9-BC}
    \ee
A simple consequence of this condition is
    \be
    \int_{-\infty}^\infty |\Psi(\zeta(s))|^2 ds<\infty.
    \label{sec9-BC-2}
    \ee

If we view $\C$ as a Riemannian manifold, namely $\R^2$ endowed with
the Euclidean metric tensor, and consider $\Gamma$ as a submanifold
of this manifold, we can use the embedding map $\zeta:\R\to\C$ to
induce a metric tensor $(\fg)$ on $\Gamma$. The corresponding line
element is given by
$d\ell:=\sqrt{\fg}\,ds=\sqrt{dx(s)^2+dy(s)^2}=|\zeta'(s)|\,ds$,
where $x(s):=\Re(\zeta(s))$ and $y(s):=\Im(\zeta(s))$. Therefore,
the integral measure defined by $\fg$ on $\Gamma$ is the arc-length
element $|\zeta'(s)|\,ds$. This in turn suggests the following
parametrization-invariant definition of the $L^2$-inner product on
$\Gamma$.
    \be
    \pbr\Psi|\Phi\pkt:=\int_{-\infty}^\infty
    \Psi(\zeta(s))^*\,\Phi(\zeta(s))\, |\zeta'(s)|\, ds.
    \label{sec9-L2-inn}
    \ee
If we identify $s$ with the arc-length parameter, for which
$|\zeta'(s)|=1$, and let
$L^2(\Gamma):=\left\{\Psi:\Gamma\to\C~\Big|~\pbr\Psi|\Psi\pkt
<\infty~\right\}$,
we can express (\ref{sec9-BC-2}) as
    \be
    \Psi\in L^2(\Gamma).
    \label{sec9=sq-int}
    \ee
This shows that the boundary condition (\ref{sec9-BC}) implies the
square-integrability condition (\ref{sec9=sq-int}) along $\Gamma$.
The converse is not generally true; the spectrum defined by the
boundary condition (\ref{sec9-BC}) is a subset of the point
spectrum\footnote{By definition, the spectrum $\sigma(A)$ of an
operator $A$ acting in a Banach space is the set of complex numbers
$E$ for which the operator $A-E I$ is not invertible, i.e., one or
more of the following conditions hold: (1) $A-E I$ is not
one-to-one; (2) $A-E I$ is not onto; (3) $A-E I$ is one-to-one so
that it has an inverse, but the inverse is not a bounded operator
\cite{hislop-sigal}. The \emph{point spectrum} of $A$ is the subset
of $\sigma(A)$ consisting of the eigenvalues $E$ of $A$, i.e., the
numbers $E$ for which $A-E I$ is not one-to-one \cite{reed-simon}.}
of the operator $-\frac{d^2}{dz^2}+V(z)$ viewed as acting in the
Hilbert space $L^2(\Gamma)$. In the following we shall use the term
spectrum to mean the subset of the point spectrum that is defined by
the boundary condition~(\ref{sec9-BC}).

In order to demonstrate the importance of the choice of the contour
in dealing with the spectral problem (\ref{sec9-eg-va}), consider
the imaginary cubic potential $V(z)=iz^3$. As we mentioned in
Section~\ref{sec1}, the spectrum defined by the boundary
condition~(\ref{sec9-BC}) along the real axis ($\zeta(s)=s$) is
discrete, real, and positive \cite{dorey,shin}. But, the spectrum
defined by the same boundary condition along the imaginary axis is
empty. To see this, we parameterize the imaginary axis according to
$z=\zeta(s)=is$ with $s\in\R$. Then the operator
$-\frac{d^2}{dz^2}+iz^3$ takes the form $\frac{d^2}{ds^2}+s^3$, and
we can respectively express the eigenvalue equation
(\ref{sec9-eg-va}) and the boundary condition (\ref{sec9-BC}) as
    \be
    \left[\frac{d^2}{ds^2}+s^3\right]\psi(s)=E\,\psi(s),
    \label{sec9-eg-va-2}
    \ee
and
    \be
    |\psi(s)|\to 0~~~{\rm exponentially~as}~~~s\to\pm\infty,
    \label{sec9-eg-va-2-boundary}
    \ee
where $\psi(s):=\Psi(is)$ for all $s\in\R$.\footnote{This was
pointed out to me by Prof.~Yavuz Nutku.} But, it is well-known that
(\ref{sec9-eg-va-2}) does not have any solution fulfilling
(\ref{sec9-eg-va-2-boundary}) for either real or complex values of
$E$.\footnote{The point spectrum of $\frac{d^2}{ds^2}+s^3$ is $\C$,
i.e., (\ref{sec9-eg-va-2}) admits square-integrable solutions for
all $E\in\C$ (This was pointed out to me by Prof.\ Patrick Dorey.)
These solutions do not however satisfy
(\ref{sec9-eg-va-2-boundary}). They do not represent physically
acceptable bound states, because they do not belong to the domain of
the observables such as position, momentum or some of their powers.}

The imaginary cubic potential belongs to the class of potentials of
the form
    \be
    V_\nu(x)=\lambda\:x^2(ix)^\nu,~~~~~~\nu\in\R,~\lambda\in\R^+.
    \label{sec9-x-nu}
    \ee
As shown in \cite{dorey,shin}, for $\nu\geq 0$ these potentials
share the spectral properties of the imaginary cubic potential, if
we impose the boundary condition (\ref{sec9-BC}) along a contour
$\Gamma_\nu$ that lies asymptotically in the union of the Stokes
wedges \cite{bender-prl}:
     \be
    S_\nu^\pm:=\Big\{r\,e^{-i(\theta_\nu^\pm+
    \varphi)}~\Big|~r\in[0,\infty),~\varphi\in(
    -\delta_\nu,\delta_\nu)\:
    \Big\},
    \label{sec9-stokes}
    \ee
where
    \be
    \theta_\nu^+:=\frac{\pi\nu}{2(\nu+4)}=:\theta_\nu,~~~~~~
    \theta_\nu^-:=\pi-\theta_\nu,~~~~~~
    \delta_\nu:=\frac{\pi}{\nu+4}.
    \label{sec9-thetas}
    \ee
    \begin{figure}
    \vspace{0.0cm} \hspace{0.00cm}
    \centerline{\includegraphics[scale=.75,clip]{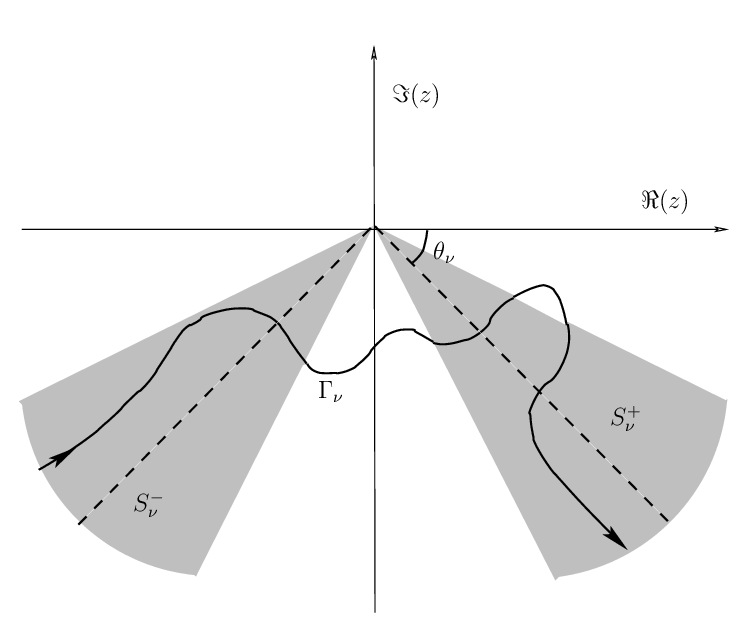}}
    \centerline{\parbox{16cm}{\caption{A contour $\Gamma_\nu$ (the solid curve) lying asymptotically in the
    union of the Stokes wedges (the grey region). The dashed lines
    are the bisectors of the Stokes wedges $S^\pm_\nu$. The angle
    $\theta_\nu$ between the positive real axis and the bisector of
    $S^+_\nu$ is also depicted.}}}
    \label{fig1}
    \vspace{0.0cm}
    \end{figure}%
Here by asymptotic inclusion of $\Gamma_\nu$ in $S_\nu^-\cup
S_\nu^+$, we mean that if $\Gamma_\nu=\{\zeta_\nu(s)|s\in\R\}$ for a
piecewise smooth one-to-one function $\zeta_\nu:\R\to\C$, then there
must exist a positive integer $M$ such that for all $s\in\R$ the
condition $\pm s>M$ implies $\zeta(s)\in S_\nu^\pm$. Figure~1 shows
Stokes wedges and a typical contour lying in $S_\nu^-\cup S_\nu^+$
asymptotically.

For $\nu=2$, (\ref{sec9-x-nu}) gives the wrong-sign quartic
potential,
    \be
    V_2(x)=-\lambda\,x^4,~~~~~~~\lambda\in\R^+,
    \label{sec9-quartic}
    \ee
which is known to have an empty spectrum (defined by (\ref{sec9-BC})
along the real axis.) Setting $\nu=2$ in (\ref{sec9-stokes}) and
(\ref{sec9-thetas}), we have
$S_2^-=\Big\{r\,e^{i\theta}~\Big|~r\in[0,\infty),~
\theta\in(-\pi,-\frac{2\pi}{3})\:\Big\}$ and
$S_2^+=\Big\{r\,e^{i\theta}~\Big|~r\in[0,\infty),~
\theta\in(-\frac{\pi}{3},0)\:\Big\}$. Therefore the condition that
$\Gamma_2$ must lie asymptotically inside $S_2^-\cup S_2^+$ excludes
the real axis as a possible choice for $\Gamma_2$. It is not
difficult to see that the same holds for all $\nu\geq 2$.

\subsection{Equivalent spectral problems defined on $\R$}
\label{sec9-equiv}

The fact that the spectrum of the potentials (\ref{sec9-x-nu})
defined by the above mentioned boundary condition along $\Gamma_\nu$
is discrete, real and positive is by no means obvious, and its proof
is quite complicated \cite{dorey,shin}. In this subsection we will
outline a transformation scheme that maps the spectral problem for
these and similar potentials to an equivalent spectral problem that
is defined on the real line \cite{jpa-2005a}. This scheme provides
an intuitive understanding of the spectral properties of the
potentials (\ref{sec9-x-nu}) and in particular allows for a
straightforward treatment of the wrong-sign quartic potential
(\ref{sec9-quartic}) that we shall consider in the following
subsection.

Given an extended contour $\Gamma$, we can use $x:=\Re(z)$ to
parameterize it. We do this by setting $\Gamma=\{\zeta(x)|x\in\R\}$
where $\zeta(x):=x+if(x)$ for all $x\in\R$, and $f:\R\to\R$ is a
piecewise smooth function. This implies that along $\Gamma$,
$dz=d\zeta(x)=[1+if'(x)]dx$ and the eigenvalue
equation~(\ref{sec9-eg-va}) takes the form
    \be
    \left[-g(x)^2\frac{d^2}{dx^2}+ig(x)^3f''(x)
    \frac{d}{dx}+\tilde v(x)\right]\tilde\psi(z)=E\,\tilde\psi(z),
    \label{sec9-eg-va-x}
    \ee
where for all $x\in\R$,
    \be
    g(x):=\frac{1}{1+if'(x)},~~~~\tilde{v}(x):=V(x+if(x)),~~~~
    \tilde\psi(x):=\Psi(x+if(x)),
    \label{sec9-trans-0}
    \ee
and a prime stands for the derivative of the corresponding function.

Next, we examine the consequences of using the arc-length
parametrization of $\Gamma$. If we define $F:\R\to\R$ by $F(x):=
\int_0^x \sqrt{1+f'(u)^2}\,du$, we can express the arc-length
parameter along $\Gamma$, which we denote by $\rx$, as $\rx:=F(x)$.
Under the transformation $x\to\rx$, the eigenvalue equation
(\ref{sec9-eg-va-x}) takes the form
    \be
    e^{-2i\xi({\rm x})} \left[-\frac{d^2}{d\rx^2}+ia(\rx)
    \frac{d}{d\rx}+v(\rx)\right]\psi(\rx)=E\,\psi(\rx),
    \label{sec9-eg-va-X}
    \ee
where\footnote{Because $F$ is a monotonically increasing function,
it is one-to-one. In particular it have an inverse that we denote by
$F^{-1}$.}
    \bea
    \xi(\rx)&:=&\tan^{-1}(f'(x))
    \Big|_{x=F^{-1}({\rm x})},
    ~~~~~~~~
    a(\rx):=\xi'(\rx)=\left. \frac{f''(x)}{[1+f'(x)^2]^{\frac{3}{2}}}
    \:\right|_{x=F^{-1}({\rm x})},
    \label{sec9-trans1}\\
    v(\rx)&:=&e^{2i\xi({\rm x})}\tilde v(F^{-1}(\rx)),~~~~~~~~~~~
    \psi(\rx):=\tilde\psi(F^{-1}(\rx)).
    \label{sec9-trans}
    \eea

The arc-length parametrization of $\Gamma$ is achieved by the
function $G:\R\to\Gamma$ defined by
$G(\rx):=F^{-1}(\rx)+if(F^{-1}(\rx))$ for all $\rx\in\R$.
This is the invertible function that maps the real line $\R$ onto
the contour $\Gamma$ in such a way that the (Euclidean) distance is
preserved, i.e., if $\rx_1,\rx_2\in\R$ are respectively mapped to
$z_1:=G(\rx_1)$ and $z_2:=G(\rx_2)$, then the length of the segment
of $\Gamma$ that lies between $z_1$ and $z_2$ is given by
$|\rx_1-\rx_2|$. In other words, $G:\R\to\Gamma$ is an isometry. In
light of (\ref{sec9-trans-0}) and (\ref{sec9-trans}), $G$ relates
the solutions $\Psi$ and $\psi$ of the eigenvalue equations
(\ref{sec9-eg-va}) and (\ref{sec9-eg-va-X}) according to
$\psi(\rx)=\Psi(G(\rx))$.
We can use $G$ to express $v$ in terms of the potential $V$
directly:
    \be
    v(\rx)=e^{2i\xi({\rm x})}V(G(\rx)).
    \label{sec9-ceq0}
    \ee
Furthermore, recalling that the arc-length parametrization of
$\Gamma$ corresponds to identifying $s$ and $\zeta(s)$ of
(\ref{sec9-param}) -- (\ref{sec9-BC-2}) respectively with $\rx$ and
$G(\rx)$, we have
    \be
    \pbr\Psi|\Psi\pkt=
    \int_{-\infty}^\infty |\Psi(G(\rx))|^2 d\rx=
    \int_{-\infty}^\infty |\psi(\rx)|^2 d\rx=\br\psi|\psi\kt.
    \label{sec9-BC-4}
    \ee
This observation has two important consequences. Firstly, it implies
that $\Psi\in L^2(\Gamma)$ if and only if $\psi\in L^2(\R)$.
Secondly, it allows for the introduction of an induced unitary
operator $G_*:L^2(\Gamma)\to L^2(\R)$, namely
    $G_*(\Psi):=\psi$ if $\psi(\rx)=\Psi(G(\rx))$
    for all $\rx\in\R$.\footnote{(\ref{sec9-BC-4}) shows that $G_*$ preserves the
norm. This is sufficient to conclude that it is a unitary operator,
for in an inner product space the norm uniquely determines the inner
product \cite[\S 6.1]{kato}.}

The above constructions show that a pseudo-Hermitian quantum system
that is defined by a Hamiltonian operator of the form
$-\frac{d^2}{dz^2}+V(z)$ acting in the reference Hilbert space
$L^2(\Gamma)$ is unitary-equivalent to the one defined by the
Hamiltonian operator
    \be
     H=e^{-2i\xi({\rm x})} \left[-\frac{d^2}{d\rx^2}+ia(\rx)
    \frac{d}{d\rx}+v(\rx)\right]
    \label{sec9-H-R}
    \ee
that is defined in the reference Hilbert space $L^2(\R)$. In terms
of the unitary operator $G_*$, we have
$-\frac{d^2}{dz^2}+V(z)=G_*^{-1}H\,G_*$.

A particularly simple choice for a contour is a wedge-shaped
contour: $\Gamma^{(\theta)}:=\{ x+if(x)|x\in\R\}$ where $f(x):=-\tan
\theta \:|x|$ and $\theta\in[0,\mbox{$\frac{\pi}{2}$})$.
A typical example is the contour obtained by adjoining the bisectors
of the Stokes wedges (the dashed lines in Figure~1.) For such a
contour,
    \bea
    &&\rx=F(x)=\sec\theta \:x,~~~~~
    x=F^{-1}(\rx)=\cos\theta\:\rx,
    \label{sec9-ceq1}\\
    &&G(\rx)=\cos\theta\:\rx-i\sin \theta\:|\rx|=\left\{
    \begin{array}{ccc}
    e^{i\theta}\,\rx&{\rm for}&\rx<0,\\
    e^{-i\theta}\,\rx&{\rm for}&\rx\geq 0,\end{array}\right.
    \label{sec9-ceq2}
    \eea
and in view of (\ref{sec9-trans1}), (\ref{sec9-trans}), and
(\ref{sec9-ceq0}) the Hamiltonian operator (\ref{sec9-H-R}) takes
the form
    \be
     H=-e^{2i\theta\,\sg({\rm x})} \left[\frac{d^2}{d\rx^2}+2i\theta\,
    \delta(\rx)\frac{d}{d\rx}\right]+
    V(\cos\theta\:\rx-i\sin \theta\:|\rx|).
    \label{sec9-H-we}
    \ee

The presence of delta-function in (\ref{sec9-H-we}) has its root in
the non-differentiability of $\Gamma^{(\theta)}$ at the origin. One
can smooth out $\Gamma^{(\theta)}$ in an arbitrarily small open
neighborhood of the origin and show that this delta-function
singularity amounts to the imposition of a particular matching
condition at $\rx=0$ for the solutions of the corresponding
eigenvalue problem. As shown in \cite{jpa-2005a}, these are given by
    \be
    \psi(0^+)=\psi(0)=\psi(0^-),~~~~~~~
    e^{-2i\theta}\psi'(0^-)=e^{2i\theta}\psi'(0^+),
    \label{sec9-match}
    \ee
where for every function $\phi:\R\to\R$, $\phi(0^\pm):= \lim_{{\rm
x}\to 0^\pm}\phi(\rx)$.

In view of (\ref{sec9-H-we}), we can express the eigenvalue equation
for $H$ in the form
    \be
    H_\pm \psi_\pm(\rx)=E\psi_\pm(\rx),~~~~~{\rm for}~~~~~\rx\in\R^\pm,
    \label{sec9-eg-va-pm}
    \ee
where
    \be
    H_\pm:=-e^{\pm 2i\theta} \frac{d^2}{d\rx^2}+
    V(e^{\mp i\theta}\,\rx),
    \label{sec9-ceq4}
    \ee
and $\psi_-:(-\infty,0]\to\C$, $\psi_+:[0,\infty)\to\C$ are defined
by $\psi_\pm(\rx):=\psi(\rx)$ for all $\rx\in\R^\pm$, $\psi_\pm(0):=
\psi(0^\pm)$, and $\psi'_\pm(0):=\psi'(0^\pm)$.

In summary, the eigenvalue problem for the Schr\"odinger operator
$-\frac{d^2}{dz^2}+V(z)$ that is defined by the boundary condition
(\ref{sec9-BC}) along $\Gamma^{(\theta)}$ is equivalent to finding a
pair of functions $\psi_\pm$ satisfying
    \bea
    &&-e^{\pm 2i\theta} \psi_\pm''(\rx)+
    V(e^{\mp i\theta}\,\rx)\:\psi_\pm(\rx)=E\:\psi_\pm(\rx)
    ~~~~~{\rm for}~~~~~\rx\in\R^\pm,
    \label{sec9-eg-va-1}\\
    &&\psi_-(0)=\psi_+(0),~~~~~~~~~~~~~~~~~~
    e^{-2i\theta}\psi_-'(0)=e^{2i\theta}\psi_+'(0),
    \label{sec9-eg-va-3}\\
    && \psi_\pm(\rx)\to 0~~~{\rm
    exponentially~as}~~~\rx\to\pm\infty.
    \label{sec9-BS-pm}
    \eea
To elucidate the practical advantage of this formulation, we explore
its application for the potentials $V_\nu(z)=\lambda\,z^2(iz)^\nu$
with $\lambda\in\R^+$.

As we explained in the preceding subsection, we need to choose a
contour that belongs to the union of the Stokes wedges $S^\pm_\nu$
asymptotically. We shall choose the wedge-shaped contour
$\Gamma^{(\theta_\nu)}$ that consists of the bisectors of
$S^\pm_\nu$. Setting $V=V_\nu$ and $\theta=\theta_\nu$ in
(\ref{sec9-ceq4}) and using (\ref{sec9-thetas}) we find the
following most surprising result.
    \be
    H_\pm=e^{\pm 2i\theta_\nu}
    \left[-\frac{d^2}{d\rx^2}+\lambda\,|\rx|^{\nu+2}\right].
    \label{sec9-ceq5}
    \ee
Similarly, (\ref{sec9-eg-va-1}) becomes
    \be
    -\psi_\pm''(\rx)+\lambda\,|\rx|^{\nu+2}\psi_\pm(\rx)=
    E\,e^{\mp 2i\theta_\nu}\psi_\pm(\rx)
    ~~~~~{\rm for}~~~~~\rx\in\R^\pm.
    \label{sec9-ceq5-1}
    \ee
The appearance of the real confining potential
$\lambda\,|\rx|^{\nu+2}$ in (\ref{sec9-ceq5}) and
(\ref{sec9-ceq5-1}) allows for an alternative proof of the
discreteness of the spectrum of the potentials $z^2(iz)^\nu$. See
\cite[Appendix]{jpa-2005a} for details.\footnote{It would be
interesting to see if this approach can lead to an alternative proof
of the reality of the spectrum.}

\subsection{Wrong-sign quartic potential}
\label{sec-quartic}

In the preceding section we showed how one can transform the
spectral problem for a potential defined along a complex contour to
one defined along $\R$. The form of the transformed Hamiltonian
operator depends on the choice of the contour and its
parametrization. This raises the natural question whether one can
choose an appropriate parameterized contour so that the transformed
Hamiltonian admits an easily constructible metric operator. The
wrong-sign quartic potential $V_2(z)=-\lambda\,z^4$, with
$\lambda\in\R^+$, is a remarkable example for which the answer to
this question is in the affirmative.

Let $\Gamma_2=\{\zeta(s)|s\in\R\}$ be the contour defined by
\cite{jm}
    \be
    \zeta(s):=-2i\sqrt{1+is},~~~~~~~\mbox{for all $s\in\R$}.
    \label{sec9-zeta=}
    \ee
If we parameterize $\Gamma_2$ using $x=\Re(\zeta(s))$, we find
$\Gamma_2=\{x+if(x)|x\in\R\}$ where $f$ is given by
    $f(x):=-\sqrt{x^2+1}$ for all $x\in\R$.
This shows that $\Gamma_2$ is a hyperbola with asymptotes
$\ell_\pm:=\{ \pm r e^{-i\frac{\pi}{4}}|r\in\R^+\}$. Because
$\ell_\pm$ lies in the Stokes wedge $S_2^\pm$, (\ref{sec9-zeta=})
defines an admissible contour for the potential $V_2$.

If we perform the change of variable $z\to\zeta(s)$, we can express
the eigenvalue equation:
$\left[-\frac{d^2}{dz^2}-\lambda\,z^4\right]\Psi(z)=E\Psi(z)$,
in the form
    \be
    \left[-(1+is)\frac{d^2}{ds^2}-\frac{i}{2}\,\frac{d}{ds}-16
    \lambda\,(1+is)^2\right]\phi(s)=E\,\phi(s),
    \label{sec-9-H-quartic-0}
    \ee
where $\phi(s):=\Psi(\zeta(s))$. We can identify $s$ with a usual
(Hermitian) position operator, introduce the corresponding wave
number operator $\fK$ as $\br s|\fK:=-i\frac{d}{ds}\br s|$, and
express (\ref{sec-9-H-quartic-0}) as the eigenvalue equation for the
Hamiltonian
    \be
    H:=(1+is)\fK^2+\frac{\fK}{2}-16\lambda\,(1+is)^2
    \label{sec-9-H-quartic-1}
    \ee
that acts in $L^2(\R)$.

As shown in \cite{jm}, the application of the perturbative scheme of
Subsection~\ref{sec-pert} yields the following exact expressions for
a metric operator and the corresponding equivalent Hermitian
Hamiltonian, respectively.
    \bea
    \eta_+&=&\exp\left(\frac{ \fK^3}{48\lambda}-2
    \fK\right),
    \label{sec-9-metric=}\\
    h&=&
    \frac{ \fK^4}{64\lambda}-\frac{\fK}{2}+16\lambda\,s^2.
    \label{sec-9-equi-h=}
    \eea
We shall offer an alternative derivation of these formulas
momentarily.

Because $h$ is a Hermitian operator that is isospectral to $H$, the
spectrum of $H$ and consequently the operator
$-\frac{d^2}{dz^2}-\lambda\,z^4$ defined along $\Gamma_2$, is real.
It is also easy to show that the common spectrum of all these
operators is positive and discrete. To see this, we express $h$ in
its $\fK$-representation where eigenvalue equation $h\Phi=E\Phi$
reads
    \be
    \left[-16\lambda\,\frac{d^2}{d\fK^2}+
    \frac{\fK^4}{64\lambda}-\frac{\fK}{2}\right]\tilde\Phi(\fK)=E
    \,\tilde\Phi(\fK),
    \label{sec9-fourier}
    \ee
and
$\tilde\Phi(\fK):=\br\fK|\Phi\kt=(2\pi)^{-1/2}\int_{-\infty}^\infty
ds\: e^{-i\fK s}\Phi(s)$. The operator in the square bracket in
(\ref{sec9-fourier}) is a Schr\"odinger operator with a confining
quartic polynomial potential. Therefore its spectrum is positive and
discrete \cite{messiah}.

We can also use the same approach to treat the quartic anharmonic
oscillator, $V(z)=\omega^2z^2-\lambda z^4$. Using the
parametrization (\ref{sec9-zeta=}), we find the following
generalizations of (\ref{sec-9-H-quartic-1}) --
(\ref{sec9-fourier}).
    \bea
    &&H:=(1+is)\fK^2+\frac{\fK}{2}-16\lambda\,(1+is)^2-4\omega^2(1+is)
    \label{sec-9-H-quartic-1-an}\\
    &&\eta_+=\exp\left[\frac{ \fK^3}{48\lambda}-(2+\frac{
    \omega^2}{4\lambda})\fK\right],
    \label{sec-9-metric=an}\\
    &&h=
    \frac{(\fK^2-4\omega^2)^2}{64\lambda}-\frac{\fK}{2}+16\lambda\,s^2.
    \label{sec-9-equi-h=an}\\
    &&\left[-16\lambda\,\frac{d^2}{d\fK^2}+
    \frac{(\fK^2-4\omega^2)^2}{64\lambda}-
    \frac{\fK}{2}\right]\tilde\Phi(\fK)=E
    \,\tilde\Phi(\fK).
    \label{sec9-fourier-an}
    \eea
Therefore, by scaling the eigenvalues according to $E\to\gamma E$
where $\gamma:=1/(16\lambda)$, we can identify the spectrum of the
operator $-\frac{d^2}{dz^2}+\omega^2z^2-\lambda z^4$ (defined along
$\Gamma_2$) with that of the operator
    \be
    -\frac{d^2}{dx^2}+
    \frac{\gamma}{4}\left[\gamma(x^2-4\omega^2)^2-2x\right]
    \label{sec9-linear-4-pot}
    \ee
that acts in $L^2(\R)$, \cite{jm}. This observation has previously
been made in \cite{buslaev-grecchi}. As discussed in
\cite{andrianov-2007}, the approach of \cite{andrianov-1982} also
leads to the same conclusion.

Next, we outline an alternative and more straightforward method of
constructing a metric operator for the Hamiltonian
(\ref{sec-9-H-quartic-1-an}).

If we separate the Hermitian and anti-Hermitian parts of $H$, we
find
    \be
    H=16\lambda\,s^2+\fK^2-\frac{\fK}{2}-(16\lambda+4\omega^2)+
    \frac{i}{2}\big\{s\,,\,\fK^2-4\omega^2-32\lambda\big\}.
    \label{sec9-H-find-eta0}
    \ee
We can combine the first and last terms on the right-hand side of
this equation to express $H$ in the form
    \be
    H=16\lambda\left[s+
    \frac{i(\fK^2-4\omega^2-32\lambda)}{32\lambda}\right]^2
    +\frac{(\fK^2-4\omega^2)^2}{64\lambda}-\frac{\fK}{2}.
    \label{sec9-H-find-eta1}
    \ee
As seen from this relation, the term responsible for the
non-Hermiticity of $H$ may be removed by a translation of $s$,
namely
    \be
    s\to s-\frac{i(\fK^2-4\omega^2-32\lambda)}{32\lambda}.
    \label{sec9-sim-zero}
    \ee
It is not difficult to see that such a translation is affected by a
$\fK$-dependent similarity transformation of the form $s\to
e^{g(\fK)}s\,e^{-g(\fK)}$. Recalling that for any analytic function
$g:\R\to\R$,
    \be
    e^{g(\fK)}s\,e^{-g(\fK)}=s-ig'(\fK),
    \label{sec9-translation}
    \ee
and comparing this equation with (\ref{sec9-sim-zero}), we find
    \be
    g(\fK)=\frac{\fK^3}{96\lambda}-(1+\frac{\omega^2}{8\lambda})+c.
    \label{sec9-f=}
    \ee
Here $c$ is an integration constant that we can set to zero without
loss of generality. In view of (\ref{sec9-translation}), we can map
$H$ to a Hermitian Hamiltonian $h$ according to
    \be
    H\to h:=e^{g(\fK)}H\,e^{-g(\fK)}=
    16\lambda\,s^2+
    \frac{(\fK^2-4\omega^2)^2}{64\lambda}-\frac{\fK}{2}.
    \label{sec9-h=derive}
    \ee
This is precisely the equivalent Hermitian Hamiltonian given by
(\ref{sec-9-equi-h=an}). Moreover, comparing (\ref{sec9-h=derive})
with the defining relation for the equivalent Hermitian Hamiltonian,
namely $h:=\rho\,H\rho^{-1}$, and recalling that
$\rho:=\sqrt\eta_+$, where $\eta_+$ is a metric operator associated
with the Hamiltonian $H$, we find $\eta_+=e^{2g(\fK)}$. In light of
(\ref{sec9-f=}), this coincides with the metric operator given by
(\ref{sec-9-metric=an}).

Next, we wish to explore the underlying classical system for the
pseudo-Hermitian quantum system defined by the potential
$V(z)=\omega^2z^2-\lambda z^4$ along $\Gamma_2$. If we view this
potential as an analytic continuation of $V(x)=\omega^2x^2-\lambda
x^4$, with $x$ denoting the Hermitian position operator, we should
identify the Hamiltonian operator for the system with
    \be
    H_{\Gamma_2}:=\frac{P_Z^2}{2m}+\Omega^2\,Z^2-\Lambda Z^4,
    \label{sec9-H=class4}
    \ee
where $\Omega\in\R$ and $\Lambda\in\R^+$ are dimensionful coupling
constants, and $Z$ and $P_Z$ are the dimensionful coordinate and
momentum operators along $\Gamma_2$. Using an arbitrary length scale
$\ell$, we can introduce the corresponding dimensionless quantities:
    \be
    \lambda:=\frac{2m\ell^6\Lambda}{\hbar^2},~~~
    \omega:=\frac{\sqrt{2m}\,\ell^2\Omega}{\hbar},~~~
    p_z:=\frac{\ell P_Z}{\hbar},~~~z:=\frac{Z}{\ell}.
    \label{sec9-trans=11}
    \ee
In terms of these the eigenvalue equation $H_{\Gamma_2}\Psi={\cal
E}\Psi$ takes the form
    \be
    \left[-\frac{d^2}{dz^2}+\omega^2z^2-\lambda\,z^4\right]\Psi(z)=
    E\Psi(z),
    \label{sec9-eg-va-dim}
    \ee
where $E:=2m\ell^2{\cal E}/\hbar^2$.

Now, we are in a position to apply our earlier results. Setting
$z=\zeta(s):=-2i\sqrt{1+is}$, we can identify (\ref{sec9-eg-va-dim})
with the eigenvalue equation for the pseudo-Hermitian Hamiltonian
(\ref{sec-9-H-quartic-1-an}). One might argue that because $z$ and
$p_z$ represent dimensionless position and momentum operators, the
same should also hold for $s$ and $\fK$, respectively. This suggests
to identify the dimensionful position ($x$) and momentum $(p)$
operators as
    \be
    x=\alpha\,\ell\, s,~~~~~~~p=\frac{\hbar\fK}{\alpha\,\ell},
    \label{sec9-trans=12}
    \ee
where $\alpha\in\R^+$ is an arbitrary constant. In view of
(\ref{sec-9-equi-h=an}), (\ref{sec9-trans=11}),
(\ref{sec9-H-find-eta1}), and (\ref{sec9-trans=12}), we obtain the
following expressions for the dimensionful pseudo-Hermitian and
equivalent Hermitian Hamiltonians.
    \bea
    H'&:=&\frac{\hbar^2\,H}{2m\ell^2}=
    \frac{(p^2-8m\,\tilde\ell^2\Omega^2)^2}{64\Lambda
    \tilde\ell^4}-
    \frac{\hbar\,p}{4m\tilde\ell}+16\Lambda\left[
    \tilde\ell\,x+\frac{i(\tilde\ell^{-2}p^2-8m\Omega^2
    -64m\ell^2\Lambda)}{64m\Lambda}\right]^2,
    \label{sec9-H-ful}\\
    h'&:=&\frac{\hbar^2\, h}{2m\ell^2}=
    \frac{(p^2-8m\,\tilde\ell^2\Omega^2)^2}{64\Lambda
    \tilde\ell^4}-
    \frac{\hbar\,p}{4m\tilde\ell}+16\,\Lambda\,\tilde\ell^2 x^2,
    \label{sec9-eq-h=}
    \eea
where $\tilde\ell:=\ell/\alpha$. Note that unlike $h'$ that only
involves the length scale $\tilde\ell$, $H'$ depends on both $\ell$
and $\tilde\ell$. The same is true for the metric operator $\eta_+$.
This shows that the pseudo-Hermitian quantum systems defined by the
Hamiltonian $H'$ and the metric operator $\eta_+$ with different
values of the parameter $\alpha$ are unitary-equivalent to a
Hermitian quantum system that depends on a single length scale
($\tilde\ell$). The latter is not, however, fixed by the Hamiltonian
(\ref{sec9-H=class4}) and the contour $\Gamma_2$ along which it is
defined.\footnote{If $\Omega\neq 0$, we can choose $\alpha$ such
that $\tilde\ell=\Omega/\sqrt\Lambda$.}

Supposing that $\tilde\ell$ is independent of $\hbar$, we can take
$\hbar\to 0$, $x\to x_c$, and $p\to p_c$ in (\ref{sec9-eq-h=}) to
obtain the underlying classical Hamiltonian. The result is
    \be
    H'_c=\frac{(p_c^2-8m\,\tilde\ell^2\Omega^2)^2}{64\Lambda
    \tilde\ell^4}-16\,\tilde\ell^2\Lambda\, x_c^2.
    \label{sec9-eq-class-h=}
    \ee
We can also introduce the pseudo-Hermitian position and momentum
operators (\ref{X-P}):
    \be
    X:=\eta_+^{-1/2}x\,\eta_+^{1/2}=x+
    \frac{i(\tilde\ell^{-2}p^2-8m\Omega^2
    -64m\ell^2\Lambda)}{64m\Lambda\tilde\ell},~~~~~
    P:=\eta_+^{-1/2}p\,\eta_+^{1/2}=p.
    \label{sec9-X-P}
    \ee
As expected, in terms of $X$ and $P$, the Hamiltonian $H'$ takes the
form
    \be
    H'=\frac{(P^2-8m\,\tilde\ell^2\Omega^2)^2}{64\Lambda
    \tilde\ell^4}-
    \frac{\hbar\,P}{4m\tilde\ell}+16\,\Lambda\,\tilde\ell^2 X^2.
    \label{sec9-H=XP}
    \ee
Furthermore, the $\eta_+$-pseudo-Hermitian canonical quantization of
the classical Hamiltonian (\ref{sec9-eq-class-h=}) yields
(\ref{sec9-H=XP}) except for the linear term in $P$.

The fact that the pseudo-Hermitian quantum systems defined by the
Hamiltonian (\ref{sec9-H=class4}) depend on an arbitrary length
scale has its root in our identification of $s$ with the relevant
dimensionless position operator. We will next consider an
alternative approach that incorporates the spectral equivalence of
the Hamiltonians (\ref{sec9-H=class4}) and
(\ref{sec9-linear-4-pot}). It is based on treating $\fK$ as the
appropriate dimensionless position operator. More specifically, it
involves replacing (\ref{sec9-trans=12}) with $x=\beta\,\ell \fK$
and $p=-\hbar\,s/(\beta\,\ell)$,
where $\beta$ is an arbitrary dimensionless real parameter. If we
set $\beta:=1/(4\sqrt\lambda)=(32 m\Lambda)^{-1/2}\ell^{-3}\hbar$,
we find the following $\ell$-independent expressions for the
equivalent Hermitian Hamiltonian $h'$ and the underlying classical
Hamiltonian $H'_c$: \footnote{Eq.~(\ref{xzxz1}) was previously
obtained in \cite{bbcjmo-prd} where the term in (\ref{xzxz1}) is
attributed to an anomaly in a path-integral quantization of the
complex classical Hamiltonian
$\frac{p_c^2}{2m}+\Omega^2z_c^2-\Lambda z_c^4$ along the contour
$\Gamma_2$. For a more careful treatment of this problem, see
\cite{jmr-prd}.}
    \bea
    h'&=&\frac{p^2}{2m}+
    4\Lambda\left(x^2-\frac{\Omega^2}{4\Lambda}\right)^2-
    \hbar\sqrt{\frac{2\Lambda}{m}}\,x,
    \label{xzxz1}\\
    H'_c&=&\frac{p_c^2}{2m}+
    4\Lambda\left(x_c^2-\frac{\Omega^2}{4\Lambda}\right)^2.
    \label{sec9-H-class=}
    \eea
Note, however, that $H'$ still depends on $\ell$:
    \be
    H'=\frac{1}{2m}\left[p-i\sqrt{\frac{m}{2}}\left(
    4\sqrt\Lambda\,x^2-\frac{\Omega^2}{\sqrt\Lambda}-
    8\sqrt\Lambda\,\ell^2\right)\right]^2+
    4\Lambda\left(x^2-\frac{\Omega^2}{4\Lambda}\right)^2-
    \hbar\sqrt{\frac{2\Lambda}{m}}\,x.
    \label{sec9-H-L-dep}
    \ee
This is also true for $\eta_+$. Again the quantum systems determined
by $H'$ and $\eta_+$ with deferent values of $\ell$ are
unitary-equivalent to a system that is independent of $\ell$.
Similarly to our earlier analysis, we can obtain an
$\ell$-independent expression for $H'$ in terms of the
pseudo-Hermitian position and momentum operators: $X=x$ and
$P=p-i\sqrt{\frac{m}{2}}\left(4\sqrt\Lambda\,x^2-
\frac{\Omega^2}{\sqrt\Lambda}-8\sqrt\Lambda\,\ell^2\right)$.
The result is $H'=\frac{P^2}{2m}+
    4\Lambda\left(X^2-\frac{\Omega^2}{4\Lambda}\right)^2-
    \hbar\sqrt{\frac{2\Lambda}{m}}\,X$.

\section{Complex Classical Mechanics versus Pseudo-Hermitian QM}
\label{sec6-Classical}

\subsection{Classical-Quantum Correspondence and Observables}
\label{sec6-CQ}

In Subsection~\ref{sec-quasi} we outlined a procedure that assigns
an underlying classical system for a given pseudo-Hermitian quantum
system with reference Hilbert space $L^2(\R^d)$. According to this
procedure, that we employed in Subsections~\ref{sec-cubic-osc},
\ref{sec-cubic-imag} and \ref{sec-quartic}, the classical
Hamiltonian $H_c$ may be computed using (\ref{class-H-2}). In other
words, to obtain $H_c$, we replace the standard position and
momentum operators with their classical counterparts in the
expression for the equivalent Hermitian Hamiltonian $h$ and evaluate
its $(\hbar\to 0)$-limit. We can quantize $H_c$ to yield $h$, if we
employ the standard canonical quantization scheme. We obtain the
pseudo-Hermitian Hamiltonian $H$, if we use the pseudo-Hermitian
canonical quantization scheme (\ref{ph-quantize}).\footnote{Clearly
this is true up to factor-ordering ambiguities/terms proportional to
positive powers of $\hbar$.}

By definition, a classical observable $O_c$ is a real-valued
function of the classical states $(\vec x_c,\vec p_c)$, i.e., points
of the phase space $\R^{2d}$. If we apply the usual (Hermitian)
canonical quantization program, the operator associated with a
classical observable $O_c(\vec x_c,\vec p_c)$ is given by
$o:=O_c(\vec x,\vec p)$ where $\vec x$ and $\vec p$ are the usual
Hermitian position and momentum operators. If we apply the
pseudo-Hermitian quantization, we find instead $O:=O_c(\vec X,\vec
P)$ where $\vec X$ and $\vec P$ are the pseudo-Hermitian position
and momentum operators, respectively (\ref{X-P}). The common feature
of both these quantization schemes is that they replace the usual
classical Poisson bracket,
    $\{A_c,B_c\}_{\rm PB}:=\sum_{j=1}^d \left(
    \frac{\partial A_c}{\partial{x_{cj}}}
    \frac{\partial B_c}{\partial{p_{cj}}}-
    \frac{\partial B_c}{\partial{x_{cj}}}
    \frac{\partial A_c}{\partial{p_{cj}}}\right)$,
of any pair of classical observables $A_c$ and $B_c$ with
$(i\hbar)^{-1}$ times the commutator of the corresponding operators
$A$ and $B$, $
    \{A_c,B_c\}_{\rm PB}\longrightarrow(i\hbar)^{-1}[A,B]$.
Let us also recall that given a pseudo-Hermitian quantum system
specified by the reference Hilbert space $\cH$, a quasi-Hermitian
Hamiltonian operator $H:\cH\to\cH$, and an associated metric
operator $\eta_+:\cH\to\cH$, the observables of the system are by
definition $\eta_+$-pseudo-Hermitian operators $O$ acting in $\cH$,
    \be
    O^\dagger=\eta_+O\,\eta_+^{-1}.
    \label{sec6-ph}
    \ee
It is absolutely essential to note that such an operator acquires
its physical meaning through the classical-to-quantum
correspondence:
    \be
    O_c\longrightarrow O,
    \label{sec6-quantize2}
    \ee
where $O_c$ is the classical observable corresponding to the
operator $O$. For example in conventional quantum mechanics, we
identify the operator $p:L^2(\R)\to L^2(\R)$ defined by $p\psi
=-i\hbar\psi'$, with the momentum of a particle moving on $\R$,
because $p$ corresponds to the classical momentum $p_c$ of the
underlying classical system. Without this correspondence, $p$ is
void of a physical meaning. It is merely a constant multiple of the
derivative operator acting in a function space.

The situation is not different in pseudo-Hermitian quantum
mechanics. Again the ($\eta_+$-pseudo-Hermitian) operators that
represent observables derive their physical meaning from their
classical counterparts through the pseudo-Hermitian version of the
classical-to-quantum correspondence. This is also of the form
(\ref{sec6-quantize2}), but the operator $O$ is now selected from
among the $\eta_+$-pseudo-Hermitian operators. As we discussed in
Subsection~\ref{sec-quasi}, if $o:\cH\to\cH$ denotes the Hermitian
observable associated with a classical observable $O_c$, the
corresponding pseudo-Hermitian observable is given by
Eq.~(\ref{O=ror}), i.e., $O:=\rho^{-1} o\, \rho$ where
$\rho:=\sqrt\eta_+$.

Whenever one deals with a symmetric, $\cP\cT$-symmetric,
diagonalizable Hamiltonian $H$ with a real spectrum, one can define
the observables as operators $O$ fulfilling the condition
\cite{bbj-erratum}
    \be
    O^T=\cC\cP\cT\,O\,\cC\cP\cT,
    \label{sec6-revised}
    \ee
where
    \be
    O^T:=\cT\,O^\dagger\cT
    \label{sec6-transpose}
    \ee
stands for the transpose of $O$, and $\cC$, $\cP$, $\cT$ are
respectively the charge, parity, and time-reversal operators that
are assumed to satisfy
    \be
    \cC^2=\cP^2=\cT^2=I,~~~~[\cC,\cP\cT]=[\cP,\cT]=[\cC,H]=0.
    \label{sec6-cpt}
    \ee
As we explained in Subsection~\ref{sec-sym}, we can relate $\cC$ to
an associated metric operator $\eta_+$ according to
    \be
    \cC=\eta_+^{-1}\cP.
    \label{sec6-c=1}
    \ee
Because $\cC^2=\cP^2=I$, we also have
    \be
    \cC=\cP\eta_+.
    \label{sec6-c=2}
    \ee

Inserting (\ref{sec6-transpose}) in (\ref{sec6-revised}) and making
use of (\ref{sec6-cpt}), we obtain
$\cT\,O^\dagger\cT=\cC\cP\cT\,O\,\cC\cP\cT=\cP\cT\cC\,O\,
\cC\cP\cT=\cT\cP\cC\,O\,\cC\cP\cT.$ In view of (\ref{sec6-cpt}),
(\ref{sec6-c=1}), and (\ref{sec6-c=2}), this relation is equivalent
to $O^\dagger=\cP\cC\,O\,\cC\cP=\cP(\cP\eta_+)\,O\,
(\eta_+^{-1}\cP)\cP=\eta_+ O\,\eta_+^{-1}$. Therefore
(\ref{sec6-revised}) implies the $\eta_+$-pseudo-Hermiticity of $O$.
The converse is also true for the cases that (\ref{sec6-revised})
can be applied consistently. This is actually not always the case.
For example, the application of (\ref{sec6-revised}) for the
Hamiltonian operator that commutes with $\cC\cP\cT$ gives $H^T=H$.
Therefore, unlike the $\eta_+$-pseudo-Hermiticity conditions
(\ref{sec6-ph}), (\ref{sec6-revised}) cannot be employed for
non-symmetric Hamiltonians. This shows that (\ref{sec6-revised}) has
a smaller domain of application than (\ref{sec6-ph}).

Another advantage of the requirement of $\eta_+$-pseudo-Hermiticity
(\ref{sec6-ph}) over the condition (\ref{sec6-revised}) is that it
makes the dynamical consistency of the definition of observables
more transparent. Recall that the main motivation for the
introduction of (\ref{sec6-revised}) in \cite{bbj-erratum} is that
its original variant \cite{bbj,bbj-ajp}, namely the requirement of
$\cC\cP\cT$-symmetry of $O$, i.e., $O=\cC\cP\cT\,O\,\cC\cP\cT$,
conflicts with the Schr\"odinger time-evolution in the Heisenberg
picture; in general the Heisenberg-picture operators
$O_H(t):=e^{-itH/\hbar}O e^{itH/\hbar}$ do not commute with
$\cC\cP\cT$ for $t\neq 0$, even if they do for $t=0$,
\cite{comment}. The reason why (\ref{sec6-revised}) does not suffer
from this problem is that it is a restatement of the
$\eta_+$-pseudo-Hermiticity of $O$. To see why the
Heisenberg-picture operators satisfy the latter condition for all
$t$, we first recall that because $H$ is $\eta_+$-pseudo-Hermitian,
it is a Hermitian operator acting in the physical Hilbert space
$\cH_{\rm phys}$ defined by the inner product
$\br\cdot|\cdot\kt_{_{\eta_+}}$. This implies that the
time-evolution operator $e^{-itH/\hbar}$ is a unitary operator
acting in $\cH_{\rm phys}$. Therefore, $e^{-itH/\hbar}O
e^{itH/\hbar}$ acts in $\cH_{\rm phys}$ as a Hermitian operator for
all $t$, i.e., it is $\eta_+$-pseudo-Hermitian for all $t$.
Alternatively, we could argue that because $H$ is
$\eta_+$-pseudo-Hermitian, $e^{-itH/\hbar}$ is
$\eta_+$-pseudo-unitary
\cite{ahmed-jain-pre,ahmed-jain-jpa,jmp-2005} and $e^{-itH/\hbar}O
e^{itH/\hbar}$ is $\eta_+$-pseudo-Hermitian.

\subsection{Complex Classical Systems and Compatible Poisson Brackets}

In their pioneering article \cite{bender-prl}, Bender and Boettcher
perform an asymptotic analysis of the spectral properties of the
complex potentials $V_\nu(z)=z^2(iz)^\nu$ that makes use of the
complex WKB-approximation. This involves the study of a certain type
of complex classical dynamical system $\cS_{BBM}$ that Bender,
Boettcher, and Meisenger (BBM) \cite{bender-jmp} identify with the
underlying classical system for the quantum system defined by
$V_\nu$. This approach, which has been the focus of attention in a
number of publications
\cite{nanayakkara,bcdm-2006,bhh-2007a,bhh-2007b}, is fundamentally
different from the prescription we used in
Subsections~\ref{sec-quasi} and \ref{sec6-CQ} to associate a
classical system ${\cal S}$ with a pseudo-Hermitian quantum system
$S$. In this subsection, we examine the structure of the complex
classical system $\cS_{BBM}$. For simplicity we consider complex
potentials $V$ that depend on a single complex variable
$\fz$.\footnote{The use of complex phase-space variables in standard
classical mechanics is an old idea \cite{marsden-ratiu}. See also
\cite{strocchi}. The subject of the present study is to consider
complex configuration variables.}

According to \cite{bender-jmp} the dynamics of $\cS_{BBM}$ is
determined by the Newton's equation
    \be
    m\ddot \fz=-V'(\fz),
    \label{sec6-newton}
    \ee
where $m\in\R^+$, each overdot stands for a time-derivative, and a
prime marks the differentiation with respect to $\fz$. We can
express (\ref{sec6-newton}) as a pair of first order differential
equations:
    \be
    m\dot\fz=\fp,~~~~~~~\dot\fp=-V'(\fz).
    \label{sec6-first-order}
    \ee
This is the Hamiltonian formulation of the dynamics of $\cS_{BBM}$;
introducing the complex Hamilton function
$\bfh:=\frac{\fp^2}{2m}+V(\fz)$,
we can express (\ref{sec6-first-order}) as the following pair of
Hamilton equations.
    \be
    \dot\fz=\frac{\partial\bfh}{\partial\fp},~~~~~~~~~~~
    \dot\fp=-\frac{\partial\bfh}{\partial\fz}.
    \label{sec6-H-eqs}
    \ee
The variables $\fz$ and $\fp$ are the coordinates of the phase space
of the system $\fP$ which is as a set identical to $\C^2$ and
$\R^4$. This observation suggests that, similarly to the quantum
systems defined along a complex contour, the complex classical
system $\cS_{BBM}$ might admit a formulation that involves real
variables. This is actually quite straightforward. We can define the
real variables
    \bea
    &&x:=\Re(\fz),~~~~~~y:=\Im(\fz),~~~~~~p:=\Re(\fp),~~~~~~q:=\Im(\fp),
    ~~~~~~V_r:=\Re(V),~~~~~~V_i :=\Im(V),
    \label{sec6-real-var}\\
    &&H_r:=\Re(\bfh)=\frac{p^2-q^2}{2m}+V_r(x,y),~~~~~~~~~~~
    H_i :=\Im(\bfh)=\frac{pq}{m}+V_i(x,y),
    \label{sec6-real-HH}
    \eea
and use the well-known relations
$\frac{\partial}{\partial\fz}=\frac{1}{2}\,\big(
    \frac{\partial}{\partial x}-i\frac{\partial}{\partial y}\big)$ and
    $\frac{\partial}{\partial\fp}=\frac{1}{2}\,\big(
    \frac{\partial}{\partial p}-i\frac{\partial}{\partial q}\big)$,
to turn the complex Hamilton equations (\ref{sec6-H-eqs}) to a
system of four real equations.

It turns out that the resulting system of equations and consequently
the complex Hamilton equations (\ref{sec6-H-eqs}) are not consistent
with the standard symplectic structure (Poisson bracket) on the
phase space $\C^2=\R^4$, \cite{CM-jmp-2007,pla-2006}. To see this,
let us also introduce $w_1:=x$, $w_2:=p$, $w_3:=y$, and $w_4:=q$.
Then the standard Poisson bracket on $\R^4$ takes the form
    \be
    \{A,B\}_{PB}=\sum_{j,k=1}^4 J^{(\rm st)}_{jk}\:
    \frac{\partial A}{\partial w_j}\,\frac{\partial B}{\partial w_j},
    \label{sec6-PB}
    \ee
where $J^{(\rm st)}_{jk}$ are the entries of the standard symplectic
matrix
    \be
    J^{(\rm st)}:=\left(\begin{array}{cccc}
    0 & 1 & 0 & 0 \\
    -1 & 0 & 0 & 0\\
    0 & 0 & 0 & 1\\
    0 & 0 & -1 & 0\end{array}\right),
    \label{sec6-J-standard}
    \ee
and $A$ and $B$ are a pair of classical observables (real-valued
functions of $w_j$).\footnote{$\Omega:=\sum_{j,k=1}^4
J_{jk}\,dw_j\wedge dw_k$ is the standard symplectic form on $\R^4$,
\cite{marsden-ratiu}.} Recall that given a Hamilton function $H$ on
the four-dimensional phase space obtained by endowing $\R^4$ with
the symplectic structure corresponding to the standard Poisson
bracket (\ref{sec6-PB}), we can express the Hamilton equations in
the form $\dot w_j=\{w_j,H\}_{\rm PB}$. If we express
(\ref{sec6-PB}) in terms of the complex variables $\fz$ and $\fp$,
we find that $\{\fz,\bfh\}_{\rm PB}=\{\fp,\bfh\}_{\rm PB}=0$,
\cite{pla-2006}. Therefore, it is impossible to formulate the
dynamics defined by (\ref{sec6-H-eqs}) using the standard symplectic
structure on $\C^2$.

This observation raises the problems of the existence, uniqueness,
and classification of the symplectic structures on $\C^2=\R^4$ that
are compatible with the dynamical equations (\ref{sec6-H-eqs}).
Ref.~\cite{CM-jmp-2007} gives a family of dynamically compatible
symplectic structures. Ref.~\cite{pla-2006} offers a complete
classification of such structures. The most general compatible
symplectic structure is defined by the following non-standard
Poisson bracket
    \be
    \lpb A,B\rpb=\sum_{j,k=1}^4 J_{jk}\:
    \frac{\partial A}{\partial w_j}\,\frac{\partial B}{\partial w_j},
    \label{sec6-gen-PB}
    \ee
where $J_{ij}$ are the entries of the symplectic matrix
    \be
    J:=\frac{1}{2}\left(\begin{array}{cccc}
    0 & 1+c & -a & -d\\
    -(1+c) & 0 & -d & -b\\
    a & d & 0 & -1+c\\
    d & b & 1-c & 0\end{array}\right),
    \label{rj=}
    \ee
and $a,b,c,d$ are arbitrary real parameters satisfying
$c^2+d^2-ab\neq 1$. Regardless of the choice of these parameters, we
have $\dot\fz=\lpb\fz,\bfh\rpb$ and $\dot\fp=\lpb\fp,\bfh\rpb$. A
particularly, simple choice is $a=b=c=d=0$ that yields
    \be
    J=J_0:=
    \frac{1}{2}\left(\begin{array}{cccc}
    0 & 1 & 0 & 0 \\
    -1 & 0 & 0 & 0\\
    0 & 0 & 0 & -1\\
    0 & 0 & 1 & 0\end{array}\right).
    \label{rj=zero}
    \ee

\subsection{Real Description of a Complex Classical System}

Among the basic results of classical mechanics is the uniqueness
theorem for symplectic structures on the phase space $\R^{2d}$,
\cite{arnold}. In order to explain the content of this theorem,
first we recall that a symplectic structure on $\R^{2d}$ is
determined by the corresponding Poisson bracket. Choosing a system
of coordinates $w_j$, we can express the latter in the form $\{
A,B\}_\cJ=\sum_{j,k=1}^{2d} \cJ_{jk}\frac{\partial A}{\partial
w_j}\,\frac{\partial B}{\partial w_j}$, where $\cJ$ is a real,
antisymmetric, nonsingular $2d\times 2d$ matrix. The above-mentioned
theorem states that there is always a system of (so-called Darboux)
coordinates in which $\{ A,B\}_\cJ$ takes the form of the standard
Poisson bracket. Application of this theorem for the Poisson bracket
(\ref{sec6-gen-PB}) yields a description of the dynamics defined by
the complex Hamiltonian $\bfh$ in terms of a real Hamiltonian $K$.

The construction of the Darboux coordinates associated with the most
general symplectic matrix (\ref{rj=}) is described in
\cite{pla-2006}. These coordinates take the following particularly
simple form for $a=b=c=d=0$.
    \be
    x_1:=\sqrt 2\, w_1=\sqrt 2\, x,~~~
    p_1:=\sqrt 2\, w_2=\sqrt 2\, p,~~~
    x_2:=\sqrt 2\, w_4=\sqrt 2\, q,~~~
    p_2:=\sqrt 2\, w_3=\sqrt 2\, y.
    \label{sec6-darboux}
    \ee
These are precisely the phase space coordinates used in \cite{xa} to
study the complex trajectories appearing in the semiclassical
treatment of the propagator for a quartic anharmonic oscillator.
They are subsequently employed in the description of
$\cP\cT$-symmetric models \cite{kaushal-singh}.

The use of the coordinates (\ref{sec6-darboux}) together with the
assumption that the potential $V$ is an analytic function, i.e., the
Cauchy-Riemann conditions, $\frac{\partial v_r}{\partial x}-
\frac{\partial v_i}{\partial y}=\frac{\partial v_r}{\partial y}+
\frac{\partial v_i}{\partial x}=0$, hold, lead to the following
remarkable observations.
    \begin{itemize}
    \item The equivalent real Hamiltonian $K$ that describes the
    dynamics is twice the real part of the complex Hamiltonian
    $\bfh$, \cite{xa,pla-2006},
        \be
        K=\frac{p_1^2-x_2^2}{2m}+2\, V_r\big(\mbox{$\frac{x_1}{2},
        \frac{p_2}{2}$}\big)=2H_r.
        \label{sec6-K=}
        \ee
    \item The imaginary part of $\bfh$, i.e.,
        \be
        H_i=\frac{x_2p_1}{2m}+V_i\big(\mbox{$\frac{x_1}{2},
        \frac{p_2}{2}$}\big)
        \label{sec6-int-mov}
        \ee
    is an integral of motion; $\{H_i,K\}_{\rm PB}=0$, \cite{pla-2006}.
    \end{itemize}
This implies that the dynamical system defined by the Hamilton
equations~(\ref{sec6-H-eqs}) is equivalent to a classical
Hamiltonian system with phase space $\R^4$ and the Hamiltonian $K=2
H_r$. Furthermore, in view of Liouville's theorem on integrable
systems, because the phase space is four-dimensional and $H_i$ is an
integral of motion that is functionally independent of $K$, this
system is completely integrable \cite{arnold,vilasi}.

The integral of motion $H_i$ generates a certain class of
transformations in the phase space that leave the dynamics
invariant. The infinitesimal form of these symmetry transformations
is given by \cite{pla-2006}:
    \bea
    x_1\to x_1+\delta x_1, && ~~~~~\delta x_1:=
    \xi \{x_1,H_i\}_{\rm PB}=\frac{\xi\,x_2}{2m},
    \label{sym1}\\
    x_2\to x_2+\delta x_2, && ~~~~~\delta x_2:=
    \xi \{x_2,H_i\}_{\rm PB}=\xi~\frac{\partial}
    {\partial{x_1}}\, V_r\big(\mbox{$\frac{x_1}{2},
        \frac{p_2}{2}$}\big),
    \label{sym2}\\
    p_1\to p_1+\delta p_1, && ~~~~~\delta p_1:=
    \xi \{p_1,H_i\}_{\rm PB}=\xi~\frac{\partial}
    {\partial p_2}\, V_r\big(\mbox{$\frac{x_1}{2},
        \frac{p_2}{2}$}\big),
    \label{sym3}\\
    p_2\to p_2+\delta p_2, && ~~~~~\delta p_2:=
    \xi \{p_2,H_i\}_{\rm PB}=-\frac{\xi\,p_1}{2m},
    \label{sym4}
    \eea
where $\xi$ is an infinitesimal real parameter.

The existence of these symmetry transformations is related to the
fact that the system involves a first class constraint. Choosing a
particular value $C$ for $H_i$, i.e., imposing the constraint
$\Phi:=H_i-C=0$, and moding out the above symmetry transformations
by identifying each orbit of these transformations with a single
point, we can construct a reduced dynamical system $\fS$ that has a
two-dimensional real phase space \cite{pla-2006}. This procedure has
been implemented for a class of monomial potentials in
\cite{smilga1} where the above symmetry transformations have been
examined in the Lagrangian formulation and the difficult issue of
the quantization of these systems has been addressed in some
detail.\footnote{The application of this approach for some
multi-dimensional systems is studied in \cite{ghosh-majhi}.} In
general the resulting quantum system depends on whether one imposes
the constraint before or after the quantization. In the former case
the prescription used to obtain the reduced classical system also
affects the resulting quantum system. For the imaginary cubic
potential that allows for a fairly detailed analysis, one obtains a
variety of quantum systems \cite{smilga1}. But none of these
coincides with the pseudo-Hermitian quantum system we studied in
Subsection~\ref{sec-cubic-imag}.

In summary the identification of the complex classical system
$\cS_{\rm BBM}$ with the classical limit of the pseudo-Hermitian
quantum system $S$ that is defined by a complex potential $V$ meets
two serious difficulties. Firstly, while $S$ has a single real
degree of freedom (one-dimensional real configuration space and
two-dimensional real phase space), $\cS_{\rm BBM}$ has two real
degrees of freedom (a four-dimensional real phase space). Secondly,
the Hamilton equations (\ref{sec6-H-eqs}) that determine the
dynamics of $\cS_{\rm BBM}$ are not consistent with the standard
symplectic structure on the phase space of $\cS_{\rm BBM}$. This in
particular means that under the naive correspondence $\cS_{\rm
BBM}\rightarrow S$, the Hamilton equations (\ref{sec6-H-eqs}) are
not mapped to the Heisenberg equations for $S$. These problems
persist regardless of whether the Hamiltonian operator for $S$ is
defined on the real line or a complex contour. In fact, the study of
the systems defined on a complex contour reveals another difficulty
with the naive correspondence $\cS_{\rm BBM}\rightarrow S$, namely
that while the definition of $S$ requires making a proper choice for
a contour, $\cS_{\rm BBM}$ is independent of such a
choice.\footnote{The phase-space path integral formulation of the
wrong-sign quartic potential defined on the contour
(\ref{sec9-zeta=}) is consistent with the standard Hilbert-space
formulation, if one restricts the complex classical Hamiltonian
$\bfh$ to the contour (\ref{sec9-zeta=}), \cite{bbcjmo-prd,jmr-prd}.
Whether imposing this restriction would lead to a particular reduced
classical system that is identical with the classical system $\cS$
defined by the Hamiltonian~(\ref{sec9-H-class=}) is worthy of
investigation.}

A proper treatment of the complex dynamical systems $\cS_{\rm BBM}$
requires the investigation of a dynamically compatible symplectic
structure. Once such a structure is selected one can adopt a
corresponding set of Darboux coordinates and use them to obtain a
standard (real) description of $\cS_{\rm BBM}$. The use of the real
description reveals the curious fact that this system admits an
integral of motion that is functionally independent of the
Hamiltonian. This has two important consequences:
    \begin{enumerate}
    \item The system is completely integrable;
    \item The system has a first class constraint.
    \end{enumerate}
The choice $H_i=0$ that is adopted in
\cite{bcdm-2006,bhh-2007a,bhh-2007b} is just one way of imposing the
constraint. It corresponds to restricting the dynamics to a
three-dimensional subspace of the phase space. On this subspace
there act the symmetry transformations (\ref{sym1}) -- (\ref{sym4})
that leave the dynamics invariant. Moding out these transformations,
one finds a reduced dynamical system $\fS$ with a two-dimensional
real phase space. It is the latter that can, in principle, be
related to the quantum system $S$. So far the existence and nature
of such a relationship could not be ascertained. For the simple
polynomial potentials that could be studied carefully, the various
known ways of constructing and quantizing $\fS$ lead to quantum
systems that differ from $S$.

Another important issue is that the above procedure of constructing
complex dynamical systems $\cS_{\rm BBM}$ and the corresponding
reduced systems $\fS$ may be carried through for any complex
analytic potential. But, not every such potential defines a unitary
pseudo-Hermitian quantum system. A typical example is the
exponential potential $V(x)=e^{i\kappa x}$ that is defined on $\R$.
It is well-known that the spectrum of the Hamiltonian operator
$H=p^2/(2m)+\epsilon\, e^{i\kappa x}$, with $m\in\R^+$ and
$\epsilon,\kappa\in\R-\{0\}$, includes spectral singularities
\cite{gasymov}. This shows that this operator cannot be mapped to a
Hermitian operator by a similarity transformation, i.e., it is not
quasi-Hermitian. Hence $H$ is not capable of defining a unitary
quantum system. Although the potential $V(x)=e^{i\kappa x}$ defines
a complex dynamical system $\cS_{\rm BBM}$, it cannot be related to
a unitary quantum system.\footnote{For a study of the classical
dynamics generated by this exponential potential, see
\cite{CM-jmp-2007}.}

We conclude this section by underlining a rather interesting
parallelism between the quantum and classical mechanics of complex
(analytic) potentials. Supposing that a complex (analytic) potential
$V$ defines a unitary pseudo-Hermitian quantum system, we showed in
Section~\ref{sec-phqm} that this system admits an equivalent
Hermitian representation. The above discussion reveals a classical
analogue of this equivalence; the complex dynamical system defined
by the complex Hamiltonian $H=p^2/(2m)+V$ admits an equivalent
description involving a real Hamiltonian. What differentiates the
pseudo-Hermitian and Hermitian representations of quantum mechanics
is the choice of the inner product (equivalently a metric operator)
on the space of state-vectors. What differentiates the complex and
real representations of the classical mechanics is the choice of the
symplectic structure (equivalently Poisson bracket) on the phase
(state) space. Mathematically the equivalence of the
pseudo-Hermitian and Hermitian representations of quantum mechanics
stems from the uniqueness theorem for separable Hilbert spaces. The
classical counterpart of this theorem that is responsible for the
above-mentioned equivalence of the complex and real representations
of the classical mechanics is the uniqueness theorem for symplectic
manifolds diffeomorphic to $\R^{2d}$.

\section{Time-Dependent Hamiltonians and Path-Integral Formulation}

\subsection{Time-dependent Quasi-Hermitian Hamiltonians}
\label{sec-time-dep}

Time-dependent Hamiltonian operators arise in a variety of
applications of conventional quantum mechanics. Their
time-dependence does not cause any difficulties except that for the
cases that the eigenvectors of the Hamiltonian are time-dependent,
the time-evolution operator takes the form of a time-ordered
exponential involving the Hamiltonian \cite{gp-book}.\footnote{A
rather common misconception in dealing with time-dependent
Hamiltonians is to think that the time-reversal operator $\cT$
changes the sign of the time variable $t$ in the Hamiltonian, i.e.,
$H(t)\to \cT H(t)\cT=H(-t)$ which in view of the definition of
$\cT$, namely for all $\psi\in L^2(\R)$, $(\cT\psi)(x):=\psi(x)^*$,
is generally false. See for example \cite{yuce}, where the author
considers a trivial non-Hermitian time-dependent Hamiltonian that is
obtained from a constant $\cP\cT$-symmetric Hamiltonian through a
time-dependent point transformation and a time-reparametrization.}
The situation is quite different when one deals with time-dependent
quasi-Hermitian Hamiltonians.\footnote{Time-dependent
quasi-Hermitian Hamiltonians arise naturally in the application of
pseudo-Hermitian quantum mechanics in quantum cosmology
\cite{cqg,ap}. See also \cite{pla-2004,FF}.} As the following no-go
theorem shows the observability of the Hamiltonian and the unitarity
of the time-evolution put a severe restriction on the way a
quasi-Hermitian Hamiltonian can depend on time \cite{plb-2007}.
    \begin{center}
    \parbox{15.5cm}{\textbf{Theorem~2:} \emph{Let $T\in\R^+$ and
    for all $t\in[0,T]$, $H(t)$ be a time-dependent quasi-Hermitian
    operator acting in a reference Hilbert space $\cH$. Suppose that
    $H(t)$ serves as the Hamiltonian operator for a pseudo-Hermitian
    quantum system with physical Hilbert space $\cH_{\rm phys}$.
    If the time-evolution of the system, that is determined by
    the Schr\"odinger equation: $i\hbar\dot\psi(t)=H(t)\psi(t)$,
    is unitary and $H(t)$ is an observable for
    all $t\in[0,T]$, then the metric operator defining
    $\cH_{\rm phys}$ does not depend on time, i.e., there must exist
    a time-independent metric operator $\eta_+$ such that $H(t)$
    is $\eta_+$-pseudo-Hermitian for all $t\in[0,T]$.}}
    \end{center}

Following \cite{plb-2007} we call a time-dependent quasi- (pseudo-)
Hermitian Hamiltonian admitting a time-independent metric
(pseudo-metric) operator \emph{quasi-stationary}. Theorem~2 states
that in pseudo-Hermitian quantum mechanics we are bound to use
quasi-stationary Hamiltonians. To demonstrate the severity of this
restriction, consider two-level quantum systems where the
Hamiltonian $H(t)$ may be represented by a $2\times 2$ complex
matrix $\underline{H(t)}$ with possibly time-dependent entries. The
requirement that $H(t)$ is quasi-Hermitian implies that
$\underline{H(t)}$ involves 6 independent real-valued functions
(because its eigenvalues are real). The additional requirement that
$H(t)$ is quasi-stationary reduces this number to 4.\footnote{This
can be easily inferred from the results of \cite{tjp}.} This is also
the same as the maximum number of independent real-valued functions
that a general time-dependent Hermitian Hamiltonian can include.

A simple implication of Theorem~2 is that the inner product of the
physical Hilbert space cannot depend on time, unless one defines the
dynamics of the quantum system by an operator that is not observable
or allows for nonunitary time-evolutions. In other words, insisting
on observability of the Hamiltonian operator and requiring unitarity
prohibit scenarios involving switching Hilbert spaces as proposed in
\cite{bbj-2007}.

\subsection{Path-Integral Formulation of Pseudo-Hermitian QM}
\label{path-integral}

Among the original motivations to consider $\cP\cT$-symmetric
quantum mechanical models is the potential applications of their
relativistic and field theoretical generalizations
\cite{bbj,bbj-ajp} in elementary particle physics. A necessary first
step in trying to explore the relativistic and field theoretical
generalizations of $\cP\cT$-symmetric or more generally
pseudo-Hermitian QM is a careful examination of its path-integral
formulation. In this section we use the approach of \cite{prd-2007}
to elucidate the role of the metric operator in the path-integral
formulation of pseudo-Hermitian QM and demonstrate the equivalence
of the latter with the path-integral formulation of the conventional
QM.

We shall first review the emergence of path integrals in dealing
with a simple conventional (Hermitian) quantum system. This requires
a brief discussion of the trace of a linear operator.

Let $L:\cH\to\cH$ be a linear operator acting in a separable Hilbert
space $\cH$ with inner product $\br\cdot|\cdot\kt$. Then the
\emph{trace} of $L$ is defined by
    \be
    {\rm tr}(L):=\sum_{n=1}^N \br\xi_n|L\xi_n\kt,
    \label{trace=0}
    \ee
where $\{\xi_n\}$ is an arbitrary orthonormal basis of $\cH$,
\cite{reed-simon}. Obviously, for $N=\infty$, the right-hand side of
(\ref{trace=0}) may not converge, and $\tr(L)$ may not exist.

Suppose that $K$ and $L$ are linear operators for which ${\rm
tr}(KL)<\infty$. Then invoking the completeness relation for $\xi_n$
and using Dirac's bra-ket notation, we can show that
    \bea
    {\rm tr}(LK)&=&
    \sum_{m,n=1}^N \br\xi_n|L|\xi_m\kt\br\xi_m|K|\xi_n\kt
    =\sum_{m,n=1}^N \br\xi_m|K|\xi_n\kt\br\xi_n|L|\xi_m\kt
    =\sum_{m=1}^N \br\xi_m|KL|\xi_m\kt={\rm tr}(KL).
    \label{trace-id}
    \eea
A simple implication of this identity is that the right-hand side of
(\ref{trace=0}) is independent of the choice of the orthonormal
basis $\{\xi_n\}$.\footnote{To see this, let $\{\zeta_n\}$ be
another orthonormal basis of $\cH$. Then as we described in
Section~\ref{sec-unitary}, $\zeta_n$ are related to $\xi_n$ by a
unitary operator $U:\cH\to\cH$, $\zeta_n=U\xi_n$. This in turn
implies $\sum_{n=1}^N \br\zeta_n|L\zeta_n\kt= \sum_{n=1}^N \br
U\xi_n|L U\xi_n\kt= \sum_{n=1}^N \br\xi_n|U^\dagger (L U)\xi_n\kt=
\sum_{n=1}^N \br\xi_n|(LU)U^\dagger\xi_n\kt= \sum_{n=1}^N
\br\xi_n|L\xi_n\kt,$ where we have used (\ref{trace-id}) and
$UU^\dagger=I$.}

For a linear operator $L$ acting in $L^2(\R)$, we can use the
position basis $\{|x\kt\}$ to compute $\tr(L)$. To demonstrate how
this is done, let $\{\xi_n\}$ be an orthonormal basis of $L^2(\R)$.
Using (\ref{trace=0}), the completeness relation for $|x\kt$ and
$\xi_n$, and Dirac's bra-ket notation, we have
    \bea
    \tr(L)&=&
    \sum_{n=0}^\infty \int_{-\infty}^\infty \int_{-\infty}^\infty
    dx\,dx' \br\xi_n|x\kt\br x|L
    |x'\kt\br x'|\xi_n\kt=\int_{-\infty}^\infty \int_{-\infty}^\infty dx\,dx'
    \br x|L|x'\kt\sum_{n=0}^\infty \br x'|\xi_n\kt\br\xi_n|x\kt
    \nn\\
    &=&
    \int_{-\infty}^\infty \int_{-\infty}^\infty dx\,dx',
    \br x|L|x'\kt~\br x'|x\kt
    =\int_{-\infty}^\infty dx \:\br x|L|x\kt.
    \label{trace-x-basis}
    \eea

Now, consider a quantum system defined by the Hilbert space
$\cH=L^2(\R)$ and a Hermitian Hamiltonian $H$ that is an analytic
(or piecewise analytic) function of the usual (Hermitian) position
operator ($x$), momentum operator ($p$), and possibly time ($t$).
The generating functional (also called partition function) for the
$n$-point (correlation) functions of the system is given by
    \be
    Z[J]=\tr\left(
    \fT\exp\left\{-\frac{i}{\hbar}\int_{t_1}^{t_2}dt\:[H-Jx]\right\}\right),
    \label{partition-fn}
    \ee
where ``$\fT\exp$'' denotes the time-ordered exponential, $t_1$ and
$t_2$ are respectively the initial and final times for the evolution
of the system that are taken to be $-\infty$ and $\infty$ in the
scattering setups used particularly in quantum field theory, and $J$
stands for the (possibly time-dependent) coupling constant for the
source terms $Jx$, \cite{das,weinberg}. The latter is by definition
an observable \cite{bryce-qft}. In view of the fact that $x$ is also
an observable, this implies that $J$ must be real-valued.

One can easily justify the condition of the observability of the
source term by noting that $Z[J]$ is used to compute the $n$-point
functions according to \cite{das}:
    $\big\langle \fT [x(\tau_1)x(\tau_2)\cdots x(\tau_n)]\big\rangle=\left.
    \frac{(-i\hbar)^n}{Z[J]}\,\frac{\delta^n Z[J]}{\delta J(\tau_1)\delta J(\tau_2)
    \cdots\delta J(\tau_n)}\right|_{J=0}$,
where $\tau_1,\tau_2,\cdots,\tau_n\in[t_1,t_2]$, $x(\tau)$ denotes
the position operator in the Heisenberg picture, i.e., $x(\tau):=
U_{_J}(\tau,t_1)\, x\, U_{_J}(t_1,\tau)$, and $U_{_J}$ is the
time-evolution operator associated with the interacting system; for
all $\ft_1,\ft_2\in\R$,
    \be
    U_{_J}(\ft_1,\ft_2):=\fT\exp\left\{-\frac{i}{\hbar}\int_{\ft_1}^{\ft_2}
    dt\:[H-Jx]\right\}.
    \label{sec7-evolution-op}
    \ee
The $n$-point functions are essentially the expectation values of
the observables $\fT [x(\tau_1)\cdots x(\tau_n)]$ (in the ground
state of the system if $t_2=-t_1\to\infty$.) Therefore, the
observability of the source term $Jx$ is linked to the observability
of the Heisenberg-picture position operators $x(\tau_i)$. The latter
is ensured by the unitarity of the time-evolution operator
$U_{_J}(\ft_1,\ft_2)$ and the observability of the
Schr\"odinger-picture position operator $x$.

In view of (\ref{trace-x-basis}), we can express the partition
function (\ref{partition-fn}) in the form
    \be
    Z[J]=\int_{-\infty}^\infty dx\,\br x|
    \fT\exp\left\{-\frac{i}{\hbar}\int_{t_1}^{t_2}dt\:[H-Jx]\right\}|x\kt=
    \int_{-\infty}^\infty dx\, \br x,t_1|x,t_2\kt,
    \label{partition-fn2}
    \ee
where
    \be
    |x,t\kt:=U_{_J}(0,t)^\dagger|x\kt,
    \label{sec7-xt-ket}
    \ee
are the (generalized) eigenfunctions of the Heisenberg-picture
position operator $x(t)$. In light of (\ref{sec7-evolution-op}) and
(\ref{sec7-xt-ket}), we also have, for all $x_1,x_2\in\R$, $\br
x_1,t_1|x_2,t_2\kt=\br x_1|U_{_J}(t_1,t_2)|x_2\kt=\br x_1|
\fT\exp\left\{-\frac{i}{\hbar}\int_{t_1}^{t_2}dt\:[H-Jx]\right\}|x_2\kt$.
Computing this quantity as a phase-space path integral and
substituting the result in (\ref{partition-fn2}), we find the
following phase-space path integral expression for the generating
functional.
    \be
    Z[J]=\int\int\cD(x)\,\cD(p)\;
    e^{-\frac{i}{\hbar}\int_{t_1}^{t_2}dt\,
    [H(x,p;t)-J(t)x]}.
    \label{partition-fn3}
    \ee
If $H$ is a quadratic polynomial in $p$, we can perform the momentum
path integral in (\ref{partition-fn3}) and convert it into a
configuration-space (Lagrangian) path integral. This yields
    \be
    Z[J]=\int\cD(x)\:
    e^{\frac{i}{\hbar}\int_{t_1}^{t_2}dt\,L_{_J}(x,\dot x;t)},
    \label{partition-fn4}
    \ee
where $L_{_J}(x,\dot x;t):=\dot x\,p-H(x,p;t)+J(t)x$,
and $p$ is to be identified with its expression obtained by solving
$\dot x=\partial H(x,p;t)/\partial p$ for $p$ as a function of $x$,
$\dot x$, and $t$.

Next, we consider the extension of the above constructions to a
system defined by the reference Hilbert space $\cH=L^2(\R)$, a
metric operator $\etap:L^2(\R)\to L^2(\R)$, and an
$\etap$-pseudo-Hermitian Hamiltonian $H:L^2(\R)\to L^2(\R)$ that is
again a (piecewise) analytic function of $x$, $p$, and possibly $t$.
According to the Theorem~2, in order for $H$ to be an observable
that generates a unitary time-evolution, it must be
quasi-stationary. As discussed in \cite{plb-2007}, this puts a
severe restriction on the form of allowed time-dependent
Hamiltonians.\footnote{In relativistic field theories, $H$ is
obtained  by integrating the Hamiltonian density over a space-like
hypersurface. This makes the time-dependence of $H$ quite arbitrary
and renders the imposition of the condition of quasi-stationarity of
$H$ an extremely difficult task.}

We can certainly work in the Hermitian representation $(\cH,h)$ of
the system where $h:=\rho\, H\,\rho^{-1}$ (with $\rho:=\sqrt\etap$)
is the equivalent Hermitian Hamiltonian (\ref{h-Hermitian}). In this
representation the generating functional has the form
    \be
    Z[J]=\tr\left(
    \fT\exp\left\{-\frac{i}{\hbar}\int_{t_1}^{t_2}dt\:[h-Jx]\right\}\right).
    \label{partition-fn-H}
    \ee
As we showed above this quantity admits a phase-space path integral
expression. But even for the case that $H$ is a quadratic polynomial
in $p$, the equivalent Hermitian Hamiltonian $h$ does not share this
property, and we cannot convert the right-hand side of
(\ref{partition-fn-H}) into a Lagrangian path integral in general.
This provides a concrete motivation for the derivation of a the
path-integral expression for the generating functional in the
pseudo-Hermitian representation of the system, i.e., $(\pH,H)$.

In \cite{bcm-2006} the authors use the
expression~(\ref{partition-fn4}) (with $t_2=-t_1\to\infty$) to
perform a perturbative calculation of the generating functional and
the one-point function for the $\cP\cT$-symmetric cubic anharmonic
oscillator (\ref{sec8-pt-sym-3}).\footnote{See also
\cite{bbmw-2002}.} As a result they find an imaginary value for the
one-point function. This is simply because the one-point function
they calculate corresponds to the ground state expectation value of
the usual position operator that is indeed not an observable of the
system. The physically meaningful generating functional is
\cite{prd-2007}
    \be
    Z[J]=\tr_\etap\!\left(
    \fT\exp\left\{-\frac{i}{\hbar}\int_{t_1}^{t_2}dt\:[H-JX]\right\}\right),
    \label{partition-fn-PH-etap}
    \ee
where for every linear operator $K:\cH\to\cH$,
    \be
    {\rm tr}_{\etap}(K):=\sum_{n=1}^N\br\psi_n| K\psi_n\kt_\etap=
    \sum_{n=1}^N\br\psi_n|{\etap}K\psi_n\kt,
    \label{ph-trace}
    \ee
$\{\psi_n\}$ is an arbitrary orthonormal basis of $\pH$, and $X$ is
the $\etap$-pseudo-Hermitian position operator (\ref{X-P}). The
$n$-point functions generated by (\ref{partition-fn-PH-etap})
correspond to the expectation value of time-ordered products of the
physical position operators and the resulting numerical values are
necessarily real.

It is not difficult to show that $\tr_\etap=\tr$. To do this, first
we recall that because $\rho:=\sqrt\eta:\pH\to\cH$ is a unitary
operator, it maps orthonormal bases of $\pH$ onto orthonormal bases
of $\cH$. In particular, $\xi_n:=\rho\,\psi_n$ form an orthonormal
basis of $\cH$. This together with $\rho^2={\etap}$,
$\rho^\dagger=\rho$, (\ref{ph-trace}), (\ref{trace=0}), and
(\ref{trace-id}) imply
    \be
    {\rm tr}_{\eta_+}(K)=
    \sum_{n=1}^N\br\psi_n|\rho^2K\psi_n\kt=
    \sum_{n=1}^N\br\rho\,\psi_n|\rho K\psi_n\kt=
    \sum_{n=1}^N\br\xi_n|\rho K\rho^{-1}\xi_n\kt=
    {\rm tr}(\rho K\rho^{-1})={\rm tr}(K).
    \label{tr=tr-identity}
    \ee
This relation allows us to express (\ref{partition-fn-PH-etap}) in
the form
    \be
    Z[J]=\tr\left(
    \fT\exp\left\{-\frac{i}{\hbar}\int_{t_1}^{t_2}dt\:[H-JX]\right\}\right).
    \label{partition-fn-PH}
    \ee

Next, we employ the definitions of $h$ and $X$, namely $h:=\rho\,
H\,\rho^{-1}$ and $X:=\rho^{-1}x\,\rho$, and the fact that $\etap$
and consequently $\rho$ do not dependent on time, to establish
$\fT\exp\left\{-\frac{i}{\hbar}\int_{t_1}^{t_2}[h-Jx]\right\}=
    \rho\;\fT\exp\left\{-\frac{i}{\hbar}\int_{t_1}^{t_2}[H-JX]\right\}
    \rho^{-1}$.
In view of this relation and (\ref{trace-id}) the right-hand sides
of (\ref{partition-fn-H}) and (\ref{partition-fn-PH}) coincide. This
is another manifestation of the physical equivalence of Hermitian
and pseudo-Hermitian representations of the system.

As we emphasized in the preceding sections the metric operator plays
a fundamental role in the operator formulation of pseudo-Hermitian
quantum mechanics. The same is true about the path-integral
formulation
 of this theory. To elucidate this point we examine the nature of the
dependence of the generating functional on the choice of a metric
operator $\etap$.

A simple consequence of (\ref{trace-x-basis}) and
(\ref{partition-fn-PH}) is
    \be
    Z[J]=\int_{-\infty}^\infty fx\:\br
    x|\fT\exp\left\{-\frac{i}{\hbar}\int_{t_1}^{t_2}dt\:[H-JX]\right\}|x\kt.
    \label{partition-fn-PH-x-basis}
    \ee
Clearly, $Z[0]$ does not depend on $\etap$, \cite{prd-2007}. This
explains the results of \cite{jakubsky-2007} pertaining the
metric-independence of thermodynamical quantities associated with
non-interacting pseudo-Hermitian statistical mechanical models.
However, in contrast to the view expressed in \cite{jr-2007}, the
metric-independence of $Z[0]$ does not extend to $Z[J]$ with $Z\neq
0$. This is actually to be expected because the knowledge of $Z[J]$
allows for the calculation of the $n$-point functions that are
expectation values of the time-ordered products of the
Heisenberg-picture $\etap$-pseudo-Hermitian position operators
$X(\tau_i)$.

The dependence of $Z[J]$ on the choice of $\etap$ is rather
implicit. In the Hermitian representation, $\etap$ enters the
expression for $Z[J]$ through the equivalent Hermitian Hamiltonian
$h$. In the pseudo-Hermitian representation, this is done through
the source term $JX$. The presence of $X$ in
(\ref{partition-fn-PH-x-basis}) prevents one from obtaining a
Lagrangian path integral for $Z[J]$ even for the cases that $H$ is a
quadratic polynomial in $p$. Therefore, contrary to claims made in
\cite{bcm-2006}, in general, the pseudo-Hermitian representation is
not practically superior to the Hermitian representation. There are
certain calculations that are performed more easily in the
pseudo-Hermitian representation, and there are others that are more
straightforward to carry out in the Hermitian representation
\cite{cjp-2004b}.

\section{Geometry of the State-Space and the Quantum Brachistochrone}

\subsection{State-Space and Its geometry in Conventional QM}
\label{geom-conv-QM}

In conventional quantum mechanics the states are not elements of the
Hilbert space $\cH$, but the rays (one-dimensional subspaces) of the
Hilbert space.\footnote{Throughout this article the word ``state''
is used to mean ``pure state''.} The space of all rays that is
usually called the \emph{projective Hilbert space} and denoted by
$\cP(\cH)$ has the structure of a manifold. For an $N$-dimensional
Hilbert space $\cH$, $\cP(\cH)$ is the complex projective space $\C
P^{N-1}$. This is a compact manifold for finite $N$ and a well-known
infinite-dimensional manifold with very special and useful
mathematical properties for infinite $N$, \cite{gp-book}.

The projective Hilbert space $\cP(\cH)$ is usually endowed with a
natural geometric structure that is of direct relevance to physical
phenomena such as geometric phases \cite{page} and optimal-speed
unitary evolutions in quantum mechanics \cite{anandan-aharonov}. To
describe this structure, we need an appropriate representation of
the elements of $\cP(\cH)$. This is provided by the projection
operators associated with the states.

Consider a state $\lambda_\psi$ represented by a state-vector
$\psi\in\cH-\{0\}$, i.e., $\lambda_\psi=\{c\psi| c\in\C\}$, and the
projection operator
    \be
    \Lambda_\psi:=\frac{|\psi\kt\br\psi|}{\br\psi|\psi\kt}.
    \label{state=}
    \ee
Clearly the relation between states $\lambda_\psi$ and state-vectors
$c\psi$ is one to (infinitely) many. But the relation between the
states $\lambda_\psi$ and the projection operators $\Lambda_\psi$ is
one-to-one. This suggests using the latter to identify the elements
of the projective Hilbert space $\cP(\cH)$. This parametrization of
$\cP(\cH)$ has the advantage of allowing us to use the algebraic
properties of the projection operators (\ref{state=}) in the study
of states.

An important property of (\ref{state=}) is that it is a positive
operator having a unit trace. The positivity of $\Lambda_\psi$ is a
simple consequence of the identities
    \be
    \Lambda_\psi^\dagger=\Lambda_\psi,~~~~~~~~~
    \Lambda_\psi^2=\Lambda_\psi.
    \label{proj-ids}
    \ee

We recall from Subsection~\ref{path-integral} that the trace of a
linear operator $J:\cH\to\cH$ is defined by
    \be
    {\rm tr}(J):=\sum_{n=1}^N \br\xi_n|J\xi_n\kt,
    \label{trace=}
    \ee
where $\{\xi_n\}$ is an arbitrary orthonormal basis of $\cH$,
\cite{reed-simon}. If $L:\cH\to\cH$ is a linear operator such that
${\rm tr}(L^\dagger L)<\infty$, $L$ is said to be a
\emph{Hilbert-Schmidt operator}. In view of (\ref{state=}),
(\ref{proj-ids}), (\ref{trace=}) and (\ref{norm2}), we have
    \be
    {\rm tr}(\Lambda_\psi^\dagger\Lambda_\psi)=
    {\rm tr}(\Lambda_\psi^2)={\rm tr}(\Lambda_\psi)=
    \sum_{n=1}^N
    \frac{\br\xi_n|\psi\kt\br\psi|\xi_n\kt}{\br\psi|\psi\kt}
    =\frac{\sum_{n=1}^N |\br\xi_n|\psi\kt|^2}{\br\psi|\psi\kt}
    =\frac{\parallel\psi\parallel^2}{\br\psi|\psi\kt}=1.
    \label{unit-trace}
    \ee
Therefore, $\Lambda_\psi$ is a Hilbert-Schmidt operator with unit
trace.

The set $\fB_2(\cH)$ of Hilbert-Schmidt operators forms a subspace
of the vector space of bounded linear operators acting in $\cH$. We
can use ``${\rm tr}$'' to define the following inner product on
$\fB_2(\cH)$.
    \be
    ( L|J ):={\rm tr}(L^\dagger J)~~~~~~~~\mbox{for all}
    ~~~~L,J\in\fB_2(\cH).
    \label{trace-inn}
    \ee
This is called the \emph{Frobenius} or \emph{Hilbert-Schmidt inner
product} \cite{horn-johnson2,reed-simon}. It has the appealing
property that given an orthonormal set $\{\chi_n\}$ of state-vectors
the corresponding set of projection operators $\{\Lambda_{\chi_n}\}$
forms an orthonormal subset of $\fB_2(\cH)$;
$\br\chi_m|\chi_n\kt=\delta_{mn}$ implies
$(\Lambda_{\chi_n}|\Lambda_{\chi_m})=\delta_{mn}$.

The set $\fH_2(\cH)$ of Hermitian Hilbert-Schmidt operators to which
the projection operators $\Lambda_\psi$ belong is a subset of
$\fB_2(\cH)$ that forms a real vector space with the usual addition
of linear operators and their scalar multiplication. It is not
difficult to see, with the help of (\ref{trace-id}), that
(\ref{trace-inn}) reduces to a real inner product on $\fH_2(\cH)$,
namely
    \be
    ( L|J ):={\rm tr}(L J)~~~~~~~~\mbox{for all}
    ~~~~L,J\in\fH_2(\cH).
    \label{trace-inn2}
    \ee
Therefore endowing $\fH_2(\cH)$ with this inner product produces a
real inner product space.

By identifying states $\lambda_\psi$ with the projection operators
$\Lambda_\psi$ we can view the projective Hilbert space $\cP(\cH)$
as a subset of $\fH_2(\cH)$ and use the inner product
(\ref{trace-inn2}) to endow $\cP(\cH)$ with a natural metric tensor.
The corresponding line element $ds$ at $\Lambda_\psi$ is given by
    \be
    ds^2(\Lambda_\psi):= \frac{1}{2}\,( d\Lambda_\psi|d\Lambda_\psi )=
    \frac{\br\psi|\psi\kt\br d\psi|d\psi\kt-|
    \br\psi|d\psi\kt|^2}{\br\psi|\psi\kt^2},
    \label{Fubini-Study=}
    \ee
where we have inserted a factor of $\frac{1}{2}$ to respect a
mathematical convention and used (\ref{state=}), (\ref{trace=}),
(\ref{trace-inn2}), and the fact that $\{\xi_n\}$ is an orthonormal
basis of $\cH$.

For $N<\infty$, we can identify $\psi$ with a nonzero complex column
vector $\vec\fz$ with components $\fz_1,\fz_2,\cdots,\fz_N$. In
terms of these we can express (\ref{Fubini-Study=}) in the form
$ds^2=\sum_{a,b=1}^N g_{ab^*}d\fz_ad\fz_b^*$ where
    \be
    g_{ab^*}:=\frac{|\vec\fz|^2\delta_{ab}-\fz_a^*\fz_b}{
    |\vec\fz|^4}.
    \label{metric-global}
    \ee
This is precisely the Fubini-Study metric on the complex projective
space $\C P^{N-1}$, \cite{egh}.

As a concrete example, consider the case that $\cH$ is
two-dimensional, i.e., $N=2$. Then using the standard basis
representation of operators acting in $\C^2$, and noting that in
this case all operators are Hilbert-Schmidt, we can infer that
$\fH_2(\cH)$ is equivalent to the set of all Hermitian matrices.
This is a four-dimensional real vector space which we can identify
with $\R^4$. Specifically, we can represent each $J\in\fH_2(\cH)$
using its standard matrix representation:
    \be
    \underline J=\left(\begin{array}{cc}
    z & x-iy\\
    x+iy & w\end{array}\right),
    \label{R4=}
    \ee
and observe that these matrices are in one-to-one correspondence
with $(x,y,z,w)\in\R^4$.

The projective Hilbert space $\cP(\cH)$ is a two-dimensional subset
of the four-dimensional real vector space $\fH_2(\cH)$. To see this
let us choose an arbitrary state-vector $\psi\in\C^2-\{\vec
0\}$. Then $\psi=\mbox{\scriptsize$\left(\begin{array}{c} \fz_1\\
\fz_2\end{array}\right)$}$ for some $\fz_1,\fz_2\in\C$ such that
$|\fz_1|^2+|\fz_2|^2\neq 0$, and in view of (\ref{state=}) we can
represent $\Lambda_\psi$ by
    \be
    \underline{\Lambda_\psi}=\frac{1}{|\fz_1|^2+|\fz_2|^2}
    \left(\begin{array}{cc}
    |\fz_1|^2 & \fz_1 \fz_2^*\\
    \fz_1^* \fz_2 & |\fz_2|^2\end{array}\right).
    \label{state-rep=}
    \ee
Using the parametrization (\ref{R4=}), we find that for
$J=\Lambda_\psi$,
    \be
    x=\frac{\fz_1 \fz_2^*+\fz_1^* \fz_2}{
    2(|\fz_1|^2+|\fz_2|^2)},~~~~
    y=\frac{i(\fz_1 \fz_2^*-\fz_1^* \fz_2)}{
    2(|\fz_1|^2+|\fz_2|^2)},~~~~
    z=\frac{|\fz_1|^2}{|\fz_1|^2+|\fz_2|^2},~~~~
    w=\frac{|\fz_2|^2}{|\fz_1|^2+|\fz_2|^2}.
    \label{xyzw=}
    \ee
Therefore, as expected $w=1-z$, so that
    \be
    \underline{\Lambda_\psi}=
    \left(\begin{array}{cc}
    z & x-iy\\
    x+iy & 1-z\end{array}\right),
    \label{state-rep=xyz}
    \ee
and the condition $\Lambda_\psi^2=\Lambda_\psi$ takes the form
    \be
    x^2+y^2+(z-\frac{1}{2})^2=\frac{1}{4}.
    \label{two-sphere-1}
    \ee
This defines a two-dimensional sphere $S^2$ that we can use to
represent $\cP(\cH)$.

If we endow $\R^3$, that is parameterized by the Cartesian
coordinates $(x,y,z)$, with the Euclidean metric, we can identify
$S^2$ with a round sphere of unit diameter. We will next obtain an
expression for the metric induced on $S^2$ by the embedding
Euclidean space $\R^3$.

Let $\mathbf{N}$ and $\mathbf{S}$ respectively denote the north and
south poles of $S^2$, i.e., $\mathbf{N}:=(x=0,y=0,z=1)$ and
$\mathbf{S}:=(x=0,y=0,z=0)$, and consider the stereographic
projection of $S^2$ onto the tangent plane $\Pi_\mathbf{N}$ at
$\mathbf{N}$, as depicted in Figure~2.
    \begin{figure}
    \vspace{0.0cm} \hspace{0.00cm}
    \centerline{\includegraphics[scale=.75,clip]{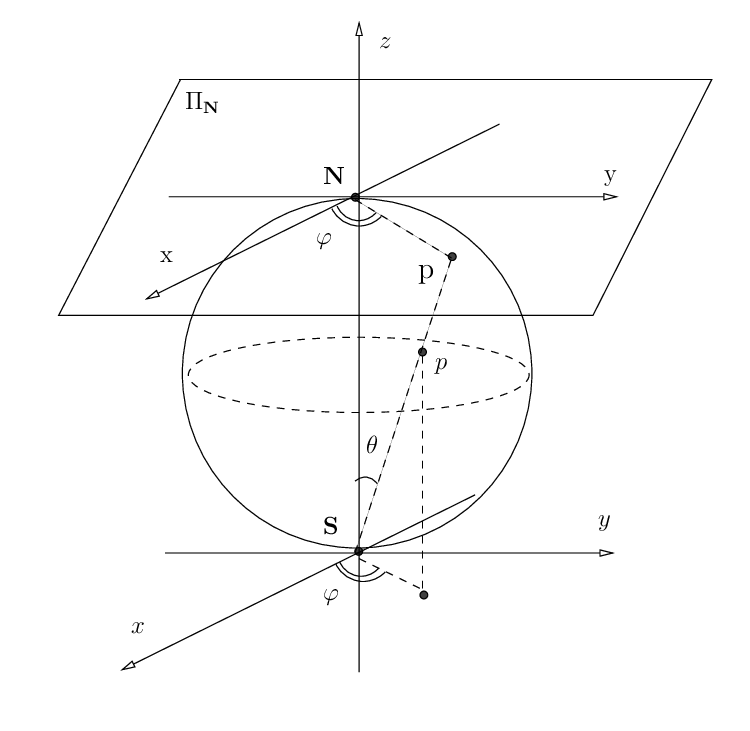}}
    \centerline{\parbox{16cm}{\caption{Stereographic projection of the sphere $S^2$ defined by
    $x^2+y^2+(z-\frac{1}{2})^2=\frac{1}{4}$: $\mathbf{N}$ and
    $\mathbf{S}$ are respectively the north and south poles of
    $S^2$. $\Pi_{\mathbf{N}}$ is the tangent plane at $\mathbf{N}$.
    $p$ is a point on $S^2-\{\mathbf{S}\}$, and $\rp$ is its
    stereographic projection on $\Pi_{\mathbf{N}}$.}}}
    \label{fig2}
    \vspace{0.0cm}
    \end{figure}
The line connecting $\mathbf{S}$ to a an arbitrary point $p=(x,y,z)$
on $S^2-\{\mathbf{S}\}$ intersects $\Pi_\mathbf{N}$ at a point
$\rp$. If we set up a Cartesian coordinate system in
$\Pi_\mathbf{N}$ with $\mathbf{N}$ as its origin and axes parallel
to the $x$- and $y$-axes, and denote by $(\rx,\ry)$ the coordinates
of $\rp$ in this coordinates system, we can uniquely identify the
points $p\in S^2-\{\mathbf{S}\}$ with $(\rx,\ry)\in\R^2$. Using
simple methods of analytic geometry, we can easily verify that
    \be
    \rx=\frac{x}{z},~~~~\ry=\frac{y}{z},~~~~
     x=\frac{\rx}{1+\rx^2+\ry^2},~~~~
     y=\frac{\ry}{1+\rx^2+\ry^2}, ~~~~
     z=\frac{1}{1+\rx^2+\ry^2}.
     \label{xyz-to-xy}
     \ee
We can employ the last three of these relations to compute the line
element over the sphere in the $(\rx,\ry)$-coordinates. A rather
lengthy but straightforward calculation yields
    \be
    ds^2=dx^2+dy^2+dz^2=\frac{d\rx^2+d\ry^2}{(1+\rx^2+\ry^2)^2}.
    \label{FS=derive1}
    \ee
This relation together with $ds^2=\sum_{i,j=1}^2 g^{(FS)}_{ij}
d\rx^id\rx^j$, where $\rx^1:=\rx$ and $\rx^2:=\ry$, gives the
following local coordinate expression for the Fubini-Study metric
tensor \cite{egh},
    \be
    g^{(FS)}_{ij}=\frac{\delta_{ij}}{\left[1+(\rx^1)^2+
    (\rx^2)^2\right]^2}.
    \label{FS-metric=}
    \ee
Expressing $\rx$ and $\ry$ in terms of the spherical coordinates,
    \be
    \varphi:=\tan^{-1}\big(\mbox{$\frac{y}{x}$}\big),~~~~~~~~
    \theta:=\cos^{-1}(2z-1),
    \label{spherical-coor}
    \ee
of $S^2$, we find
    \be
    \rx=\tan\left(\mbox{$\frac{\theta}{2}$}\right)\cos\varphi,
    ~~~~~\ry=\tan\left(\mbox{$\frac{\theta}{2}$}\right)\sin\varphi.
    \label{xy=tf}
    \ee
Substituting these in (\ref{FS=derive1}) leads to the following
familiar relation for the line element of a sphere of unit diameter.
    \be
    ds^2=\frac{1}{4}\,\left(d\theta^2+\sin^2\theta\,
    d\varphi^2\right).
    \label{FS-explicit}
    \ee

In order to make the relationship between (\ref{Fubini-Study=}) and
(\ref{FS=derive1}) more transparent, we recall that the south pole
$\mathbf{S}$ of $S^2$ corresponds to the projection operator
$\Lambda_{e_2}$ represented by
    \be
    \underline{\Lambda_{e_2}}=\left(\begin{array}{cc}
    0 & 0\\
    0 & 1\end{array}\right)
    \label{state-minus}
    \ee
and the state $\lambda_{e_2}:=\{c\, e_2\,|\,c\in\C\}$, where
$e_2:=${\scriptsize$\left(\begin{array}{c}0\\1\end{array}\right)$}.
The coordinates $(\rx,\ry)$ parameterize all the states except
$\lambda_{e_2}$. These are represented by the state-vectors
    \be
    \psi=\left(\begin{array}{c} \fz_1\\
    \fz_2\end{array}\right)~~~{\rm with}~~~\fz_1\neq 0.
    \label{psi=FS}
    \ee
Introducing $\rz:=\frac{\fz_2}{\fz_1}$, we can simplify the
expression (\ref{state-rep=}) for the corresponding projection
operator. In terms of $\rz$, the coordinates $x$, $y$, $z$ appearing
in (\ref{state-rep=xyz}) take the form
    $x=\frac{\Re(\rz)}{1+|\rz|^2}$, $y=\frac{\Im(\rz)}{1+|\rz|^2}$,
    $z=\frac{1}{1+|\rz|^2}$.
Comparing these with (\ref{xyz-to-xy}) reveals
    \be
    \rx=\Re(\rz),~~~~~~~~\ry=\Im(\rz).
    \label{x-y=re-im-z}
    \ee
Now, we are in a position to compute the line element
(\ref{Fubini-Study=}). Inserting (\ref{psi=FS}) in
(\ref{Fubini-Study=}), setting $\fz_2=\fz_1 \rz$, and using
(\ref{x-y=re-im-z}) and (\ref{xy=tf}), we find
    \be
     ds^2=\frac{d\rz^* d\rz}{(1+|\rz|^2)^2}=
    \frac{d\rx^2+d\ry^2}{(1+\rx^2+\ry^2)^2}=
    \frac{1}{4}\,\left(d\theta^2+\sin^2\theta\,
    d\varphi^2\right).
    \label{twice-FS=}
    \ee
Therefore, as a Riemannian manifold the state-space $\cP(\cH)$ is
identical to a two-dimensional (round) sphere of unit
diameter.\footnote{The above calculation of the metric tensor on
$S^2$ makes use of the stereographic projection of
$S^2-\{\mathbf{S}\}$ onto the plane $\Pi_\mathbf{N}$ which is a copy
of $\R^2=\C$. We could also consider the stereographic projection of
$S^2-\{\mathbf{N}\}$ onto the tangent plane $\Pi_\mathbf{S}$ at
$\mathbf{S}$. Using both the projections we are able to describe all
the points on $S^2$. This is a manifestation of the manifold
structure of $S^2$.}

\subsection{State-Space and Its geometry in Pseudo-Hermitian QM}
\label{geom-conv-PHQM}

The construction of the space of states in pseudo-Hermitian quantum
mechanics is similar to that in conventional quantum mechanics.
Again, the states are rays in the (reference) Hilbert space $\cH$
which are identical with the rays in the physical Hilbert space
$\cH_{\rm phys}$. The only difference is in the way one associates
projection operators to states and defines an appropriate notion of
distance (metric tensor) on the state-space.

In the following we shall use $\cP(\cH_{\rm phys})$ to denote the
state-space of a pseudo-Hermitian quantum system with physical
Hilbert space $\cH_{\rm phys}$. The latter is obtained by endowing
the underlying vector space of $\cH$ with the inner product
$\br\cdot|\cdot\kt_{\eta_+}$, where ${\etap}:\cH\to\cH$ is a given
metric operator rendering the Hamiltonian of the system
${\etap}$-pseudo-Hermitian. We shall also introduce the following
notation that will simplify some of the calculations: For all
$\xi,\zeta\in\cH$, $|\zeta\pkt:=|\zeta\kt=\zeta$,
$\pbr\zeta|:=\br\zeta|{\etap}$,
$\pbr\xi|\zeta\pkt:=\br\xi|{\etap}|\zeta\kt=\br\xi|{\etap}\zeta\kt=
    \br\xi|\zeta\kt_{\eta_+}$.

First we define, for each pair of linear operators $L,J:\cH\to\cH$,
    \be
    (L|J)_{\eta_+}:={\rm tr}_{\eta_+}(L^\sharp J)=\tr(L^\sharp J),
    \label{ph-HS-inn-prod-1}
    \ee
where  $L^\sharp$ stands for the ${\etap}$-pseudo-adjoint of $L$:
    \be
    L^\sharp:={\etap}^{-1}L^\dagger{\etap},
    \label{eta-pseudo-adj}
    \ee
$\tr_\etap$ is defined by (\ref{ph-trace}), and we have used
(\ref{tr=tr-identity}). The linear operators $A:\pH\to\pH$ for which
$(A|A)_{\eta_+}$ is finite together with (\ref{ph-HS-inn-prod-1})
form the inner product space $\fB_2(\pH)$ of Hilbert-Schmidt
operators acting in $\cH_{\rm phys}$. Substituting
(\ref{eta-pseudo-adj}) in (\ref{ph-HS-inn-prod-1}) and using
$\rho^2={\etap}$ and $\rho^\dagger=\rho$, we also have
    \be
    (L|J)_{\eta_+}={\rm tr}({\etap}^{-1}
    L^\dagger{\etap} J)={\rm tr}(\rho^{-1}
    L^\dagger\rho^2 J \rho^{-1})={\rm tr}((\rho
    L\rho^{-1})^\dagger\rho J \rho^{-1})=(\rho
    L\rho^{-1}|\rho J \rho^{-1}).~~~
    \label{ph-HS-inn-prod-2}
    \ee
These calculations show that $\rho:\pH\to\cH$ induces a unitary
operator $U_\rho:\fB_2(\pH)\to\fB_2(\cH)$ defined by
    \be
    U_\rho(L):=\rho\, L\, \rho^{-1},~~~~~~\mbox{for all $L\in
    \fB_2(\pH)$}.
    \label{induced-unitary}
    \ee

Now, consider a state $\lambda_\psi:=\{c\psi|c\in\C\}$ for some
$\psi\in\cH_{\rm phys}-\{0\}$. Because $\pbr\cdot|\cdot\pkt$ is the
inner product of $\cH_{\rm phys}$, the orthogonal projection
operator onto $\lambda_\psi$ is given by
    \be
    \Lambda^{(\eta_+)}_\psi:=\frac{|\psi\pkt\pbr\psi|}{
    \pbr\psi|\psi\pkt}=\frac{|\psi\kt\br\psi|{\etap}}{
    \br\psi|{\etap}\psi\kt}=\frac{\br\psi|\psi\kt\,\Lambda_\psi\,
    {\etap}}{\br\psi|{\etap}\psi\kt}.
    \label{ph-proj-op}
    \ee
A quick calculation shows that
${\Lambda^{({\etap})}_\psi}^2=\Lambda^{({\etap})}_\psi$.
Furthermore, using the arguments leading to (\ref{unit-trace}), we
have
    \[{\rm
    tr}_{\eta_+}(\Lambda^{({\etap})\sharp}_\psi\Lambda^{({\etap})}_\psi)=
    {\rm tr}_{\eta_+}(\Lambda^{({\etap})}_\psi)=\frac{\sum_{n=1}^N
    \pbr\psi_n|\psi\pkt\pbr\psi|\psi_n\pkt}{\pbr\psi|\psi\pkt}
    =\frac{\sum_{n=1}^N
    |\pbr\psi_n|\psi\pkt|^2}{\pbr\psi|\psi\pkt}=1.\]
This shows that $\Lambda^{({\etap})}_\psi\in \fB^{(\eta_+)}_2$, and
also, because ${\rm tr}_{\eta_+}={\rm tr}$,
    \be
    {\rm tr}(\Lambda^{({\etap})}_\psi)=1.
    \label{ph-unit-trace2}
    \ee

Another direct implication of (\ref{ph-proj-op}) is
$\Lambda^{({\etap})\dagger}_\psi ={\etap}
\Lambda^{({\etap})}_\psi{\etap}^{-1}$. Hence
$\Lambda^{({\etap})}_\psi$ belongs to the subset $\fH_2(\pH)$ of
${\etap}$-pseudo-Hermitian elements of $\fB_2(\pH)$. We can view
this as a real vector space. (\ref{ph-HS-inn-prod-1}) defines a real
inner product on this space, and the restriction of
(\ref{induced-unitary}) onto $\fH_2(\pH)$ that we also denote by
$U_\rho$ yields a unitary operator that maps $\fH_2(\pH)$ onto
$\fH_2(\cH)$. In fact $\fH_2(\pH)$ and $\fH_2(\cH)$ are real
separable Hilbert spaces and the existence of the unitary operator
$U_\rho:\fH_2(\pH)\to\fH_2(\cH)$ is a manifestation of the fact that
real separable Hilbert spaces of the same dimension are
unitary-equivalent.

The set of the projection operators (\ref{ph-proj-op}) that is in
one-to-one correspondence with the projective Hilbert space
$\cP(\pH)$ is a proper subset of $\fH_2(\pH)$. Similarly to the case
of conventional quantum mechanics, we can define a natural metric on
this space whose line element has the form
    \bea
    ds_{\eta_+}^2(\Lambda_\psi):=\frac{1}{2}
    (d\Lambda^{({\etap})}_\psi|d\Lambda^{({\etap})}_\psi )_{{\etap}}
    &=&\frac{\pbr\psi|\psi\pkt\pbr d\psi|d\psi\pkt-|
    \pbr\psi|d\psi\pkt|^2}{\pbr\psi|\psi\pkt^2}\nn\\
    &=&
    \frac{\br\psi|{\etap}\psi\kt\br d\psi|{\etap} d\psi\kt-|
    \br\psi|{\etap}d\psi\kt|^2}{\br\psi|{\etap}\psi\kt^2}.
    \label{ph-Fubini-Study=}
    \eea
It is important to note that as smooth manifolds $\cP(\cH)$ and
$\cP(\pH)$ are the same, but as Riemannian manifolds they are
different. While $\cP(\cH)$ is endowed with the Fubini-Study metric,
$\cP(\pH)$ is endowed with the metric corresponding to
(\ref{ph-Fubini-Study=}). For $N<\infty$ we can obtain a global
expression for the latter in terms of the coordinates
$\fz_1,\fz_2,\cdots,\fz_N$ of the state-vectors
$\psi=\vec\fz\in\C^N$. This yields the following generalization of
(\ref{metric-global}) that satisfies $ds_{\eta_+}^2=\sum_{a,b=1}^N
g^{({\etap})}_{ab^*}\,d\fz_a\fz_b^*$.
    \be
    g^{({\etap})}_{ab^*}:=
    \frac{\sum_{c,d=1}^N(\eta_{_+cd}\:\eta_{_+ba}-
    \eta_{_+ca}\:\eta_{_+bd})\,\fz^*_c\fz_d}{
    \left(\sum_{r,s=1}^N\eta_{_+rs}\:\fz_r^*\fz_s\right)^2}.
    \label{ph-metric-global}
    \ee
Here $\eta_{_+ab}$ are the entries of the standard representation
$\underline{{\etap}}$ of ${\etap}$, \cite{prl-2007}.

For two-level systems where $N=2$, we can easily obtain an explicit
expression for the line element (\ref{ph-Fubini-Study=}).  In
general the metric operator ${\etap}$ is represented by
    \be
    \underline{{\etap}}=\left(\begin{array}{cc}
    a & b_1-ib_2\\
    b_1+i b_2 & c\end{array}\right),
    \label{underline-eta=}
    \ee
where $a,b_1,b_2,c\in\R$ are such that
    \be
    a+c={\rm tr}(\underline{{\etap}})>0,~~~~~~~~~
    \rd:=ac-(b_1^2+b_2^2)=\det(\underline{{\etap}})>0.
    \label{constants=}
    \ee
In view of (\ref{ph-proj-op}) we can again parameterize
$\underline{\Lambda^{({\etap})}_\psi}$ using the cartesian
coordinates $x,y,z$ of the sphere $S^2$ defined by
(\ref{two-sphere-1}). For the states differing from $\lambda_{e_2}$,
we can alternatively choose the coordinates $\rx$ and $\ry$ of
(\ref{x-y=re-im-z}) and show using (\ref{psi=FS}),
(\ref{ph-proj-op}), and (\ref{underline-eta=}) that
    \be
    ds_{{\etap}}^2=\frac{\rd(d\rx^2+d\ry^2)}{\left[a+2
    (b_1\rx+b_2\ry)+c(\rx^2+\ry^2)\right]^2}.
    \label{prl-8-cor}
    \ee
It is easy to see that for ${\etap}=I$ where $a=c=1$ and
$b_1=b_2=0$, (\ref{prl-8-cor}) reproduces (\ref{twice-FS=}).

To gain a more intuitive understanding of (\ref{prl-8-cor}), we next
express its right-hand side in terms of the spherical coordinates
(\ref{spherical-coor}). Inserting (\ref{xy=tf}) in (\ref{prl-8-cor})
and carrying out the necessary calculations, we find
    \be
    ds_{{\etap}}^2=\frac{k_1(d\theta^2+\sin^2\theta\,
    d\varphi^2)}{\left[1+k_2\cos\theta+g(\varphi)
    \sin\theta\right]^2}=
    \frac{k_1\:(d\theta^2+\sin^2\theta\,
    d\tilde\varphi^2)}{\left[1+k_2\cos\theta+
    k_3\cos\tilde\varphi\,\sin\theta\right]^2}
    \label{prl-8-cor-sph}
    \ee
where we have introduced
    \bea
    &&k_1:=\frac{\rd}{(a+c)^2}=
    \frac{{\det(\underline{{\etap}})}}{
    {\rm tr}(\underline{{\etap}})^2},~~~~~~~~
    k_2:=\frac{a-c}{a+c},~~~~~~~~
    k_3:=\frac{2\sqrt{b_1^2+b_2^2}}{a+c},
    \\
    &&g(\varphi):=\frac{2(b_1\cos\varphi+b_2\sin\varphi)}{a+c},~~~~~~~~
    \tilde\varphi:=\varphi-\beta,~~~~~~~~
    \beta:=\tan^{-1}\big(\frac{b_2}{b_1}\big).
    \eea
Note that the change of coordinate $\varphi\to\tilde\varphi$
corresponds to a constant rotation about the $z$-axis, and because
of (\ref{constants=}) we have $k_1>0$, $-1<k_2<1$ and $0\leq k_3<1$.

The projective Hilbert space $\cP(\pH)$ is the Riemannian manifold
obtained by endowing the sphere $S^2$ with the metric
$\mathbf{g^{({\etap})}}$ corresponding to the line
element~(\ref{prl-8-cor-sph}).

Next, consider the general case where $N$ need not be two. In order
to compare the geometric structure of $\cP(\pH)$ and $\cP(\cH)$, we
recall that $\cP(\pH)\subset\fB_2(\pH)$,
$\cP(\cH)\subset\fB_2(\cH)$, and that the linear operator $U_\rho$
of Eq.~(\ref{induced-unitary}) maps $\fB_2(\pH)$ onto $\fB_2(\cH)$.
It is easy to check that the restriction of $U_\rho$ on $\cP(\pH)$,
i.e., the function $u_\rho:\cP(\pH)\to\cP(\cH)$ defined by
    \be
    u_\rho(\Lambda^{({\etap})}_\psi):=\rho\, \Lambda^{({\etap})}_\psi
    \rho^{-1},~~~~~~~~~\mbox{for all $\Lambda^{({\etap})}_\psi\in
    \cP(\pH)$,}
    \label{u-rho-restrict}
    \ee
is a diffeomorphism. Furthermore, in view of (\ref{state=}),
(\ref{ph-proj-op}), and $\rho^\dagger=\rho=\sqrt{\etap}$, we have
    \be
    u_\rho(\Lambda^{({\etap})}_\psi)=
    \frac{\rho|\psi\kt\br\psi|{\etap}\rho^{-1}}{\br\psi|\rho^2\psi\kt}
    =\frac{\rho|\psi\kt\br\psi|\rho}{\br\rho\,\psi|\rho\,\psi\kt}=
    \frac{|\Psi\kt\br\Psi|}{\br\Psi|\Psi\kt}=\Lambda_\Psi,
    \label{u-rho-restrict2}
    \ee
where $\Psi:=\rho\,\psi\in\cH$. A straightforward consequence of
(\ref{Fubini-Study=}), (\ref{trace-inn2}), (\ref{ph-HS-inn-prod-2}),
(\ref{ph-Fubini-Study=}), (\ref{u-rho-restrict}) and
(\ref{u-rho-restrict2}) is
    \bea
    ds^2(u_\rho(\Lambda^{({\etap})}_\psi))&=&
    ds^2(\Lambda_\Psi)=\frac{1}{2}\,(d\Lambda_\Psi|d\Lambda_\Psi)=
    \frac{1}{2}\,(\rho\, d\Lambda^{({\etap})}_\psi\rho^{-1}|
    \rho\, d\Lambda^{({\etap})}_\psi\rho^{-1})\nn\\
    &=&\frac{1}{2}\,(d\Lambda^{({\etap})}_\psi|d\Lambda^{({\etap})}_\psi)_{{\etap}}=
    ds_{{\etap}}^2(\Lambda^{({\etap})}_\psi).
    \eea
This shows that $u_\rho:\cP(\pH)\to\cP(\cH)$ is an \emph{isometry},
i.e., it leaves the distances invariant. Therefore, $\cP(\pH)$ and
$\cP(\cH)$ have the same geometric structure. In particular, for
$N=2$, we can identify $\cP(\pH)$ with a sphere of unit diameter
embedded in $\R^3$ with its standard geometry.

This result is another manifestation of the fact that
pseudo-Hermitian quantum mechanics is merely an alternative
representation of the conventional quantum mechanics. Because the
geometry of the state-space may be related to physical quantities
such as geometric phases, we should not have expected to obtain
different geometric structures for $\cP(\pH)$ and $\cP(\cH)$.

\subsection{Optimal-Speed Evolutions}

Let $H:\cH\to\cH$ be a possibly time-dependent Hermitian Hamiltonian
with a discrete spectrum. Suppose that we wish to use $H$ to evolve
an initial state $\lambda_{\psi_I}\in\cP(\cH)$ into a final state
$\lambda_{\psi_F}\in\cP(\cH)$ in some time $\tau$. Then the evolving
state-vector $\psi(t)\in\cH$ satisfies
    \be
    H\psi(t)=i\hbar\dot\psi(t),~~~~~
    \psi(0)=\psi_I,~~~~~\psi(\tau)=\psi_F,
    \label{sch-eq-QBP}
    \ee
and the corresponding state $\lambda_{\psi(t)}$ traverses a curve in
the projective Hilbert space $\cP(\cH)$.

The instantaneous speed for the motion of $\lambda_{\psi(t)}$ in
$\cP(\cH)$ is
    \be
    v_\psi:=\frac{ds}{dt},
    \label{speed1=}
    \ee
where $ds$ is the line element given by (\ref{Fubini-Study=}). In
view of this equation,
    \be
    v_\psi^2=\frac{\br\psi(t)|\psi(t)\kt\br\dot\psi(t)|\dot\psi(t)\kt-
    |\br\psi(t)|\dot\psi(t)\kt|^2}{\br\psi(t)|\psi(t)\kt^2}
    =\frac{\Delta E_\psi(t)^2}{\hbar^2},
    \label{speed2=}
    \ee
where
    \be
    \Delta E_\psi(t)^2:=
    \frac{\br\psi(t)|H^2\psi(t)\kt}{\br\psi(t)|\psi(t)\kt}
    -\frac{|\br\psi(t)|H\psi(t)\kt|^2}{\br\psi(t)|\psi(t)\kt^2},
    \label{uncertainty=}
    \ee
is the square of the energy uncertainty, and we have used
(\ref{sch-eq-QBP}) and the Hermiticity of $H$. We can employ
(\ref{speed1=}) and (\ref{speed2=}) to express the length of the
curve traced by $\lambda_{\psi(t)}$ in $\cP(\cH)$ in the form
\cite{anandan-aharonov}:
    \be
    s=\frac{1}{\hbar}\int_0^\tau \Delta E_\psi(t)\, dt.
    \label{length=uncertainty}
    \ee
Because $\Delta E_\psi$ is non-negative, $s$ is a monotonically
increasing function of $\tau$. This makes $\tau$ a monotonically
increasing function of $s$. Therefore, the shortest travel time is
achieved for the paths of the shortest length, i.e., the geodesics
of $\cP(\cH)$.

For a time-independent Hamiltonian $H$ we have
$\psi(t)=e^{-itH/\hbar}\psi_I$ and as seen from
(\ref{uncertainty=}), $\Delta E_\psi$ is also time-independent. In
this case, (\ref{length=uncertainty}) yields
    \be
    \tau=\frac{\hbar s}{\Delta E_\psi},
    \label{tau=QBP}
    \ee
and the minimum possible travel time is achieved when $s$ is
identified with the geodesic distance between $\psi_I$ and $\psi_F$.

Because of the particular structure of $\cP(\cH)$ the geodesic(s)
connecting any two states $\lambda_{\psi_I}$ and $\lambda_{\psi_F}$
lie entirely on a two-dimensional submanifold of $\cP(\cH)$ that is
actually the projective Hilbert space $\cP(\cH_{I,F})$ for the
subspace $\cH_{I,F}$ of $\cH$ spanned by $\psi_I$ and $\psi_F$. If
the time evolution generated by the Hamiltonian minimizes the travel
time the evolving state $\lambda_{\psi_I}$ should stay in
$\cP(\cH_{I,F})$ during the evolution. This means that the problem
of determining minimum-travel-time evolutions
\cite{fleming,Vaidman,Margolus,brody,carlini1} reduces to the case
that $\cH$ is two-dimensional \cite{brody-hook}. As we saw in
Subsection~\ref{geom-conv-QM}, in this case $\cP(\cH)$ is a round
sphere of unit diameter and the geodesics are the large circles on
this sphere.

Next, we study, without loss of generality, the case $N=2$. It is
easy to show the existence of a time-independent Hamiltonian that
evolves $\lambda_{\psi_I}$ to $\lambda_{\psi_F}$ along a geodesic.
We will next construct such a Hamiltonian.

Consider a time-independent Hamiltonian $H$ acting in a
two-dimensional Hilbert space. We can always assume that $H$ has a
vanishing trace so that its eigenvalues have opposite sign,
$E_2=-E_1=:E$.\footnote{This is true for general possibly
time-dependent Hamiltonians $H(t)$. Under the gauge transformation
$\psi(t)\to \psi'(t)\to e^{i\alpha(t)/\hbar}\psi(t)$ with
$\alpha(t)=N^{-1}\int_0^t {\rm tr}[H(s)]ds$, the Hamiltonian $H(t)$
transforms into the traceless Hamiltonian $H'(t):=H(t)-N^{-1}{\rm
tr}[H(t)]I$.} Because $\Delta E_\psi$ is time-independent we can
compute it at $t=0$. Expanding $\psi(0)=\psi_I$ in an orthonormal
basis $\{\psi_1,\psi_2\}$ consisting of a pair of eigenvectors of
$H$, i.e., writing it in the form
    \be
    \psi_I=c_1\psi_1+c_2\psi_2,~~~~~c_1,c_2\in\C,
    \label{psi-expand-ini}
    \ee
we find \cite{p87}
    \be
    \Delta E_\psi=E \sqrt{1-\left(\frac{|c_1|^2-|c_2|^2}{
    |c_1|^2+|c_2|^2}\right)^2}\leq E.
    \label{delta-E=}
    \ee
Therefore, the travel time $\tau$ satisfies
    \be
    \tau\geq \tau_{\rm min}:=\frac{\hbar s}{E},
    \label{bound}
    \ee
where $s$ is the geodesic distance between $\lambda_{\psi_I}$ and
$\lambda_{\psi_F}$ in $\cP(\cH)$. (\ref{bound}) identifies
$\tau_{\rm min}$ with a lower bound on the travel time. Next, we
shall construct a Hamiltonian $H_\star$ with eigenvalues $\pm E$ for
which $\tau=\tau_{\rm min}$. This will, in particular, identify
$\tau_{\rm min}$ with the minimum travel time.

In order to determine $H_\star$ we only need to construct a pair of
its linearly independent eigenvectors $\psi_1$ and $\psi_2$ and use
its spectral resolution:
    \be
    H_\star=E\big(-|\psi_1\kt\br\psi_1|+|\psi_2\kt\br\psi_2|\big).
    \label{H-star=1}
    \ee
As seen from (\ref{delta-E=}), to saturate the lower bound on
$\tau$, we must have $|c_1|=|c_2|$. In view of time-independence of
$\Delta E_\psi$ we could also use $\psi_F$ to compute this quantity.
If we expand the latter as
    \be
    \psi_F=d_1\psi_1+d_2\psi_2,~~~~~d_1,d_2\in\C,
    \label{psi-expand-fin}
    \ee
we find that $\Delta E_\psi$ satisfies (\ref{delta-E=}) with
$(c_1,c_2)$ replaced by $(d_1,d_2)$. Therefore, by the same argument
we find $|d_1|=|d_2|$. Therefore, there must exist
$\beta_I,\beta_{F}\in\R$ such that
    \be
    c_2=e^{\beta_I}c_1,~~~~~~~d_2=e^{\beta_F}d_1.
    \label{cc-dd}
    \ee
Inserting these relations in (\ref{psi-expand-ini}) and
(\ref{psi-expand-fin}) gives
    \be
    \psi_1+e^{i\beta_I}\psi_2=c_1^{-1}\psi_I,~~~~~~
    \psi_1+e^{i\beta_F}\psi_2=d_1^{-1}\psi_F.
    \label{psi-eqns}
    \ee
Solving these for $\psi_1$ and $\psi_2$, we obtain
    \be
    \psi_1=\frac{\sqrt 2\big(\hat\psi_I-
    e^{\frac{i\vartheta}{2}}\hat\psi_F\big)}{1-e^{i\vartheta}},
    ~~~~~~~~~~~
    \psi_2=\frac{\sqrt 2\:e^{-i\beta_F}\big(-\hat\psi_I+
    e^{\frac{-i\vartheta}{2}}\hat\psi_F\big)}{1-e^{i\vartheta}},
    \label{psi12=QBP}
    \ee
where we have introduced
    \be
    \vartheta:=\beta_I-\beta_F,~~~~~~~
    \hat\psi_I:=\frac{\psi_I}{\sqrt 2\, c_1},~~~~~~~
    \hat\psi_F:=\frac{e^{\frac{i\vartheta}{2}}\psi_F}{
    \sqrt 2\, d_1}.
    \label{vartheta=}
    \ee

Clearly $\hat\psi_n$ determines the same state as $\psi_n$ for
$n\in\{1,2\}$. Also in view of (\ref{H-star=1}), the presence of
$\beta_F$ in (\ref{psi12=QBP}) does not affect the expression for
$H_\star$. The only parameter that has physical significance is the
angle $\vartheta$. The orthonormality of $\psi_1$ and $\psi_2$
implies that $\hat\psi_I$ and $\hat\psi_F$ have unit norm and more
importantly that $\vartheta$ fulfils
    \be
    \br\hat\psi_I|\hat\psi_F\kt=\cos(\mbox{$\frac{\vartheta}{2}$}).
    \label{angle-beta1}
    \ee
In terms of $\psi_I$ and $\psi_F$ this equation takes the form
    \be
    \br \psi_I| \psi_F\kt=c_1^*d_1\big(1+
    e^{-i\vartheta}\big).
    \label{angle-beta11}
    \ee
Note that because $\hat\psi_I$ and $\hat\psi_F$ have unit norm,
according to (\ref{vartheta=}), $\br\psi_I|\psi_I\kt=2|c_1|^2$ and
$\br\psi_F|\psi_F\kt=2|d_1|^2$. These relations together with
(\ref{vartheta=}) and (\ref{angle-beta1}) imply
    \be
    \cos^2(\mbox{$\frac{\vartheta}{2}$})=
    \frac{|\br\psi_I|\psi_F\kt|^2}{\br\psi_I|\psi_I\kt\,
    \br\psi_F|\psi_F\kt}.
    \label{cosvartheta=}
    \ee
Therefore whenever $\psi_I$ and $\psi_F$ are orthogonal,
$\vartheta=\pi$. Furthermore, as discussed in
\cite{anandan-aharonov}, (\ref{cosvartheta=}) shows that $\vartheta$
is related to the geodesic distance $s$ between $\lambda_{\psi_I}$
and $\lambda_{\psi_F}$ according to\footnote{Note that the metric on
$\cP(\cH)$ that is used in \cite{anandan-aharonov} differs from our
metric by a factor of $\sqrt 2$.}
    \be
    \vartheta=2s.
    \label{angle-beta4}
    \ee
In the case that $\psi_I$ and $\psi_F$ are orthogonal,
$\lambda_{\psi_I}$ and $\lambda_{\psi_F}$ are \emph{antipodal}
points on $\cP(\cH)$. Therefore their geodesic distance $s$ is half
of the perimeter of a large circle. Because $\cP(\cH)$ is a round
sphere of unit diameter, we have $s=\pi/2$, which is consistent with
(\ref{angle-beta4}).

Having calculated the eigenvectors $\psi_1$ and $\psi_2$, we can use
(\ref{H-star=1}) to obtain an explicit expression for the
Hamiltonian $H_\star$ that evolves $\lambda_{\psi_I}$ into
$\lambda_{\psi_F}$ in time $\tau_{\rm min}$. Substituting
(\ref{psi12=QBP}) in (\ref{H-star=1}) and using (\ref{vartheta=}),
(\ref{angle-beta1}) and (\ref{angle-beta4}), we find
\cite{carlini1,brody-hook,p87}
    \be
    H_\star=
    \frac{iE\big(|\hat\psi_F\kt\br\hat\psi_I|-
    |\hat\psi_I\kt\br\hat\psi_F|\big)}{4\sin(\frac{\vartheta}{2})}=
    \frac{iE\cot(s)}{4}
    \left(\frac{|\psi_F\kt\br\psi_I|}{\br\psi_I|\psi_F\kt}-
    \frac{|\psi_I\kt\br\psi_F|}{\br\psi_F|\psi_I\kt}\right).
    \label{H-star=expl}
    \ee
The last equation shows that $H_\star$ depends only on the states
$\lambda_{\psi_I}$ and $\lambda_{\psi_F}$ and not on the particular
state-vectors one uses to represent these states. Note also that
this equation is valid generally; it applies for quantum systems
with an arbitrary finite- or infinite-dimensional $\cH$.

This completes our discussion of the quantum Brachistochrone problem
in conventional quantum mechanics. We can use the same approach to
address this problem within the framework of pseudo-Hermitian
quantum mechanics \cite{prl-2007}. This amounts to making the
following substitutions in the above analysis: $|\psi_n\kt \to
|\psi_n\pkt$, $\br\psi_n| \to \:\pbr\psi_n|$, and $s\to
s_{{\etap}}$. In particular, the minimum travel time is given by
    \be
    \tau_{\rm min}^{({\etap})}=\frac{\hbar s_{{\etap}}}{E},
    \label{ph-bound}
    \ee
and the ${\etap}$-pseudo-Hermitian Hamiltonian that generates
minimal-travel-time evolution between $\lambda_{\psi_I}$ and
$\lambda_{\psi_F}$ has the form
    \be
    H^{({\etap})}_\star=
    \frac{iE\cot(s_{{\etap}})}{4}
    \left(\frac{|\psi_F\pkt\pbr\psi_I|}{\pbr\psi_I|\psi_F\pkt}-
    \frac{|\psi_I\pkt\pbr\psi_F|}{\pbr\psi_F|\psi_I\pkt}\right),
    \label{ph-H-star=expl}
    \ee
where
    \be
    \cos^2(s_{{\etap}})=
    \frac{|\pbr\psi_I|\psi_F\pkt|^2}{\pbr\psi_I|\psi_I\pkt\,
    \pbr\psi_F|\psi_F\pkt}.
    \label{ph-vartheta=}
    \ee
Eq.~(\ref{ph-H-star=expl}) gives the expression for the most general
time-independent optimal-speed quasi-Hermitian Hamiltonian operator
that evolves $\psi_I$ into $\psi_F$. Similarly to its Hermitian
counterpart, it applies irrespective of the dimensionality of the
Hilbert space.

In \cite{bbj-2007}, the authors show using a class of
quasi-Hermitian Hamiltonians that one can evolve an initial state
$\lambda_{\psi_I}$ into a final state $\lambda_{\psi_F}$ in a time
$\tau$ that violates the condition $\tau\geq\tau_{\rm min}$. They
actually show that by appropriately choosing the form of the
quasi-Hermitian Hamiltonian one can make $\tau$ arbitrarily small.
This phenomenon can be easily explained using the above treatment of
the problem. The minimum travel time for an
${\etap}$-pseudo-Hermitian Hamiltonian is given by (\ref{ph-bound}).
In view of (\ref{prl-8-cor-sph}), depending on the value of
$k_1=\det(\underline{{\etap}})/{\rm tr}(\underline{{\etap}})^2$ one
can make $s_{{\etap}}$ and consequently $\tau_{\rm min}^{({\etap})}$
as small as one wishes. This observation does not, however, seem to
have any physically significant implications, because a physical
process that involves evolving $\lambda_{\psi_I}$ into
$\lambda_{\psi_F}$ using an ${\etap}$-pseudo-Hermitian Hamiltonian
$H$ can be described equally well by considering the evolution of
$\lambda_{\rho\psi_I}$ into $\lambda_{\rho\psi_F}$ using the
Hermitian Hamiltonian $h:=\rho\, H\,\rho^{-1}$. In light of the
existence of the isometry $u_\rho:\cP(\pH)\to\cP(\cH)$, the distance
between $\lambda_{\psi_I}$ and $\lambda_{\psi_F}$ in $\cP(\pH)$ is
equal to the distance between $\lambda_{\rho\psi_I}$ and
$\lambda_{\rho\psi_F}$ in $\cP(\cH)$. It is also easy to show that
the travel time for both the evolutions are identical. Therefore as
far as the evolution speed is concerned there is no advantage of
using the ${\etap}$-pseudo-Hermitian Hamiltonian $H$ over the
equivalent Hermitian Hamiltonian $h$, \cite{prl-2007}.

We wish to emphasize that the existence of a lower bound on travel
time is significant, because it limits the speed with which one can
perform unitary transformations dynamically. Such transformations
play a central role in quantum computation. For example the
construction of efficient NOT-gates involves unitary transformations
that map a state into its antipodal state. The distance between
antipodal states in $\cP(\pH)$ is the same as in $\cP(\cH)$. Hence,
for such states $\tau^{({\etap})}_{\rm min}=\tau_{\rm min}$.

The situation is quite different if we consider the evolution
generated by the ${\etap}$-pseudo-Hermitian Hamiltonian
$H_\star^{(\etap)}$ in the standard projective Hilbert space
$\cP(\cH)$. In this case, we can indeed obtain arbitrarily fast
evolutions, but they will not be unitary \cite{fring-jpa-2007}. The
possibility of infinitely fast non-unitary evolutions is actually
not surprising. What is rather surprising is that one can achieve
such evolutions using quasi-Hermitian Hamiltonians; \emph{there are
arbitrarily fast quasi-unitary evolutions} \cite{p87}.

A scenario that is also considered in \cite{bbj-2007} is to use both
Hermitian and quasi-Hermitian Hamiltonians to produce an arbitrarily
fast evolution of $\lambda_{\psi_I}$ into $\lambda_{\psi_F}$. This
is done in three stages. First, one evolves the initial state
$\lambda_{\psi_I}$ into an auxiliary state $\lambda_{\psi'_I}$ using
a Hermitian Hamiltonian $h_1$ in time $\tau_1$, then one evolves
$\lambda_{\psi'_I}$ into another auxiliary state $\lambda_{\psi'_F}$
using a quasi-Hermitian Hamiltonian $H$ in time $\tau'$, and finally
one evolves $\lambda_{\psi'_F}$ into the desired final state
$\lambda_{\psi_F}$ using another Hermitian Hamiltonian $h_2$ in time
$\tau_2$. By choosing the intermediate states $\lambda_{\psi'_I}$
and $\lambda_{\psi'_F}$ appropriately one can make $\tau_1$ and
$\tau_2$ as small as one wishes. By choosing $H$ to be an
${\etap}$-pseudo-Hermitian operator of the form
(\ref{ph-H-star=expl}) with the parameter $k_1$ of ${\etap}$
sufficiently small one can make the total travel time
$\tau:=\tau_1+\tau'+\tau_2$ smaller than $\tau_{\rm min}$. In this
scenario both the initial and final states belong to $\cP(\cH)$, but
to maintain unitarity of the evolution one is bound to switch (the
defining metric of) the physical Hilbert space at $t=\tau_1$ and
$t=\tau_1+\tau'$. Therefore, this scheme involves a physical Hilbert
space with a time-dependent inner product. As discussed in
Subsection~\ref{sec-time-dep}, the latter violates the condition
that the Hamiltonian is an observable. Therefore, \emph{there seems
to be no legitimate way of lowering the bound on travel time between
two states of a given distance except allowing for non-unitary
(possibly quasi-unitary) evolutions.}

\section{Physical Applications}

Since its inception in the form of $\cP\cT$-symmetric models in the
late 1990's and later as a consistent quantum mechanical scheme
\cite{jpa-2004b} pseudo-Hermitian QM has been the subject of
extensive research. The vast majority of the publications on the
subject deal with issues related to formalism or various (quantum
mechanical as well as field theoretical) toy models with mostly
obscure physical meaning. There are however a number of exceptions
to this general situation where concrete problems are solved using
the methods developed within the framework of pseudo-Hermitian QM.
In this section we outline the basic ideas upon which these recent
developments rest. Before engaging into a discussion of these,
however, we wish to list some of the applications that predate the
recent activities in the field.

\subsection{Earlier Applications}

\subsubsection{Dyson Boson Mapping}

Among the earliest manifestations of pseudo-Hermitian operators is
the one appearing in the context of the Dyson mapping of Hermitian
Fermionic Hamiltonians to equivalent quasi-Hermitian bosonic
Hamiltonians \cite{dyson-1956}. Dyson mapping has subsequently found
applications in nuclear physics \cite{janssen-1971} and provided the
basic idea for the formulation of quasi-Hermitian QM \cite{geyer}.
For a brief review of the Dyson mapping method see
\cite{geyer-cjp-2004}.

\subsubsection{Complex Scaling and Resonances}

Consider the one-parameter family of operators:
$ru_\alpha=\exp\left(\frac{i\alpha}{2\hbar}\:\{x,p\}\right)$ with
$\alpha\in\C$, that act in the Hilbert space $L^2(\R)$. We can
easily use the Backer-Campbell-Hausdorff
identity~(\ref{bch-identity})  together with the canonical
computation relation $[x,p]=i\hbar$ to show that $\ru_\alpha$
induces a scaling of the position and momentum operators: $x\to
\ru_\alpha\; x\;\ru_\alpha^{-1}=e^\alpha x$ and $p\to \ru_\alpha\;
p\;\ru_\alpha^{-1}=e^{-\alpha} p$. For $\alpha\in\R$, $\ru_\alpha$
is a unitary transformation, and one can use the latter property to
show that
$(\ru_\alpha\psi)(x)=e^{-\frac{\alpha}{2}}\psi(e^{\alpha}x)$ for all
$\psi\in L^2(\R)$.

For $\alpha\in\C-\R$, the transformation $\psi\to\ru_\alpha\psi$ is
called a \emph{complex scaling} transformation. In this case,
$\ru_\alpha$ is no longer a unitary operator. In fact, neither
$\ru_\alpha$ nor its inverse is bounded. This implies that its
action on a Hermitian Hamiltonian $H$, namely $H\to H'=\ru_\alpha
H\ru_\alpha^{-1}$, that (neglecting the unboundedness of
$\ru_\alpha$ and $\ru_\alpha^{-1}$) maps $H$ into a quasi-Hermitian
Hamiltonian $H'$, can have dramatic effects on the nature of its
continuous spectrum. This observation has applications in the
treatment of resonances (where one replaces $x$ with the radial
spherical coordinate in $\R^3$). The main idea is to perform an
appropriate complex scaling transformation so that the
non-square-integrable wave functions representing resonant states of
$H$ are mapped to square-integrable eigenfunctions of $H'$. For
details, see \cite{simon,lowdin,antoniou} and references therein.

\subsubsection{Vortex Pinning in Superconductors}

Consider the Hamiltonian: $H_g=\frac{(p+ig)^2}{2m}+v(x)$, where $g$
is a real constant and $v$ is a real-valued potential. This
Hamiltonian can be mapped to the Hermitian Hamiltonian $H_0$ by the
similarity transformation:
    \be
    H_g\to e^{-gx/\hbar}\,H_g\,e^{gx/\hbar}=H_0.
    \label{sec9-HN-2}
    \ee
This in turn implies that $H_g$ is $\etap$-pseudo-Hermitian for the
metric operator $\etap:=e^{-2gx/\hbar}$. This is to be expected,
because (\ref{sec9-HN-2}) is an example of the quasi-Hermitian
Hamiltonians of the form (\ref{sec9-HN-3}) that admit $x$-dependent
metric operators. In \cite{hn-prl-1996,hn-prb-1997-8} the
Hamiltonian $H_g$ (with random potential $v$) is used in modeling a
delocalization phenomenon relevant for the vortex pinning in
superconductors. A review of the ensuing developments is provided in
\cite{hatano-pa-1998}.

\subsection{Relativistic QM, Quantum Cosmology, and QFT}
\label{rqm-cq-qft}

The issue of constructing an appropriate inner product for the
defining Hilbert space of a quantum mechanical system, which we
shall refer to as the \emph{Hilbert-space problem}, is almost as old
as quantum mechanics itself. Probably the first serious encounter
with this problem is Dirac's attempts to obtain a probabilistic
interpretation of the (first-quantized) Klein-Gordon fields in the
late 1920's. The same problem arises in the study of other bosonic
fields and particularly in the application of Dirac's method of
constrained quantization for systems with first class constraints
\cite{dirac-book}. This method defines the ``physical space'' $\cV$
of the state-vectors as the common null space (kernel) of the
constraints, but it does not specify the inner product necessary to
make $\cV$ into a Hilbert space. Often the inner product induced
from the auxiliary Hilbert space of the unconstrained system is not
physically admissible, and one must find an alternative method of
constructing an appropriate inner.

In trying to deal with the Hilbert-space problem for Klein-Gordon
fields, Dirac was led to the discovery of the wave equation for
massive spin-half particles and the antimatter that earned him the
1933 Nobel prize in physics. Another major historical development
that has its root in attempts to address the Hilbert-space problem
is the discovery of the method of second quantization and eventually
relativistic quantum field theories. These developments did not
bring a definitive resolution for the original problem, but
diminished the interest in its solution considerably.

In the 1960's the discovery of the Hamiltonian formulation of the
General Theory of Relativity \cite{ADM} provided the necessary means
to apply Dirac's method of constrained quantization to gravity. This
led to the formulation of canonical quantum gravity and quantum
cosmology \cite{dewitt-1967,wheeler-1968} and brought the
Hilbert-space problem to forefront of research in fundamental
theoretical physics for the second time. In this context it emerges
as the problem of finding an appropriate inner product on the space
of solutions of the Wheeler-DeWitt equation. Without such an inner
product these solutions, that are often called the
 ``wave functions of the universe,'' are void of a physical meaning. The
lack of a satisfactory solution to this  problem has been one of the
major obstacles in transforming canonical quantum gravity and
quantum cosmology into genuine physical theories
\cite{kuchar-1992,isham-1993}.

A widely used approach in dealing with the Hilbert-space problem for
Klein-Gordon and Proca fields is to use the ideas of
indefinite-metric quantum theories. These fields admit a conserved
current density whose integral over space-like hypersurfaces yields
a conserved scalar charge. This is however not positive-definite and
as a result cannot be used to define a positive-definite inner
product and make the space of all fields $\cV$ into a genuine
Hilbert space directly. This makes one pursue the following
well-known scheme \cite{wald}. First, one uses the conserved charge
to define an indefinite inner product on $\cV$, and then restricts
this indefinite inner product to the (so-called positive-energy)
subspace of $\cV$ where the inner product is positive-definite. The
common practice is to label this subspace as ``physical'' and define
the Hilbert space using the fields belonging to this ``physical
space.''

This approach is not quite satisfactory, because even for physical
fields the above-mentioned conserved current density can take
negative values \cite{bk-2003}. Therefore it cannot be identified
with a probability density. There are also other problems related
with the observables that mix ``physical fields'' with ``unphysical
fields'' or ``ghosts.''

The application of pseudo-Hermitian QM in dealing with the Hilbert
space problem in relativistic QM and quantum cosmology
\cite{cqg,ap,ijmpa-2006,ap-2006a,js-2006,zm-2008}, and the removal
of ghosts in certain quantum field theories
\cite{bender-lee-model,jones-prd-2008} relies on the construction of
an appropriate (positive-definite) inner product on the space of
solutions of the relevant field equation.\footnote{This should be
distinguished with the treatment of the Pais-Uhlenbeck oscillator
proposed in \cite{bender-mannheim}, because the latter involves
changing the boundary conditions on the field equation which in turn
changes the vector space of fields.}

The basic idea behind the application of pseudo-Hermitian QM in
dealing with the Hilbert-space problem in relativistic QM and
quantum cosmology is that the relevant field equations whose
solutions constitute the state-vectors of the desired quantum theory
are second order differential equations in a ``time''
variable.\footnote{This is the physical time variable in an inertial
frame in relativistic QM or a fictitious evolution parameter in
quantum cosmology which may not be physically admissible
\cite{ap}.}. These equations have the following general form.
    \be
    \frac{d^2}{dt^2}\,\psi(t)+D\psi(t)=0,
    \label{sec9-f-eq}
    \ee
where $t$ denotes a dimensionless time variable, $\psi:\R\to\cL$ is
a function taking values in some separable Hilbert space $\cL$, and
$D:\cL\to\cL$ is a positive-definite operator that may depend on
$t$.

We can express (\ref{sec9-f-eq}) as a two-component Schr\"odinger
equation \cite{feshbach-villars},
    \be
    i\frac{d}{dt}\Psi(t)=H\psi(t),
    \label{sec9-2-comp}
    \ee
where $\Psi:\R\to\cL^2$ and $H:\cL^2\to\cL^2$ are defined by
\cite{jpa-1998,jmp-1998}
    \be
    \Psi(t):=\left(\begin{array}{c}
    \psi(t)+i\dot\psi(t)\\
    \psi(t)-i\dot\psi(t)\end{array}\right),~~~~~
    H:=\frac{1}{2}
    \left(\begin{array}{cc}
    D+1 & D-1\\
    -D+1 & -D-1\end{array}\right),
    \label{sec9-2-comp2}
    \ee
$\cL^2$ stands for the Hilbert space $\cL\oplus\cL$, and a dot
denotes a $t$-derivative. The Hamiltonian (\ref{sec9-2-comp2}) can
be easily shown to be quasi-Hermitian \cite{cqg}.

In Subsection~\ref{sec-two-level}, we examined in detail the quantum
system defined by the Hamiltonian (\ref{sec9-2-comp2}) for the case
that $\cL$ is $\C$ with the usual Euclidean inner product and $D$ is
multiplication by a positive number. In this case, the field
equation~(\ref{sec9-f-eq}) is the classical equation of motion for a
(complex) harmonic oscillator with frequency $\sqrt D$. It turns out
that most of the practical and conceptual difficulties of addressing
the Hilbert-space problem for Klein-Gordon, Proca, and
Wheeler-DeWitt fields can be reduced to and dealt with in the
context of this simple oscillator.\footnote{For a discussion of this
particular quantization of the classical harmonic oscillator, see
\cite{jmp-2005}.} In particular, the cases in which $D$ is
$t$-dependent (that arises in quantum cosmological models) require a
more careful examination. We will not deal with these cases here.
Instead, we refer the interested reader to \cite{ap} where a
comprehensive discussion of these issues and their ramifications is
provided.

Following the approach taken in Subsection~\ref{sec-two-level}, one
can construct a metric operator $\eta_+:\cL^2\to\cL^2$ and a new
inner product $\br\cdot|\cdot\kt_\etap$ on $\cL^2$ that renders $H$
Hermitian. This defines a physical Hilbert space $\cK$ of
two-component fields $\Psi(t)$. Because $H:\cK\to\cK$ is Hermitian,
it generate a unitary time-evolution in $\cK$. In particular, for
every initial time $t_0\in\R$, every pair $\Psi_1$ and $\Psi_2$ of
solutions of the Schr\"odinger equation (\ref{sec9-2-comp}), and all
$t\in\R$,
    \be
    \br\Psi_1(t)|\Psi_2(t)\kt_\etap=
    \br\Psi_1(t_0)|\Psi_2(t_0)\kt_\etap.
    \label{sec9-unitary}
    \ee

As a vector space $\cK$ (and $\cL^2$) are isomorphic to the space of
solutions of the single-component field equation (\ref{sec9-f-eq}),
i.e., $\cV:=\{\psi:\R\to\cL~|~\ddot\psi(t)+D\psi(t)=0~{\rm
for~all}~t\in\R~\}$. We can obtain an explicit realization of this
isomorphism as follows. Let $t_0$ be an initial time and
$\cU_{t_0}:\cV\to\cK$ be defined by
    \be
    \cU_{t_0}(\psi):=\Psi(t_0).
    \label{sec9-U=}
    \ee
According to (\ref{sec9-2-comp2}) and (\ref{sec9-U=}), the effect of
$\cU_{t_0}$ on solutions $\psi$ of the field equation
(\ref{sec9-f-eq}) is to map them to the corresponding initial
conditions $\psi(t_0)$ and $\dot\psi(t_0)$. Because the field
equation is linear and second order, this mapping is a linear
bijection. Therefore, $\cU_{t_0}$ is a vector space isomorphism.
This is an important observation, because it allows us to use
$\cU_{t_0}$ to induce a positive-definite inner product
$(\cdot,\cdot)_\etap$ on $\cV$ form the inner product
$\br\cdot|\cdot\kt_\etap$ on $\cK$. The induced inner product is
defined by
    \be
    (\psi_1,\psi_2)_\etap:=\br\cU_{t_0}(\psi_1)|\cU_{t_0}(\psi_2)\kt_\etap,
    ~~~~{\rm for~all}~~~~\psi_1,\psi_2\in\cV.
    \label{sec9-inn-prod}
    \ee
Note that in view of (\ref{sec9-U=}) and (\ref{sec9-unitary}), the
right-hand side of (\ref{sec9-inn-prod}) is independent of the value
of $t_0$. This makes $(\cdot,\cdot)_\etap$ into a well-defined inner
product on $\cV$ and gives it the structure of a Hilbert
space.\footnote{Strictly speaking one must also perform a Cauchy
completion of the inner product space obtained by endowing $\cV$
with $(\cdot,\cdot)_\etap$.}

In principle different choices for $\etap$ give rise to different
inner products $(\cdot,\cdot)_\etap$, but the resulting Hilbert
spaces $\cH_\etap:=(\cV,(\cdot,\cdot)_\etap)$ are
unitary-equivalent. The arbitrariness in the choice of $\etap$ can
be restricted by imposing additional physical conditions. For
example in the case of Klein-Gordon and Proca fields, the
requirement that $(\cdot,\cdot)_\etap$ be Lorentz-invariant reduces
the enormous freedom in the choice of $\etap$ to a finite number of
free numerical parameters \cite{cqg,zm-2008}.

The most general Lorentz-invariant and positive-definite inner
product on the space of (real or complex) Klein-Gordon fields has
the following form \cite{ap-2006a}
    \be
    (\psi_1,\psi_2)_\etap:=-\frac{i\hbar\kappa}{2mc}\int_{\Sigma}
    d\sigma^\mu\left[\psi_1(x)^*
    \stackrel{\leftrightarrow}{\partial_\mu}\cC\psi_2(x)+
    a\,\psi_1(x)^*\stackrel{\leftrightarrow}{\partial_\mu}\psi_2(x)\right],
    \label{kg-inn=}
    \ee
where $\Sigma$ is a spacelike Cauchy hypersurface in the Minkowski
spacetime, $x:=(x^0,x^1,x^2,x^3)$ are the spacetime coordinates in
an inertial frame, $\psi_1$ and $\psi_2$ are a pair of solutions of
the Klein-Gordon equation:
$\hbar^2\left[-\partial_0^2+\nabla^2\right]\psi(x)=m^2c^2\psi(x)$,
such that for all $x^0\in\R$, $\psi(x^0,\vec x)$ and
$\partial_0\psi(x^0,\vec x)$ define square-integrable functions of
$\vec x:=(x^1,x^2,x^3)$, $\partial_\mu:=\partial/
\partial x^\mu$, $\nabla^2:=\partial_1^2+\partial_2^2+\partial_3^2$,
for any pair of differentiable functions $f$ and $g$,
$f\!\stackrel{\leftrightarrow}{\partial_\mu}\!g:= f\partial_\mu
g-g\partial_\mu f$, $\kappa\in\R^+$ and $a\in(-1,1)$ are arbitrary
dimensionless free parameters demonstrating the arbitrariness in the
choice of $\etap$, and $\cC$ is the grading operator defined by
    \be
    (\cC\psi)(x):=i\left(-\nabla^2+\frac{m^2c^2}{\hbar^2}\right)^{\!-1/2}\!\!
    \psi(x)=\int_{\R^3}\int_{\R^3} dk^3 dy^3\;
    \frac{e^{i\vec k\cdot(\vec x-\vec y)}\psi(x^0,\vec y)}{
    \sqrt{\vec k^2+\frac{m^2c^2}{\hbar^2}}}.
    \label{sec9-charge}
    \ee
Note that $-\nabla^2+\hbar^2/(m^2c^2)$ is a positive operator acting
in $\cL=L^2(\R^3)$ and that $\cC$ is Lorentz-invariant
\cite{ap-2006a}.

According to (\ref{sec9-inn-prod}), as a linear operator mapping
$\cH_\etap$ to $\cK$, $\cU_{t_0}$ is a unitary operator. Similarly
$\rho:=\sqrt\etap$ is a unitary operator mapping $\cK$ to $\cL^2$.
Therefore $\rho\:\cU_\etap:\cH_\etap\to\cL^2$ is also unitary.
Usually $\cL$ is an $L^2$-space with well-known self-adjoint
operators. This allows for a simple characterization of the
self-adjoint operators $o$ acting in $\cL^2$. We can use these
operators and the unitary operator $\rho\:\cU_\etap$ to construct
the self-adjoint operators $O:\cH_\etap\to\cH_\etap$ that serve as
the observables of the desired quantum theory. This is done using
    \be
    O=(\rho\:\cU_\etap)^{-1}\,o\:\rho\:\cU_\etap.
    \label{sec9-obs}
    \ee
The application of this construction for Klein-Gordon
\cite{ijmpa-2006,ap-2006a} and Proca \cite{zm-2008} fields yields
explicit expressions for the corresponding relativistic position
operators and localized states, a problem that has been a subject of
ongoing research since the 1940's \cite{pryce-1948,newton-wigner}. A
natural consequence of these developments is the construction of a
set of genuine relativistic coherent states for Klein-Gordon fields
interacting with a constant magnetic field \cite{ap-2006b}.

\subsection{Electromagnetic Wave Propagation}

An interesting application of pseudo-Hermitian QM is its role in
dealing with the centuries-old problem of the propagation of
electromagnetic waves in linear dielectric media \cite{epl}. Unlike
the applications we discussed in the preceding section, here it is
the spectral properties of quasi-Hermitian operators and their
similarity to Hermitian operators that plays a key role.

Consider the propagation of the electromagnetic waves inside a
source-free dispersionless (linear) dielectric medium with
dielectric and permeability tensors $\Ep=\Ep(\vec x)$ and
$\MU=\MU(\vec x)$ that may depend on space $\vec x\in\R^3$ but not
on time $t\in\R$. Maxwell's equations in such a medium read
\cite{jackson}
    \bea
    &&\vec\nabla\cdot\vec D=0,~~~~~~~~~~~~~~\vec\nabla\cdot\vec B=0,
    \label{sec9-max-1}\\
    &&\dot{\vec B}+\vec\nabla\times\vec E=0,~~~~~~~
    \dot{\vec D}-\vec\nabla\times\vec H=0,
    \label{sec9-max-2}
    \eea
where $\vec E$ and $\vec B$ are the electric and magnetic fields, a
dot means a time-derivative, and
    \be
    \vec D:=\Ep\,\vec E,~~~~~~~~~~~~~ \vec H:=\MU^{-1}\vec B.
    \label{sec9-D-H}
    \ee

Eqs.~(\ref{sec9-max-1}) and (\ref{sec9-max-2}) are respectively
called the constraint and dynamical equations. The former may be
viewed as conditions on the initial values of the electromagnetic
field, because once they are satisfied for some initial time the
dynamical equations ensure their validity for all time.

Similarly to Klein-Gordon equation, we can express the dynamical
Maxwell equations (\ref{sec9-max-2}) as first order ordinary
differential equations for state-vectors belonging to a separable
Hilbert space. To achieve this we introduce the complex vector space
$\cV$ of vector fields $\vec F:\R^3\to\C^3$ and endow it with the
inner product $\br\vec F_1|\vec F_2\kt:=\int_{\R^3}d^3x\:\vec
F_1(\vec x)^*\cdot\vec F_2(\vec x),~ {\rm for~all}~ \vec F_1,\vec
F_2\in\cV$, to define the Hilbert space of square-integrable vector
fields: $\cH:=\{\vec F:\R^3\to\C^3~|~\br\vec F|\vec F\kt<\infty~\}$.
The operation of computing the curl of the (differentiable) elements
of this Hilbert space turns out to define a linear Hermitian
operator $\fD:\cH\to\cH$ according to $(\fD\vec F)(\vec
x):=\vec\nabla\times\vec F(\vec x)$. We can use $\fD$ to write
(\ref{sec9-max-2}) in the form: $\dot{\vec B}(t)+\fD\vec E(t)=0$ and
$\dot{\vec D}(t)-\fD\vec H(t)=0$. Evaluating the time-derivative of
both sides of the second of these equations and using the first of
these equations and (\ref{sec9-D-H}), we find
    \be
    \ddot{\vec E}(t)+\Omega^2\vec E(t)=0,
    \label{sec9-max-4}
    \ee
where $\Omega^2:\cH\to\cH$ is defined by
$\Omega^2:=\Ep^{-1}\fD\MU^{-1}\fD$.

In view of the fact that $\Ep$, $\MU$, and consequently $\Omega^2$
are time-independent, we can integrate (\ref{sec9-max-4}) to obtain
the following formal solution
    \be
    \vec E(t)=\cos(\Omega t)\vec E_0+
    \Omega^{-1}\sin(\Omega t)\dot{\vec E}_0,
    \label{sec9-E=}
    \ee
where $\vec E_0:=\vec E(0)$, $\dot{\vec E}_0:=\dot{\vec E}(0)=
\Ep^{-1}\fD\MU^{-1}\vec B(0)$, and
    \be
    \cos(\Omega t):=\sum_{n=0}^\infty
    \frac{(-1)^n}{(2n)!}\;(t^2\Omega^2)^n,
    ~~~~~~~~~
    \Omega^{-1}\sin(\Omega t):=t\sum_{n=0}^\infty
    \frac{(-1)^n}{(2n+1)!}\;(t^2\Omega^2)^n.
    \label{sec9-cos-sin}
    \ee
Given initial values of the electric and magnetic fields $\vec E(0)$
and $\vec B(0)$, we can use (\ref{sec9-E=}) and (\ref{sec9-cos-sin})
to obtain a series expansion for the evolving electric field. One
can select specific initial fields so that this expansion involves a
finite number of nonzero terms, but these are of little physical
significance. In general the resulting solution is an infinite
derivative series expansion that is extremely difficult to sum or
provide reliable estimates for. A crucial observation which
nonetheless makes this expansion useful is that, for the cases that
$\Ep$ and $\MU$ are Hermitian, the operator $\Omega^2:\cH\to\cH$ is
$\Ep$-pseudo-Hermitian: ${\Omega^2}^\dagger=
\Ep\:\Omega^2\:\Ep^{-1}$. In particular, for lossless material where
$\Ep$ is a positive $\vec x$-dependent matrix, $\Omega^2$ is a
quasi-Hermitian operator\footnote{As an operator acting in $\cH$ the
dielectric tensor $\Ep$ plays the role of a metric operator. This is
one of the rare occasions where a metric operator has a concrete
physical meaning.} that can be mapped to a Hermitian operator $h$ by
a similarity transformation, namely $h=\rho\:\Omega^2\:\rho^{-1}$
where $\rho:=\Ep^{\frac{1}{2}}$.

In terms of $h$ the solution~(\ref{sec9-E=}) takes the form: $
    \vec E(t)=\rho^{-1}[\cos(h^{\frac{1}{2}}t)\rho\:\vec E_0+
    h^{-\frac{1}{2}}\sin(h^{\frac{1}{2}} t)\rho\:\dot{\vec E}_0]$.
Therefore,
    \be
    \vec E(\vec x,t)=\br\vec x|\vec E(t)\kt=
    \rho^{-1}(\vec x)\int_{\R^3}d^3y~
    \left[\stackrel{\leftrightarrow}{C}\!\!(\vec x,\vec y;t)
    \rho(\vec y)\vec E_0(\vec y)+
          \stackrel{\leftrightarrow}{S}\!\!(\vec x,\vec y;t)
          \rho(\vec y)\dot{\vec E}_0(\vec y)\right],
    \label{sec9-E=h2}
    \ee
where
    \be
    \stackrel{\leftrightarrow}{C}\!\!(\vec x,\vec y;t):=
    \br\vec x|\cos(h^{\frac{1}{2}}t)|\vec y\kt,~~~~~
    \stackrel{\leftrightarrow}{S}\!\!(\vec x,\vec y;t):=
    \br\vec x|h^{-\frac{1}{2}}\sin(h^{\frac{1}{2}} t)|\vec y\kt.
    \label{sec9-kernel}
    \ee

The fact that $h$ is a Hermitian operator acting in $\cH$ makes it
possible to compute the kernels
$\stackrel{\leftrightarrow}{C}\!\!(\vec x,\vec y;t)$ and
$\stackrel{\leftrightarrow}{S}\!\!(\vec x,\vec y;t)$ of the
operators $\cos(h^{\frac{1}{2}}t)$ and
$h^{-\frac{1}{2}}\sin(h^{\frac{1}{2}} t)$ using the spectral
representation of $h$:
    \be
    h=\sum_{n=1}^N \sum_a E_n~|\psi_{n,a}\kt\br\psi_{n,a}|,
    \label{sec9-spec-res}
    \ee
where the sum over the spectral label $n$ should be identified with
an integral or a sum together with an integral whenever the spectrum
of $h$ has a  continuous part, $E_n$ and $|\psi_{n,a}\kt$ denote the
eigenvalues and eigenvectors of $h$ respectively, and $a$ is a
degeneracy label. In view of (\ref{sec9-spec-res}), for every
analytic function $F$ of $h$, such as $\cos(h^{\frac{1}{2}}t)$ and
$h^{-\frac{1}{2}}\sin(h^{\frac{1}{2}} t)$, we have $\br\vec
x|F(h)|\vec y\kt=\sum_{n=1}^N\sum_a F(E_n)~\psi_{n,a}(\vec
x)\psi_{n,a}(\vec y)^*$. In the scattering setups where $\Ep$ and
$\MU$ tend to constant values, $h$ has a continuous spectrum and one
finds integral representations for the kernels (\ref{sec9-kernel})
that reduce the solution~(\ref{sec9-E=h2}) of Maxwell's equations
into performing certain integrals (after solving the eigenvalue
problem for $h$.)

Ref.~\cite{epl} outlines the application of this method for the
cases that the medium is isotropic, the initial fields as well as
the dielectric and permeability constants change only along the
$z$-direction, and the WKB approximation is applicable in dealing
with the eigenvalue problem for $h$. Under these conditions, one can
compute the kernels (\ref{sec9-kernel}) analytically. This allows
for the derivation of the following closed form expression for the
propagating electromagnetic field in terms of the initial fields
$\vec E_0$, $\dot{\vec E}_0$, and the $z$-dependent dielectric and
permeability constants $\varepsilon(z)$ and $\mu(z)$.
    \bea
    \vec
    E(z,t)&=&\frac{1}{2}
    \left[\frac{\mu(z)}{\varepsilon(z)}\right]^{\frac{1}{4}}
    \left\{\left[\frac{\varepsilon(w_-(z,t))}{\mu(w_-(z,t))}
    \right]^{\frac{1}{4}}\right.\!\!
    \vec E_0(w_-(z,t))+
    \vspace{.5cm}\left[\frac{\varepsilon(w_+(z,t))}{\mu(w_+(z,t))}
    \right]^{\frac{1}{4}}\!\!
    \vec E_0(w_+(z,t))+\nn\\
    &&\vspace{1cm}\left.\int_{w_-(z,t)}^{w_+(z,t)} dw~
    \mu(w)^{\frac{1}{4}}\varepsilon(w)^{\frac{3}{4}}
    \dot{\vec E}_0(w)\right\},
    \nn
    \eea
where $w_\pm(z,t):=u^{-1}(u(z)\pm t)$ and $u(z):=\int_0^z
d\fz\:\sqrt{\varepsilon(\fz)\mu(\fz)}$, and $u^{-1}$ stands for the
inverse function for $u$, \cite{epl}.

The possibility of the inclusion of dispersion effects in the above
approach of solving Maxwell's equations is considered in
\cite{pla-2010a}.

\subsection{Other Applications and Physical Manifestations}

The following are some other areas where pseudo-Hermitian operators
arise and/or the methods of pseudo-Hermitian QM are used in dealing
with specific physics problems.

\subsubsection{Atomic Physics and Quantum Optics}

Effective quasi-Hermitian scattering Hamiltonians arise in the study
of the bound-state scattering from spherically symmetric short range
potentials. As shown by Matzkin in \cite{matzkin}, the use of the
machinery of pseudo-Hermitian QM in the study of these Hamiltonians
leads to a more reliable quantitative description of the scattering
problem. It also provides a better understanding of the
approximation schemes used in this context in the past and allows
for their improvement.

The relevance of pseudo-Hermitian operators to two-level atomic and
optical systems has been noted in
\cite{ben-aryeh-jpa-2004,ben-aryeh-jmo-2008,ss-jpa-2008}, and their
application in describing squeezed states is elucidated in
\cite{deb-epjd-2005,ben-aryeh-job-2005}. The optical systems provide
an important arena for manufacturing non-Hermitian and in particular
pseudo- and quasi-Hermitian effective Hamiltonians. Recent
experimental studies of $\cP\cT$-symmetric periodic potentials that
make use of $\cP\cT$-symmetric optical lattices is based on this
observation \cite{mec-prl-2008}. See also \cite{berry-2008}.

\subsubsection{Open Quantum Systems}

The emergence of non-Hermitian effective Hamiltonians in the
description of the resonant states, that is based on Feshbach's
projection scheme \cite{feshback-1962} is a very well-known
phenomenon \cite{muga-2004}. The application of a similar idea, that
replaces the projection scheme with an averaging scheme, for open
quantum systems also leads to a class of non-Hermitian effective
Hamiltonians (usually called Liouvillian or Liouville's super
operator) \cite{breur-petruccione}. These Hamiltonians that
determine the dynamics of the reduced density operators can, under
certain conditions, be pseudo-Hermitian or even quasi-Hermitian. In
\cite{stenholm-ap-2002,jakob-stenholm-pra-2004,stenholm-jakob-ap-2004},
Stenholm and Jakob explore the application of the properties of
pseudo- and quasi-Hermitian operators in the study of open quantum
systems. The key development reported in these articles is the
construction of a metric operator, that uses the spectral method we
discussed in Subsection~\ref{sec-spectral}, and the identification
of the corresponding norm with a viable candidate for a generalized
notion of entropy.

\subsubsection{Magnetohydrodynamics}

Pseudo-Hermitian effective Hamiltonians arise in the study of the
dynamo effect in magnetohydrodynamics
\cite{gs-jmp-2003,gsz-jmp-2005}. These Hamiltonians are typically
non-quasi-Hermitian and involve exceptional points. Therefore, they
can only be treated in the framework of indefinite-metric theories
and using the properties of Krein spaces \cite{azizov}.

\subsubsection{Quantum Chaos and Statistical Mechanics}

In \cite{djm-pra-1995}, Date el al study the spectrum of the
Hamiltonian operator: $H=\frac{1}{2}\left(p_x+\frac{\alpha
y}{r^2}\right)^2+\frac{1}{2}\left(p_y-\frac{\alpha
y}{r^2}\right)^2$, where $\alpha$ is a real coupling constant and
$r:=\sqrt{x^2+y^2}$. This Hamiltonian that is Hermitian and
$\cP\cT$-symmetric describes a rectangular Aharonov-Bohm billiard.
Here Note that the spectrum is obtained by imposing Dirichlet
boundary condition on the boundary of the rectangular configuration
space, that is defined by $|x|\leq a$ and $|y|\leq b$ for some
$a,b\in\R^+$, and also at the location of the flux line, namely
$x=y=0$. The main result of \cite{djm-pra-1995} is that the nearest
neighbor spacing distribution for this system has a transition that
interpolates between the Poisson (level clustering) and Wigner
(level repulsion) distributions. In an attempt to obtain a random
matrix model with this kind of behavior, Ahmed and Jain constructed
and studied certain pseudo-Hermitian random matrix models in
\cite{ahmed-jain-pre,ahmed-jain-jpa}.

\subsubsection{Biophysics}

In \cite{ee-jcp-2008}, Eslami-Moossallam and Ejtehadi have
introduced the following effective Hamiltonian for the description
of the dynamics of an anisotropic DNA molecule.
    \be
    H:=\frac{J_1^2}{2A_1}+\frac{J_2^2}{2A_2}+\frac{J_3^2}{2C}+
    i\omega_0J_3-\tilde f\cos\beta,
    \label{sec9-DNA}
    \ee
where $A_1,A_2,C,\omega_0,$ and $\tilde f$ are real coupling
constants, $\alpha,\beta,\gamma$ are Euler angles, and
$J_1,J_2,J_3$, that satisfy the commutation relations for angular
momentum operators, are defined by
$J_1:=-i\left(-\frac{\cos\gamma}{\sin\beta}\:\frac{\partial}{\partial\alpha}
+\sin\gamma\:\frac{\partial}{\partial\beta}+\cot\beta\cos\gamma\:
\frac{\partial}{\partial\gamma}\right)$,
$J_2:=-i\left(\frac{\sin\gamma}{\sin\beta}\:\frac{\partial}{\partial\alpha}
    +\cos\gamma\:\frac{\partial}{\partial\beta}-\cot\beta\sin\gamma\:
    \frac{\partial}{\partial\gamma}\right)$, and
    $J_3:=-i\:\frac{\partial}{\partial\gamma}$.
Clearly the Hamiltonian (\ref{sec9-DNA}) is non-Hermitian, but it is
at the same time real, i.e., it commutes with the time-reversal
operator $\cT$. In light of the fact that $\cT$ is an antilinear
operator, this implies that $H$ is a pseudo-Hermitian operator
\cite{p3}. It would be interesting to see if this observation has
any physically interesting implications, besides the restriction it
puts on the spectrum of $H$.

\section{Summary and Conclusions}
\label{conclusions}

In this article, we addressed various basic problems related to what
we call pseudo-Hermitian quantum mechanics. Our starting point was
the observation that a given quantum system admits an infinity of
unitary-equivalent representations in terms of Hilbert
space-Hamiltonian operator pairs. This freedom in the choice of
representation can be as useful as gauge symmetries of elementary
particle physics.

We have surveyed a variety of mathematical concepts and tools to
establish the foundations of pseudo-Hermitian quantum on a solid
ground and to clarify the shortcomings of the treatment of the
subject that is based on the so-called charge operator $\cC$. We
showed that it is the metric operator $\etap$ that plays the central
role in pseudo-Hermitian quantum mechanics. Although, one can in general
introduce a $\cC$ operator and express $\etap$ in terms of $\cC$,
the very construction of observables of the theory and the
calculations of the physical quantities requires the knowledge
of $\etap$. This motivates addressing the problem of the computation
of a metric operator for a given quasi-Hermitian operator. We have
described different approaches to this problem.

We have discussed a number of basic issues related to the
classical-to-quantum correspondence to elucidate the status of the
classical limit of pseudo-Hermitian quantum mechanics. We have also
elaborated on the surprising limitation on the choice of
time-dependent quasi-Hermitian Hamiltonians, the role of the metric
operator in path-integral formulation of the theory, a treatment
of the systems defined on complex contours, and a careful study of
the geometry of the space of states that seems to be indispensable
for clarifying the potential application of quasi-Hermitian
Hamiltonians in generating fast quantum evolutions.

Finally, we provided a discussion of various known applications and
manifestations of pseudo-Hermitian quantum mechanics.

Among the subjects that we did not cover and suffice to provide a few
references for are pseudo-supersymmetry and its extensions
\cite{p4,sr-jmp-2005,ss-jpa-2005,sr-jpa-2006}, weak pseudo-Hermiticity
\cite{solombrino-2002,bq-pla-2002,znojil-pla-2006,jmp-2006b}, and
the generalizations of $\cP\cT$-symmetry \cite{bbm-jpa-2002,jpa-2008a}.
This omission was particularly because of our intention not to treat
the results or methods with no direct or concrete implications for the
development of pseudo-Hermitian quantum mechanics. We particularly
avoided discussing purely formal results and speculative ideas.

\section*{Acknowledgments}

I would like to express my gratitude to Patrick Dorey and his
colleagues, Clare Dunning and Roberto Tate, for bringing to my
attention an error in an earlier version of this manuscript. I am
also indebted to Emre Kahya and Hossein Mehri-Dehnavi to help me
find and correct a number of typos. This project was supported by
the Turkish Academy of Sciences (T\"UBA).


\begin{appendix}

\section*{Appendix: Reality of Expectation values Implies
Hermiticity of the Observables} \label{appendix-1}

\noindent \textbf{Theorem~3:} \emph{Let ${\cal H}$ be a Hilbert
space with inner product $\br\cdot|\cdot\kt$ and $A:{\cal H}\to{\cal
H}$ be a (densely-defined, closed) linear operator satisfying ${\cal
D}(A)={\cal D}(A^\dagger)$, i.e., $A$ and its adjoint $A^\dagger$
have the same domain ${\cal D}(A)$. Then $A$ is a Hermitian operator
if and only if $\br\psi|A\psi\kt$ is real for all $\psi\in{\cal
D}(A)$.}

\noindent {\bf Proof:} If $A$ is Hermitian, we have
$\br\phi|A\psi\kt=\br A\phi|\psi\kt$ for all $\psi,\phi\in{\cal
D}(A)$. Then according to property (ii) of Subsection~\ref{sec-inn},
$\br\psi|A\psi\kt\in\R$ for all $\psi\in{\cal D}(A)$. Next, suppose
that for all $\psi\in{\cal D}(A)$, $\br\psi|A\psi\kt\in\R$. We will
show that this condition implies the Hermiticity of $A$ in two
steps.

\noindent \emph{Step 1}: Let $A_+:=\frac{1}{2}\,(A+A^\dagger)$ and
$A_-:=\frac{1}{2i}\,(A-A^\dagger)$. Then $A=A_++iA_-$, ${\cal
D}(A_\pm)={\cal D}(A)$, and $A_\pm$ are Hermitian operators. In view
of the first part of the theorem, this implies that
    \be
    \br\psi|A_\pm\psi\kt\in\R,~~~~~~\mbox{for all $\psi\in{\cal D}(A)$}.
    \label{app2.1}
    \ee
Furthermore, according to $A=A_++iA_-$ and the hypothesis of the
second part of the theorem,
$\br\psi|A_+\psi\kt+i\br\psi|A_-\psi\kt=\br\psi|A\psi\kt\in\R$. This
relation and (\ref{app2.1}) show that
    \be
    \br\psi|A_-\psi\kt=0~~~~\mbox{for all $\psi\in{\cal D}(A)$}.
    \label{app-3}
    \ee

\noindent  \emph{Step 2}: Let $\phi,\psi$ be arbitrary elements of
${\cal D}(A)$, $\xi_\pm:=\phi\pm\psi$, and $\zeta_\pm=\phi\pm
i\psi$. Then a direct calculation, using the property (\emph{iii})
of Subsection~\ref{sec-inn}, shows that
$\br\phi|A_-\psi\kt=\frac{1}{4}\left(\br\xi_+|A_-\xi_+\kt
    -\br\xi_-|A_-\xi_-\kt-i\br\zeta_+|A_-\zeta_+\kt+i
    \br\zeta_-|A_-\zeta_-\kt\right)=0$,
where the last equality follows from (\ref{app-3}) and the fact that
$\xi_\pm,\zeta_\pm\in{\cal D}(A)$. This establishes
$\phi|A_-\psi\kt=0$ for all $\phi,\psi\in{\cal D}(A)$. In particular
setting $\phi=A_-\psi$, we find $\br A_-\psi|A_-\psi\kt=0$ which in
view of the property (\emph{i}) of Subsection~\ref{sec-inn} implies
$A_-\psi=0$ for all $\psi\in{\cal D}(A)$. Hence $A_-=0$, and
according to $A=A_++iA_-$, we finally have $A=A_+$. But $A_+$ is
Hermitian.~~~$\square$

\end{appendix}



\ed